\documentclass[11pt,a4paper]{article}
\pdfoutput=1
\usepackage{jheppub}
\usepackage[T1]{fontenc}
\usepackage[utf8]{inputenc}
\usepackage{mathbbol}
\usepackage{bbm}
\usepackage{mathtools}
\usepackage{t1enc}
\usepackage{dsfont}
\usepackage{slashed}
\usepackage{booktabs}
\usepackage{longtable}
\usepackage{makecell}
\usepackage{multirow}
\usepackage{microtype}
\usepackage{rotfloat}
\usepackage{rotating}
\usepackage{cmap}
\usepackage{xfrac}

\newcommand{\dd}{\textmd{d}}
\DeclareMathOperator{\Tr}{Tr}
\DeclareMathOperator{\arccosh}{arccosh}
\newcommand{\MS}{\overline{\text{MS}}}
\newcommand{\unitj}{\ensuremath{\hat{\jmath}}}
\newcommand{\uniti}{\ensuremath{\hat{\imath}}}

\long\def\symbolfootnote[#1]#2{\begingroup%
\def\thefootnote{\fnsymbol{footnote}}\footnote[#1]{#2}\endgroup} 

\title{\boldmath{Scale setting and the light baryon spectrum in $N_f=2+1$ QCD with Wilson
fermions}}
\collaborationImg{\includegraphics{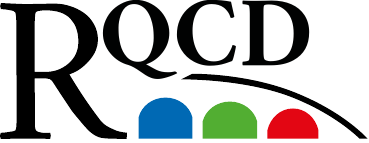} Collaboration}
\author[a]{Gunnar~S.~Bali,}
\author[a]{Sara~Collins,}
\author[a]{Peter~Georg,}
\author[a]{Daniel~Jenkins,}
\author[b]{Piotr~Korcyl,}
\author[a]{Andreas~Schäfer,}
\author[a]{Enno~E.~Scholz,}
\author[a]{Jakob~Simeth,}
\author[a]{Wolfgang~Söldner,}
\author[a]{Simon~Weishäupl}

\affiliation[a]{Institut für Theoretische Physik, Universität Regensburg,\\93040 Regensburg, Germany.}
\affiliation[b]{Institute of Theoretical Physics, Jagiellonian University,\\ul.~Łojasiewicza 11, 30--348 Kraków, Poland.}

\emailAdd{gunnar.bali@ur.de}
\emailAdd{sara.collins@ur.de}
\emailAdd{peter.georg@ur.de}
\emailAdd{daniel.jenkins@ur.de}
\emailAdd{piotr.korcyl@uj.edu.pl}
\emailAdd{andreas.schaefer@ur.de}
\emailAdd{enno.scholz@ur.de}
\emailAdd{jakob.simeth@ur.de}
\emailAdd{wolfgang.soeldner@ur.de}
\emailAdd{simon.weishaeupl@ur.de}

\abstract{We determine the light baryon spectrum on ensembles generated
by the Coordinated Lattice Simulations (CLS) effort, employing
$N_f=2+1$ flavours of non-perturbatively improved Wilson fermions. The hadron
masses are interpolated and
extrapolated within the quark mass plane, utilizing three distinct
trajectories, two of which intersect close to the
physical quark mass point and the third one approaching the
$\textmd{SU(3)}$ chiral limit.
The results are extrapolated to the continuum limit,
utilizing six different lattice spacings ranging from
$a\approx 0.10\,\textmd{fm}$ down to below 0.04\,\textmd{fm}.
The light pion mass varies from $M_{\pi}\approx 429\,\textmd{MeV}$
down to 127\,\textmd{MeV}. In general, the spatial extent
is kept larger than four times the inverse pion mass
and larger than $2.3\,\textmd{fm}$, with
additional small and large volume ensembles to investigate
finite size effects. We determine the Wilson flow scales
$\sqrt{t_{0,\text{ph}}}=0.1449^{(7)}_{(9)}\,\textmd{fm}$~\cite{Luscher:2010iy}
and $t_0^*\approx t_{0,\text{ph}}$~\cite{Bruno:2016plf}
from the octet cascade ($\Xi$ baryon). Determining the light baryon
spectrum in the continuum limit, we
find the nucleon mass $m_N=941.7^{(6.5)}_{(7.6)}\,\textmd{MeV}$
and the other stable baryon masses to agree
with their experimental values within sub-percent level uncertainties.
Moreover, we determine $\textmd{SU(3)}$ and $\textmd{SU(2)}$ chiral
perturbation theory low energy
constants, including the octet and the $\Omega$ baryon
sigma~terms $\sigma_{\pi N}=43.9(4.7)\,\textmd{MeV}$,
$\sigma_{\pi\Lambda}=28.2^{(4.3)}_{(5.4)}\,\textmd{MeV}$,
$\sigma_{\pi\Sigma}=25.9^{(3.8)}_{(6.1)}\,\textmd{MeV}$,
$\sigma_{\pi\Xi}=11.2^{(4.5)}_{(6.4)}\,\textmd{MeV}$ and
$\sigma_{\pi\Omega}=6.9^{(5.3)}_{(4.3)}\,\textmd{MeV}$,
as well as various parameters, renormalization factors
and improvement coefficients that are relevant for simulations
with our lattice action.}


\begin{document}
\maketitle

\section{Introduction}
Lattice QCD calculations have become indispensable for the theoretical
understanding of Standard Model processes and beyond that involve
reactions with quarks that are bound inside hadrons. An essential step
in the analysis is the scale setting, assigning a physical value to the
lattice spacing $a$.
This is achieved by equating a lattice observable, computed at the
physical point in terms of the quark masses, to the corresponding experimental
value. Ideally, both the experimental value and the results of the
lattice simulations should be known as precisely as possible.
Some of the most accessible of such observables are hadron masses and decay
constants. In the past, for instance the $\Omega$ baryon
mass~\cite{PACS-CS:2009sof,Capitani:2011fg,Borsanyi:2012zs,Blum:2014tka,Wilson:2019wfr,Miller:2020evg}, the $\Xi$ baryon mass~\cite{Durr:2008zz,PACS:2019ofv},
the nucleon mass~\cite{Bali:2012qs,Alexandrou:2018egz},
the average octet baryon mass~\cite{Bornyakov:2015eaa},
bottomonium mass splittings~\cite{HPQCD:2011qwj}, the pion
decay constant~\cite{Dowdall:2013rya,Alexandrou:2018egz,ExtendedTwistedMass:2021qui},
the kaon decay constant~\cite{Borsanyi:2013bia}
and combinations of the pion and kaon decay
constants~\cite{MILC:2015tqx,Bruno:2016plf,FermilabLattice:2018est,Strassberger:2021tsu} were employed to set the scale.

Basing the scale setting on pseudoscalar decay constants assumes
knowledge of the Cabibbo--Kobayashi--Maskawa (CKM) matrix elements
$V_{ud}$ or $V_{us}$. Moreover, extracting decay constants from experimental
decay rates of electrically charged
particles requires some understanding of the role of electromagnetic
interactions and soft photons (for a discussion in the context
of Lattice QCD, see, e.g., refs.~\cite{Carrasco:2015xwa,Lubicz:2016xro,Patella:2017fgk}). Pseudoscalar decay constants can, however, be determined
with high statistical accuracy. Regarding the spectrum of hadrons
that do not undergo strong decay, the experimental input is
cleaner. However, for heavy-light meson or quarkonium masses
also the heavy quark mass needs to be matched to experiment
and, to achieve a controlled continuum limit, very fine lattice spacings are
required, whereas for light baryon
masses the statistical errors are a limiting factor.

Often results are not available exactly at the physical point so that an
extrapolation or an interpolation is required. In view of this,
it is useful to introduce an intermediate scale parameter
to relate between different lattice spacings,
that can be determined very precisely, shows little
dependence on the quark masses and can be computed at a reference point in the
quark mass (hyper)plane where simulations are computationally
more affordable than at the physical point.
For an overview of different reference scale parameters, see
ref.~\cite{Sommer:2014mea}.
One such quantity is the Wilson flow scale
$t_0$~\cite{Luscher:2010iy} and, in particular, its value
$t_0^*$~\cite{Bruno:2016plf}, obtained at equal quark masses for
a reference value of the squared pion mass in units
of $t_0$. One can use the combination
$a^2/t_0^*$ to translate between different lattice spacings, however,
to determine the physical scale and the correct quark mass values,
experimental input is still required. Here we determine the combination
$\sqrt{t_{0,\text{ph}}}m_{\Xi}$ at the
experimental values of $M_{\pi}/m_{\Xi}$ and $M_K/m_{\Xi}$
(where $m_{\Xi}$, $M_{\pi}$ and $M_K$ are the masses 
of the $\Xi$ baryon, the pion and the kaon, respectively)
in the continuum limit. This procedure assigns physical units to the
lattice scale $t_{0,\text{ph}}$ and, extrapolating the ratio
$t_0^*/t_{0,\text{ph}}$ to the continuum limit, also to $t_0^*$.

Apart from setting the scale, the light baryon spectrum in itself
and its dependence on the light and strange quark masses are of great
interest: comparison with experiment serves
as a check that all systematics for the lattice setup are under control,
before more complicated observables are considered, while from
the quark mass dependence the validity ranges
of chiral effective theories can be estimated and the related
low energy constants (LECs) determined.
There exists a long history of lattice studies of the light hadron
spectrum including baryons, starting with calculations in the quenched approximation~\cite{Hamber:1981zn,Weingarten:1981jy,Bernard:1982hh,deForcrand:1987gx,Cabasino:1990zu,Guagnelli:1992zq,Lombardo:1992uc,Allton:1993ue,Butler:1994em,Akemi:1994zq,Bhattacharya:1995fz,Iwasaki:1995cm,Allton:1996yv,Lee:1997bq,Aoki:1999yr,Aoki:2002fd}, with
$N_f=4$~\cite{Born:1989iv,Altmeyer:1992dd} and $N_f=2$~\cite{Bitar:1990cb,Bitar:1992dk,Fukugita:1992hr,Bitar:1993xd,Eicker:1998sy,Allton:1998gi,AliKhan:2001xoi,Aoki:2002uc,Alexandrou:2008tn,Alexandrou:2009qu,Chowdhury:2012wa,Engel:2013ig,Alexandrou:2017xwd} mass-degenerate sea quark flavours,
with $N_f=2+1$ flavours~\cite{Bernard:2001av,WalkerLoud:2008bp,Aoki:2008sm,Lin:2008pr,Bazavov:2009bb,Ishikawa:2009vc,Durr:2008zz,PACS-CS:2009sof,Bietenholz:2011qq,Beane:2011pc,Horsley:2012fw,PACS:2019ofv,Francis:2022hyr},
with $N_f=2+1+1$ flavours~\cite{Alexandrou:2014sha,Miller:2020evg,Miller:2022vcm} and including
electromagnetic and mass isospin breaking effects~\cite{Borsanyi:2013lga,Borsanyi:2014jba,Horsley:2015eaa}. Motivated by the fact that effects due to a charm
sea quark can essentially be integrated out~\cite{Bruno:2014ufa,Athenodorou:2018wpk}, in this first high statistics study with a large number of independent gauge ensembles, covering a significant region of the parameter space in terms of the lattice spacing, quark mass combinations and the volume, we restrict ourselves to $N_f=2+1$ sea quark flavours.

In terms of observables, in this article we compute the spectrum of the
light baryons and flavour non-singlet pseudoscalar mesons at the
physical point. In particular, we determine the masses of
all positive parity octet and decuplet baryons, i.e.\ the
$N$, the $\Sigma$, the $\Lambda$ and the $\Xi$, as well as the
$\Delta(1232)$, the $\Sigma^*(1385)$, the $\Xi^*(1530)$ and the $\Omega$.
We remark that our results for the $\Delta$, the $\Sigma^*$ and the $\Xi^*$,
that strongly decay into $N\pi$, $\Lambda\pi/\Sigma\pi$ and $\Xi\pi$ in nature,
demonstrate the limitations of the conventional approach. A more refined study
of strongly decaying baryons would require
a finite volume scattering analysis, also including
baryon-meson-type operators into the interpolator basis.
In addition to determining the spectrum, we map out the dependence of
the baryon masses on the pion and
kaon masses in the continuum limit and determine
chiral perturbation theory (ChPT) LECs as well as the baryon
$\sigma$~terms.

Lattice simulations require an extrapolation to the continuum and infinite
volume limits. In particular, a controlled
extrapolation to the continuum limit is challenging, as has been
emphasized recently in ref.~\cite{Husung:2019ytz}. Clearly,
several simulation points are necessary, all employing inverse
lattice spacings that are larger than any of the physical
scales that need to be resolved.
The present computations are carried out
for a multitude of quark mass combinations
at six different values of the lattice spacing, ranging from
$a\lesssim 0.098\,\textmd{fm}$ down to $a<0.039\,\textmd{fm}$, covering
a factor larger than 6 in terms of $a^2$.
Simulating at our smallest lattice spacing at the physical pion mass
while maintaining a sufficiently large volume would require a linear spatial
lattice dimension larger than 150 points. In view of the
computational effort, near-physical quark masses are only realized at
$a\approx 0.064\,\textmd{fm}$ and $a\approx 0.085\,\textmd{fm}$.
Values of the lattice spacing even smaller than $a=0.039\,\textmd{fm}$
may be desirable in future studies, e.g., of heavy quark physics or
of nucleon structure observables that require large momenta.

Irrespective of the computational cost, in general it is difficult
to tune the simulation parameters to exactly match
the physical quark mass point.
Therefore, usually an interpolation or extrapolation,
reweighting~\cite{PACS-CS:2009sof,RBC:2010qam,Aoki:2012st,Finkenrath:2013soa}
or a Taylor expansion (computing derivatives with respect to the
quark masses~\cite{Bruno:2016plf}) is carried out.
We implement the first strategy. One novelty of our simulations is
the excellent coverage of the plane spanned by the light and strange
quark masses. Most simulations involve
the light quark mass being reduced while the strange quark
mass is kept almost constant. Here we
combine two trajectories that intersect close to the physical point, one
keeping the average quark mass constant~\cite{Bietenholz:2010jr,Bruno:2014jqa}
and one keeping the strange quark mass approximately
constant~\cite{Bali:2016umi}, which tightly constrains the extrapolation.
As a by-product, we
also obtain the strange and light quark masses, which will
be subject of a separate publication.
Additional ensembles along the symmetric $m_s=m_{\ell}$ line are
realized that approach the $\textmd{SU(3)}$ chiral limit. These
are essential for the determination of ChPT LECs.

We employ $N_f=2+1$ flavours of non-perturbatively order $a$ improved
Wilson fermions and the tree-level Symanzik improved gauge action. For details
on the action, see ref.~\cite{Bruno:2014jqa}. To avoid freezing of the
topological charge at small lattice spacings~\cite{Schaefer:2010hu},
most ensembles utilize open boundary
conditions in time~\cite{Luscher:2011kk}, and, in particular, all
ensembles at our smallest two values of $a$.
In addition to the baryon spectrum and ChPT LECs,
we determine a number of observables like $t_0^*/a^2$,
the critical hopping parameter and combinations of coefficients of
order $a$ improvement terms as
functions of the lattice coupling. These are
important for the planning of future simulation points.

The article is organized as follows. In section~\ref{sec:ensembles}, an
overview of the gauge ensembles analysed is provided. Then, in
section~\ref{sec:t0kappa}, combinations of renormalization
constants and improvement coefficients are determined, employing
global fits, updating earlier results~\cite{Bali:2016umi} and adding new ones.
In addition, the scale parameter $t_0/a^2$ is computed and
interpolating formulae are given for a number of related quantities.
In section~\ref{sec:ex}, the procedure of extrapolating
the baryon masses to the physical limit is explained: details
of the continuum limit extrapolations (maintaining full
order $a$ improvement) are described in section~\ref{sec:extrap}, followed by
section~\ref{sec:chiral}, where the extrapolation and interpolation
strategy in the quark mass plane is explained, and section~\ref{sec:finite},
where finite volume corrections are considered.
In section~\ref{sec:physical}, the relevant continuum limit
expectations and parametrizations in terms of $\textmd{SU(3)}$ LECs
as well as polynomial expansions are introduced.
Subsequently, in section~\ref{sec:results}, extrapolations to physical quark
masses in the infinite volume continuum limit are illustrated and the
lattice scale as well as the baryon spectrum are determined and
the systematics are quantified.
Moreover, values of the $\sigma$~terms and various LECs are computed
and the results discussed.
The main results are then highlighted in section~\ref{sec:summary}, before we
conclude in section~\ref{sec:conclusions}.

For the non-specialist reader the figures and tables
of section~\ref{sec:ensembles} are of interest as is the general
extrapolation strategy of section~\ref{sec:ex}. On first reading,
the reader may wish to skip section~\ref{sec:physical}, which
details the continuum limit parametrizations,
as these are referred to in the results section~\ref{sec:results}.
All the main results are summarized in section~\ref{sec:summary}
with references to where to find these in the body of the article.

Several appendices are provided:
expectations for hadron masses in an
isospin symmetric world are given in appendix~\ref{sec:isospin}.
In appendix~\ref{sec:finitebaryon} the parametrizations that are
used for the finite volume
effects of the baryon masses are presented, whereas
in appendix~\ref{sec:sigmadef} the $\sigma$~terms are related
to the dependence of the baryon masses on the quark masses and
the pseudoscalar meson masses. Moreover, some NLO
LECs of mesonic $\textmd{SU(3)}$
ChPT are determined. In appendix~\ref{sec:su2lec}
we describe how we obtain $\textmd{SU(2)}$ (H)BChPT
LECs from their $\textmd{SU(3)}$ (H)BChPT counterparts.
In appendix~\ref{sec:spectrum}, the extraction of the masses from
the relevant two-point functions is discussed in detail,
illustrative examples are provided and the masses are
tabulated. All the statistical methods employed are
discussed in appendix~\ref{sec:statistical}, where
autocorrelations in Monte Carlo time as well as correlations
between different masses within each individual ensemble are addressed.
These methods are used to extract the hadron and quark masses from
two-point functions
and to determine fit parameters, which describe, e.g.,
the dependence of baryon masses on the pseudoscalar masses, the volume and
the lattice spacing.
Some of the technical Hybrid Monte Carlo (HMC) simulation parameters
are presented in
appendix~\ref{sec:reweight}, where we also
discuss how the results, obtained from simulations
with a small twisted mass term, are reweighted to the target action.

\section{Overview of the ensembles}
\label{sec:ensembles}
As mentioned above, we employ $N_f=2+1$ flavours of
non-perturbatively order $a$ improved Wilson
fermions~\cite{Sheikholeslami:1985ij,Bulava:2013cta}
and the tree-level Symanzik improved gauge action~\cite{Weisz:1982zw}.
For details
on the action and the simulation, see ref.~\cite{Bruno:2014jqa}.
Since that publication many new
CLS ensembles have been generated and we discuss the present status
below.\footnote{For an up-to-date CLS configuration
status, see \url{https://www-zeuthen.desy.de/alpha/public-cls-nf21/}.}
A few ensembles are also included, which are not part of the CLS effort
since these have been generated using the {\sc BQCD}
code~\cite{Nakamura:2010qh}, all with periodic boundary conditions in
time and equal quark masses. These are labelled as ``rqcd0{\tt mn}'' below.
Six values of the inverse coupling constant $\beta=6/g^2$ are
realized, corresponding to lattice spacings ranging from
$a\approx 0.098\,\textmd{fm}$ down to $a\approx 0.039\,\textmd{fm}$.
The scale $t_0^*$~\cite{Bruno:2016plf}, defined
in section~\ref{sec:t0def} below, was used
for the conversion into physical units.
Our result $\sqrt{8t_0^*}\approx0.4097\,\textmd{fm}$
is presented in section~\ref{sec:t0sum}.

For Wilson fermions the so-called vector Ward identity
(or lattice) quark mass of a flavour $j$ is related to the
corresponding hopping parameter $\kappa_j$ that appears in the action
as follows:
\begin{equation}
  \label{eq:quark}
m_j=\frac{1}{2a}\left(\frac{1}{\kappa_j}-\frac{1}{\kappa_{\text{crit}}}\right),
\end{equation}
where $m_1=m_2=m_{\ell}$ and $m_3=m_s$ are the light and strange quark
masses, respectively, and $\kappa_{\text{crit}}$ is the critical
hopping parameter. One can also use the axial Ward identity (AWI)
to define unrenormalized non-singlet AWI masses:
\begin{equation}
  \label{eq:awimass}
\widetilde{m}_j+\widetilde{m}_k=
\frac{i\partial_0\langle 0|A_0^{jk}|\pi^{jk}\rangle}
{\langle 0|P^{jk}|\pi^{jk}\rangle},
\end{equation}
where $A_{\mu}^{jk}=
A_{\mu}^{jk,0}-iac_A\partial_{\mu}P^{jk}$ with $A_{\mu}^{jk,0}=\overline{\psi}_j\gamma_\mu\gamma_5\psi_k$ is the order $a$
improved non-singlet axial current for flavours $j\neq k$ and
$P^{jk}={\overline{\psi}_j}i\gamma_5\psi_k$ is
the corresponding pseudoscalar current.
The improvement constant $c_A(g^2)$ has been determined
in ref.~\cite{Bulava:2016ktf}. The critical value of the hopping
parameter $\kappa_{\text{crit}}(g^2)$, that appears in eq.~\eqref{eq:quark}, is
defined by the requirement that
the lattice quark mass along the symmetric line $m_s=m_\ell$ vanishes at
the same point as the AWI mass $\widetilde{m}_s=\widetilde{m}_{\ell}$.
The above quark masses can be converted into renormalized
quark masses $\widehat{m}_j$ in a standard continuum scheme,
e.g., the $\MS$ scheme at the scale $\mu=2\,\textmd{GeV}$. For the
AWI masses, the conversion (including order $a$ improvement) reads
\begin{equation}
  \label{eq:massdef}
  \widehat{m}_j=\frac{Z_A}{Z_P}\left\{1+a\left[3\left(\widetilde{b}_A-
    \widetilde{b}_P\right)\overline{m}+\left(b_A-b_P\right)m_j\right]\right\}\widetilde{m}_j,
\end{equation}
where
\begin{equation}
  \label{eq:maver}
  \overline{m}\coloneqq\frac13\left(2m_{\ell}+m_s\right)=\frac13\Tr M
\end{equation}
denotes the average sea quark mass and $\Tr M$ the trace of the quark
mass matrix.
The ratio of axial over pseudoscalar renormalization factors
$Z_A(g^2)/Z_P(\mu a,g^2)$ was, for instance, determined in
ref.~\cite{Campos:2018ahf} and the mass-dependent improvement coefficients
$b_A(g^2)$, $b_P(g^2)$, $\widetilde{b}_A(g^2)$ and $\widetilde{b}_P(g^2)$ in
ref.~\cite{Korcyl:2016ugy}. Since these order $a$ correction terms
are numerically small, keeping $\widetilde{m}_s$ constant will
result in an almost constant $\widehat{m}_s$.
Our strategy for keeping the AWI strange quark mass near its physical value
was introduced in ref.~\cite{Bali:2016umi}.

\setlength\LTcapwidth{\columnwidth}
\begin{footnotesize}
\begin{longtable}{cccp{4.5cm}r@{ $\cdot$ }lrrrr}
  \caption{\label{tab:parameters}\small Parameters of the analysed CLS
    and RQCD ensembles. Mass plane trajectory, ensemble name, 
    open (o) or (anti-)periodic (p) boundary conditions (bc),
    hopping parameter $\kappa$, the number of lattice points
    $N_t\cdot N_s^3$, the number of molecular dynamics units (MDUs) used in
    the spectrum analysis, $N_{\text{MD}}$, and the number of MDUs between
    measurements, $\Delta_{\text{MD}}$. For the determination of
    $t_0/a^2$ and its autocorrelation time $\tau_{t_0,\text{int}}$,
    in some cases a larger number of MDUs and a different $\Delta_{\text{MD}}$
    (in brackets) was employed. Italics indicate that
    the autocorrelation time is only estimated,
    due to a short Monte Carlo time series.
    The resulting $t_0/a^2$-values are listed in tables~\ref{tab:mass_sym}
    and~\ref{tab:mass_mes}.
    Ensembles A653, U103, H101, B450, H200, N202, N300 and
    J500 are both on
    the $\overline{m}=m_{\text{symm}}$ and the $m_s=m_{\ell}$ lines while
    D150 and E250 are approximately on both the $\overline{m}=m_{\text{symm}}$ and
    the $\widehat{m}_s\approx \widehat{m}_{s,\text{ph}}$ lines. For H102 there
    exist
    two runs, H102a and H102b with slightly different algorithmic parameters
    (H102r001 and H102r002 in table~2 of ref.~\cite{Bruno:2014jqa}).
  }\\\toprule
trajectory&id& bc &\multicolumn{1}{l}{$(\kappa_{\ell}, \kappa_s)$}&$N_t$ & $N_s^3$&$N_{\text{MD}}$&$\Delta_{\text{MD}}$&$\tfrac{\tau_{t_0,\text{int}}}{\text{MDU}}$\\\hline\endfirsthead
\caption{\small Parameters of analysed ensembles (continued).}\\\toprule
trajectory&id& bc &\multicolumn{1}{l}{$(\kappa_{\ell}, \kappa_s)$}&$N_t$ & $N_s^3$&$N_{\text{MD}}$&$\Delta_{\text{MD}}$&$\tfrac{\tau_{t_0,\text{int}}}{\text{MDU}}$\\\hline\endhead
\hline
\multicolumn{8}{r}{\textit{Continued on next page}} \\
\endfoot
\hline
\endlastfoot
\multicolumn{8}{c}{$\beta=3.34$}\\\hline
\multirow{3}{*}{$m_s=m_{\ell}$}
           & A651    & p & $(0.1365, 0.1365)$                         & $48$  & $24^3$ & 20400  &  4 & $78(_{12}^{15})$ \\
           & A652    & p & $(0.1365695, 0.1365695)$                   & $48$  & $24^3$ & 19980  &  4 & $48(_{6}^{7})$ \\
           & A650    & p & $(0.1366, 0.1366)$                         & $48$  & $24^3$ & 18624  &  4 & $64(_{9}^{12})$ \\\hline
\multirow{2}{*}{$\overline{m}=m_{\text{symm}}$}
           & A653    & p & $(0.1365716, 0.1365716)$                   & $48$  & $24^3$ & 20200  &  4 & $44(_{5}^{6})$ \\
           & A654    & p & $(0.13675, 0.136216193)$                   & $48$  & $24^3$ & 20268  &  4 & $99(_{19}^{27})$ \\\hline
\multicolumn{8}{c}{$\beta=3.4$}              & \\\hline
\multirow{3}{*}{$m_s=m_{\ell}$}
           & rqcd019 & p & $(0.1366, 0.1366)$                         & $32$  & $32^3$ & 1686   &  1 & ${\it 18(_{4}^{7})}$ \\
           & rqcd021 & p & $(0.136813, 0.136813)$                     & $32$  & $32^3$ & 1541   &  1 & $0.8(_{0.2}^{0.2})$ \\
           & rqcd017 & p & $(0.136865, 0.136865)$                     & $32$  & $32^3$ & 1849   &  1 & $7(_{2}^{2})$ \\\hline
\multirow{11}{*}{$\overline{m}=m_{\text{symm}}$}
           & U103    & o & $(0.13675962, 0.13675962)$                 & $128$ & $24^3$ & 19800  &8(4)& $44(_{5}^{6})$ \\
           & H101    & o & $(0.13675962, 0.13675962)$                 & $96$  & $32^3$ & 8000   &  4 & $45(_{8}^{9})$ \\
           & U102    & o & $(0.136865, 0.136549339)$                  & $128$ & $24^3$ & 17680  &8(4)& $53(_{8}^{10})$ \\
           & H102a   & o & $(0.136865, 0.136549339)$                  & $96$  & $32^3$ & 3720   &  4 & $34(_{8}^{10})$ \\
           & H102b   & o & $(0.136865, 0.136549339)$                  & $96$  & $32^3$ & 3948   &  4 & $33(_{6}^{8})$ \\
           & U101    & o & $(0.13697, 0.13634079)$                    & $128$ & $24^3$ & 6624   &  4 & $40(_{8}^{10})$ \\
           & H105    & o & $(0.13697, 0.13634079)$                    & $96$  & $32^3$ & 7944   &  4 & $40(_{5}^{6})$ \\
           & N101    & o & $(0.13697, 0.13634079)$                    & $128$ & $48^3$ & 5824   &  4 & $50(_{9}^{12})$ \\
           & S100    & o & $(0.13703, 0.136222041)$                   & $128$ & $32^3$ & 3932   &  4 & $44(_{12}^{19})$ \\
           & C101    & o & $(0.13703, 0.136222041)$                   & $96$  & $48^3$ & 9368   &  4 & $28(_{4}^{4})$ \\
           & D101    & o & $(0.13703, 0.136222041)$                   & $128$ & $64^3$ & 1292   &  4 & ${\it 144(_{2}^{2})}$ \\
           & D150    & p & $(0.137088, 0.13610755)$                   & $128$ & $64^3$ & 2408   &  4 & ${\it 35(_{11}^{20})}$ \\\hline
\multirow{3}{*}{$\widetilde{m}_s=\widetilde{m}_{s,\text{ph}}$}
           & H107    & o & $(0.13694566590798,$ $0.136203165143476)$  & $96$  & $32^3$ & 6256   &  4 & $46(_{8}^{11})$ \\
           & H106    & o & $(0.137015570024, 0.136148704478)$         & $96$  & $32^3$ & 6212   &  4 & $32(_{5}^{5})$ \\
           & C102    & o & $(0.13705084580022,$ $0.13612906255557)$   & $96$  & $48^3$ & 6000   &  4 & $35(_{6}^{7})$ \\\hline
\multicolumn{8}{c}{$\beta=3.46$}                                     & \\\hline
\multirow{3}{*}{$m_s=m_{\ell}$}
           & rqcd029 & p & $(0.1366, 0.1366)$                         & $64$  & $32^3$ & 1476   &  1 & ${\it 22(_{7}^{18})}$ \\
           & rqcd030 & p & $(0.1369587, 0.1369587)$                   & $64$  & $32^3$ & 1224   &  1 & ${\it 20(_{7}^{18})}$ \\
           & X450    & p & $(0.136994, 0.136994)$                     & $64$  & $48^3$ & 1600   &  4 & ${\it 17(_{5}^{7})}$ \\\hline
\multirow{3}{*}{$\overline{m}=m_{\text{symm}}$}
           & B450    & p & $(0.13689, 0.13689)$                       & $64$  & $32^3$ & 6448   &  4 & $56(_{12}^{17})$ \\
           & S400    & o & $(0.136984, 0.136702387)$                  & $128$ & $32^3$ & 11488  &  4 & $40(_{6}^{7})$ \\
           & N401    & o & $(0.1370616, 0.1365480771)$                & $128$ & $48^3$ & 4376   &  4 & $32(_{6}^{7})$ \\
           & D450    & p & $(0.137126, 0.136420428639937)$            & $128$ & $64^3$ & 2488   &  4 & ${\it 98(_{2}^{2})}$ \\\hline
\multirow{3}{*}{$\widetilde{m}_s=\widetilde{m}_{s,\text{ph}}$}
           & B451    & p & $(0.136981435679729,$ $0.136408545268417)$ & $64$  & $32^3$ & 7996   &  4 & $44(_{8}^{12})$ \\
           & B452    & p & $(0.1370455, 0.136378044)$                 & $64$  & $32^3$ & 7772   &  4 & $29(_{4}^{6})$ \\
           & N450    & p & $(0.1370986, 0.136352601)$                 & $128$ & $48^3$ & 4524   &  4 & $43(_{10}^{16})$ \\
           & D451    & p & $(0.13714, 0.136337761)$                   & $128$ & $64^3$ & 1828   &  4 & ${\it 38(_{13}^{26})}$ \\\hline
  \multicolumn{8}{c}{$\beta=3.55$}                                   & \\\hline
\multirow{3}{*}{$m_s=m_{\ell}$}
           & B250    & p & $(0.1367, 0.1367)$                         & $64$  & $32^3$ & 1776   &  4 & ${\it 34(_{11}^{17})}$ \\
           & X250    & p & $(0.13705, 0.13705)$                       & $64$  & $48^3$ & 1380   &  4 & ${\it 70(_{35}^{161})}$ \\
           & X251    & p & $(0.1371, 0.1371)$                         & $64$  & $48^3$ & 1744   &  4 & ${\it 53(_{22}^{64})}$ \\\hline
\multirow{6}{*}{$\overline{m}=m_{\text{symm}}$}
           & H200    & o & $(0.137, 0.137)$                           & $96$  & $32^3$ & 8000   &  4 & $33(_{5}^{6})$ \\
           & N202    & o & $(0.137, 0.137)$                           & $128$ & $48^3$ & 3536   &  4 & $63(_{16}^{26})$ \\
           & N203    & o & $(0.13708, 0.136840284)$                   & $128$ & $48^3$ & 6172   &  4 & $23(_{3}^{3})$ \\
           & N200    & o & $(0.13714, 0.13672086)$                    & $128$ & $48^3$ & 6848   &  4 & $34(_{5}^{6})$ \\
           & S201    & o & $(0.13714, 0.13672086)$                    & $128$ & $32^3$ & 8372   &  4 & $22(_{2}^{3})$ \\
           & D200    & o & $(0.1372, 0.136601748)$                    & $128$ & $64^3$ & 7996   &  4 & $28(_{4}^{5})$ \\
           & E250    & p & $(0.137232867, 0.136536633)$               & $192$ & $96^3$ & 1956   &  4 & ${\it 67(_{30}^{114})}$ \\\hline
\multirow{3}{*}{$\widetilde{m}_s=\widetilde{m}_{s,\text{ph}}$}
           & N204    & o & $(0.137112, 0.136575049)$                  & $128$ & $48^3$ & 6000   &  4 & $45(_{8}^{11})$ \\
           & N201    & o & $(0.13715968, 0.136561319)$                & $128$ & $48^3$ & 6000   &  4 & $38(_{6}^{7})$ \\
           & D201    & o & $(0.1372067, 0.136546844)$                 & $128$ & $64^3$ & 4312   &  4 & $34(_{7}^{9})$ \\\hline
  \multicolumn{8}{c}{$\beta=3.7$}                                      & \\\hline
\multirow{1}{*}{$m_s=m_{\ell}$}
           & N303    & o & $(0.1368, 0.1368)$                         & $128$ & $48^3$ & 2000   &  4 & ${\it 26(_{7}^{10})}$ \\\hline
\multirow{3}{*}{$\overline{m}=m_{\text{symm}}$}
           & N300    & o & $(0.137, 0.137)$                           & $128$ & $48^3$ & 6080   &  4 & $44(_{9}^{11})$ \\
           & N302    & o & $(0.137064, 0.1368721791358)$              & $128$ & $48^3$ & 8804   &  4 & $41(_{6}^{7})$ \\
           & J303    & o & $(0.137123, 0.1367546608)$                 & $192$ & $64^3$ & 7992   &  8 & $77(_{16}^{20})$ \\
           & E300    & o & $(0.137163, 0.1366751636177327)$           & $192$ & $96^3$ & 1992   &8(4)& $38(_{10}^{15})$ \\\hline
\multirow{3}{*}{$\widetilde{m}_s=\widetilde{m}_{s,\text{ph}}$}
           & N305    & o & $(0.137025, 0.136676119)$                  & $128$ & $48^3$ & 8000   &  4 & $37(_{7}^{9})$ \\
           & N304    & o & $(0.137079325093654,$ $0.136665430105663)$ & $128$ & $48^3$ & 6136   &  4 & $39(_{8}^{11})$ \\
           & J304    & o & $(0.13713, 0.1366569203)$                  & $192$ & $64^3$ & 6076   &  4 & $56(_{14}^{22})$ \\\hline
  \multicolumn{8}{c}{$\beta=3.85$}                                     & \\\hline
\multirow{1}{*}{$m_s=m_{\ell}$}
           & N500    & o & $(0.13672514, 0.13672514)$                 & $128$ & $48^3$ & 3760   &  4 & ${\it 57(_{18}^{35})}$ \\\hline
\multirow{2}{*}{$\overline{m}=m_{\text{symm}}$}
           & J500    & o & $(0.136852, 0.136852)$                     & $192$ & $64^3$ & 6008   &  8 & $75(_{12}^{15})$ \\
           & J501    & o & $(0.1369032, 0.136749715)$                 & $192$ & $64^3$ & 5988   &  4 & $79(_{16}^{23})$ 
\end{longtable}
\end{footnotesize}

\setlength\LTcapwidth{\columnwidth}
\begin{small}
\begin{longtable}{cccccc}
  \caption{\label{tab:physparams}\small Overview of the physical parameters
    of the analysed CLS and RQCD ensembles.
    Mass plane trajectory, ensemble name, spatial lattice extent $L$ in physical
    units and in units of the pion mass as well as the pion and the kaon masses
    (which are volume corrected, see section~\ref{sec:finite}).
    The physical units have been assigned using
    $\sqrt{8t_0^*}=0.4097\,\textmd{fm}$. Statistical errors
    are not shown in this overview table. The lattice spacings
    including errors are given in table~\ref{tab:t0star}, see also
    section~\ref{sec:t0} for details, while the raw
    pion and kaon mass data including errors can be found in
    tables~\ref{tab:mass_sym} and~\ref{tab:mass_mes}.
  }\\\hline
  trajectory&id& $L/\textmd{fm}$& $L M_\pi$ & $M_\pi/\textmd{MeV}$ & $M_K/\textmd{MeV}$ \\\hline
  \endfirsthead
  \caption{\small Physical parameter values (continued).}\\\hline
  trajectory&id& $L/\textmd{fm}$& $L M_\pi$ & $M_\pi/\textmd{MeV}$ & $M_K/\textmd{MeV}$ \\\hline
  \endhead
\hline \multicolumn{6}{r}{\textit{Continued on next page}} \\
\endfoot
\hline
\endlastfoot

\hline
\multicolumn{6}{c}{$\beta=3.34, a = 0.098\,\mathrm{fm}$ }\\
\hline
\multirow{3}{*}{$m_s=m_{\ell}$} 
 & A651 & 2.34 & 6.6 & 556 & 556 \\
 & A652 & 2.34 & 5.13 & 432 & 432 \\
 & A650 & 2.34 & 4.39 & 371 & 371 \\
\hline
\multirow{2}{*}{$\overline{m}=m_{\text{symm}}$} 
 & A653 & 2.34 & 5.09 & 429 & 429 \\
 & A654 & 2.34 & 4.0 & 338 & 459 \\
\hline
\multicolumn{6}{c}{$\beta=3.4, a = 0.085\,\mathrm{fm}$ }\\
\hline
\multirow{3}{*}{$m_s=m_{\ell}$} 
 & rqcd019 & 2.72 & 8.4 & 608 & 608 \\
 & rqcd021 & 2.72 & 4.7 & 340 & 340 \\
 & rqcd017 & 2.72 & 3.26 & 236 & 236 \\
\hline
\multirow{12}{*}{$\overline{m}=m_{\text{symm}}$} 
 & U103 & 2.04 & 4.35 & 420 & 420 \\
 & H101 & 2.72 & 5.85 & 423 & 423 \\
 & U102 & 2.04 & 3.7 & 357 & 445 \\
 & H102a & 2.72 & 4.95 & 359 & 444 \\
 & H102b & 2.72 & 4.89 & 354 & 442 \\
 & U101 & 2.04 & 2.81 & 271 & 464 \\
 & H105 & 2.72 & 3.88 & 281 & 468 \\
 & N101 & 4.09 & 5.82 & 281 & 467 \\
 & S100 & 2.72 & 2.95 & 214 & 476 \\
 & C101 & 4.09 & 4.6 & 222 & 476 \\
 & D101 & 5.45 & 6.13 & 222 & 476 \\
 & D150 & 5.45 & 3.51 & 127 & 482 \\
\hline
\multirow{3}{*}{$\widetilde{m}_s=\widetilde{m}_{s,\text{ph}}$} 
 & H107 & 2.72 & 5.09 & 368 & 550 \\
 & H106 & 2.72 & 3.77 & 273 & 520 \\
 & C102 & 4.09 & 4.62 & 223 & 504 \\
\hline
\multicolumn{6}{c}{$\beta=3.46, a = 0.075\,\mathrm{fm}$ }\\
\hline
\multirow{3}{*}{$m_s=m_{\ell}$} 
 & rqcd029 & 2.41 & 8.72 & 713 & 713 \\
 & rqcd030 & 2.41 & 3.9 & 319 & 319 \\
 & X450 & 3.62 & 4.86 & 265 & 265 \\
\hline
\multirow{4}{*}{$\overline{m}=m_{\text{symm}}$} 
 & B450 & 2.41 & 5.15 & 421 & 421 \\
 & S400 & 2.41 & 4.33 & 354 & 445 \\
 & N401 & 3.62 & 5.27 & 287 & 464 \\
 & D450 & 4.82 & 5.28 & 216 & 480 \\
\hline
\multirow{4}{*}{$\widetilde{m}_s=\widetilde{m}_{s,\text{ph}}$} 
 & B451 & 2.41 & 5.16 & 422 & 577 \\
 & B452 & 2.41 & 4.31 & 352 & 548 \\
 & N450 & 3.62 & 5.26 & 287 & 528 \\
 & D451 & 4.82 & 5.35 & 219 & 507 \\
\hline
\multicolumn{6}{c}{$\beta=3.55, a = 0.064\,\mathrm{fm}$ }\\
\hline
\multirow{3}{*}{$m_s=m_{\ell}$} 
 & B250 & 2.04 & 7.37 & 713 & 713 \\
 & X250 & 3.06 & 5.43 & 350 & 350 \\
 & X251 & 3.06 & 4.16 & 268 & 268 \\
\hline
\multirow{7}{*}{$\overline{m}=m_{\text{symm}}$} 
 & H200 & 2.04 & 4.36 & 422 & 422 \\
 & N202 & 3.06 & 6.42 & 414 & 414 \\
 & N203 & 3.06 & 5.39 & 348 & 445 \\
 & S201 & 2.04 & 3.0 & 290 & 471 \\
 & N200 & 3.06 & 4.43 & 286 & 466 \\
 & D200 & 4.08 & 4.18 & 202 & 484 \\
 & E250 & 6.12 & 4.05 & 131 & 493 \\
\hline
\multirow{3}{*}{$\widetilde{m}_s=\widetilde{m}_{s,\text{ph}}$} 
 & N204 & 3.06 & 5.48 & 353 & 549 \\
 & N201 & 3.06 & 4.44 & 287 & 527 \\
 & D201 & 4.08 & 4.14 & 200 & 504 \\
\hline
\multicolumn{6}{c}{$\beta=3.7, a = 0.049\,\mathrm{fm}$ }\\
\hline
$m_s=m_{\ell}$
 & N303 & 2.36 & 7.75 & 646 & 646 \\
\hline
\multirow{4}{*}{$\overline{m}=m_{\text{symm}}$} 
 & N300 & 2.36 & 5.1 & 425 & 425 \\
 & N302 & 2.36 & 4.17 & 348 & 455 \\
 & J303 & 3.15 & 4.14 & 259 & 479 \\
 & E300 & 4.73 & 4.22 & 176 & 496 \\
\hline
\multirow{3}{*}{$\widetilde{m}_s=\widetilde{m}_{s,\text{ph}}$} 
 & N305 & 2.36 & 5.14 & 428 & 584 \\
 & N304 & 2.36 & 4.24 & 353 & 558 \\
 & J304 & 3.15 & 4.18 & 261 & 527 \\
\hline
\multicolumn{6}{c}{$\beta=3.85, a = 0.039\,\mathrm{fm}$ }\\
\hline
$m_s=m_{\ell}$
 & N500 & 1.85 & 5.69 & 604 & 604 \\
\hline
\multirow{2}{*}{$\overline{m}=m_{\text{symm}}$} 
 & J500 & 2.47 & 5.19 & 413 & 413 \\
 & J501 & 2.47 & 4.21 & 336 & 448 \\
\hline
\end{longtable}
\end{small}

\begin{figure}[htp]
  \centering
  \resizebox{0.32\textwidth}{!}{\includegraphics[width=\textwidth]{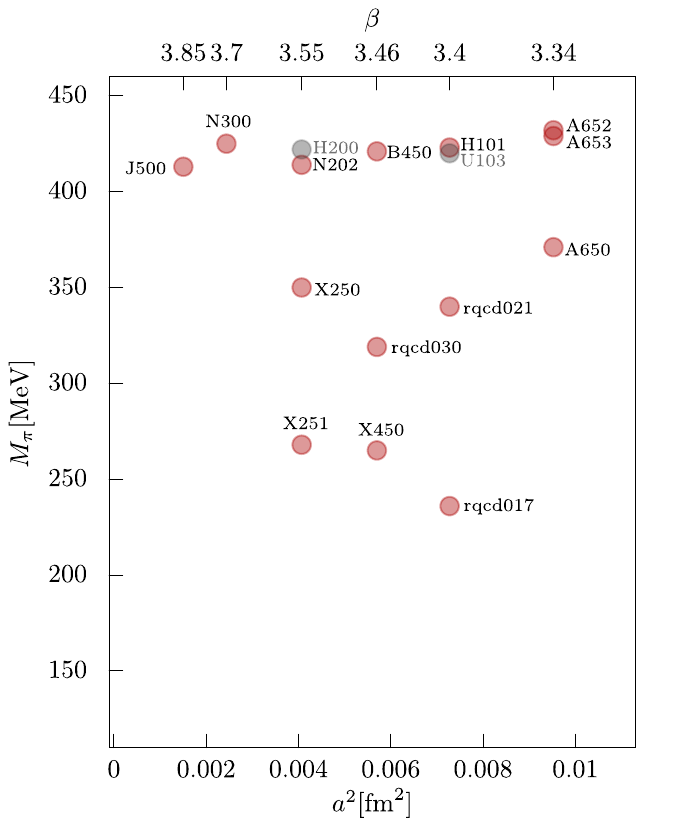}}
  \resizebox{0.32\textwidth}{!}{\includegraphics[width=\textwidth]{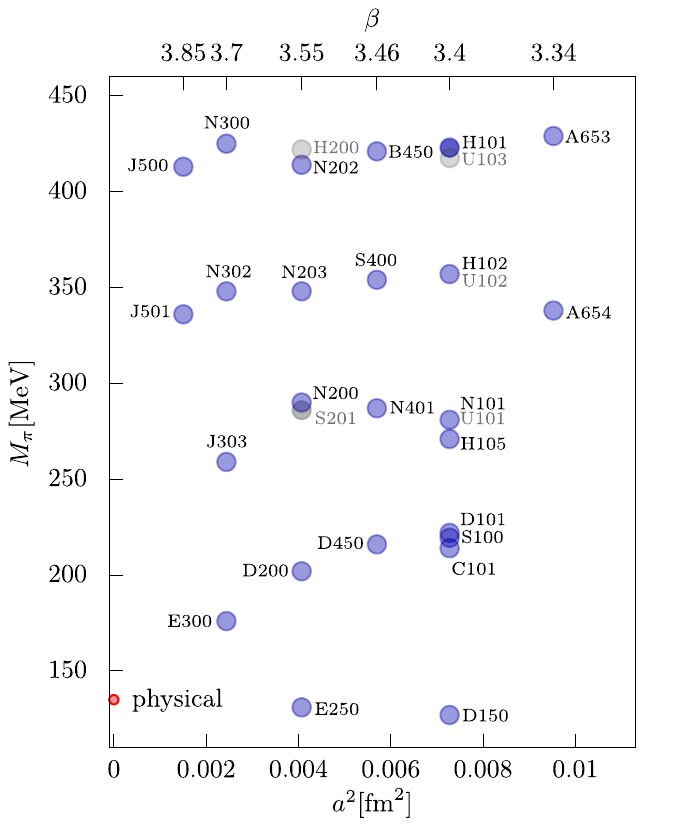}}
  \resizebox{0.32\textwidth}{!}{\includegraphics[width=\textwidth]{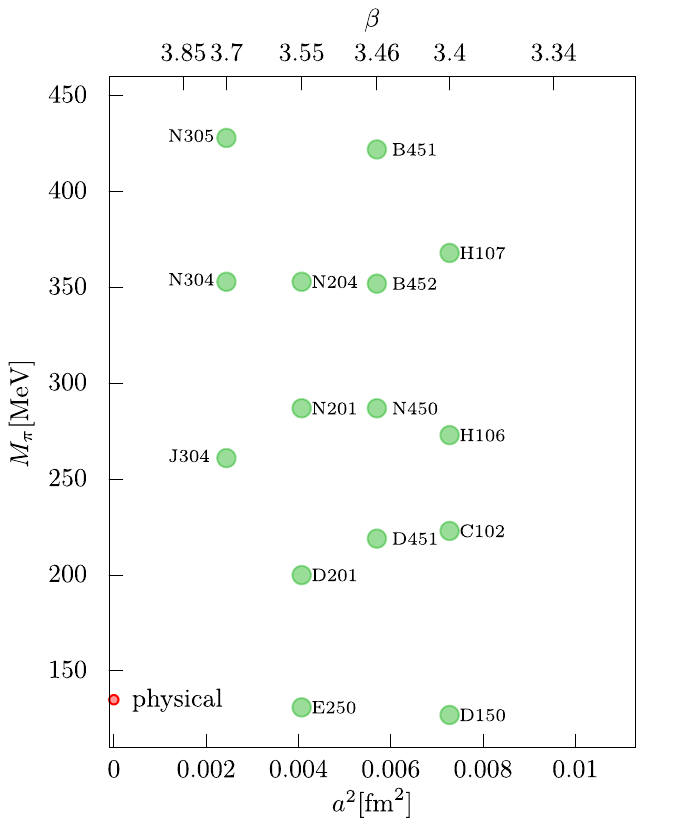}}
  \caption{
        Overview of the analysed ensembles: three different quark mass
    trajectories (left: $m_s=m_{\ell}$, centre: $\Tr M=\text{const}$, right: $\widehat{m}_s\approx\text{const}$) have been analysed at six
    (four for $\widehat{m}_s\approx\text{const}$)
    different lattice
    spacings. On the $m_s=m_{\ell}$ trajectory, six additional ensembles
    with $M_{\pi}>450\,\textmd{MeV}$ exist
    (A651, rqcd019, rqcd029, B250, N303 and N500, see
    tables~\ref{tab:parameters} and~\ref{tab:physparams}),
    which do not enter our hadron spectroscopy
    analysis. We also omit ensembles with
    $L<2.3\,\textmd{fm}$ (grey circles).\label{fig:ensembles}}
\end{figure}

\begin{figure}[htp]
  \centering
  \resizebox{0.8\textwidth}{!}{\includegraphics[width=\textwidth]{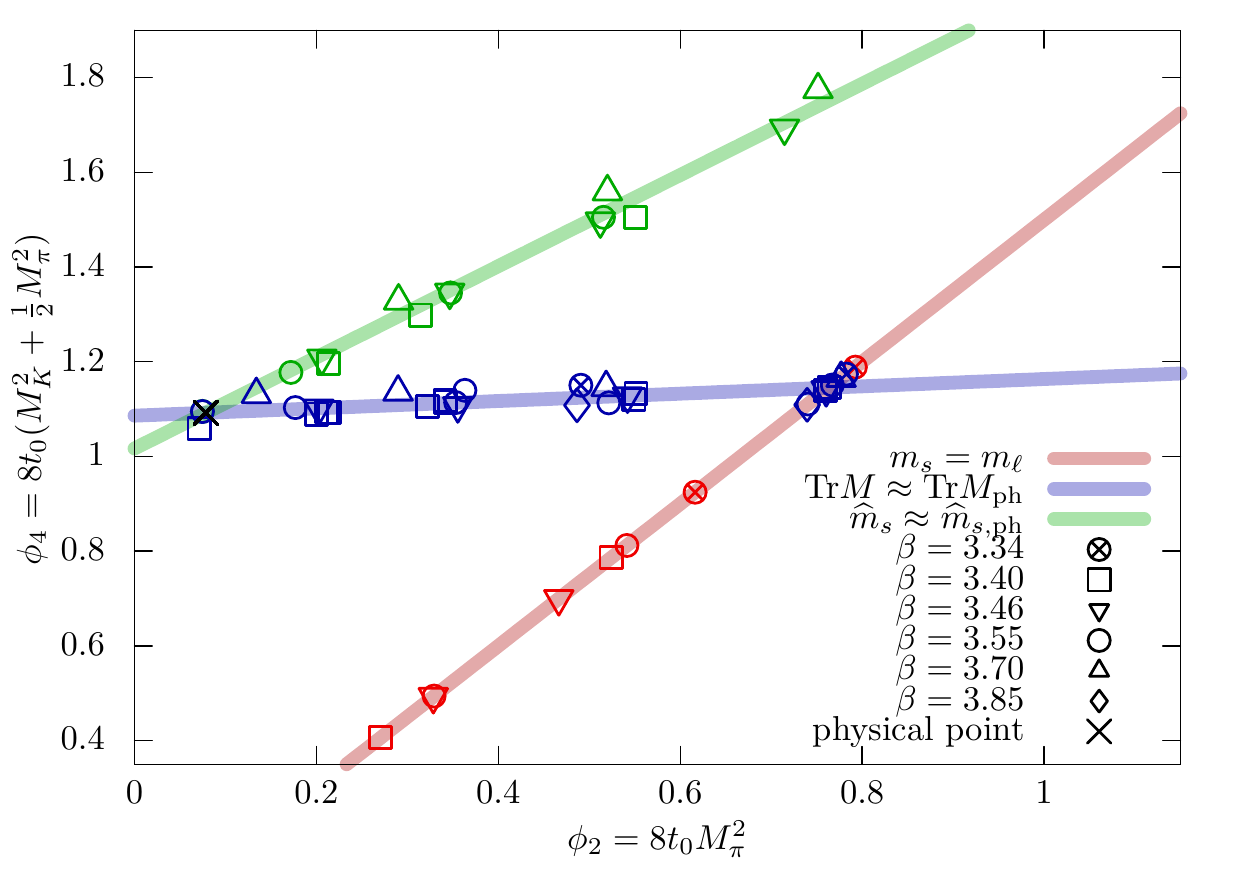}}
  \caption{Overview of the analysed ensembles in the quark mass plane.
    The ordinate is (approximately) proportional
    to $\widehat{m}_{\ell}$, the abscissa to
    $2\widehat{m}_{\ell}+\widehat{m}_s$. The
    $\widehat{m}_s\approx\text{const}$ and the $\Tr M=\text{const}$ trajectories
    intersect close to the physical point (black cross), while the latter
    trajectory starts from the point on the symmetric
    $m_s=m_{\ell}$ line, where $M_{\pi}=M_K\approx 411\,\textmd{MeV}$.\label{fig:closeup}}
\end{figure}

At each value of $\beta$ the
simulations are carried out along three trajectories in the quark mass plane:
\begin{itemize}
  \item The symmetric line: $m_s=m_{\ell}$, i.e.\ $\widehat{m}_s=\widehat{m}_{\ell}$.
  \item The $\Tr M=\text{const}$ line: $a(2m_{\ell}+m_s)=3am_{\text{symm}}$, i.e.\
    $2\widehat{m}_{\ell}+\widehat{m}_s=\text{const} +\mathcal{O}(a)$. The constant
    is chosen such
    that the combination $(2M_K^2+M_\pi^2)t_0^*$ is close to its physical value,
    assuming $\sqrt{8t_0^*}=0.413\,\textmd{fm}$~\cite{Bruno:2016plf}.
    The strategy of keeping the sum of quark masses constant was
    pioneered by QCDSF/UKQCD~\cite{Bietenholz:2010jr}.
  \item The line of fixed strange quark mass: the renormalized
    strange quark mass is kept near its physical value~\cite{Bali:2016umi}.
\end{itemize}
We analyse a large number of ensembles along these three trajectories,
most of which have open boundary conditions in time~\cite{Luscher:2011kk}
in order to circumvent critical
slowing down towards the continuum limit, due to the freezing of
the topological charge~\cite{Schaefer:2010hu}. The relevant simulation
parameters are listed in table~\ref{tab:parameters},\footnote{The time $\tau_{t_0,\text{int}}$ is the largest
autocorrelation time for the observables we have studied.
It may be close to the exponential autocorrelation time of the system,
see appendices~\ref{sec:bin} and~\ref{sec:autocorr}.}
whereas the physical values of the lattice
spacings, spatial lattice volumes and pseudoscalar
meson masses are given in table~\ref{tab:physparams}.
The $t_0/a^2$-values are listed in
tables~\ref{tab:mass_sym} and~\ref{tab:mass_mes} of
appendix~\ref{sec:masstables}. For some of the ensembles these
have been determined previously~\cite{Bruno:2014jqa,Bali:2016umi,Bruno:2016plf}. In all these cases, within
statistical errors, our determination agrees with the previous ones.
An overview of the ensembles is
shown in figure~\ref{fig:ensembles}. Figure~\ref{fig:closeup} illustrates
our coverage of the quark mass plane within the region of
interest.

\begin{figure}[htp]
  \centering
  \resizebox{0.49\textwidth}{!}{\includegraphics[width=\textwidth]{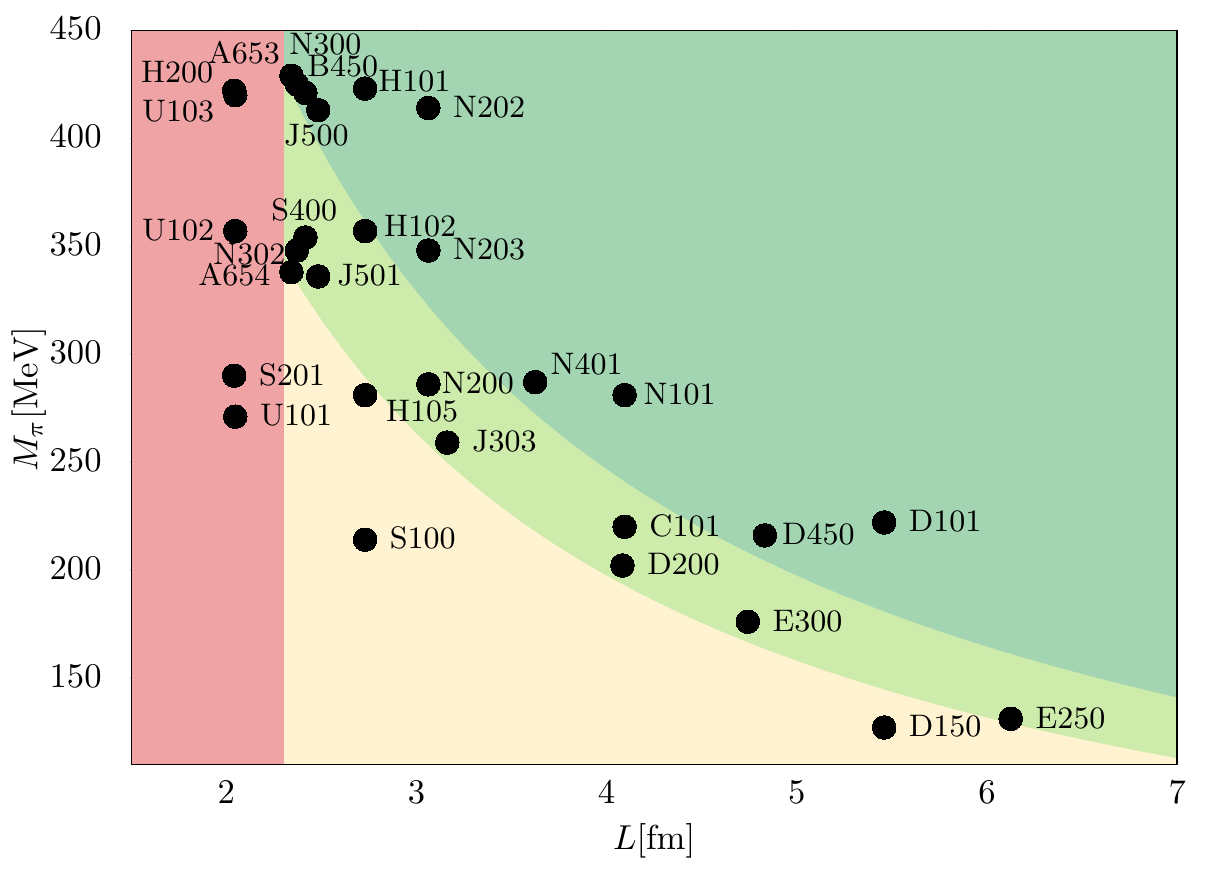}}
  \resizebox{0.49\textwidth}{!}{\includegraphics[width=\textwidth]{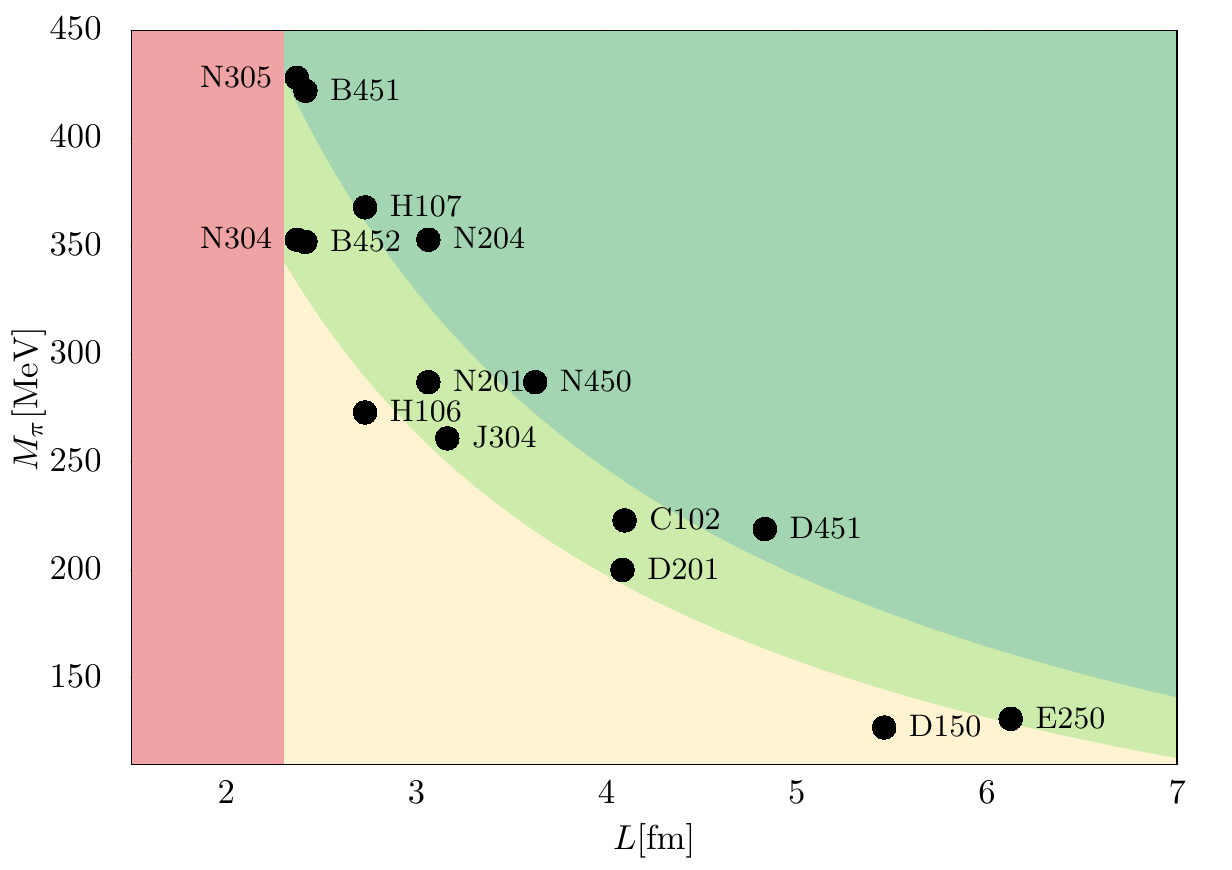}}
  \caption{The spatial lattice extents $L$ and the pion masses $M_{\pi}$
    for the two quark mass trajectories leading to the physical point.
    The coloured regions distinguish between \(LM_{\pi} \leq 4\) (yellow), \(4 <
    LM_{\pi} \leq 5\) (light green), \(5 < LM_\pi\) (dark green) and
    \(L<2.3\,\textmd{fm}\) (red).\label{fig:volumes}}
\end{figure}

With three exceptions (rqcd017 for $m_s=m_{\ell}$, D150 for $\Tr M=\text{const}$
and H106 for $\widehat{m}_s\approx\text{const}$), at least one ensemble
exists at each simulation point with a spatial lattice extent
$L=N_s a>\max\{4/M_{\pi},2.3\,\textmd{fm}\}$.
In some cases additional volumes were generated to enable the study of
finite volume effects. Figure~\ref{fig:volumes} provides an overview
regarding this: the dark green areas correspond to $LM_{\pi}>5$, light green to
$5\geq LM_\pi>4$, yellow to $4\geq LM_\pi$ and red to $L<2.3\,\textmd{fm}$.

As mentioned above, details on the simulations performed by CLS
using the {\sc openQCD} code~\cite{Luscher:2012av}\footnote{Publicly
available at \url{http://luscher.web.cern.ch/luscher/openQCD}.}
can be found in ref.~\cite{Bruno:2014jqa}. In appendix~\ref{sec:reweight}
we discuss the technical parameters for some of the more recently performed
simulations.

\section{Wilson flow scales and the determination of action-specific parameters}
\label{sec:t0kappa}
We start with a brief discussion of some of the subtleties related to
order $a$ improvement, before we determine the
critical hopping parameter values and combinations
of some of the relevant renormalization constants and improvement parameters.
We then determine a reference point in the quark mass
plane and compute different observables related to the scale
parameter $t_0$~\cite{Luscher:2010iy}.
We give interpolating formulae for the dependence of all
these quantities on $g^2$ and determine the continuum limit
dependence of $t_0$ on the pseudoscalar masses.
For each lattice spacing, the $\Tr M=\text{const}$ trajectory starts
from a point where $m_s=m_{\ell}$ and we determine the optimal
start value for this trajectory to intersect the physical point.
Finally, we determine the values $\kappa^*$ and $am^*$ associated
with a reference point on the symmetric line, where
$t_0^*$ is defined.

\subsection{Order $\boldsymbol{a}$ improvement of the coupling constant}
\label{sec:coupordera}
The simulations are carried out at
fixed values of the bare coupling $g^2=6/\beta$, however,
the coupling undergoes order $a$
improvement~\cite{Jansen:1995ck,Luscher:1996sc},
\begin{equation}
  \label{eq:improvecoup}
  \tilde{g}^2=\tilde{g}^2(g^2,a\overline{m})\coloneqq g^2\left[1+b_g(g^2)a\overline{m}\right],
\end{equation}
with an as yet unknown improvement coefficient function $b_g(g^2)$.
In order to implement order $a$ Symanzik improvement, when varying
the average quark mass $\overline{m}$,
naively one would keep $\tilde{g}^2$ fixed, as is assumed,
e.g., in ref.~\cite{Bhattacharya:2005rb}.
Since $b_g>0$ (at least in perturbation theory)
this means $g^2$ should be reduced as $\overline{m}$ is increased.
We find this impractical and instead keep $g^2$ fixed, thereby changing the
improved coupling $\tilde{g}^2$ as
$a\overline{m}$ is varied. As a result of this choice, the lattice
spacing --- although defined in the
$N_f=3$ chiral limit --- acquires a dependence on $a\overline{m}$:
$a(\tilde{g}^2)=a(\tilde{g}^2(g^2,a\overline{m}))$.
We can determine the order $a$ difference between this
lattice spacing, corresponding to a fixed value of the
improved coupling $\tilde{g}^2$, and that
corresponding to a constant $g^2$, $a(g^2)$, by expanding
\begin{equation}
  \label{eq:atilde}
  a(\tilde{g}^2)=a\left(g^2\left(1+b_g a\overline{m}\right)\right)
  =a(g^2)\left(1+b_a a\overline{m}+\ldots\right).
\end{equation}

Integrating the $\beta$-function gives\footnote{$\Lambda_{\text{L}}$ is the QCD $\Lambda$ parameter in the lattice scheme defined by our action
and we use the normalization
\[
  \beta_0=11-\frac23 N_f=9,\quad
  \beta_1=102-\frac{38}{3}N_f=64.
  \]
}
\begin{equation}
  \label{eq:hdef}
  a(g^2)\Lambda_{\text{L}}=h(g^2)=\exp\left[-\frac{8\pi^2}{\beta_0g^2}
  -\frac{\beta_1}{2\beta_0^2}\ln\frac{\beta_0g^2}{16\pi^2}+\mathcal{O}(g^2)\right].
\end{equation}
Plugging $a(\tilde{g}^2)$ into this equation
allows us to relate $b_a$ and $b_g$:
\begin{equation}
  b_a(g^2)=\frac{\dd \ln h(g^2)}{\dd g^2}g^2b_g(g^2).
\end{equation}
The one-loop result $b_g^{(1)}=0.012000(2)N_f g^2$~\cite{Luscher:1996sc}, setting
$N_f=3$, translates into
\begin{equation}
  \label{eq:bapert}
  b^{(1)}_a=\frac{8\pi^2}{\beta_0g^2}b^{(1)}_g=0.31583(5).
\end{equation}

For simplicity of notation, below we will refer to
$a(\tilde{g}^2)$
as $a$ while we refer to the lattice spacing that we will approach in the
chiral limit when keeping $g^2$ fixed as $a_0$, i.e.\
\begin{equation}
  \label{eq:azero}
  a_0(g^2)=a\left(\tilde{g}^2(g^2,a\overline{m})\right)\left(1-b_a a\overline{m}\right).
\end{equation}
Hadron masses determined in lattice units $Ma$ are
subject to quark mass-dependent order $a$ effects since we do not simulate at
fixed values of $\tilde{g}^2$, whereas combinations $Ma_0
=Ma(1-b_aa\overline{m})$ are free of such contributions.
In the analysis we will use dimensionless combinations of physical
observables to circumvent this complication. This is necessary to
maintain order $a$ improvement when including points in the quark mass
plane that are not on the $\Tr M=\text{const}$ trajectory. Another
subtle issue concerns the definition of quark masses in the lattice scheme
eq.~\eqref{eq:quark}. In this case $\kappa_{\text{crit}}$ as a
function of the improved coupling $\tilde{g}^2$ should be used.
Nevertheless, we employ
$\kappa_{\text{crit}}(g^2)$. The difference is
of order $g^4a\overline{m}$ and can be absorbed into the definition of the
improvement coefficients $\tilde{b}_m$ and $\tilde{d}_m$~\cite{Bali:2016umi}.

Finally, we remark that keeping $g^2$ fixed rather than $\tilde{g}^2$
means that we use the renormalization constant $Z_J(g^2)$ to renormalize
a current $J(g^2,am_{\ell},am_s)$,
instead of $Z_J(\tilde{g}^2(g^2,a\overline{m}))$ in equations
such as eq.~\eqref{eq:massdef}.
This substitution can be implemented
consistently but alters some of the $\mathcal{O}(a)$ terms,
as discussed in refs.~\cite{Bali:2016umi,Korcyl:2016ugy,inprep}.
Again, we remark that $Z_J$ remains
the same function of $g^2$ but, unlike ref.~\cite{Bhattacharya:2005rb},
we keep its argument constant as the quark masses are varied.

\subsection{The critical hopping parameter and combinations of renormalization
  constants and improvement coefficients}
\label{sec:kappaImpr}
Following ref.~\cite{Bali:2016umi}, we parameterize the dependence of the
AWI quark masses on the hopping parameter values as follows,
\begin{align}
a\widetilde{m}_s-a\widetilde{m}_{\ell}&=\frac{Z}{2}\left(\frac{1}{\kappa_s}-
\frac{1}{\kappa_{\ell}}\right)\left[1-\frac{\mathcal{A}}{12}\left(
\frac{1}{\kappa_s}-\frac{1}{\kappa_{\ell}}\right)
-\mathcal{B}_0a\overline{m}\right],
\label{eq:fitdiff}\\
a\overline{\widetilde{m}}&=r_m Z\left[a\overline{m}-\frac{\mathcal{C}_0}{36}
  {\left(\frac{1}{\kappa_s}-\frac{1}{\kappa_{\ell}}\right)}^2
  -\frac{\mathcal{D}_0}{2}{(a\overline{m})}^2\right].
\label{eq:fitsum}
\end{align}
The two AWI quark mass combinations on the left hand sides, at each
value of $\beta$, depend on six parameters:
$Z$, $r_m Z$, $\kappa_{\text{crit}}$ (implicit in the average lattice quark mass),
$\mathcal{B}_0$, $\mathcal{C}_0$ and $\mathcal{D}_0$, while $\mathcal{A}$
is already known non-perturbatively~\cite{Korcyl:2016ugy,inprep}.
The combination of flavour non-singlet renormalization constants $Z$ is
defined as
\begin{equation}
  \label{eq:z1loop}
  Z=\frac{Z_m Z_P}{Z_A}=1+0.05274\,C_F g^2+\mathcal{O}(g^4),
\end{equation}
where the one-loop result was obtained in
refs.~\cite{Taniguchi:1998pf,Constantinou:2014fka} and $C_F=4/3$.
While flavour non-singlet combinations of lattice quark masses renormalize
with $Z_m=Z_S^{-1}$, the average sea quark mass renormalizes with $Z_m^s=r_m Z_m$,
for details, see, e.g., refs.~\cite{Bhattacharya:2005rb,Bali:2016umi}.

\begin{table}[htb]
\caption{\label{tab:fitresults}
Results of fits to our AWI quark mass
data according to eqs.~\eqref{eq:fitdiff} and~\eqref{eq:fitsum}.
The $\mathcal{A}$-values were determined in ref.~\cite{inprep}.
In addition to the fit parameters obtained separately at each
$\beta$-value, we list the results from the global interpolations
eqs.~\protect\eqref{eq:pade},~\protect\eqref{eq:zinter},~\protect\eqref{eq:rminter} and~\protect\eqref{eq:kappainter} as ``(int)''. In the
cases where the reduced $\chi^2$-value turned out to be larger than one,
we multiplied our errors with its square root. Due to the
smaller number of ensembles available, determining the parameters
$B_0$ and $C_0$ via such a fit was not possible at $\beta=3.85$, while
$B_0$ could not be predicted at $\beta=3.34$.
Therefore, in these cases the result from the interpolation,
obtained at the remaining five or four lattice spacings, was
used as an input. At $\beta=3.85$, in addition, we constrained
$D_0=1\pm 10$. The values for $\beta=3.4$ and 3.55 below
supersede those that we published in ref.~\cite{Bali:2016umi}.
We also include an independent determination
of $r_m$~\cite{Heitger:2021bmg}, interpolated to the same
lattice spacings, for comparison.}
\begin{center}
  {\scriptsize
    \begin{tabular}{ccccccc}
      \toprule
$\beta$&3.34 & 3.4 &            3.46&            3.55&          3.7&           3.85\\\midrule
$\chi^2/N_{\text{DF}}$&1.8/2&41.1/13&32.6/11&20.9/12&13.0/8&3.9/1\\         
$Z$        &0.8061(218)&0.8705(127)&0.9186(84)&0.9819(41)&1.0514(19)&1.0843(23)\\
$Z$ (int)  &0.7924(167)&0.8681 (88)&0.9246(56)&0.9857(41)&1.0493(26)&1.0861(28)\\
$r_m$      &3.818(958) &2.625(156) &1.848(65) &1.550(15) &1.300(14) &1.170(117)\\
$r_m$ (int)&4.594(551) &2.437 (68) &1.879(23) &1.541(12) &1.317 (9) &1.216(35)\\
$r_m$~\cite{Heitger:2021bmg}& &2.335 (31) &1.869(19) &1.523(14) &1.267(16) &1.149(33)\\
$\kappa_{\text{crit}}$&0.1366944(218)&0.1369112(45)&0.1370657(28)&0.1371709(11)&0.1371532(14)&0.1369768(84)\\
$\kappa_{\text{crit}}$ (int)&0.1366938(45)&0.1369153 (9)&0.1370613(10)&0.1371715(10)&0.1371530 (9)&0.1369767(26)\\
$\mathcal{A}$&2.058(49)&2.026(47)&1.995(45)&1.952(43)&1.886(40)&1.828(37)\\      
$\mathcal{B}_0$&   &$-$1.11(3.20)&$-$1.56(1.62)&$-$1.17(83)(1)&0.11(36)&   \\
$\mathcal{B}_0$ (int)&$-$4.41(6.78)&$-$2.42(2.26)&$-$1.44(95)&$-$0.65(42)&$-$0.01(33)&0.32(31)\\      
$\mathcal{C}_0$&5.42(69)&4.03(30)&2.57(42)&2.45(20)&2.10(33)&   \\
$\mathcal{C}_0$ (int)&5.60(58)&3.80(19)&3.04(16)&2.47(14)&2.02(11)&1.79(9)\\
$\mathcal{D}_0$&$-$31(542)&6(14)&$-$5(7)&2.7(1.3)&4.8(2.1)&   \\\bottomrule
  \end{tabular}}
\end{center}
\end{table}

The parameters
$\mathcal{A}$, $\mathcal{B}_0$, $\mathcal{C}_0$ and $\mathcal{D}_0$
are normalized such that these are unity in the free field case. They
correspond to combinations of improvement coefficients that are defined in
ref.~\cite{Bhattacharya:2005rb} (for the difference between
$\widetilde{b}_P$ and $\overline{b}_P$ etc., see refs.~\cite{Bali:2016umi,inprep}):
\begin{align}\label{eq:adef}
\mathcal{A}&=b_P-b_A-2b_m,\\
\mathcal{B}_0&=
-(r_m+1)(b_P-b_A)-2b_m-3(\tilde{b}_P-\tilde{b}_A+\tilde{b}_m),\\
\mathcal{C}_0&=-\frac{1}{2r_m}(b_P-b_A)-2d_m,\\
\mathcal{D}_0&=-2(b_P-b_A+d_m)-6(\tilde{b}_P-\tilde{b}_A+\tilde{d}_m).
\label{eq:ddef}
\end{align}
The combination $\mathcal{A}$ has been determined in ref.~\cite{inprep}
and is not fitted here. Also $\mathcal{B}_0$ was computed in this
reference, however, we choose to re-determine this here.
We repeat the analysis of ref.~\cite{Bali:2016umi}
and obtain the values displayed in table~\ref{tab:fitresults}.
The larger set of ensembles at our disposal enables us to fit all
the parameters. In general $\mathcal{D}_0$ is not
well constrained by the data.
In order to discriminate $Z$ from the combination $Z\mathcal{B}_0$,
at least two ensembles with $m_s\neq m_{\ell}$ at different
values of $2m_\ell+m_s$ are necessary. Such sets of ensembles are not
available at $\beta=3.34$ and $\beta=3.85$, where the number of
different ensembles is smaller than at the intermediate four
lattice spacings. Therefore, in these cases
we estimate $\mathcal{B}_0$, extrapolating
from the other lattice spacings as described below. At $\beta=3.85$,
in addition $\mathcal{C}_0$ is obtained from an extrapolation. We then
use these values and their errors as input.
At $\beta=3.85$ we also vary $\mathcal{D}_0=1\pm 10$.
The input parameter variations are implemented as pseudo-bootstrap
samples in the cases of $\mathcal{A}$, $\mathcal{B}_0$ and
$\mathcal{D}_0$ while we can add the constraint on $\mathcal{C}_0$
as a prior, without overly biasing the result. Different combinations
of extrapolating some parameters and fitting the remaining ones were
carried out, with consistent results.

We will use the interpolated results
``(int)'' of table~\ref{tab:fitresults} in our analysis.
The parameters $\mathcal{B}_0$ and $\mathcal{C}_0$ start out at very small
and very large values, respectively, but steadily approach unity as
$\beta$ is increased.
In the case of $\mathcal{D}_0$ the $\beta<3.55$ results are compatible with zero
within their large errors as the data are not very sensitive with respect
to this parameter. Only at $\beta\geq 3.55$ we are able to obtain positive,
non-zero values.

\begin{figure}[htp]
  \centering
  \resizebox{0.49\textwidth}{!}{\includegraphics[width=\textwidth]{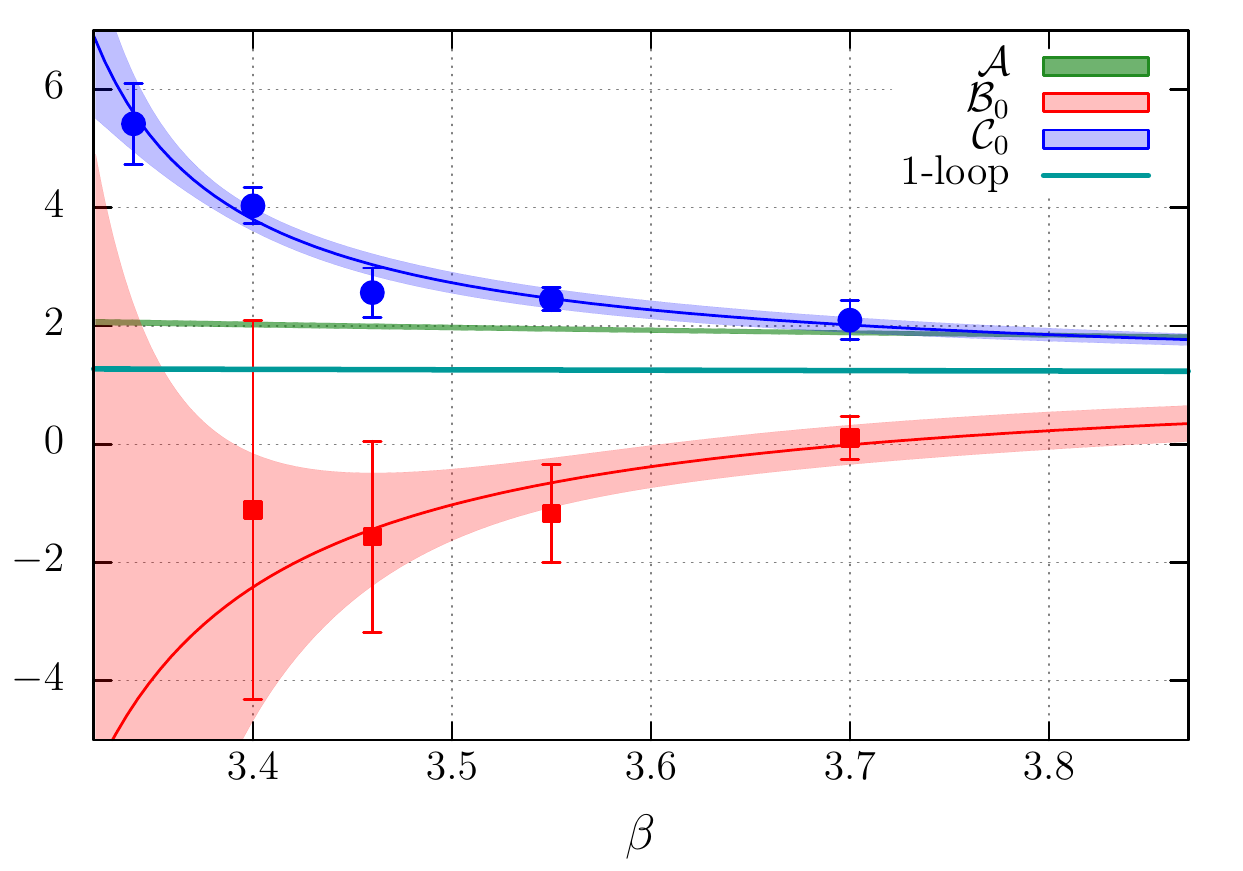}}
  \resizebox{0.49\textwidth}{!}{\includegraphics[width=\textwidth]{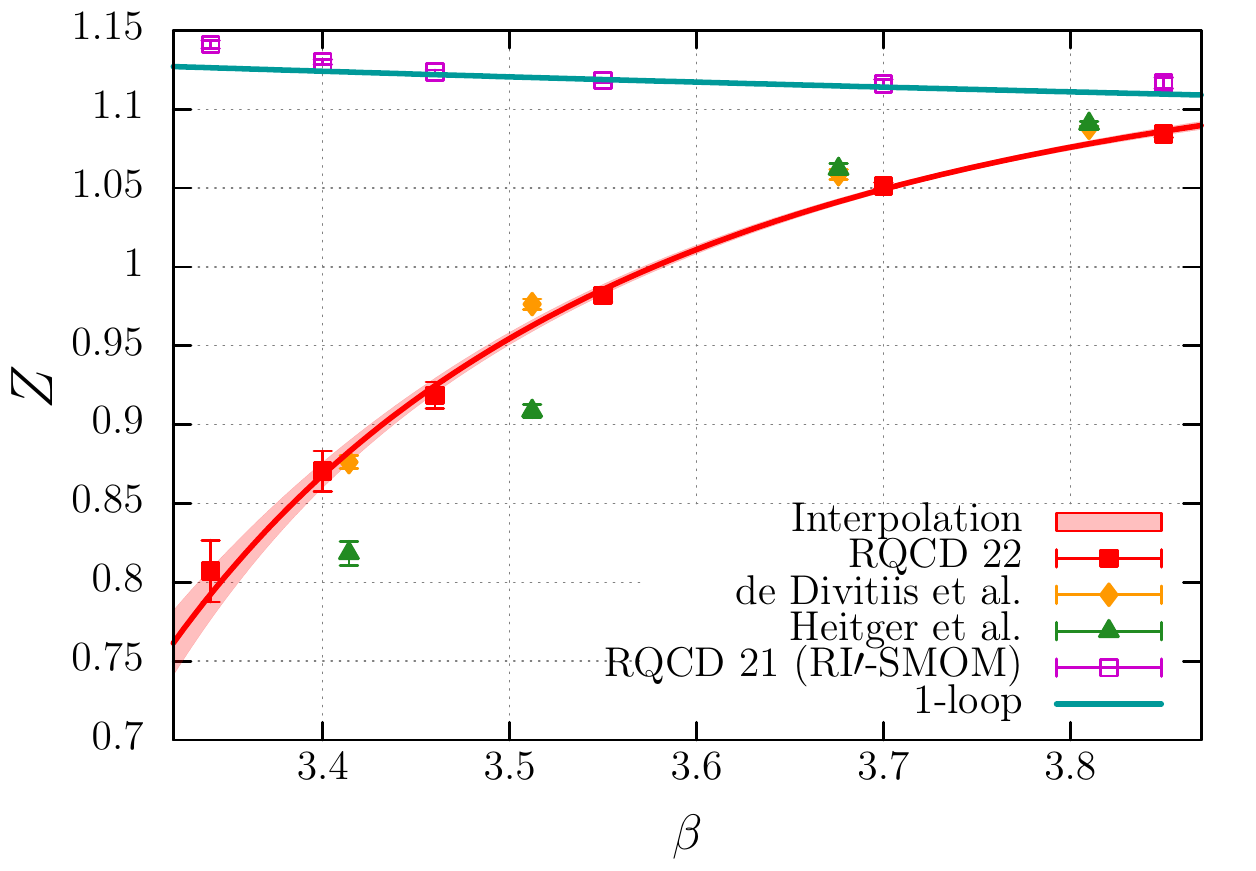}}
  \caption{\label{fig:ZBC}Left: $\mathcal{A}$~\cite{inprep},
    $\mathcal{B}_0$ and
    $\mathcal{C}_0$ along with the one-loop expectations~\eqref{eq:aper}--\eqref{eq:cper} and the
    parametrization eq.~\eqref{eq:pade} with the parameter values of
    table~\ref{tab:fitresults2}. Right: the renormalization constant combination
    $Z=Z_m Z_P/Z_A$, together with the one-loop expectation and the
    parametrization eq.~\eqref{eq:zinter}. Also shown are the ``LCP-0''
    results for $Z$ of ref.~\cite{deDivitiis:2019xla} (de~Divitiis et al.)
    and the ``$Z^{(T/3)}$'' definition of ref.~\cite{Heitger:2021bmg} (Heitger et al.) as well
    as the RI'-SMOM determination of ref.~\cite{Bali:2020lwx} (RQCD~21),
    using ``$Z_A'$'', with leading lattice artefact subtraction and
    the fixed scale method.
}
\end{figure}

For our action to one-loop order~\cite{Taniguchi:1998pf} the
above combinations of improvement coefficients
read~\cite{Bali:2016umi}\footnote{There is one subtlety here:
  it turns out that
  the effect of $b_g$ already propagates at $\mathcal{O}(g^2)$
  into the $\tilde{b}_J$ and $\tilde{d}_J$
  improvement coefficients for currents with an
  anomalous dimension, see Ref.~\cite{inprep} for details.
  However, since $\tilde{b}_m=-\tilde{b}_S$ and
  $2d_m+6\tilde{d}_m=-b_S-3\tilde{b}_S-3\tilde{d}_S$,
  the anomalous dimension contributions cancel from the combinations
  $\mathcal{B}_0$ and $\mathcal{D}_0$, as they should.}
\begin{align}
  \mathcal{A}  &=1+0.1538(2)\,g^2,\label{eq:aper}\\
  \mathcal{B}_0&=\mathcal{D}_0=1+0.1501(4)\,g^2,\label{eq:bper}\\
  \mathcal{C}_0&=1+0.1520(2)\,g^2.\label{eq:cper}
\end{align}
This motivates fits of $\mathcal{O}\in\{\mathcal{B}_0,\mathcal{C}_0\}$
to the interpolating ansatz:
\begin{equation}
  \label{eq:pade}
  \mathcal{O}=1+b_{\mathcal{O}}^{\text{one-loop}}g^2\frac{1+\gamma_{\mathcal{O}}g^2}{1+\delta_{\mathcal{O}}g^2}.
\end{equation}
The resulting fit parameters are displayed in table~\ref{tab:fitresults2}.
For the $\mathcal{B}_0$ interpolation
we obtain $\chi^2/N_{\text{DF}}=0.67/2$ and for $\mathcal{C}_0$
$\chi^2/N_{\text{DF}}=1.99/3$.
In figure~\ref{fig:ZBC} we show the data for $\mathcal{B}_0$ and
$\mathcal{C}_0$, along with the interpolating parametrizations
eq.~\eqref{eq:pade} and the one-loop expectations, which are
indistinguishable on the scale of the figure.
We also include the parametrization
$\mathcal{A}=1+0.1538\,g^2+0.242(15)\,g^4$ of ref.~\cite{inprep}.
The non-perturbatively determined values differ substantially
from the one-loop expectations (grey curves).

\begin{table}
  \caption{Parameters of fits of the $3.4\leq\beta\leq 3.7$ data
    to eq.~\eqref{eq:pade}, using
    the one-loop coefficients of ref.~\protect\cite{Taniguchi:1998pf},
    see eqs.~\eqref{eq:bper} and~\eqref{eq:cper}.
\label{tab:fitresults2}}
\begin{center}
  \begin{tabular}{ccccc}
    \toprule
    Coefficient&$b_{\mathcal{O}}^{\text{one-loop}}$&$\gamma_{\mathcal{O}}$&$\delta_{\mathcal{O}}$&cov($\gamma_{\mathcal{O}}$,$\delta_{\mathcal{O}}$)\\
    \midrule
    $\mathcal{B}_0$&0.1501&$-0.946(223)$&$-0.537(34)$&$-0.918$\\
    $\mathcal{C}_0$&0.1520&$-0.308 (48)$&$-0.542(4)$&0.928\\\bottomrule
  \end{tabular}
\end{center}
\end{table}

In the right panel of figure~\ref{fig:ZBC} we show our data for $Z=Z_mZ_P/Z_A$
in comparison to different literature results~\cite{deDivitiis:2019xla,Bali:2020lwx,Heitger:2021bmg}. Up to order $a^2$ effects, the different sets
should approach each other towards large $\beta$-values. For instance,
in figure~20 of ref.~\cite{Bali:2020lwx} it has been demonstrated
that the ratios between the results for $Z$ from refs.~\cite{Bali:2020lwx}
and~\cite{deDivitiis:2019xla} are consistent with this expectation,
when plotted as a function of $a^2$. We note that
the former non-perturbative set of results, obtained using the RI'-SMOM
scheme (RQCD~21~\cite{Bali:2020lwx}), is quite close
to the one-loop expectation but differs substantially
from the three other sets within our window of $\beta$-values.
For $3.34\leq\beta\leq 3.85$, our data for $Z$ can be parameterized with
$\chi^2/N_{\text{DF}}=3.79/3$ as follows:
\begin{equation}
  \label{eq:zinter}
  Z(g^2)=1+0.07032\,g^2\frac{1+0.4896\,g^2-0.6473\,g^4}{1-0.4857\,g^2}.
\end{equation}
We refrain from stating the covariance matrix and the errors of the fit
parameters, however, the accuracy of the interpolation
(about 1\% at $\beta=3.4$ and
2\textperthousand\ at $\beta=3.7$) can be read off
table~\ref{tab:fitresults}.
We also show the above interpolation in the figure.

\begin{figure}[htp]
  \centering
  \resizebox{0.49\textwidth}{!}{\includegraphics[width=\textwidth]{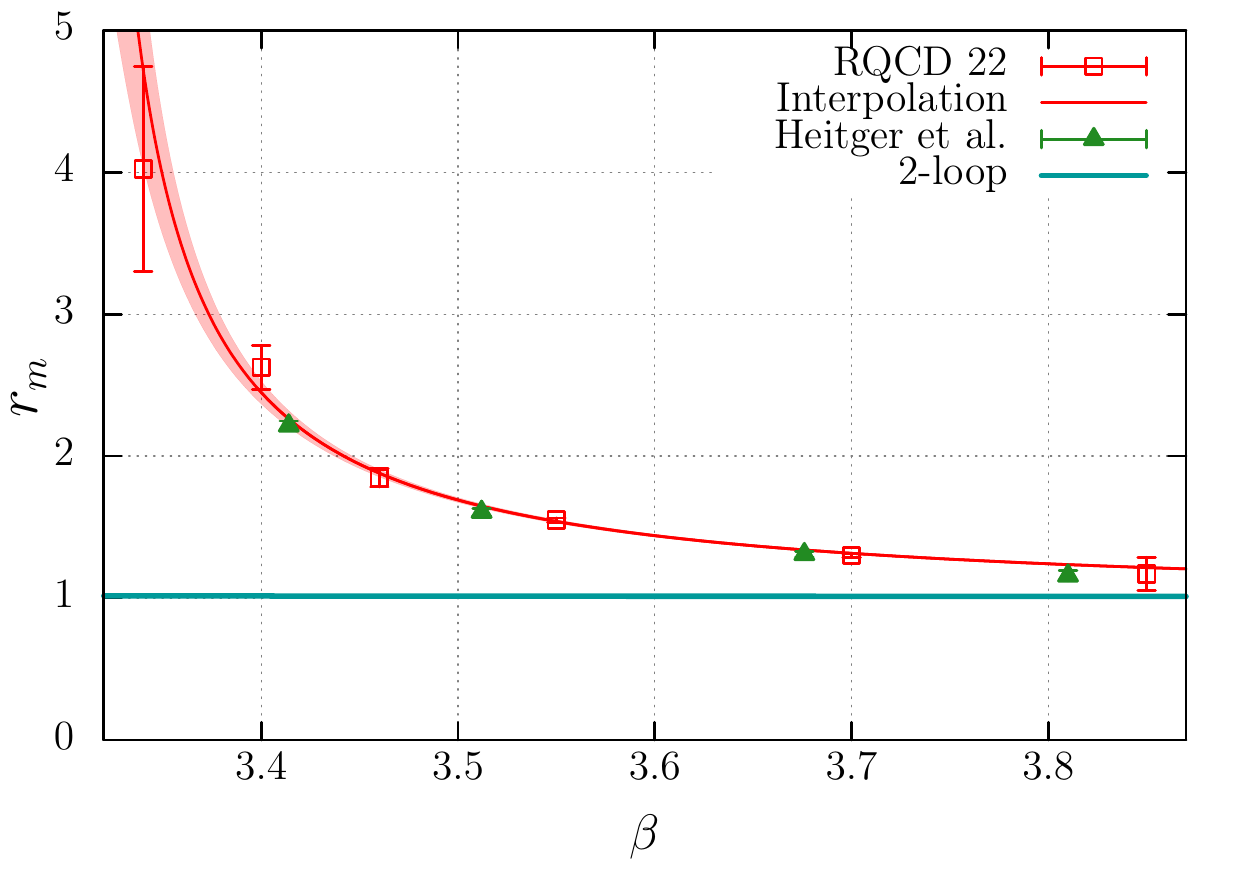}}
  \resizebox{0.49\textwidth}{!}{\includegraphics[width=\textwidth]{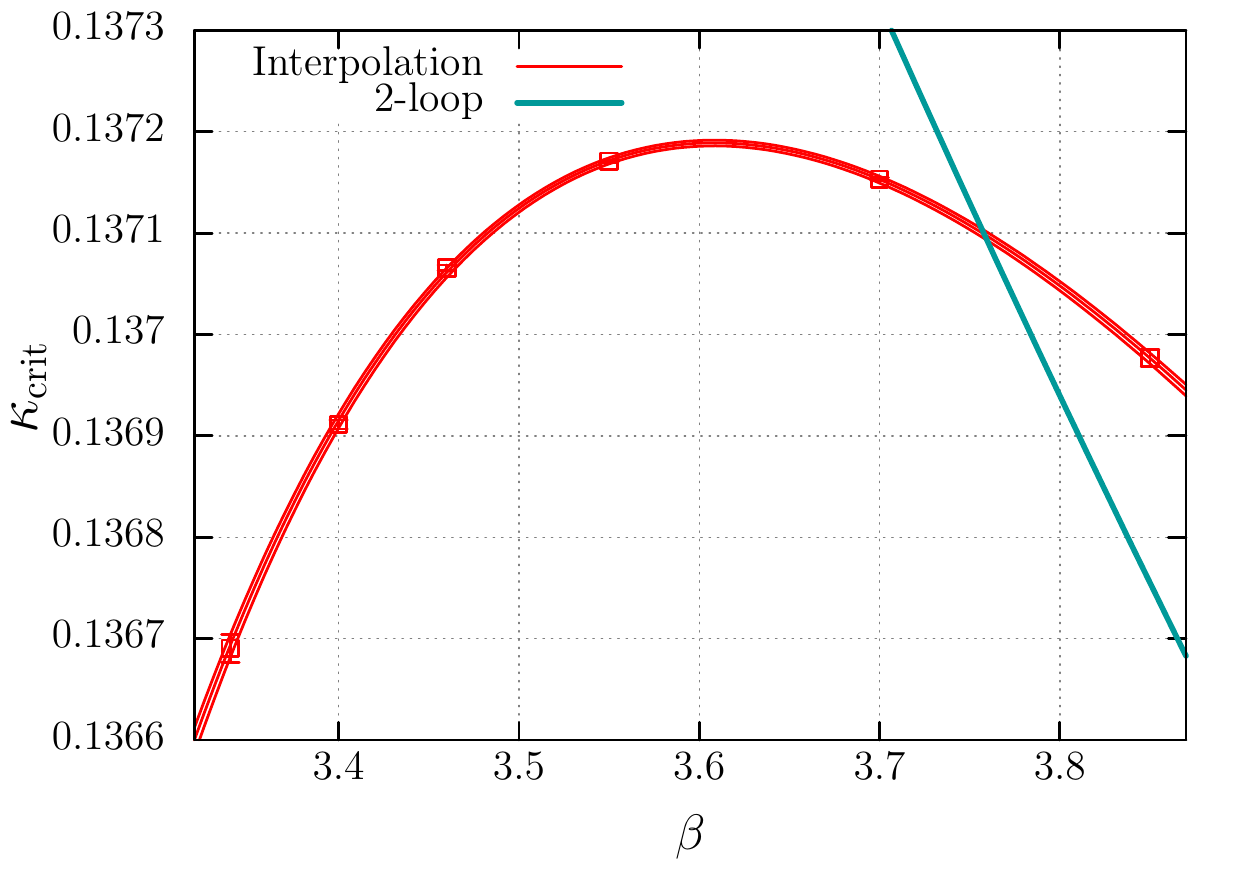}}
  \caption{\label{fig:rmkappa}Left: the ratio of the singlet over the
    non-singlet quark mass renormalization constant $r_m=Z_m^{s}/Z_m$,
    along with the parametrization eq.~\eqref{eq:rminter} and the two-loop
    expectation. Also shown is the recent determination of
    ref.~\cite{Heitger:2021bmg}, using their ``$\nu,Z,T/3$'' prescription.
    Right: the critical hopping parameter $\kappa_{\text{crit}}$,
    together with the interpolation eq.~\eqref{eq:kappainter} and the
    two-loop expectation eq.~\eqref{eq:kappa2loop}.}
\end{figure}

Regarding $r_m$, we obtain the interpolation
\begin{equation}
  \label{eq:rminter}
  r_m(g^2)=1+0.004630\,g^4\frac{1+0.128(72)\,g^2}{1-0.5497(13)\,g^2},
\end{equation}
where the two-loop coefficient $0.004630(2)$
was computed in ref.~\cite{Constantinou:2016ieh}.
We obtain $\chi^2/N_{\text{DF}}=3.91/4$ and the correlation between the
two fit parameters reads 0.844. The numerical values
are compiled in table~\ref{tab:fitresults}.
The result is also shown, along with the interpolating
formula and the two-loop expectation, in the left panel of
figure~\ref{fig:rmkappa}. We compare this to the recent determination
of Heitger et al.~\cite{Heitger:2021bmg}. The two sets can
in principle, differ by order $a^2$ effects. With
the exception of $\beta\approx 3.7$, the data sets are compatible
with one another (see table~\ref{tab:fitresults}), however,
the respective interpolations intersect (not shown).
Note that --- similarly to $r_m$ ---
also the data for $Z$ of Heitger et al.\ cross our
results inbetween $\beta=3.512$ and $\beta=3.676$,
as can be seen in figure~\ref{fig:ZBC}. The error obtained
at $\beta=3.85$ for our interpolation was quite small compared to
the actually measured uncertainty at this point
(0.007 vs.\ 0.117). This is due to the perturbation theory
constraint for $\beta\rightarrow\infty$. Therefore, to be on the safe
side, we inflated the error given for the interpolation of $r_m$
at this end point by a factor of five to bring this more
in line with the statistical uncertainties.

The critical hopping parameter for our action is known to
two-loops~\cite{Skouroupathis:2008ry}: for the one-loop
coefficient we insert the values of table~II into eq.~(20) of that reference,
setting $c_{\text{sw}}=1$. At the two-loop level we combine this result
with the one-loop correction to $c_{\text{sw}}$, $0.19624449(1)g^2$~\cite{Aoki:2003sj} (see also ref.~\cite{Aoki:1998qd}), and add the two-loop
results of tables~III--VII of ref.~\cite{Skouroupathis:2008ry},
setting $c_{\text{sw}}=1$ and $c_2=0$:
\begin{equation}
  \label{eq:kappa2loop}
  \frac{1}{\kappa_{\text{crit}}}=8-0.402453622(12)\,g^2-0.024893(5)\,g^4+\mathcal{O}(g^6).
\end{equation}
We use these results to fit $\kappa_{\text{crit}}$:
\begin{equation}
  \label{eq:kappainter}
  \frac{1}{\kappa_{\text{crit}}}=8-0.402454\,g^2\frac{1+0.28955\,g^2-0.1660\,g^6}{1+0.22770\,g^2-0.2540\,g^4}.
\end{equation}
This interpolation with $\chi^2/N_{\text{DF}}=3.67/3$, also shown
in figure~\ref{fig:rmkappa},
is valid for $3.34\leq\beta\leq 3.85$ and has the two-loop asymptotic
large $\beta$ limit built in. Its relative accuracy
ranges from approximately $3\cdot 10^{-5}$ at $\beta=3.34$ to
$2\cdot 10^{-5}$ at $\beta=3.85$ and is more precise inbetween these
end points.
The relevant numbers are included in table~\ref{tab:fitresults}.
The slow approach towards the two-loop expectation is striking.

Note again that $Z$ and $r_m=r_m Z/Z$ can differ by $\mathcal{O}(a^2)$ terms if
determined following a different prescription while the improvement
coefficients $\mathcal{A}$, $\mathcal{B}_0$, $\mathcal{C}_0$ and
$\mathcal{D}_0$ have ambiguities of order $a$.

\subsection{The scale parameter $\boldsymbol{t_0}$}
\label{sec:t0def}

\begin{figure}[htp]
  \centering
  \resizebox{\textwidth}{!}{\includegraphics[width=\textwidth]{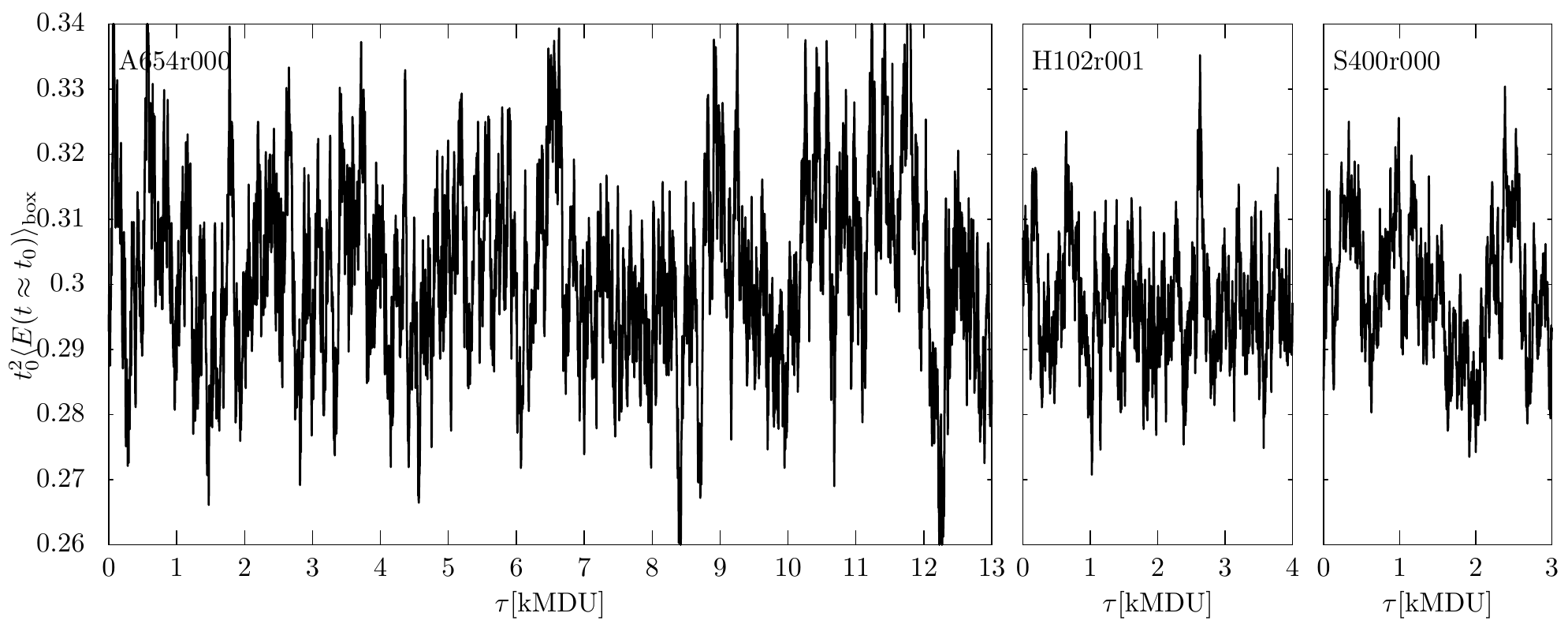}}
  \resizebox{\textwidth}{!}{\includegraphics[width=\textwidth]{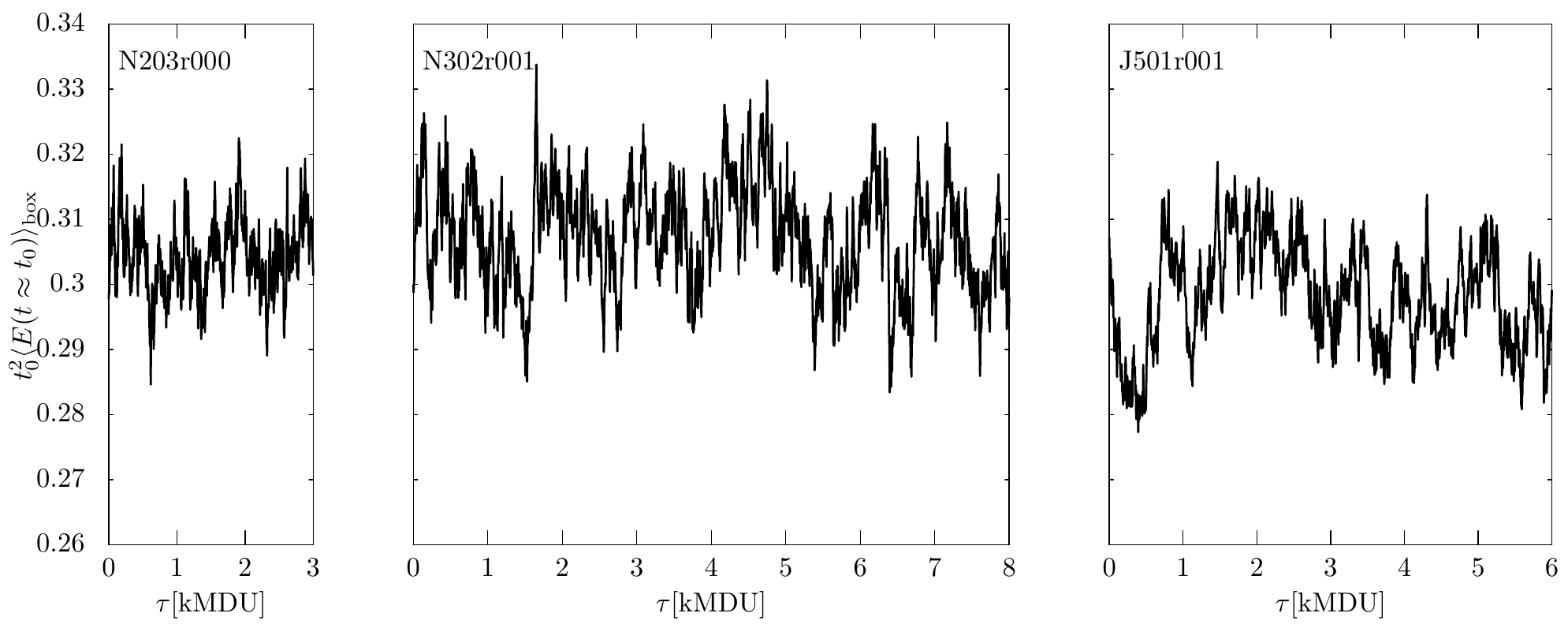}}
  \caption{History of the Wilson flow action density, multiplied
    by $t_0^2$~\cite{Luscher:2010iy},
    at a flow time close to
    \(t_0\),
        inside a sub-volume $V_4$ of approximately
        \(1\,\textmd{fm} \cdot {(a N_s)}^3\) (except for A654 with
        periodic boundary conditions, where
        we employ the whole volume), along a line of
    \(M_\pi \approx 350\,\textmd{MeV}\) (for $\Tr M=\text{const}$)
    from coarse to fine lattice
    spacings. The amplitude of the fluctuation varies, e.g., due to
    somewhat different physical volumes.
    Autocorrelations increase from top left to bottom right,
    with the exception of
    A654 at $\beta=3.34$ where we observe larger autocorrelation times
    than at $\beta=3.4$. For the cases where more than one
    Monte Carlo chain exists, only one replica is shown.
    \label{fig:wilson}
    }
\end{figure}

The parameter $t_0$, introduced in ref.~\cite{Luscher:2010iy},
corresponds to the Wilson flow time $t$ at which the equality
\begin{equation}
  \left.t^2\,E(t)\right|_{t=t_0} = 0.3,\quad
  E(t)=\frac{1}{V_4}\int_{V_4}\!\!{\dd}^4\!x\,\frac14G_{\mu\nu}^a(x,t)G_{\mu\nu}^a(x,t)
\end{equation}
holds, where we employ the clover leaf definition of the
average action density $E(t)$ and the
integration scheme of ref.~\cite{Luscher:2010iy}. For lattices with
open boundary conditions in time, $E(t)$ is only averaged over
the central temporal region of the lattice,
$V_4\approx 1\,\text{fm}\cdot(aN_s)^3$. Otherwise,
we employ the whole lattice volume $V_4=a^4N_tN_s^3$.
In figure~\ref{fig:wilson} we show the
Monte Carlo history in MDUs of $E(t)$ at $t\approx t_0$,\footnote{Note
  that the data shown are only approximately at $t=t_0$.
  In the actual scale setting analysis we interpolate $E(t)$ between available
  flow times to determine the correct value of $t_0/a^2$.}
for different ensembles with $M_{\pi}\approx 350\,\textmd{MeV}$
along the $\Tr M=\text{const}$ line, see figure~\ref{fig:ensembles}.
The lattice spacing decreases from the left ($a\approx 0.098\,\textmd{fm}$)
to the right ($a\approx 0.039\,\textmd{fm}$). This observable is known
to have very large autocorrelation times~\cite{Luscher:2011kk} and indeed
some slowing down is clearly visible. Nevertheless, even at the finest
lattice spacing we are able to sufficiently sample the action density
at this flow time.

Using an intermediate scale derived from $t_0$ to translate between
different lattice spacings is a convenient choice since all its mass dependence
is due to sea quark effects. Therefore, one would expect $t_0/a^2$
to vary only moderately at each $\beta$-value.
At the same time the $t_0/a^2$-values carry
tiny statistical errors. In section~\ref{sec:t0}
we study the quark mass dependence of $t_0/a^2$ and find that
in the continuum limit this combination remains constant
along trajectories where the sum of quark masses is kept fixed, within
our present uncertainties.

We introduce four related scales:
\begin{itemize}
\item The ratio $t_{0,\text{symm}}/a^2$ refers to $t_0$ in lattice
  units at the point along our $\Tr M=\text{const}$ lines where
  $m_s=m_{\ell}$. Note that $t_{0,\text{symm}}$ (in physical units)
  differs slightly between different $\beta$-values.
\item
  The scale $t_0^*$~\cite{Bruno:2016plf} is defined
  as the value of $t_0$ at the point in the quark
  mass plane where, along the symmetric $m_s=m_{\ell}$ line,
  \begin{equation}
    \label{eq:phi4}
    \phi_4^*=8t_0^*\left(M_K^2+\frac{M_{\pi}^2}{2}\right)=
    12t_0^*M_{\pi}^2\coloneqq 1.110.
  \end{equation}
  This is close to $t_{0,\text{symm}}$ and the combination $t_0^*/a^2$ can
  be obtained by a small interpolation or by reweighting nearby simulation
  points~\cite{Bruno:2016plf}. If at this point $2M_K^2+M_\pi^2$
  had the same value as at the physical quark mass point this definition
  would imply $\sqrt{8t_0^*}=0.413\,\textmd{fm}$. However, we stress
  that the above choice can always be made, independent of any
  such assumption.
  \item The scale $t_{0,\text{ph}}$ refers to the value of $t_0$ at the physical
  values of $M_{\pi}$ and $M_K$.
\item Finally, the scale $t_{0,\text{ch}}$ refers to $t_0$ in the chiral limit
  $M_{\pi}=M_K=0$.
\end{itemize}
We will frequently use the above sub- and superscripts also for other
quantities taken at the respective points in the quark mass plane.
$t_{0,\text{symm}}(\beta)/a^2$ is specific to our set of ensembles. These
values are in general close to $t_0^*/a^2$, however, the corresponding
$\Tr M=\text{const}$ lines do not always touch the physical point.
Unlike $t_0$ at unphysical positions in the quark mass plane,
the scale $t_{0,\text{ph}}$ can be determined
from an experimental input quantity, e.g., the mass of the cascade baryon.
ChPT LECs are defined in the chiral limit and, if dimensionful, obtain their
scale from $t_{0,\text{ch}}$. For this purpose, the ratio
$t_{0,\text{ch}}/t_{0,\text{ph}}$ is needed. Precise determinations of
$t_{0,\text{ph}}$ and of $t_{0,\text{ch}}$ require ensembles with small
pion masses. In contrast, $t_0^*/a^2$ can be determined easily with pseudoscalar
masses around $400\,\textmd{MeV}$. It is therefore an ideal intermediate
scale to relate different lattice spacings.

Below we will determine the dependence of
$t_0(\sqrt{8t_0}M_{\pi},\sqrt{8t_0}M_K,a)/a^2$ on the pion
and kaon masses as well as on the lattice spacing, which will
enable us to translate between all the above-mentioned scales. We will also
extract $t_0^*/a^2(g^2,m^*)$
as a function of $g^2=6/\beta$.

\subsection{The quark mass dependence of the $\boldsymbol{t_0}$- and the $\boldsymbol{t_0^*/a^2}$-values}
\label{sec:t0}
It turns out to be convenient to define the following pseudoscalar meson
mass combinations:
\begin{align}
  \label{eq:mass0}
\overline{M}\vphantom{M}^2\coloneqq \frac{2M_K^2+M_{\pi}^2}{3},\quad
\delta M^2\coloneqq2\left(M_K^2-M_{\pi}^2\right),
\end{align}
which we correct for finite volume effects in next-to-leading order
(NLO) ChPT as described in
section~\ref{sec:finite}. With few exceptions\footnote{For H105, H106, B452,
  N302, N304, J303 and J304 the pion mass correction becomes larger than
  half of the statistical error, but in no case does it exceed it.}
these corrections are much smaller than our statistical errors.

The dependence of the continuum limit $t_0(\overline{M},\delta M)$ on the
pseudoscalar masses has been computed to next-to-next-to
leading order (NNLO) in $\textmd{SU(3)}$ ChPT
in ref.~\cite{Bar:2013ora}. To NLO only a term proportional to $\overline{M}\vphantom{M}^2$
appears while to NNLO $M^4$ and $M^4\ln(M/\mu)$ terms
contribute. Within the present precision, at the
three finest lattice spacings we find
our $t_0$ data to be insensitive to $\delta M$:
at $\beta=3.4$ the variation
of $t_0/a^2$ along the $\overline{m}=m_{\text{symm}}$
line amounts to less than 3\% and this decreases further with
increasing $\beta$. Along the other lines in the quark mass plane
we are unable to detect any deviation from a linear dependence on
$\overline{M}\vphantom{M}^2$, the relative slope of which decreases towards the
continuum limit. Therefore, we assume the continuum limit
behaviour~\cite{Bar:2013ora}
\begin{align}
  t_0(\overline{M},\delta M) = t_{0,\text{ch}}\left( 1 + k_1\frac{3\overline{M}\vphantom{M}^2}{{(4\pi F_0)}^2} \right)
  \approx t_{0,\text{ch}}\left(1+\tilde{k}_1 8t_{0}\overline{M}\vphantom{M}^2\right),\label{eq:t0chi}
\end{align}
where $k_1 = \tilde{k}_1\cdot 8 t_{0,\text{ch}}{(4\pi F_0)}^2/3$.

\begin{figure}[htp]
  \centering
  \resizebox{0.49\textwidth}{!}{\includegraphics[width=\textwidth]{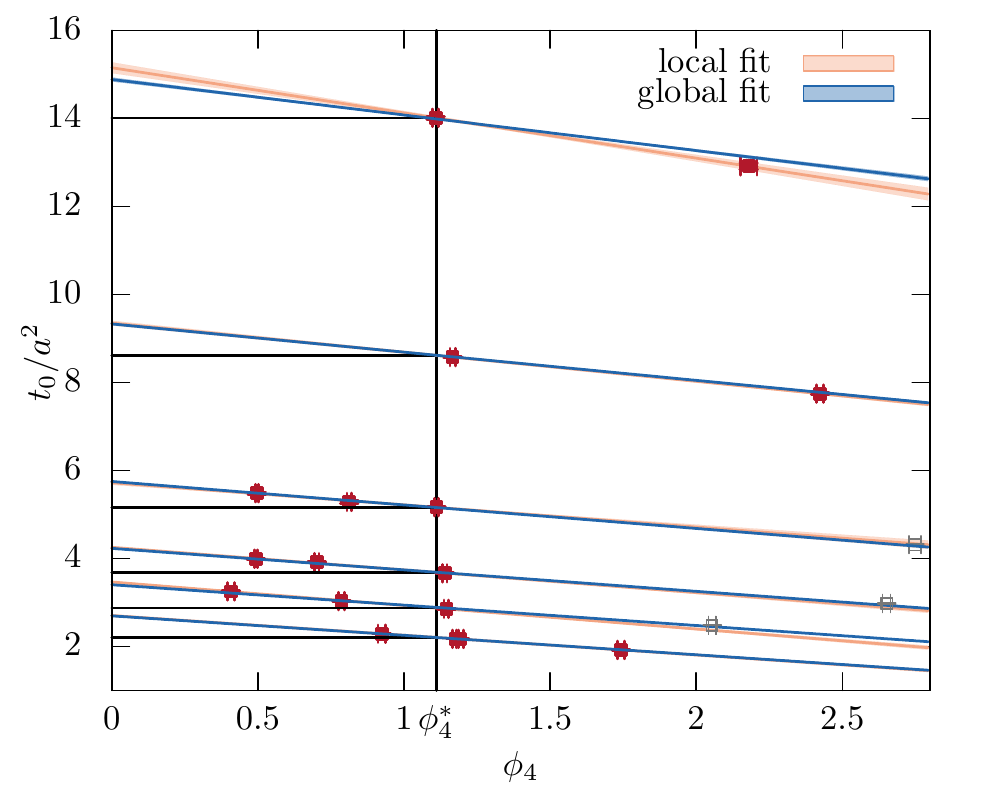}}
  \resizebox{0.49\textwidth}{!}{\includegraphics[width=\textwidth]{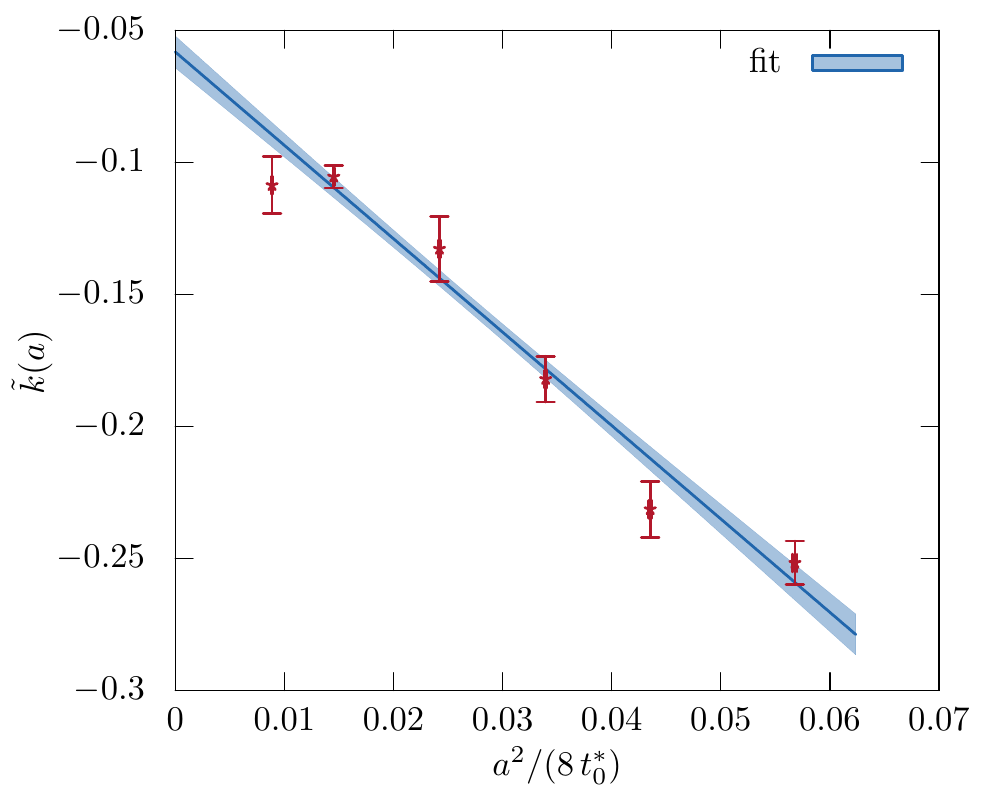}}
  \caption{Left: linear interpolation of $t_0/a^2$ for $m_s=m_{\ell}$ for
    each value of the
    coupling, from $\beta=3.34$ (bottom) to $\beta=3.85$ (top), cf.\
    eq.~\eqref{eq:localt0fit} (local fit). Also shown is the result of the
    global fit including all quark mass trajectories, see
    eqs.~\eqref{eq:t0par1}, \eqref{eq:t0par2}, \eqref{eq:globalt0fit2} and~\eqref{eq:globalt0fit3}. The vertical line marks the value of the
    $\phi_4^*$ reference point and horizontal lines the resulting values of
    $t_0^*/a^2$.
    Right: the slope $\tilde{k}$, as a function of $a^2/(8t_0^*)$ from the
    global fit, together with a
quadratic continuum limit extrapolation. 
\label{fig:t0starfit}}
\end{figure}

\subsubsection{Survey of the $\boldsymbol{t_0/a^2}$ data}
At a non-vanishing lattice spacing we will encounter
mass-dependent
order $a$ lattice corrections $t_0/a^2=(1-2b_a\,a \overline{m})t_0/a_0^2$
(see eq.~\eqref{eq:azero}) and higher order corrections. The
order $a^2$ contributions are proportional to
either a constant, $\Lambda^2$, the average squared pion mass $\overline{M}\vphantom{M}^2$
or to $\delta M^2$. As a first step,
along our $m_s = m_\ell$
lines (where $\delta M = 0$), we attempt fits of the form
\begin{equation}
  \label{eq:localt0fit}
  \frac{t_0}{a^2} \approx \frac{t_{0,{\text{ch}}}}{a_0^2}\left[1 + \tilde{k}(a) \,8 t_0(\overline{M},0) \overline{M}\vphantom{M}^2\right]
\end{equation}
with a different set of parameters $t_{0,{\text{ch}}}/a_0^2$ and $\tilde{k}(a)$
for each value of the coupling.
These fits effectively describe our data as is demonstrated in
figure~\ref{fig:t0starfit} (local fit). We list the results of this simplest way
of extracting
$t_0^*/a^2\approx (t_{0,\text{ch}}/a_0^2)[1+\tfrac{2}{3}\tilde{k}(a)\phi_4^*]$
in the first line of table~\ref{tab:t0star} as ``linear''.
To minimize a possible bias due
to higher order correction terms, in the cases where
more than three well-separated
data points were available, i.e.\ at $\beta=3.4$, 3.46 and
3.55, we excluded the heaviest pion mass from the fit.
It turns out that the resulting $t_0^*/a^2$-values are most sensitive
with respect to the value of the data point $t_0/a^2=t_{0,\text{symm}}/a^2$
closest to $t_0^*/a^2$. The results compare well with the previous determination
of ref.~\cite{Bruno:2016plf}, updated in ref.~\cite{Bruno:2017lta}, that
we show in the third row of table~\ref{tab:t0star}. We also
list the lattice spacings in this table, using the result of this article,
$\sqrt{8t_0^*}=0.4097^{(20)}_{(25)}\,\textmd{fm}$ for the conversion.
The errors of $a$ across the $\beta$-values are correlated 
because of this and are dominated
by the scale setting uncertainty. When performing continuum limit
extrapolations, we will use
the globally interpolated $t_0^*/a^2$-values of the second row of the
table instead.

\begin{table}[h!tp]
  \caption{\label{tab:t0star} Results for $t_0^*$ in lattice units from
    this work (lines 1 and 2), in comparison to the respective numbers
    of ref.~\cite{Bruno:2016plf} (updated in
    ref.~\cite{Bruno:2017lta}, line 3).
    We consider the global fit results of the second line
    as the most reliable ones. Also shown is the value of $t_0$ in the chiral
    limit, $t_{0,\text{ch}}(a)$. The continuum limit ratio $t_0^*/t_{0,\text{ch}}$
    is shown in
    eq.~\eqref{eq:t0ratio}. The first errors are statistical, the second
    errors reflect the uncertainty of the improvement coefficient $b_a$
    (that is related to $b_g$). In the last line we list the lattice spacings
    obtained through $\sqrt{8t_0^*}=0.4097^{(20)}_{(25)}\,\textmd{fm}$, see
    eq.~\eqref{eq:t0starp} below, where
    we added all errors in quadrature after symmetrizing the scale error.
  }
{
  \footnotesize
  \renewcommand{\arraystretch}{1.2}
  \setlength\tabcolsep{3 pt}
\begin{center}
\begin{tabular}{lcccccc}
\toprule
$\beta$ & 3.34 & 3.4 & 3.46 & 3.55 & 3.7 & 3.85 \\
\midrule
  $t_0^*/a^{2}$, linear fit                        & 2.204(5)     & 2.872(10)     & 3.682(12)    & 5.162(16)     & 8.613(25)    & 14.011(39) \\
$t_0^*/a^{2}$, global fit                        & 2.204(4)(4)  & 2.888(4)(7)  & 3.686(4)(10) & 5.157(5)(14)  & 8.617(7)(21) & 13.988(19)(28) \\
  $t_0^*/a^{2}$ \cite{Bruno:2016plf,Bruno:2017lta} &          & 2.862(5)     & 3.662(12)    & 5.166(15)     & 8.596(27)    & 13.880(220) \\
  $t_{0,\text{ch}}/a_0^{2}$, global fit             & 2.695(13)(2)  & 3.402(11)(1) & 4.228(10)(5) & 5.749(12)(1)  & 9.329(27)(4) & 14.885(57)(14) \\
  $a/\textmd{fm}$&0.09757(56)&0.08524(49)&0.07545(44)&0.06379(37)&0.04934(28)&0.03873(22)
\\\bottomrule
\end{tabular}
\end{center}
}
\end{table}

We plot the resulting slopes $\tilde{k}(a)$ as a function of
$a^2/(8t_0^*)$ in the right panel of figure~\ref{fig:t0starfit}.
Combining eqs.~\eqref{eq:azero}, \eqref{eq:t0chi} and~\eqref{eq:localt0fit},
we obtain
\begin{equation}
  \tilde{k}(a) = \tilde{k}_1 - \frac{2b_a a\overline{m}}{8 t_0 \overline{M}\vphantom{M}^2} + \mathcal{O}(a^2).
  \label{eq:k1tilde}
\end{equation}
However, we are unable to detect any linear contribution in the figure.
Instead, a clearly quadratic dependence on $a^2$ is visible.
Adding any other power
of $a$ to the quadratic continuum limit extrapolation results in a coefficient
that is compatible with zero.
The failure to resolve a term proportional to $a$ reflects the fact that
within our range of lattice spacings the combination
$a\overline{m}$ remains
almost constant when keeping the average renormalized quark mass
$\widehat{\overline{m}}=r_m Z_m m
\propto\overline{M}\vphantom{M}^2$ fixed because the factor $r_m$
decreases rapidly with $\beta$, as can be seen in the left panel of
figure~\ref{fig:rmkappa}. This also means that,
unlike the coefficients $r_m$, $\mathcal{A}$, $\mathcal{B}_0$,
$\mathcal{C}_0$ or $\mathcal{D}_0$, $b_a$ cannot
depend strongly on $\beta$; otherwise there would have been visible
corrections to the quadratic behaviour. Therefore, from now on
we will assume that $b_a$ coincides with its one-loop value $b_a^{(1)}$ within
a 100\% error band: $b_a=0.32(32)$. Note that $b_a^{(1)}$ is independent
of $g^2$, see eqs.~\eqref{eq:hdef}--\eqref{eq:bapert}.
The naive quadratic continuum limit extrapolation gives
\begin{equation}
  \label{eq:tildekk}
  \tilde{k}=\tilde{k}(a=0)=-0.0600(85).
\end{equation}

Motivated by the above considerations, below
we will attempt a global fit to our data
according to the effective parametrization
\begin{align}
  \label{eq:globalt0fit1}
  \frac{t_0}{a^2}& = \frac{t_{0,\text{ch}}}{a_0^2}(g^2)\left( 1 + \tilde{k}\, 8 t_0 \overline{M}\vphantom{M}^2 \right) + \bar{c}\,8 t_0\overline{M}\vphantom{M}^2 + \delta c\, 8 t_0 \delta M^2,
\end{align}
with the parameters $\bar{c}$ and $\delta c$, in addition to $\tilde{k}$.
To set the stage for this global fit, we first interpolate the
$t_0/a^2$-values at each $\beta$-value locally via a linear
fit to determine $t_0^*/a^2$.

\subsubsection{Interpolating formula for $\boldsymbol{t_0^*/a^2}$}
For very small values of the coupling $g^2$ the dependence of
$t_0^*/a^2$ on $g^2$ is controlled by the perturbative $\beta$
function. Its three-loop coefficient $\beta_2$ is at present not
known for our action. However, the ratio of $\Lambda$ parameters,
\begin{equation}
  \frac{\Lambda_{\text{L}}}{\Lambda_{\MS}}=0.2887542,
\end{equation}
was calculated in ref.~\cite{Skouroupathis:2007mq}. Combining this
with the recent determination~\cite{Bruno:2017gxd}
\begin{equation}
  \label{eq:t0ms}
  \Lambda_{\MS} \sqrt{8t_0^*} = 0.712(24)
\end{equation}
for the three flavour theory, we arrive at the expectation
\begin{equation}
  \frac{t_0^*}{a^2(g^2)}=
  \frac{t_0^*}{a_0^2(g^2)}\left[1-2 b_a(g^2)a m^*\right]=
  \left[f(g^2)+c_{t_0}+d_{t_0}f^{-1/2}(g^2)\right]\left[1-2 b_a(g^2)a m^*\right],
  \label{eq:t0par1}
\end{equation}
where
\begin{equation}
  f(g^2)=\frac{t_0^*\Lambda^2_{\text{L}}}{h^2(g^2)}=0.00528(36)\exp\left(\frac{16\pi^2}{\beta_0g^2}+\frac{\beta_1}{\beta_0^2}\ln\frac{\beta_0g^2}{16\pi^2}-b_{t_0}g^2+\cdots\right).\label{eq:t0par2}
\end{equation}  
The coefficient $b_{t_0}\approx (\beta_1^2-\beta_0\beta_2)/(16\pi^2\beta_0^3)$
effectively parameterizes higher order perturbative contributions,
$c_{t_0}$ describes the leading $\mathcal{O}(a^2)$ lattice correction to
eq.~\eqref{eq:t0ms} and $d_{t_0}$ a subleading
$\mathcal{O}(a^3)$ correction.
Setting $b_a=0$, the resulting fit parameters read
\begin{equation}
  c_{t_0}=0.18(12),\quad d_{t_0} = -1.43(16), \quad b_{t_0} = 0.9293(46),\quad \chi^2/N_{\text{DF}} = 2.1 / 3.
  \label{eq:t0par3}
\end{equation}

\subsubsection{Global interpolation of $\boldsymbol{t_0/a^2}$}
Having determined $t_0^*/a^2$ from individual fits
to data obtained at the different $\beta$-values and having
obtained an interpolating formula, we may also attempt a global
fit to all the available data, utilizing this parametrization.
Substituting $t_0^*/a^2$ for $t_{0,\text{ch}}/a_0^2$
(using eqs.~\eqref{eq:azero} and~\eqref{eq:t0chi}) and
$\tilde{k}_1$ for $\tilde{k}$ (using eq.~\eqref{eq:k1tilde}),
from eq.~\eqref{eq:globalt0fit1} we obtain
\begin{align}
  \frac{t_0}{a^2}& = \frac{t_0^*}{a^2}(g^2)\left[1 + \tilde{k}_1 A - 2 b_a \left( a \overline{m} - am^*\right)\right]  + \bar{c}A
  + \delta c\,8t_0\delta M^2\nonumber\\
   &=\left[f(g^2)+c_{t_0}+d_{t_0}f^{-1/2}(g^2)\right]
     \left(1 + \tilde{k}_1 A - 2 b_a a \overline{m} \right) + \bar{c}A
     +\delta c\, 8 t_0 \delta M^2,  \label{eq:globalt0fit2}
\end{align}
where
\begin{equation}
  A=8t_0\left(\overline{M}\vphantom{M}^2-M^{*2}\right).
  \label{eq:globalt0fit3}
\end{equation}
In the second step above we used eqs.~\eqref{eq:t0par1} and~\eqref{eq:t0par2}
to parameterize $t_0^*/a^2(g^2)$. This adds the parameters
$b_{t_0}$, $c_{t_0}$ and $d_{t_0}$, such that the total number
of parameters for this combined fit across different values
of $g^2$ is 6. Note that
when inserting this parametrization the term that is proportional to
$2b_a a m^*$ cancels from the above equation but it resurfaces within
the relation eq.~\eqref{eq:t0par1} between $t_0^*/a^2$ and $f(g^2)$.

\begin{figure}[htp]
  \centering
  \resizebox{0.49\textwidth}{!}{\includegraphics[width=\textwidth]{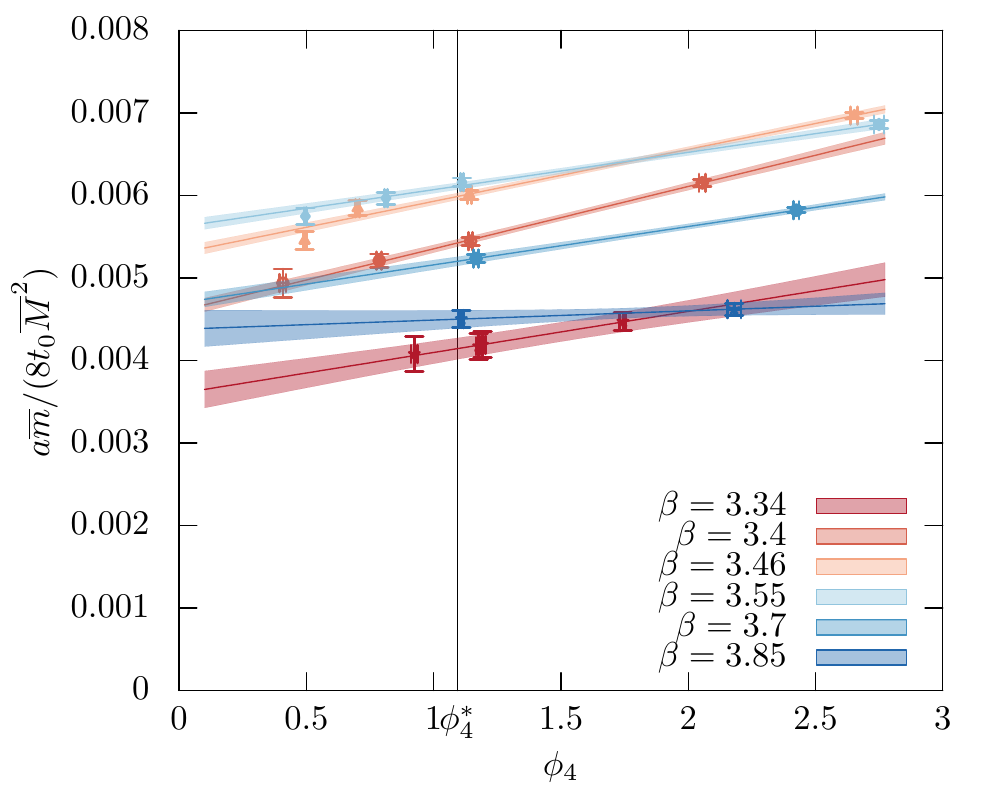}}
  \resizebox{0.49\textwidth}{!}{\includegraphics[width=\textwidth]{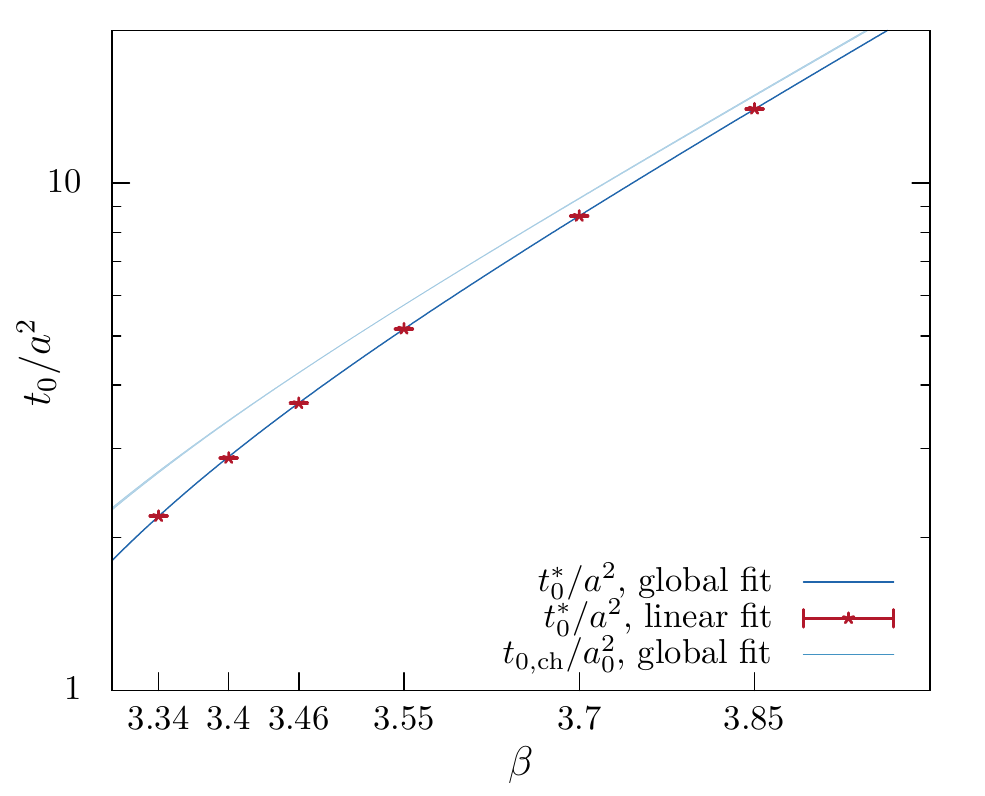}}
  \caption{\label{fig:ambarratio}Left: the ratio of the average lattice quark
    mass in lattice units over the square of the average pseudoscalar
    mass, including all three quark mass trajectories. This ratio varies
    by only $20\%$. This makes it impractical to
    discriminate between terms proportional to $a\overline{m}$ and terms
    $\propto 8\,t_0\overline{M}^2$.
    Right: global fit results (cf.\
    eq.~\eqref{eq:globalt0fit2}) for $t_0^*/a^2$ and $t_{0,\text{ch}}/a_0^2$,
    together with results for $t_0^*/a^2$ taken
    from separate linear fits (cf.\
    eq.~\eqref{eq:localt0fit}) to the $m_s = m_{\ell}$ data only.
  }
\end{figure}

As discussed above, the effect of $b_a$ cannot be isolated
within our range of lattice spacings (see figure~\ref{fig:t0starfit})
as $a\overline{m}$ is approximately proportional to
$t_0\overline{M}\vphantom{M}^2$.
This is shown in the left panel of figure~\ref{fig:ambarratio},
where we plot the ratio as a function of $\phi_4$.
The $\Tr M=\text{const}$ ensembles can be found in the vicinity of the 
vertical $\phi_4=\phi_4^*$ line. There is no detectable dependence of
this ratio on $\delta M$. The slopes with respect to
$\overline{M}\vphantom{M}^2\propto
\phi_4$ decrease with the lattice spacing, indicating that
the dominant violations of the GMOR
relation $\overline{m}/\overline{M}\vphantom{M}^2=\text{const}$ are due to
lattice artefacts. However, as already discussed
above, the ratio $a\overline{m}/[8t_0\overline{M}\vphantom{M}^2]$
at $\phi_4^*$ itself does not decrease linearly with $a$. Instead,
between $\beta=3.34$ and $\beta=3.4$ it first increases and only
from $\beta=3.46$ onwards it decreases as the parameter $r_m$
slowly approaches unity.

Since the $a\overline{m}$ dependence is hard to distinguish from
the $\overline{M}\vphantom{M}^2$ dependence, within the global fit we fix
$b_a$ to its one-loop value
eq.~\eqref{eq:bapert}, $b_a=b_a^{(1)}\approx 0.3158$.
The relation~\eqref{eq:k1tilde} between $\tilde{k}_1$ and $\tilde{k}$
will enable a cross-check with the previous fit result
eq.~\eqref{eq:tildekk} that was obtained
by extrapolating the individual slopes $\tilde{k}(a)$ of
eq.~\eqref{eq:localt0fit} to the continuum limit.
In addition to this central fit, we carry out a second fit,
setting $b_a = 0$ and interpret the difference
between the resulting parameters as a systematic uncertainty.
This second fit also allows for a comparison not only
with eq.~\eqref{eq:tildekk} but also with the earlier
results shown in eq.~\eqref{eq:t0par3}.
The difference of the $b_a$-values used in eq.~\eqref{eq:t0par1}
only slightly affects the parameters $b_{t_0}$, $c_{t_0}$ and $d_{t_0}$
and the impact on the $t_0^*/a^2$-values is even smaller.

\begin{figure}[htp]
  \centering
  \resizebox{0.49\textwidth}{!}{\includegraphics[width=\textwidth]{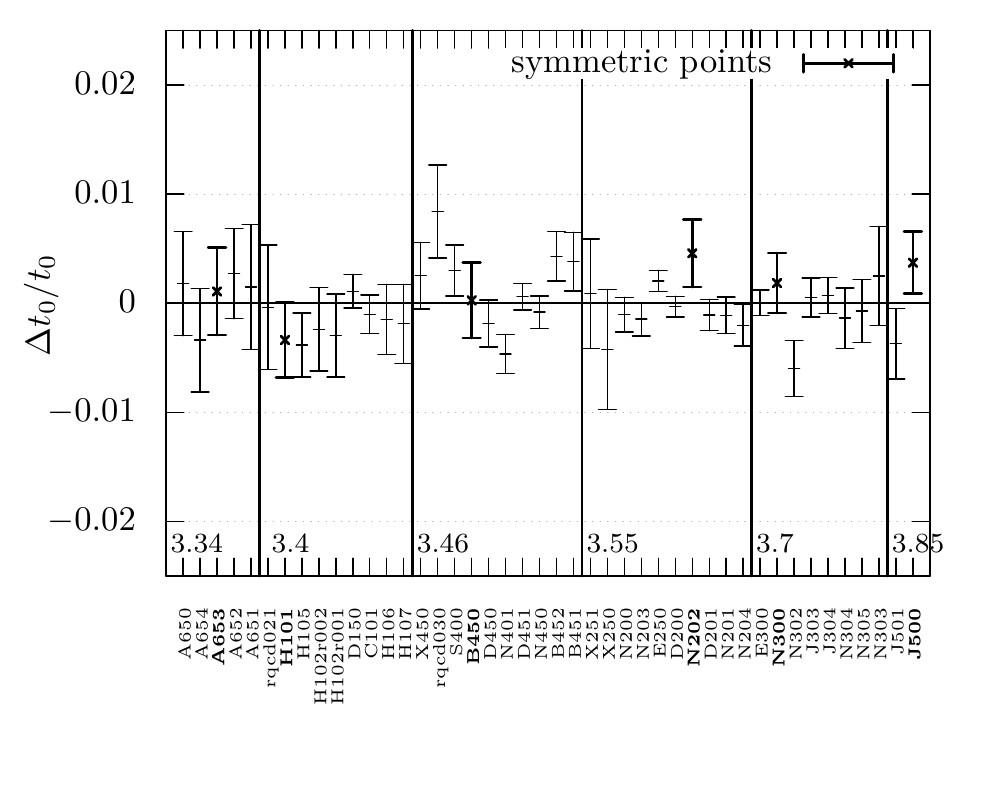}}
  \resizebox{0.49\textwidth}{!}{\includegraphics[width=\textwidth]{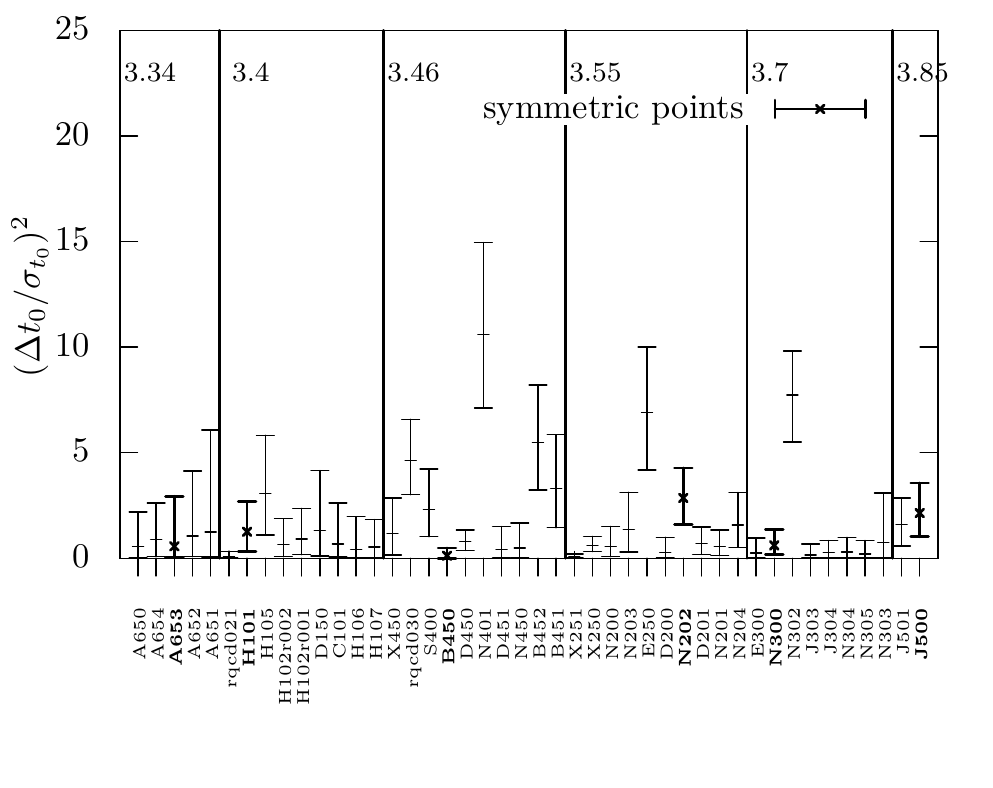}}
  \caption{Deviations $\Delta t_0 / a^2 =  t_0/a^2(g^2, \overline{M}^2,
    \delta M^2)-t_0/a^2$ between the values postdicted by the fit
    eq.~\eqref{eq:globalt0fit2} and the data
    are typically below one per cent (left) or of the order of the statistical
    error (right), with only a few exceptions. The ensembles are sorted from
    left to right in terms of decreasing lattice spacing and increasing
    values of the average $\overline{M}^2$. Boldface ensemble names correspond
    to the ``symmetric'' points, i.e.\ those closest to the position
    of $\phi_4^*$.\label{fig:globalt0fitquality}
    }
\end{figure}

Since not only $t_0/a^2$ but also the variables
$t_0 \overline{M}\vphantom{M}^2$ and $t_0\delta M^2$
carry errors, in these fits (as well as in the previous fit to
eq.~\eqref{eq:localt0fit}) we use the generalized least
squares fit method described in appendix~\ref{sec:leastsquares},
where we also take into account the correlation among the arguments.
However, we neglect correlations between these variables
and the $t_0/a^2$-values on the left hand side of the equation.
This is justified by the fact that the relative errors
of $t_0/a^2$ are much smaller than those of the
pseudoscalar meson masses.

We demonstrate in figure~\ref{fig:globalt0fitquality} that
the data are well described by the fit with $b_a=b_a^{(1)}$ and
remark that the picture looks very similar for $b_a=0$. The fit
curve itself is shown for the $m_s=m_{\ell}$ points in
figure~\ref{fig:t0starfit} (left). The respective fit parameters
read
\begin{align}
  \label{eq:globalt0fit2results}
  \tilde{k}_1(b_a= b_a^{(1)})    & = -0.0466(62), \quad & \tilde{k}_1(b_a=0) & =\tilde{k}=-0.0506(63),\nonumber\\
  \overline{c}(b_a= b_a^{(1)}) & = -0.560(27), \quad &\overline{c}(b_a=0) & = -0.560(27),\nonumber \\
  \delta c (b_a= b_a^{(1)})    & = 0.0213(28), \quad & \delta c(b_a=0) & = 0.0210(28),\nonumber\\
  b_{t_0}(b_a= b_a^{(1)})       & = 0.9336(26), \quad & b_{t_0}(b_a=0) & = 0.9340(26),\nonumber\\
  c_{t_0}(b_a= b_a^{(1)})       & = 0.286(63), \quad &c_{t_0}(b_a=0) & = 0.254(63),\nonumber\\
  d_{t_0}(b_a= b_a^{(1)})       &= -1.567(84), \quad & d_{t_0}(b_a=0) & = -1.515(84),\nonumber\\
  \chi^2/N_{\text{DF}}(b_a= b_a^{(1)}) & = 59.4/38, \quad & \chi^2/N_{\text{DF}}(b_a=0) & = 58.7/38,
\end{align}
where the errors have been obtained from the bootstrap
distributions of the parameters and scaled with $\sqrt{\chi^2/N_{\text{DF}}}$.
In the right panel of figure~\ref{fig:ambarratio} we compare the global fit
to the $t_0^*/a^2$-values obtained from the linear (local)
interpolation results above. In addition, we show the chiral limit of
this ratio.

The differences between the central values of the two columns of
eq.~\eqref{eq:globalt0fit2results} constitute
the systematic errors from varying $b_a$ from zero to its one-loop
value. As expected, the slope parameter $\tilde{k}_1$ is most affected
by this change. For $b_a=0$, one has $\tilde{k}_1=\tilde{k}$ and the
above result
still agrees within errors with our first estimate~\eqref{eq:tildekk}
of $\tilde{k}$. We remark that
the difference between the above two values is consistent
with eq.~\eqref{eq:k1tilde}: combining the $b_a=0$ result
with the typical value $am^*=0.004$ (see the left panel of
figure~\ref{fig:ambarratio} and table~\ref{tab:kappa}), we
obtain
$\tilde{k}_1\approx\tilde{k} + 3b_a am^*/\phi_4^*\approx -0.0472(63)$,
which indeed is very close to the result of the $b_a=b_a^{(1)}$
fit $\tilde{k}_1 = -0.0466(62)$.
Moreover, the parameters $b_{t_0}$, $c_{t_0}$ and $d_{t_0}$ agree
within errors with the determination~\eqref{eq:t0par3},
based on the local fit results.

We used the central value $\Lambda_{\MS} \sqrt{8t_0^*} = 0.712$
from the determination of ref.~\cite{Bruno:2017gxd} [eq.~(3.31)]
as an input. Instead, we could have included the normalization as
a free fit parameter. Carrying out such a fit, we find
$\Lambda_{\MS} \sqrt{8t_0^*} = 0.68(17)$,
in agreement with the more precise
result that was obtained employing the step scaling function within
the Schrödinger functional framework~\cite{Bruno:2017gxd}.

\subsection{Summary of the main results for $\boldsymbol{t_0}$, $\boldsymbol{a}$ and the low energy constant $\boldsymbol{k_1}$}
\label{sec:t0sum}
We summarize the $t_0^*/a^2$ results from the global fit in
the second line of table~\ref{tab:t0star}, where the first error
is statistical and the second one reflects the impact of
the uncertainty of $b_a$. Similarly, in the fourth line of the
table we list $t_{0,{\text{ch}}}/a_0^2$, where naturally the
statistical uncertainty is larger while the uncertainty of the
$b_a$-value has less of an impact.
In general, the $t_0^*/a^2$ results from the global fit
agree well with those obtained at the individual $\beta$-values.
Comparing with the results of refs.~\cite{Bruno:2016plf,Bruno:2017lta},
we only find deviations of about 2.5 and 1.5 standard deviations,
respectively, at $\beta=3.4$ and $\beta=3.46$.
Also in our case there is some difference between the linear (local)
and the global fit results at $\beta=3.4$, which is mainly due to
the $t_0/a^2$-value determined on ensemble H101. We regard the
results from the global interpolation where statistical fluctuations
average out to some extent as more robust and we will use these values.

The $t_0^*/a^2$-values are well described by the interpolating formula
\begin{align}
  \frac{t_0^*}{a^2}(g^2) & = f_{\text{eff}}(g^2) + 0.285 - 1.566 f_{\text{eff}}^{-1/2}(g^2),\quad\text{where}\nonumber\\
 f_{\text{eff}}(g^2)  & = \exp\left(17.54596\,g^{-2} - 7.507 + 0.790123\,\ln(g^2)-0.9334 g^2\right).\label{eq:t0interpol}
\end{align}
The relative errors are below 0.3\% over the entire fitted range $3.34 \leq \beta
\leq 3.85$.
This more convenient parametrization was obtained by
refitting the result, setting $b_{t_0}=0$ but keeping the $g^{-2}$
and $\ln(g^2)$ coefficients fixed. Within the errors of the $\Lambda$ parameter
of ref.~\cite{Bruno:2017gxd}, this interpolation converges towards
the two-loop running of the scale at small values of $g^2$, making
this formula particularly useful for predictions regarding
future runs at smaller lattice spacings. For the values at already
existing simulation points we
refer the reader to the second row of table~\ref{tab:t0star}.

The slope parameter~\cite{Bar:2013ora} defined in eq.~\eqref{eq:t0chi}
has the numerical value
\begin{equation}
  \label{eq:k1result}
  \tilde{k}_1=-0.0466(62),\quad
  k_1=\tilde{k}_1\cdot\frac{8t_{0,\text{ch}}{(4\pi F_0)}^2}{3}\approx -0.055(8),
\end{equation}
where we used the globally fitted $\tilde{k}_1$ with the effect of
$b_a$ and its uncertainty included in the central value and the error.
For the last conversion, we used
$\sqrt{8t_{0,\mathrm{ch}}}F_0=0.1502^{(56)}_{(29)}$~\cite{Bali:2022qja}.

Plugging the result for $\tilde{k}_1$ above as well as $\phi^*=1.11$
into eq.~\eqref{eq:t0chi} gives the
continuum limit relations
\begin{align}
  t^*_0&=\left(1+\frac{2}{3}\tilde{k}_1\phi_4^*\right)t_{0,\text{ch}}
  = 0.9655(46)t_{0,\text{ch}},
  \label{eq:t0ratio}\\
  \label{eq:t0ratio2}
  t^*_0&=\left[1+\frac{2}{3}\tilde{k}_1\left(\phi_4^*-\phi_{4,\text{ph}}\right)\right]t_{0,\text{ph}}
  = 0.99947(7)t_{0,\text{ph}}.
\end{align}
In the second equation we used the result
of this work $\phi_{4,\text{ph}}=1.093^{(10)}_{(14)}$, which is due to
$\sqrt{8t_{0,\text{ph}}}=0.4098^{(20)}_{(25)}\,\textmd{fm}<0.413\,\textmd{fm}$
(see eq.~\eqref{eq:t0result}).
This then implies
\begin{equation}
  \label{eq:t0starp}
  \sqrt{8t_0^*}=0.99974(4)\sqrt{8t_{0,\text{ph}}}
  =0.4097^{(20)}_{(25)}\,\textmd{fm}.
\end{equation}
Note that at a non-vanishing lattice spacing, the above relations between
the $t_0$-values at different points in the quark mass plane
depend on the action used and, in the case of $t_0^*/t_{0,\text{ph}}$,
also on the exact definition of the physical point. However, in
the continuum limit (up to the treatment of isospin breaking effects),
the results should be universal. Within this study,
we use $a^2/(8t_0^*)$ only to relate different lattice spacings within our
continuum limit extrapolation. In the end the scale is set by
$t_{0,\text{ph}}$, as obtained from the mass of the $\Xi$ baryon at
the physical point, in the continuum limit.
The lattice spacings of table~\ref{tab:t0star}
are computed using $a^2/(8t_0^*)$ with the physical value of $t_0^*$ set
by the continuum limit ratio $t_{0,\text{ph}}/t_0^*$.

In principle, along the line with $\Tr M=\text{const}$ there could
be a dependence of $\phi_4$ on the mass difference $\delta M^2$,
which we have neglected above.
This would then result in a correction to the above
relation between $\sqrt{8t_0^*}$ and $\sqrt{8t_{0,\text{ph}}}$.
The relative statistical uncertainty of the latter quantity
is of size 0.5\%. A correction of a comparable size to eqs.~\eqref{eq:t0ratio2}
and~\eqref{eq:t0starp} would require $\phi_4$ to vary by more than 30\% along
this line between $\delta M=0$ and the physical point, due to the smallness
of the parameter $\tilde{k}_1$. As we will see in the
following subsection, we are unable to detect any
corrections to $\phi_4(\delta M)$ in the continuum limit
that would exceed our statistical accuracy of about 1\%. Therefore,
the above value for $\sqrt{8t_0^*}$ and its error
remain unaffected.

\subsection{The symmetric point parameters}
\label{sec:t02}
Above, we have defined the parameter $\phi_4=12t_0^2\overline{M}\vphantom{M}^2$.
Several values of $\phi_4$ are of relevance:
\begin{itemize}
\item $\phi_{4,\text{ph}}$, the value at the physical point in the continuum limit,
\item $\phi_4^*=1.11$, which --- together with $M_K=M_{\pi}$ --- defines
  the reference point for the determination of $t_0^*/a^2$,
\item $\phi_{4,\text{symm}}(a)$, the starting point, where for each
  lattice spacing $a$ our actual $\Tr M=\text{const}$ trajectory branches off
  the symmetric $m_s=m_{\ell}$ line,
\item $\phi_{4,\text{opt}}(a)$, the branch point that should be chosen such
  that the $\Tr M=\text{const}$ trajectory touches the physical point
  for the lattice spacing $a$ and
\item $\phi_{4,\text{opt}}$, the corresponding starting point in the continuum limit.
\end{itemize}
Note that in general there is some degree of mistuning so that
$\phi_{4,\text{symm}}(a)\neq \phi_{4,\text{opt}}(a)$. 
The optimal values $\phi_{4,\text{opt}}(a)$ will depend somewhat on
how the physical point is defined.
Here we match the kaon and pion masses in units of $\sqrt{8t_0}$
to their experimental values, where $t_{0,\text{ph}}$ is obtained
from $m_{\Xi}$ in the continuum limit.

\begin{table}[h!tp]
  \caption{\label{tab:kappa}Results for the critical hopping parameter
    $\kappa_{\text{crit}}$ and for the hopping parameter at the $N_f=3$ symmetric
    point, where $\phi_4 = \phi_4^* = 1.11$, $\kappa^*$, together with the
    bare quark mass at this point, 
    $am^* = (\kappa^{*-1} - \kappa_{\text{crit}}^{-1})/2$.}
  {
\renewcommand{\arraystretch}{1.2}
\begin{center}
\begin{tabular}{llll}
\toprule
$\beta$ & $\kappa_{\text{crit}}$   & $\kappa^*$     & $am^*$\\\midrule
3.34    & 0.1366938(45)& 0.1365791(23)& 0.00307(10)\\
3.4     & 0.1369153 (9)& 0.1367647(11)& 0.00402 (3)\\
3.46    & 0.1370613(10)& 0.1368948(13)& 0.00444 (3)\\
3.55    & 0.1371715(10)& 0.1370013(10)& 0.00453 (3)\\
3.7     & 0.1371530 (9)& 0.1370081(13)& 0.00385 (4)\\
3.85    & 0.1369767(26)& 0.1368518(39)& 0.00333 (8)\\\bottomrule
\end{tabular}
\end{center}
}
\end{table}

Other parameters of interest are the lattice quark mass
at this point $am^*$ and the corresponding hopping parameter $\kappa^*$.
To determine $\kappa^*$ and $am^*$, at each $\beta$-value we carry out a
simple phenomenological fit to the $m_s = m_\ell$ ensembles
using the previously determined $\kappa_{\text{crit}}$-values as an input:
\begin{equation}
  am = p_1 \phi_4 + p_2 \phi_4^2.
  \label{eq:aminter}
\end{equation}
$am^*$ is then the value of this interpolation at $\phi_4 = \phi_4^*$.
The fit is depicted in the left panel of figure~\ref{fig:ambarratio},
where we plot $(3/2)am/\phi_4$ versus $\phi_4$ for the symmetric
ensembles. The results are collected in table~\ref{tab:kappa}, where we also
display the critical hopping parameter values.

Starting at the symmetric
point ($\delta M^2=0$) and reducing the pion mass, keeping the sum of
lattice quark masses constant,
results in a decreasing $\phi_4(\delta M_{\text{ph}},a)$:
in order to simulate on a trajectory
that goes through the physical point, at a given lattice spacing $a$,
one has to start from
values $\phi_{4,\text{opt}}(a)$ somewhat larger than $\phi_{4,\text{ph}}$.
Along lines of constant $\Tr M$, in the continuum limit $\phi_4$
cannot depend linearly on $\delta M^2$~\cite{Bar:2013ora},
but this is not so regarding
lattice artefacts. Therefore, for the data taken along such trajectories,
we make the ansatz:
\begin{equation}
  \phi_4(\delta M,a)=\phi_4(0,a)+\delta c_{\phi}\frac{a^2}{t_0^*}
  8t_0\delta M^2+\left[c_{\phi}+d_{\phi}\left(\phi_4(0,a)-\phi_{4,\text{ph}}\right)\right]\left(8t_0\delta M^2\right)^2.
  \label{eq:phi4fit}
\end{equation}
We added the term proportional to $d_{\phi}$ to compensate
for the effect that the starting point
$\phi_4(0,a)=\phi_{4,\text{symm}}(a)$ is not kept constant across the
six lattice spacings, see the fourth column of table~\ref{tab:kappas}.

\begin{figure}[thp]
  \centering
  \resizebox{0.49\textwidth}{!}{\includegraphics[width=\textwidth]{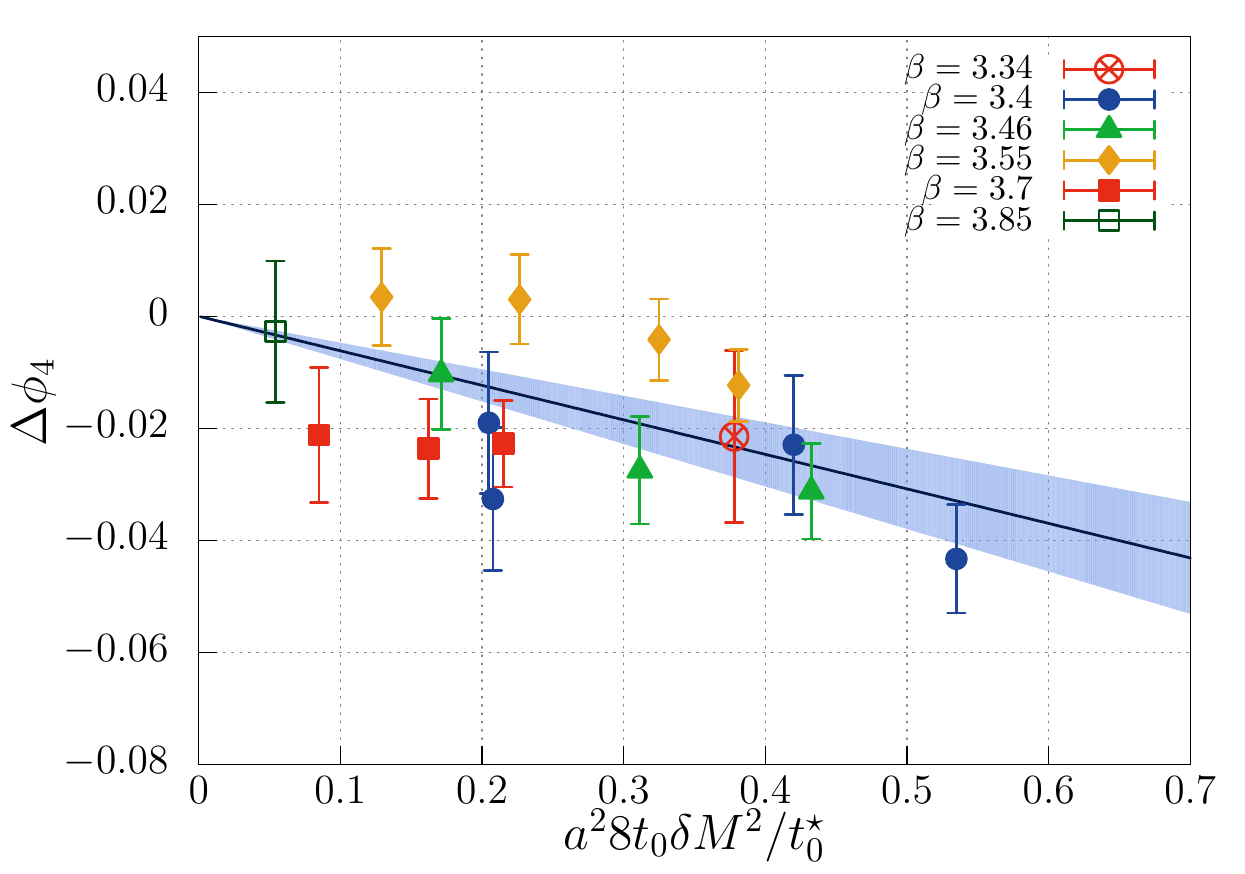}}
  \resizebox{0.49\textwidth}{!}{\includegraphics[width=\textwidth]{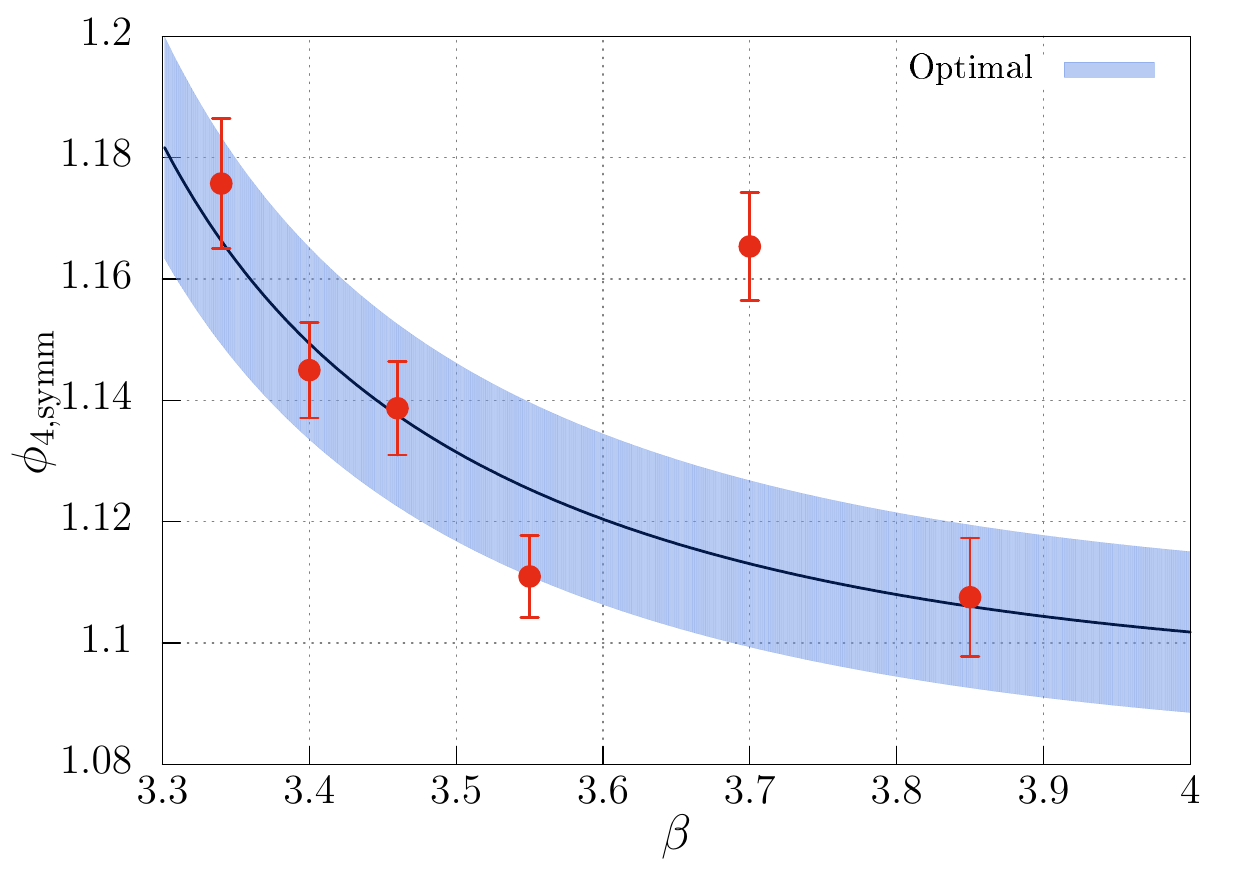}}
  \caption{Left: the differences defined in eq.~\eqref{eq:dephi4}
    as a function of $(a^2/t_0^*) 8t_0\delta M^2$.
    The curve and error band correspond to a
    two-parameter fit of the $\Tr M=\text{const}$ data with $m_s\neq m_{\ell}$
    to eq.~\eqref{eq:phi4fit}, setting $c_{\phi}=0$.
    Right: the optimal starting
    values of $\phi_4$ at the symmetric point that should be
    used in order to obtain $\phi_{4,\text{ph}}=1.093^{(11)}_{(13)}$
    at the physical point (blue error band), along with the actually
    simulated values
    of $\phi_{4,\text{symm}}(a)$.\label{fig:phi4symfit}
    }
\end{figure}

\begin{table}[tp]
  \caption{\label{tab:kappas} The values of $\phi_{4,\text{symm}}(a)$ and
    $\phi_{4,\text{opt}}(a)$ and the corresponding hopping parameter-values.}
  {
\renewcommand{\arraystretch}{1.2}
\begin{center}
\begin{tabular}{lllll}
\toprule
$\beta$ & $\kappa_{\text{symm}}$&  $\kappa_{\text{opt}}$&$\phi_{4,\text{symm}}(a)$& $\phi_{4,\text{opt}}(a)$\\
\midrule
3.34    & 0.1365715 & 0.1365725(30)& 1.1757(109)& 1.166(17)\\
3.4     & 0.13675962& 0.1367585(28)& 1.1450 (78)& 1.149(16)\\
3.46    & 0.13689   & 0.1368901(29)& 1.1387 (76)& 1.138(15)\\
3.55    & 0.137     & 0.1369988(26)& 1.1110 (66)& 1.125(14)\\
3.7     & 0.137     & 0.1370077(24)& 1.1653 (88)& 1.113(14)\\
3.85    & 0.136852  & 0.1368523(43)& 1.1075 (96)& 1.106(13)\\\bottomrule
\end{tabular}
\end{center}
}
\end{table}

In this article we find the central value of
$\phi_{4,\text{ph}}=\phi_4(\delta M_{\text{ph}},0)=
1.093^{(11)}_{(13)}$ to be
smaller than $\phi_4^*=1.110$ by almost 2\%, corresponding to
about 1.6 standard deviations,
due to $\sqrt{8t_{0,\text{ph}}}<0.413\,\textmd{fm}$. Furthermore,
we obtain $8t_0 \delta M_{\text{ph}}^2 =1.950^{(19)}_{(24)}$.
Setting $\phi_4(\delta M_{\text{ph}},a)=\phi_{4,\text{ph}}$ in
eq.~\eqref{eq:phi4fit} then gives $\phi_4(0,a)=\phi_{4,\text{opt}}(a)$:
\begin{equation}
  \label{eq:phi4opt}
  \phi_{4,\text{opt}}(a)=\phi_{4,\text{ph}}-
  \frac{\delta c_{\phi}\frac{a^2}{t_0^*}
    8t_0\delta M_{\text{ph}}^2+c_{\phi}\left(8t_0\delta M_{\text{ph}}^2\right)^2}
  {1+d_{\phi}\left(8t_0\delta M_{\text{ph}}^2\right)^2}.
\end{equation}

We carry out correlated one-, two- and three-parameter fits to
eq.~\eqref{eq:phi4fit}, setting $c_{\phi}=d_{\phi}=0$,
$c_{\phi}=0$, $d_{\phi}=0$ and leaving all parameters free, respectively.
In total we have 15 large volume simulation points along the
$\Tr M=\text{const}$ lines with $m_{\ell}\neq m_s$ (see the central
panel of figure~\ref{fig:ensembles}). We omit D150 from
this counting and the fits since $LM_{\pi}<4$ in this case.
Taking into account the fact that we have two data
points that correspond to the H102 parameters, this
then gives 15, 14, 14 and 13 degrees of freedom,
respectively, for the four fits. Regarding the one-parameter fit,
we find $\delta c_{\phi}=-0.0839(76)$ with
$\chi^2/N_{\text{DF}}=1.30$. Allowing for $d_{\phi}\neq 0$, we find
\begin{equation}
  \label{eq:phi4res}
  \delta c_{\phi}=-0.062(14),\quad d_{\phi}=-0.062(33),\quad
  \chi^2/N_{\text{DF}}=1.21.
\end{equation}
We define
\begin{equation}
  \Delta \phi_4(\delta M,a)= \phi_4(\delta M,a)-\phi_4(0,a)-d_{\phi}\left(\phi_4(0,a)-\phi_{4,\text{ph}}\right)\left(8t_0\delta M^2\right)^2
  \label{eq:dephi4}
\end{equation}
and plot the resulting fit for these shifted differences as a function
of $(a^2/t_0^*)t_0\delta M^2$ in the left panel of figure~\ref{fig:phi4symfit}.
Additionally including the parameter $c_{\phi}$ slightly decreases the fit
quality
($\chi^2/N_{\text{DF}}=1.22$) and gives $c_{\phi}=0.0029(27)$ while
$\delta c_{\phi}=-0.070(14)$ and $d_{\phi}=-0.114(54)$ remain
unchanged within errors, relative to eq.~\eqref{eq:phi4res}.
Setting $d_{\phi}=0$ but including $c_{\phi}$ gives $c_{\phi}=-0.0017(16)$
with $\chi^2/N_{\text{DF}}=1.33$. Also in this case $\delta c_{\phi}=
-0.069(17)$ agrees with the other fit results.

We conclude that we are unable to
discriminate $c_{\phi}$ from zero and we actually obtain the
best fit quality when removing this parameter. Therefore,
within errors $\phi_{4,\text{opt}}=\phi_{4,\text{ph}}$
in the continuum limit. Of particular interest is also $\phi_{4,\text{opt}}(a)$,
which can be obtained from eq.~\eqref{eq:phi4opt}. In the right panel of
figure~\ref{fig:phi4symfit} we compare this to the actual simulation points
$\phi_{4,\text{symm}}$. The error of the prediction, also included
in table~\ref{tab:kappas}, is dominated by the uncertainty of
the scale $t_{0,\text{ph}}$. It turns out that, with the exception
of $\beta=3.7$, the simulated $\Tr M=\text{const}$ trajectories are within
the target range. The continuum limit
agreement of $\phi_{4,\text{opt}}$ at the symmetric line with
$\phi_{4,\text{ph}}$ within the present errors is a universal feature
of $N_f=2+1$ QCD. However, we remark again that
$\phi_{4,\text{opt}}(a)$ as depicted in the figure will depend on the
input quantities used to define the physical point at non-vanishing values
of the lattice spacing.

Finally, we predict
the hopping parameter values that correspond to the optimal
$m_s=m_{\ell}$ starting point,
\begin{equation}
  \frac{1}{\kappa_{\text{opt}}(a)}= 2 \left[ p_1 \phi_{4,\text{opt}}(a) + p_2 \phi_{4,\text{opt}}^2(a)\right] + \frac{1}{\kappa_{\text{crit}}},
\end{equation}
where $p_1$ and $p_2$ are defined in eq.~\eqref{eq:aminter}.
These $\kappa$-values are also included in table~\ref{tab:kappas}.

\section{Physical point extrapolation strategy}
\label{sec:ex}
We detail our strategy to extrapolate the hadron masses to
the physical point in the quark mass plane, and to the continuum
and infinite volume limits. In particular,
we wish to maintain full order $a$ improvement within
the continuum limit extrapolation.
Since the scale should be set by
comparing to an experimental measurement at the physical
quark mass point, the exact position of which in turn depends
on the scale setting, some care needs to be taken.
First we explain how we retain order $a$ improvement in the
continuum limit extrapolation. Then, in section~\ref{sec:chiral},
we give the general framework of our combined chiral and continuum
limit extrapolations, before
discussing how we account for eventual
finite volume effects in section~\ref{sec:finite}.

\subsection{The continuum limit}
\label{sec:extrap}
Naive Wilson fermions have order $a$ cut-off effects. Here we
implement a complete, non-perturbative order $a$ improvement programme.
This consists of improving the action and --- for the
determination of the AWI quark masses --- also
the axial and pseudoscalar currents.
To improve the action, the Sheikholeslami-Wohlert
(clover) term~\cite{Sheikholeslami:1985ij} is added,
with its coefficient $c_{\text{sw}}$
determined in ref.~\cite{Bulava:2013cta} for the tree-level
Symanzik improved gluon action~\cite{Weisz:1982zw} that we use.
Most of our gauge
ensembles have open boundary conditions in time~\cite{Luscher:2011kk} where
the boundary terms are not order $a$ improved. However, measurements
are only taken in the bulk, far away from these boundaries, see
appendix~\ref{sec:positions}, exponentially suppressing such residual
order $a$ effects. Also the quark masses and the coupling constant
appearing in the action need to be improved. The former does not
affect the position of the
physical point in terms of the kaon and pion masses but was
relevant for the vector Ward identity quark mass determination of
section~\ref{sec:kappaImpr} (with the combinations
of improvement coefficients $\mathcal{A}$,
$\mathcal{B}_0$, $\mathcal{C}_0$ and $\mathcal{D}_0$).

As mentioned above and discussed in section~\ref{sec:coupordera}, the bare
coupling $g^2=6/\beta$ undergoes
order $a$ improvement~\cite{Jansen:1995ck,Luscher:1996sc}:
$\tilde{g}^2=g^2(1+b_g a\overline{m})$, where
$\overline{m}$ denotes the average sea quark mass eq.~\eqref{eq:maver}.
Instead of simulating at fixed values of $\tilde{g}^2$, thereby
keeping the lattice spacing $a$ constant, we simulate at specified
values of $g^2$. This will imply order $a$ cut-off effects
on lattice hadron mass values $Ma$,
unless the average quark mass is kept constant. Along two of our quark mass
plane trajectories, namely $m_s=m_{\ell}$ and $\widehat{m}_s\approx
\widehat{m}_{s,{\text{ph}}}$, we vary $\overline{m}$ and therefore care
needs to be taken.

Assuming that the combination $\phi_4$ of eq.~\eqref{eq:phi4}
remains constant along the $2/\kappa_{\ell}+1/\kappa_s =\text{const}$ line
(i.e.\ the line of a constant average lattice quark mass
$\overline{m}=m_{\text{symm}}$) and setting
$\sqrt{8t_0^*}\approx 0.413\,\textmd{fm}$
gives the value $\phi_4^*=1.11$. Previous lattice
studies gave results consistent with this value of $t_0$, either at
the $\phi_4^*$ position~\cite{Bruno:2016plf} or at the
physical point~\cite{Borsanyi:2012zs}. In section~\ref{sec:t02}
we confirm that $\phi_4$ remains constant within our present
statistical errors as long as $\Tr M$ is kept constant, up to lattice
artefacts.
However, we will find a somewhat lower value for $\sqrt{8t_0^*}$,
giving $\phi_{4,\text{ph}}=1.093^{(11)}_{(13)}$. Nevertheless, we will keep
the value $\phi_4=\phi_4^*=1.11$ in order to define the
``$*$'' reference point.
Given our good coverage of the quark mass plane,
whether this reference value exactly coincides with the value of $\phi_4$ at the
physical point is not relevant for our scale and mass
determinations.

In the end the scale needs to be set independently, i.e.\ a physical value
has to be assigned, e.g., to $t_{0,\text{ph}}$ or $t_0^*$ in the continuum limit,
that is consistent with our physical point definition.
In the absence of stable light mesons other than the $\pi$ and the
$K$, the remaining possibilities
of input quantities include baryon
masses as well as pion and kaon decay constants.\footnote{The $\phi$ meson
is reasonably narrow too, however, it has the same quantum numbers
as the lighter, less stable $\omega$ meson.
Including input from heavy hadrons, e.g., charmed baryons, while statistically
precise, would require to simultaneously fix the charm quark mass.}
In the former case statistical errors are larger, while
the accuracy of the latter observables is limited by the precision
of the determination of the renormalization factor $Z_A$,
the accuracy of the experimental input (converted into an isospin
symmetric world without soft photon effects) and using
phenomenological values for the
CKM matrix elements $V_{ud}$ and, in particular, $V_{us}$.
These values in turn depend on previous lattice QCD
determinations by other groups.

Within this subsection
we denote a hadron mass of the continuum theory as $M$.
Lattice simulations give dimensionless combinations $Ma(g^2,a\overline{m})=
Ma_0(g^2)(1+b_a a\overline{m})$, see eq.~\eqref{eq:azero}.
This $a\overline{m}$ correction term cancels from ratios of masses
determined at the same value of $\overline{m}$. These will
then approach the continuum limit without any linear dependence on $a$,
which holds for hadron masses obtained at different
positions along the $\overline{m}=m_{\text{symm}}=\text{const}$
quark mass plane trajectory. However, in general
ratios between hadron masses at different points in the
quark mass plane will not be order $a$ improved
since $M_1 a(g^2,a\overline{m}_1)/[M_2 a(g^2,a\overline{m}_2)]=
(M_1/M_2)[1+b_a a(\overline{m}_1-\overline{m}_2)]$.
The average quark mass varies along the
$m_s=m_{\ell}$ and $\widehat{m}_s\approx\text{const}$
trajectories. Order $a$ effects will, however, cancel if
dimensionless combinations $Ma\,\sqrt{8t_0}/a=M\sqrt{8t_0}$, where $t_0$ is
computed on the same ensemble as $M$, are taken.

Extrapolating $\sqrt{8t_0}M$ combinations
mixes the continuum limit quark mass dependence of $t_0$
with that of $M$. This is unproblematic since we have already
determined the continuum limit functional dependence of $t_0$
on the pseudoscalar masses
in section~\ref{sec:t0}, see eqs.~\eqref{eq:t0chi} and~\eqref{eq:k1result},
allowing us to disentangle the two effects. 
Also physical point extrapolated masses will remain unaffected.
We remark that the $\overline{m}=m_{\text{symm}}$
trajectory is a notable exception in our extrapolation
strategy since in this case the ratio of
lattice numbers $(t_0^*/a^2)/(t_{0,\text{ph}}/a^2)$ taken
at different points does not receive any order $a$ contributions,
at least if we assume that
this trajectory (that
starts out from somewhere near $\phi_4^*$) goes through the physical point.
Moreover, in this case $t_0$ only varies at
NNLO in ChPT~\cite{Bar:2013ora}.

\subsection{Parametrization of the quark mass, lattice spacing and volume dependence}
\label{sec:chiral}
It is convenient not only to define the average sea quark mass
eq.~\eqref{eq:maver} but also to introduce the
quark mass difference
\begin{equation}
\label{eq:deltam}
\delta m \coloneqq
m_s-m_{\ell}=3(\overline{m}-m_{\ell})=-\frac{3}{2}(\overline{m}-m_s).
\end{equation}
The leading order (LO) $\textmd{SU(3)}$ GMOR relations read
\begin{equation}
M_{\pi}^2=2B_0\widehat{m}_{\ell}+\mathcal{O}(\widehat{m}_{\ell}^2),\quad
M_K^2=B_0(\widehat{m}_s+\widehat{m}_{\ell})
+\mathcal{O}(\widehat{m}_s^2,\widehat{m}_{\ell}^2,\widehat{m}_s\widehat{m}_{\ell}),
\end{equation}
where $B_0=\Sigma_0/F_0^2$  and $\Sigma_0=-\langle\overline{u}u\rangle>0$
and $F_0$ are the (negative) chiral
condensate and the pion decay constant, respectively, in the limit of
$N_f=3$ massless quarks. We remind the reader that
the renormalized quark masses in a
continuum scheme are denoted as $\widehat{m}_{\ell}$ and $\widehat{m}_s$.
The GMOR relations also link the meson mass combinations of
eq.~\eqref{eq:mass0} to corresponding quark mass combinations:
\begin{align}
  \label{eq:mass12}
\overline{M}\vphantom{M}^{\,2}= \frac{2M_K^2+M_{\pi}^2}{3}\approx2B_0\widehat{\overline{m}},\quad
\delta M^2=2\left(M_K^2-M_{\pi}^2\right)\approx 2B_0\delta \widehat{m},
\end{align}
where $\delta m=m_s-m_{\ell}$, see eq.~\eqref{eq:deltam}.

In section~\ref{sec:extrap} we discussed how to avoid
order $a$ lattice artefacts. We implement this programme by employing
the fit strategy outlined below. We distinguish between
$t_0^*$, defined at the position $\phi_4=1.11$
along the $m_s=m_{\ell}$ flavour symmetric line
and $t_{0,\text{ph}}$, defined at the physical point in the quark mass
plane. The ratio of these two quantities is unity
to NLO in ChPT~\cite{Bar:2013ora}.
Note that the combination $\sqrt{8t_0^*}/a$ can be employed
to translate between lattice spacings obtained at different values of
the inverse coupling constant $\beta$, without the need to
know the corresponding $\sqrt{8t_{0,\text{ph}}}/a$-values.

The fit strategy consists of first defining parametrizations
(here for the example of the $\Xi$ mass):
\begin{equation}
  \label{eq:fit}
\sqrt{8t_0} m_{\Xi}=f_{\Xi}(\sqrt{8t_0}M_{\pi},\sqrt{8t_0}M_K,L/\sqrt{8t_0},a^2/(8t_0^*)).
\end{equation}
In principle, we could have chosen $a^2/(8t_{0,\text{ph}})$ as the last argument
on the right hand side,
however, this is not known prior to the fit while we have
already determined $a^2/(8t_0^*)$ in section~\ref{sec:t0}, see the
second line of table~\ref{tab:t0star}.
We also remark that in the last argument we replaced $a_0^2$ by
$a^2=a_0^2(1+2b_a am^*)$,
the difference being of order $a^3$.

In view of eq.~\eqref{eq:fit} it is convenient to define the
dimensionless quantities
\begin{align}
  \label{eq:defin}
  \mathbbm{m}_B=\sqrt{8t_0}\,m_B,\quad
  \overline{\mathbbm{M}}= \sqrt{8 t_0}\, \overline{M},\quad
  \delta\mathbbm{M}= \sqrt{8 t_0}\, \delta M,\quad
  \mathbbm{L}=\frac{L}{\sqrt{8t_0}},\quad
  \mathbbm{a}=\frac{a}{\sqrt{8t_0^*}},
\end{align}
where $B\in\{N,\Lambda,\Sigma,\Xi,\Delta,\Sigma^*,\Xi^*,\Omega\}$.
Other quantities
such as $\mathbbm{M}_{\pi}=\sqrt{8 t_0}M_{\pi}$ are rescaled
analogously. Note that we choose to rescale $a$ into units of
$\sqrt{8t_0^*}$ rather than $\sqrt{8t_0}$.

The above products of hadron masses and $\sqrt{8t_0}$, determined at
the same lattice coupling and quark mass values, have no linear lattice
spacing dependence. Therefore, the leading lattice spacing effects are
of orders
$a^2\Lambda^2$, $a^2\Lambda\overline{m}$,
$a^2\Lambda\delta m$, $a^2\overline{m}^2$, $a^2\overline{m}\delta m$
and $a^2\delta m^2$. We will also investigate
corrections $\propto a^3$.\footnote{In principle, the
$a^2$ effects are accompanied
by different powers $g^{2\Gamma_i}$ with the
anomalous dimensions $\Gamma_i$ determined by the
Symanzik counterterms~\cite{Husung:2019ytz}. However, for our action
it was found for the minimal dimension $\Gamma_{\min}$
that $\Gamma_{\min}\approx 0.247$ without and
$\Gamma_{\min}=-0.111$ with quark mass terms~\cite{Husung:2021mfl}.
In fact, the dominant
contribution is expected to be
$\propto g^{1.52}a^2$~\cite{Husung:2021mfl}.
Therefore, ignoring the anomalous dimensions should be
a conservative and safe assumption for our lattice action,
in particular, since the $\Gamma_i$ are positive and
$g^2$ varies only by a factor
of $1.15$, whereas the lattice spacing $a$ changes by
a factor of $2.5$.}
Omitting these latter terms, converting the quark
masses into pseudoscalar meson masses and truncating the dependence
of lattice artefacts at quadratic order in the pseudoscalar
masses,\footnote{This is justified by the fact that our parametrizations
  of the dependence on pseudoscalar masses in the continuum
  limit will not exceed $\mathcal{O}(M^4)$.
  Moreover, one would expect that $\overline{m}\lesssim m_s/3\ll\Lambda$ and
  $\delta m<m_s\ll\Lambda$.}
the fit function for the baryon octet can be factorized
as follows:
\begin{align}
  \label{eq:fit2}
  \mathbbm{m}_O(\mathbbm{M}_{\pi},\mathbbm{M}_K,\mathbbm{L},\mathbbm{a})
  =  \mathbbm{m}_O(\mathbbm{M}_{\pi},\mathbbm{M}_K,\mathbbm{L})\left[
    1+\mathbbm{a}^2\left(c_o+\bar{c}_o\,\overline{\mathbbm{M}}\vphantom{M}^2+
    \delta c_O\,\delta\mathbbm{M}^2\right)\right].
\end{align}
While $c_o$ and $\bar{c}_o$ are independent of the baryon in question,
the $\delta c_O$ will be different for different
baryons $O\in\{N,\Lambda,\Sigma,\Xi\}$.
The same applies to decuplet baryons with
the replacements $\mathbbm{m}_O\mapsto\mathbbm{m}_D$,
$c_o\mapsto c_d$, $\bar c_o\mapsto \bar{c}_d$,
$\delta c_O\mapsto\delta c_D$
and $D\in\{\Delta,\Sigma^*,\Xi^*,\Omega\}$.
We summarize possible continuum limit parametrizations of
$\mathbbm{m}_B(\mathbbm{M}_{\pi},\mathbbm{M}_K,\mathbbm{L})=
\mathbbm{m}_B(\mathbbm{M}_{\pi},\mathbbm{M}_K,\mathbbm{L},0)$ in
section~\ref{sec:physical} below.
We remark that since our lattice action breaks chiral symmetry
at any non-vanishing value of the lattice spacing,
there are no obvious $\textmd{SU(3)}$ constraints, relating the
$\delta c_B$ parameters for different baryons $B$.

Our physical point is defined as the position in the quark mass plane
where $M_{\pi,\text{ph}}=134.8(3)\,\textmd{MeV}$ and
$M_{K,\text{ph}}=494.2(3)\,\textmd{MeV}$. We
use $m_{\Xi,\text{ph}}=1316.9(3)\,\textmd{MeV}$ as the input to
set the scale in the continuum limit.
We refer to appendix~\ref{sec:isospin} for a discussion of these numbers
and specifically to table~\ref{tab:contmasses}. Using these values,
the scale $t_{0,\text{ph}}$ can be obtained via the relation
\begin{align}
  \label{eq:physical}
  \sqrt{8t_{0,\text{ph}}}=
  \frac{\mathbbm{m}_{\Xi}(\mathbbm{M}_{\pi}=0.10236(22)\mathbbm{m}_{\Xi},\mathbbm{M}_K=0.37528(24)\mathbbm{m}_{\Xi},0)}{1316.9(3)\,\textmd{MeV}}.
\end{align}
Subsequently, also at $a> 0$ we may define the physical point
as the position where
\begin{align}
  \label{eq:physpoint}
  \mathbbm{M}_K=\sqrt{8t_{0,\text{ph}}}\,494.2(3)\,\textmd{MeV},\quad
  \frac{\mathbbm{M}_{\pi}}{\mathbbm{M}_K}=
  \frac{M_{\pi,\text{ph}}}{M_{K,\text{ph}}}=0.2728(6).
\end{align}
Finally, if needed, the continuum limit ratio
$t_0^*/t_{0,\text{ph}}$ can be determined via
eqs.~\eqref{eq:k1result} and~\eqref{eq:t0ratio2}, see
eq.~\eqref{eq:t0starp}.

\subsection{Finite size effects}
\label{sec:finite}
Within section~\ref{sec:chiral} we have assumed infinite volume hadron masses.
Here we use ChPT as a guide to study the impact of potential finite size
effects. To be safely within the so-called $p$-regime of ChPT we not only
restrict the linear box size to $L\gg M_{\pi}^{-1}$ but we also require this to
be much larger than the inverse pseudo-critical
temperature. To this end, in the latest Flavour Lattice Averaging Group (FLAG)
Review~\cite{Aoki:2021kgd} an extent $L>2\,\textmd{fm}$ is advocated.
Here we set the somewhat stricter limit $L>2.3\,\textmd{fm}$. Any remaining
finite size effects are expected to be suppressed at least exponentially
with the mass gap, in proportion to $\exp(-LM_{\pi})$.

The pseudoscalar meson masses, which we are able
to determine very precisely, can, in principle, be affected by finite
size effects in a statistically significant way.
To limit the number of parameters and also to be consistent
with the order, i.e.\ $\mathcal{O}(p^3)$, that we use in the
chiral expansion of the baryon masses,
we only consider finite size effects to NLO (order $p^2$) in mesonic ChPT,
where no order $p^3$ corrections exist.
For $N_f=3$ mass-degenerate light quark flavours one obtains
in a finite volume~\cite{Gasser:1986vb,Gasser:1987zq}:
\begin{align}
\label{eq:mpi1}
M^2_{\pi}(L)&=M^2_{\pi}\left[1+\frac{1}{N_f}h(\lambda_{\pi},M_{\pi}^2)+\cdots\right],
\end{align}
where $\lambda_{\pi} = LM_{\pi}$ and
\begin{equation}
h(\lambda_{\pi},M_{\pi}^2)=\frac{4M_{\pi}^2}{(4\pi F_0)^2}
\sum_{\mathbf{n}\neq\mathbf{0}}\frac{K_1(\lambda_{\pi}
  |\mathbf{n}|)}{\lambda_{\pi}|\mathbf{n}|}.
\label{eq:h1}
\end{equation}
Above, $\mathbf{n}\in\mathbb{Z}^3$ are integer component vectors
and $K_1(x)$ is the modified Bessel function of the second kind.

\begin{figure}[htp]
  \centering
  \resizebox{0.8\textwidth}{!}{\includegraphics[width=\textwidth]{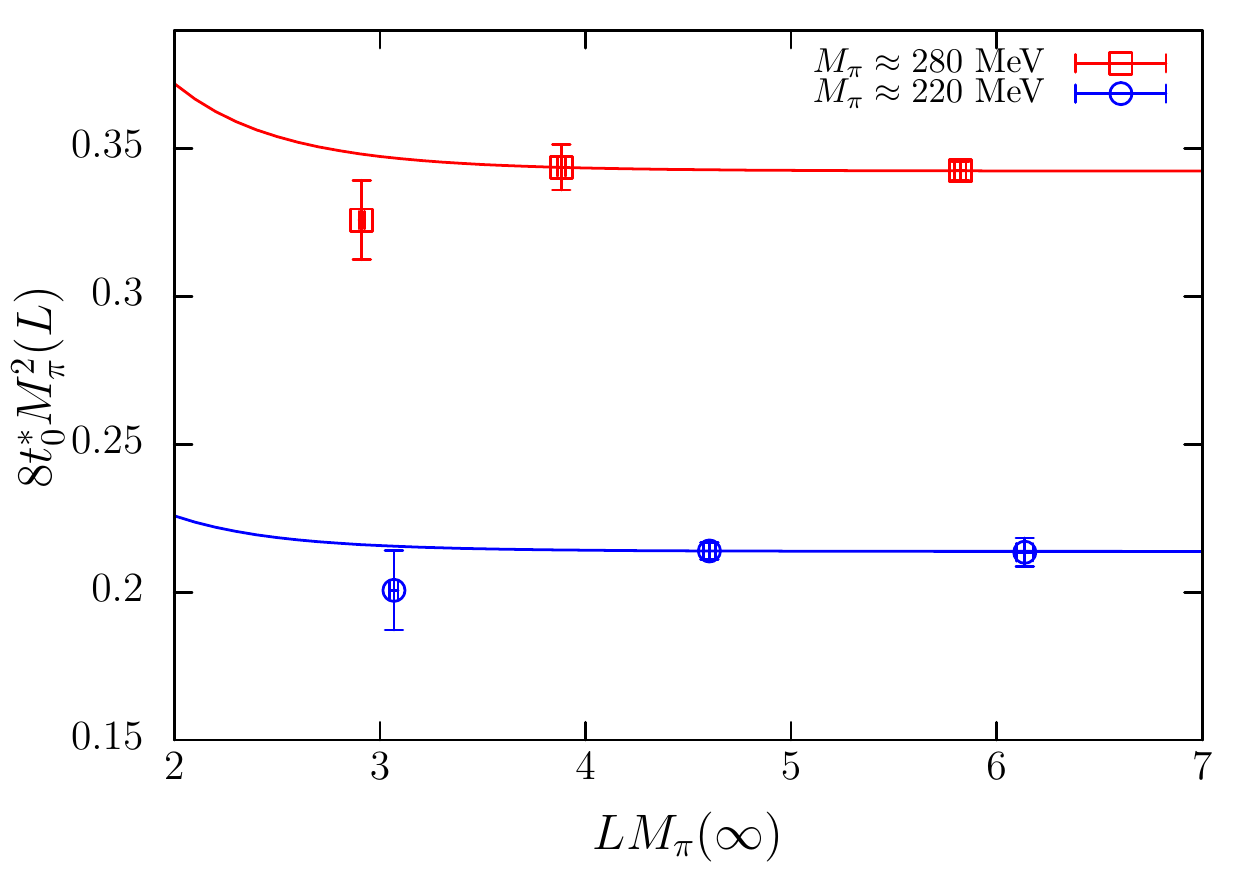}}
  \caption{Dependence of the pion mass
    on the lattice extent at $\beta=3.4$. The $M_{\pi}\approx 280\,\textmd{MeV}$
    data correspond to ensembles U101, H105 and N101 (in order
    of increasing $L$), the $M_{\pi}\approx
    220\,\textmd{MeV}$ data to ensembles
    S100, C101 and D101. The curves, that are normalized
    with respect to the most precise data point, correspond to
    the parametrization eq.~\eqref{eq:mpi2}, where we made use of
    eqs.~\eqref{eq:h1} and~\eqref{eq:meta8}.\label{fig:fsepion}}
\end{figure}

For non-degenerate quark masses, one obtains (see, e.g.,
eq.~(16) of ref.~\cite{Bijnens:2014dea}, where also the
NNLO corrections can be found)
\begin{align}
\label{eq:mpi2}
M_{\pi}^2(L)&=M_{\pi}^2\left[1+\frac12h(\lambda_{\pi},M_{\pi}^2)-\frac16h(\lambda_{\eta_8},M_{\eta_8}^2)\right],\\
\label{eq:mpi3}
M_K^2(L)&=M_K^2\left[1+\frac13h(\lambda_{\eta_8},M_{\eta_8}^2)\right],
\end{align}
where we made use of
\begin{equation}
  M_{\eta_8}^2\approx \frac{4M_K^2-M_{\pi}^2}{3}=\overline{M}\vphantom{M}^2+\frac13\delta M^2,
  \label{eq:meta8}
\end{equation}
which holds to this order in ChPT. In figure~\ref{fig:fsepion} we
compare the pion mass data to this expectation for two pion masses where
simulations at three different volumes exist. In these
cases the deviation of the smallest volume points from the large volume limit
seems to have the wrong sign, however, the deviation from the
NLO ChPT expectation is smaller than two standard deviations. Note that
the smallest volume shown (U101) does not enter our analysis since
$L<2.3\,\textmd{fm}$ in this case.
For the kaon mass we do not detect any
statistically significant finite volume effects. In this case, away from the
$m_s=m_{\ell}$ line, both the
predicted finite size effects and the relative statistical
errors are smaller than for the pion.

To be on the safe side, we correct all pion and kaon masses for the
finite volume effect eqs.~\eqref{eq:h1}--\eqref{eq:meta8}.
Since these equations only encapsulate the leading non-trivial
order, we add in quadrature
half of the difference between finite volume and infinite
volume extrapolated results as a systematic error to
the statistical error of the pion
and kaon masses. In practice, this is done by adding uncorrelated Gaussian
distributed random variables to the existing bootstrap samples.
We remark that except for some small volume
ensembles that do not enter our extrapolations, e.g., U101, the finite
size correction is always smaller than the statistical error
and, for the vast majority of ensembles, much smaller.

\begin{figure}[htp]
  \centering
  \resizebox{0.49\textwidth}{!}{\includegraphics[width=\textwidth]{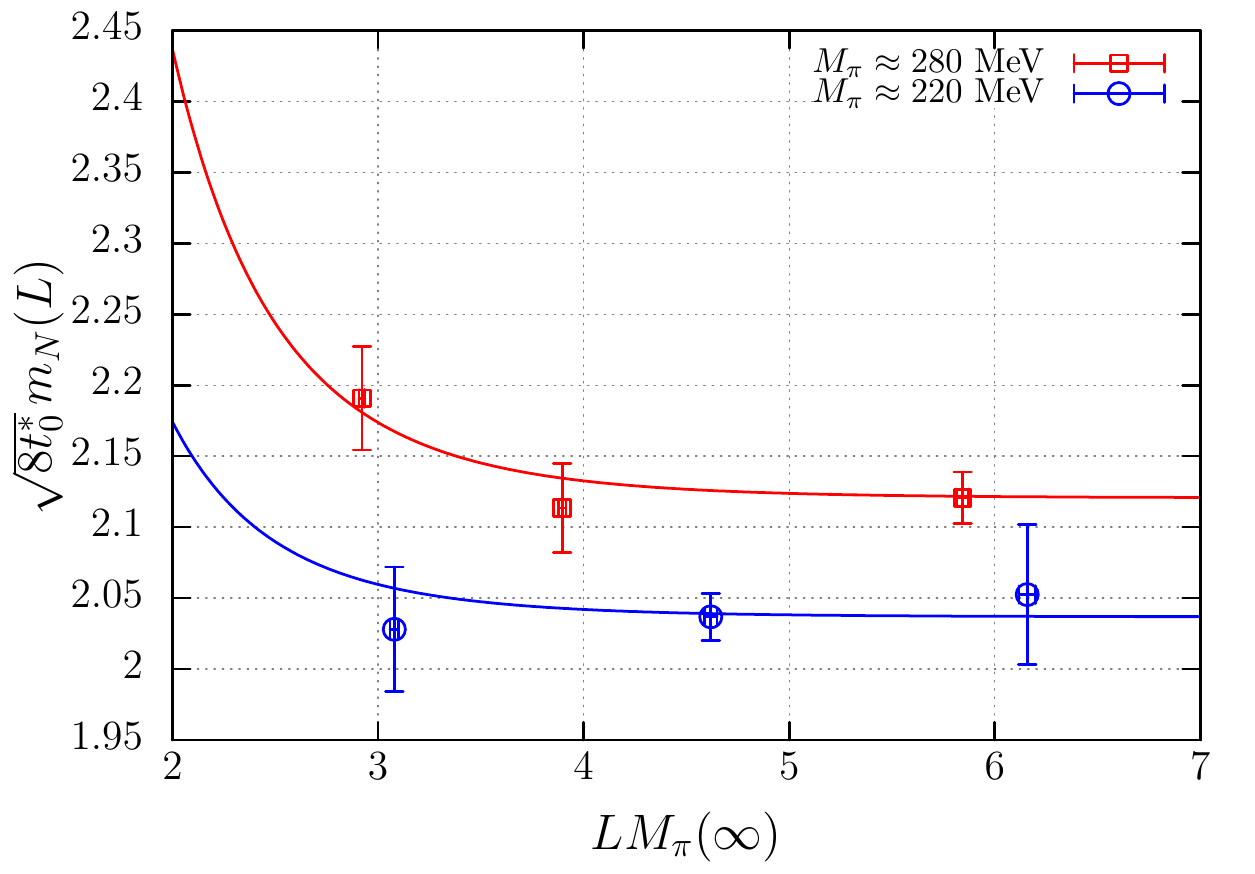}}
  \resizebox{0.49\textwidth}{!}{\includegraphics[width=\textwidth]{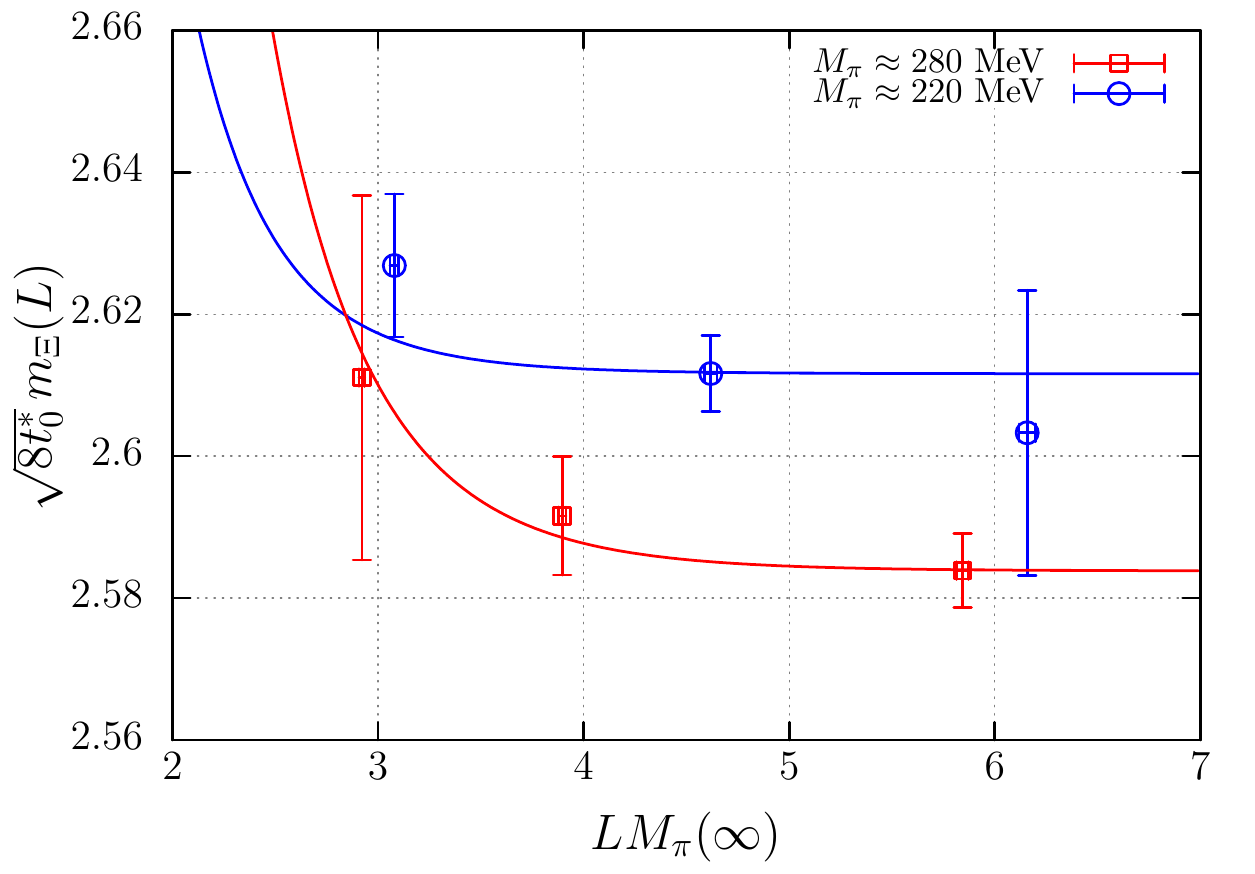}}
  \caption{Dependence of the nucleon (left) and $\Xi$ baryon (right) mass
    on the lattice extent. The masses are determined on the
    ensembles that are displayed in figure~\ref{fig:fsepion}.
    The curves, that are normalized
    with respect to the most precise data point, correspond to
    the parametrization~\eqref{eq:baryonfse}.
    \label{fig:fsebaryon}}
\end{figure}
The analytical expressions for finite volume effects of the baryon masses
are discussed in appendix~\ref{sec:finitebaryon}. These are
included into the functional form of the fit to
$\mathbbm{m}_B(\mathbbm{M}_\pi,\mathbbm{M}_K,\mathbbm{L})$.
In general, we expect finite volume effects in the baryon sector
to be much smaller than our statistical errors, at least for
$L\gtrsim\max\left\{4M_{\pi}^{-1},2.3\,\textmd{fm}\right\}$.
In figure~\ref{fig:fsebaryon} we compare
nucleon and cascade masses with the
order $p^3$ BChPT expectation eq.~\eqref{eq:baryonfse}, matching this
in each case to the first data point with $LM_{\pi}>4$. The coefficients
$g_{N,P}/(4\pi F_0)^2$ and $g_{\Xi,P}/(4\pi F_0)^2$ (see eq.~\eqref{eq:octcop})
and the scale $m_0$, all defined in the chiral limit, are
taken from our best global continuum limit BChPT fit to the octet baryon
spectrum, see section~\ref{sec:barspec} and table~\ref{tab:lecresults}.
$M_{\eta_8}$ is computed via eq.~\eqref{eq:meta8}.
The curves are subject to systematics that are due to
truncating at order $p^3$, omitting effects of decuplet
baryon loops and the uncertainties of the low energy constants.
Similar results are obtained for the other baryons and at different
simulation points.
The figure qualitatively confirms our expectation:
some of the $LM_{\pi}\approx 3$ points appear to give larger masses
than the large volume data, while
we see no significant differences between the $LM_{\pi}\gtrsim 4$ points.

\section{The continuum limit dependence of baryon masses on the meson masses}
\label{sec:physical}
In the preceding section, we discussed our combined
chiral, continuum and infinite volume limit extrapolation strategy.
The generic form of our fits is given in eq.~\eqref{eq:fit2},
see also the definitions eqs.~\eqref{eq:mass12} and~\eqref{eq:defin}.
We remark that for analysing $m_s=m_{\ell}$ ensembles alone,
the parameters $\delta c_{O|D}$ are not needed since
$\delta\mathbbm{M}=0$.
Likewise, only one of the coefficients $c_{o|d}$ or $\bar{c}_{o|d}$ is required
when considering only the $\overline{m}=\text{const}$ data, where
$\overline{\mathbbm{M}}\approx \text{const}$. If only
$\widehat{m}_s\approx \widehat{m}_{s,\text{ph}}$ data are used,
the effect of $\bar{c}_{o|d}$ can be absorbed into $c_{o|d}$
and $\delta c_{O|D}$. However, here we attempt a joint analysis of all
the ensembles and we therefore include all six discretization
coefficients both for the octet and for the decuplet baryons.

Below we define the relevant parametrizations of the
quark mass dependence in the continuum limit.
These are based on $\textmd{SU(3)}$ ChPT, i.e.\ an expansion about
$m_s=m_{\ell}=0$, as well as
on a Taylor expansion about points on the $m_s=m_{\ell}$ line.
The volume dependence (that does not contain additional LECs)
is detailed in
appendix~\ref{sec:finitebaryon}.
We start the discussion with a linear dependence of
the baryon masses on the quark masses, i.e.\ NLO ChPT.
The fits that we will carry out are based on
NNLO BChPT in extended-on-mass-shell (EOMS) regularization,
NNLO BChPT, including transitions between
octet and decuplet baryons (small
scale expansion (SSE)),
and a Taylor expansion up to quadratic order in the quark
masses about points on the symmetric line $m_s=m_{\ell}$.
The heavy baryon limit (HBChPT) is also discussed.

\subsection{Linear: NLO BChPT}
\label{sec:fit1}
Terms proportional to quark masses are quadratic in the pseudoscalar
meson masses due to the GMOR relations. Therefore, to lowest non-trivial
order the infinite volume continuum limit
expectations for octet and decuplet baryon masses
read
\begin{align}
  \label{eq:linear1}
m_O(M_\pi,M_K) &= m_{0} + \bar{b}\, \overline{M}\vphantom{M}^2 + \delta b_O\, \delta M^2,\\
m_D(M_\pi,M_K) &= m_{D0} + \bar{t}\, \overline{M}\vphantom{M}^2 + \delta t_D \, \delta M^2,\label{eq:linear2}
\end{align}
respectively,
where $O\in\{N,\Lambda,\Sigma,\Xi\}$, $D\in\{\Delta,\Sigma^*,\Xi^*,\Omega\}$
and $m_{0}$ and $m_{D0}=m_0+\delta$ are the baryon masses in the
$\textmd{SU(3)}$ chiral limit. The
remaining parameters are related to the three standard octet $\textmd{SU(3)}$
LECs $b_0$, $b_D$ and $b_F$ and the two decuplet LECs $t_{D0}$ and $t_D$
as follows (where
we deviate from the standard notation to distinguish the LEC $t_{D0}$ from the
scale parameter $t_0$),
\begin{align}
\bar{b}&=-6 b_0-4 b_D,\quad \bar{t}=3t_{D0}+3t_D,\label{eq:btbar}\\
\delta b_N&=\tfrac23(3b_F-b_D),\quad 
\delta b_\Lambda=-\tfrac43 b_D,\quad 
\delta b_\Sigma=\tfrac43 b_D,\quad 
\delta b_\Xi=-\tfrac23(3b_F+b_D),\label{su3cons1}\\
\delta t_{\Delta}&=-t_D,\quad 
\delta t_{\Sigma^*}=0,\quad 
\delta t_{\Xi^*}=t_D,\quad 
\delta t_{\Omega}=2t_D.
\label{su3cons}
\end{align}
One parameter encapsulates the dependence on the average
quark mass ($b_0$ and $t_{D0}$ or, equivalently, $\bar b$ and $\bar t$) while the
quark mass splittings
depend on two parameters  or one parameter only ($b_D$ and $b_F$ or $t_D$,
respectively), which is due to
$\textmd{SU(3)}$ constraints. Note that the above functional forms
are consistent with the Gell-Mann--Okubo
mass splitting relations~\cite{GellMann:1962xb,Okubo:1961jc}.

Based on these continuum relations, we employ the ansatz
\begin{align}
  \mathbbm{m}_O(\mathbbm{M}_\pi,\mathbbm{M}_K) &= \mathbbm{m}_{0}
  + \overline{\mathbbm{b}} \, \overline{\mathbbm{M}}\vphantom{M}^2 + \delta\mathbbm{b}_O \, \delta\mathbbm{M}^2,\\
  \mathbbm{m}_D(\mathbbm{M}_\pi,\mathbbm{M}_K) &= \mathbbm{m}_{D0}
  + \overline{\mathbbm{t}} \, \overline{\mathbbm{M}}\vphantom{M}^2 + \delta \mathbbm{t}_D \, \delta\mathbbm{M}^2
\end{align}
for the dimensionless combinations eq.~\eqref{eq:defin}
for octet and decuplet baryons, respectively, where
$\mathbbm{m}_{0}$ and $\mathbbm{m}_{D0}$ are the products
of $m_{0}$ and $m_{D0}$ with $\sqrt{8t_{0,\text{ch}}}$ in the
$N_f=3$ chiral limit.
Again, we have the $\textmd{SU(3)}$ constraints
\begin{align}\label{su3cons2}
 \delta \mathbbm{b}_N&=\tfrac23(3 \mathbbm{b}_F- \mathbbm{b}_D), &
 \delta \mathbbm{b}_\Lambda&=-\tfrac43 \mathbbm{b}_D, &
 \delta \mathbbm{b}_\Sigma&=\tfrac43 \mathbbm{b}_D, &
 \delta \mathbbm{b}_\Xi&=-\tfrac23(3\mathbbm{b}_F+\mathbbm{b}_D),\\
 \delta \mathbbm{t}_{\Delta}&=-\mathbbm{t}_D, &
 \delta \mathbbm{t}_{\Sigma^*}&=0, &
 \delta \mathbbm{t}_{\Xi^*}&=\mathbbm{t}_D,  &
 \delta \mathbbm{t}_{\Omega}&=2\,\mathbbm{t}_D.
 \label{su3cons3}
\end{align}
To this order, $t_0(M_\pi,M_K)$ in the continuum limit only depends
on $\overline{M}\vphantom{M}^2$ but not on the mass splitting,
see eq.~(4.10) of ref.~\cite{Bar:2013ora}. Therefore,
$\mathbbm{b}_F$, $\mathbbm{b}_D$ and $\mathbbm{t}_D$
can be obtained by trivially rescaling the baryonic low energy constants.
However, the slopes $\overline{\mathbbm{b}}$ and $\overline{\mathbbm{t}}$
will contain an additive term, due
to the dependence of $\sqrt{8t_0}$ on $\overline{M}\vphantom{M}^2$
(see eq.~\eqref{eq:t0chi}):
\begin{align}\label{eq:bbmshift0}
  b_F&=\sqrt{8t_{0,\text{ch}}}\,\mathbbm{b}_F,\quad
  b_D=\sqrt{8t_{0,\text{ch}}}\,\mathbbm{b}_D,\quad
  t_D=\sqrt{8t_{0,\text{ch}}}\,\mathbbm{t}_D,\\
  \bar{b}&=\sqrt{8t_{0,\text{ch}}}
  \left(\overline{\mathbbm{b}}-\frac{\tilde{k}_1}{2}\mathbbm{m}_0\right),\quad
  \bar{t}=\sqrt{8t_{0,\text{ch}}}
  \left(\overline{\mathbbm{t}}-\frac{\tilde{k}_1}{2}\mathbbm{m}_{D0}\right),  \label{eq:bbmshift}
\\
  b_0&=\sqrt{8t_{0,\text{ch}}}
  \left(\mathbbm{b}_0+\frac{\tilde{k}_1}{12}\mathbbm{m}_0\right),\quad
  t_{D0}=\sqrt{8t_{0,\text{ch}}}
  \left(\mathbbm{t}_{D0}-\frac{\tilde{k}_1}{6}\mathbbm{m}_{D0}\right),\label{eq:bbmshift2}
\end{align}
where eq.~\eqref{eq:bbmshift2} follows from eqs.~\eqref{eq:bbmshift0}
and~\eqref{eq:bbmshift} and the definitions~\eqref{eq:btbar}
of $\bar{b}$ and $\bar{t}$.

Linear octet baryon fits would contain a total of $10$ parameters:
the six parameters of eq.~\eqref{eq:fit2} describing the lattice
spacing dependence ($c_o$, $\bar c_o$, $\delta c_N$, $\delta c_\Lambda$, $\delta c_\Sigma$, $\delta c_\Xi$)
and four parameters describing the chiral behaviour
in the continuum limit ($\mathbbm{m}_0$, $\overline{\mathbbm{b}}$, $\mathbbm{b}_D$,
$\mathbbm{b}_F$). The decuplet masses only depend on
nine parameters ($\mathbbm{m}_{D0}$, $\overline{\mathbbm{t}}$, $\mathbbm{t}_D$,
$c_d$, $\bar c_d$, $\delta c_\Delta$, $\delta c_{\Sigma^*}$,
$\delta c_{\Xi^*}$, $\delta c_\Omega$).

\subsection{BChPT: NNLO BChPT}
\label{sec:fit2}
We label the $\mathcal{O}(1)$ and $\mathcal{O}(p^2)$ baryon ChPT as
LO and NLO, respectively, and the 
$\mathcal{O}(p^3)$ covariant baryon ChPT
(BChPT)~\cite{Jenkins:1990jv} as NNLO.\footnote{Note that
  in the naming conventions that are usually employed within
  HBChPT, this order is labelled as ``NLO''.
  Also note that in NNLO BChPT the baryon masses are
  accurate to $\mathcal{O}(m_q^{3/2})$ in the quark masses $m_q\sim p^2$.
  Due to the absence of such terms in the mesonic sector,
  NLO meson ChPT already contains all $\mathcal{O}(p^4)$ terms,
  while the LO GMOR relations are accurate
  at $\mathcal{O}(p^2)$. Consequently, both
  within NLO and NNLO BChPT
  it is sufficient to employ the GMOR relations to convert
  quark masses into meson masses. NLO meson ChPT is only
  needed in conjunction with BChPT at NNNLO.}
This leading one-loop order includes the usual sunset self-energy
diagrams. We regulate the loop-function according to
the EOMS scheme~\cite{Gegelia:1999gf,Fuchs:2003qc,Lehnhart:2004vi}
for removing terms that break the power counting. For a
review of different ChPT approaches, see, e.g., ref.~\cite{Geng:2013xn}.
The infinite volume mass dependence for octet baryons reads~\cite{Ellis:1999jt}
\begin{align}
  \label{eq:octetc}
m_O(M_\pi,M_K) &= m_0 + \bar b \, \overline{M}\vphantom{M}^2 + \delta b_O \, \delta M^2 \nonumber \\&\quad + \frac{m_0^3}{{(4\pi F_0)}^2}\left[g_{O,\pi} f_O\left( \frac{M_\pi}{m_0}\right) + g_{O,K} f_O\left( \frac{M_K}{m_0}\right) + g_{O,\eta_8} f_O\left( \frac{M_{\eta_8}}{m_0}\right)\right],
\end{align}
where the $\delta b_O$ are the same as above, see eq.~\eqref{su3cons1},
the $\eta_8$ mass to this order is given by eq.~\eqref{eq:meta8}
and the EOMS loop-function reads
\begin{align}
  f_O(x) &= -2x^3\left[\sqrt{1-\frac{x^2}{4}}\arccos\left(\frac{x}{2}\right)
    +\frac{x}{2}\ln(x)\right].
  \label{eq:loopf}
\end{align}
The dimensionless couplings
$g_{O,P}$ are given as\footnote{The $g_{O,P}$ satisfy the constraints
  $g_{O,\pi} + g_{O,K} + g_{O,\eta_8} = 2(5 D^2+9 F^2)/3$ and
$2 g_{N,P} + g_{\Lambda,P} + 3 g_{\Sigma,P} + 2 g_{\Xi,P} = 2c_P(5 D^2+9 F^2)/3$,
where the couplings $g_{O,P}$ already include
the meson multiplicities $c_P$ ($c_\pi=3$, $c_K=4$, $c_{\eta_8}=1$).}
  \begin{align}
g_{N,\pi}&= \tfrac{3}{2}{(D+F)}^2, & g_{N,K}&=\tfrac{5}{3} D^2 - 2D F + 3 F^2, & g_{N,\eta_8} &= \tfrac{1}{6} {(D-3F)}^2,\nonumber \\
g_{\Lambda,\pi}&=2 D^2, & g_{\Lambda,K}&=\tfrac{2}{3} D^2 + 6 F^2, & g_{\Lambda,\eta_8}&=\tfrac{2}{3} D^2,\nonumber \\
g_{\Sigma,\pi}&=\tfrac{2}{3}D^2 + 4 F^2, & g_{\Sigma,K}&=2D^2 + 2F^2, & g_{\Sigma,\eta_8}&=\tfrac{2}{3} D^2,\nonumber \\
g_{\Xi,\pi}&=\tfrac{3}{2}{(D-F)}^2, & g_{\Xi,K}&=\tfrac{5}{3} D^2 + 2D F + 3 F^2, & g_{\Xi,\eta_8}&=\tfrac{1}{6} {(D+3F)}^2,\label{eq:octcop}
\end{align}
where $D$ and $F$ are the usual $\textmd{SU(3)}$
LECs describing the LO
baryon-meson-baryon coupling. For instance,
$g_{N,\pi}=(3/2)(D+F)^2=(3/2)\mathring{g}_A^2$ is related to the axial charge
of the nucleon in the chiral limit $\mathring{g}_A^2$.
Note that truncating the loop-function~\eqref{eq:loopf} at $\mathcal{O}(x^3)$,
i.e.\ expanding about $m_0=\infty$,
results in the HBChPT~\cite{Jenkins:1990jv}
expression~\cite{Gasser:1987rb,Bernard:1992qa}
\begin{equation}
  f_O(x)=-\pi x^3+\mathcal{O}(x^4).
  \label{eq:loophb}
\end{equation}
We remark that
the NNNLO BChPT expression is also known~\cite{Ren:2012aj}.

The functional form above, when rescaled, reads
\begin{align}
\mathbbm{m}_O(\mathbbm{M}_\pi,\mathbbm{M_K}) &= \mathbbm{m}_0 + \overline{\mathbbm{b}} \; \overline{\mathbbm{M}}\vphantom{M}^2 + \delta \mathbbm{b}_O \; \delta
\mathbbm{M}^2 \nonumber \\&\quad + \mathbbm{g}_{O,\pi} f_O\left( \frac{\mathbbm{M}_\pi}{\mathbbm{m}_0}\right) + \mathbbm{g}_{O,K} f_O\left( \frac{\mathbbm{M}_K}{\mathbbm{m}_0}\right) + \mathbbm{g}_{O,\eta_8} f_O\left( \frac{\mathbbm{M}_{\eta_8}}{\mathbbm{m}_0}\right).
\label{eq:rescalem}
\end{align}
If we were to go to the next order then we would have to include
corrections to the GMOR relations when replacing
the quark masses by the meson masses, as well as
additional terms arising from the $\mathcal{O}(p^4)$ chiral expansion of
$t_0$.

The couplings $\mathbbm{g}_{O,P}$ are related to the original ones
$g_{O,P}$ by substituting $D\mapsto \mathbbm{D}, F\mapsto \mathbbm{F}$, where
\begin{align}
  D^2 = \frac{{(4 \pi \mathbbm{F}_0)}^2}{\mathbbm{m}_0^3}\mathbbm{D}^2, \quad
  F^2 = \frac{{(4 \pi \mathbbm{F}_0)}^2}{\mathbbm{m}_0^3}\mathbbm{F}^2.
  \label{eq:DFrescale}
\end{align}

In total our NNLO fit to the octet baryon masses
has only $12$ parameters: the six parameters describing the lattice
spacing dependence of eq.~\eqref{eq:fit2} ($c_o$, $\bar c_o$, $\delta c_N$, $\delta c_\Lambda$, $\delta c_\Sigma$, $\delta c_\Xi$), the four parameters describing the linear, NLO
chiral continuum limit behaviour ($\mathbbm{m}_0$, $\overline{\mathbbm{b}}$, $\mathbbm{b}_D$, $\mathbbm{b}_F$) and, in addition,
$\mathbbm{F}$ and $\mathbbm{D}$ that parameterize the 12
couplings $\mathbbm{g}_{O,P}$.

Decuplet baryon masses have been computed in $\textmd{SU(3)}$ HBChPT in
ref.~\cite{Tiburzi:2004rh}, to NNLO in covariant $\textmd{SU(2)}$
BChPT with infrared cut-off (IR BChPT)~\cite{Bernard:2005fy}
as well as in $\textmd{SU(3)}$
EOMS BChBT to NNLO in ref.~\cite{MartinCamalich:2010fp} and to
NNNLO in ref.~\cite{Ren:2013oaa}.
Restricting ourselves to self-energy diagrams that do not contain
octet baryon exchanges, the relevant infinite volume expression reads
\begin{align}
  m_D(M_\pi,M_K) &= m_{D0} + \bar{t} \,\overline{M}\vphantom{M}^2 + \delta t_D \,\delta M^2 \nonumber \\&\quad +
  \frac{m_{D0}^3}{{(4\pi F_0)}^2}\left[g_{D,\pi} f_D\left(\frac{M_\pi}{m_{D0}},\frac{M_{\pi}}{m_0}\right)
    + g_{D,K} f_D\left(\frac{M_K}{m_{D0}},\frac{M_K}{m_0}\right)\right.\nonumber\\
    &\qquad\qquad\qquad + \left.g_{D,\eta_8} f_D\left(\frac{M_{\eta_8}}{m_{D0}},\frac{M_{\eta_8}}{m_0}\right)\right].
  \label{eq:bchptdec}
\end{align}
The decuplet loop-function in the EOMS regularization~\cite{MartinCamalich:2010fp} reads:\footnote{It is derived in the IR regularization in
ref.~~\cite{Bernard:2005fy}.}
\begin{align}
  f_D(x,y)=
  -2x^3
  \left\{
    \left(1-\frac{x^2}{4}\right)^{5/2}
    \!\!\!\arccos\left(\frac{x}{2}\right)
    +\frac{x}{64}
    \left[ 17 - 2x^2 +
      2\left(30-10x^2+x^4\right)\ln(y)\right]\right\},
    \label{eq:decloopf}
\end{align}
where we use the same renormalization scale $\mu=m_0$ as in the
octet baryon case. This is the reason for the two arguments of
the loop function.
Truncating $f_D$
at $\mathcal{O}(x^3)$ again gives the HBChPT result~\cite{Tiburzi:2004rh}
\begin{align}
  f_D(x) &= -\pi x^3.
\end{align}
The decuplet couplings $g_{D,P}$ are given as\footnote{For
  the LEC $\mathcal{H}$ we use the normalization of
  ref.~\cite{Tiburzi:2004rh}. The decuplet
  couplings satisfy the $\textmd{SU(3)}$ constraints
  $g_{D,\pi} + g_{D,K} + g_{D,\eta_8} = 20\mathcal{H}^2/27$ and
  $4g_{\Delta,P} + 3g_{\Sigma^*,P} + 2g_{\Xi^*,P} + g_{\Omega,P} =
  25c_P\mathcal{H}^2/27$ where $c_\pi=3$, $c_K=4$, $c_{\eta_8}=1$.}~\cite{Tiburzi:2004rh,MartinCamalich:2010fp}
\begin{align}
  g_{\Delta,\pi}&=\tfrac{25}{54}\mathcal{H}^2,
  & g_{\Delta,K}&=\tfrac{5}{27}\mathcal{H}^2,
  & g_{\Delta,\eta_8} &=\tfrac{5}{54}\mathcal{H}^2,\nonumber \\
  g_{\Sigma^*,\pi}&=\tfrac{20}{81}\mathcal{H}^2,
  & g_{\Sigma^*,K}&=\tfrac{40}{81}\mathcal{H}^2,
  & g_{\Sigma^*,\eta_8} &=0\nonumber \\
    g_{\Xi^*,\pi}&=\tfrac{5}{54}\mathcal{H}^2,
  & g_{\Xi^*,K}&=\tfrac{5}{9}\mathcal{H}^2,
  & g_{\Xi^*,\eta_8} &=\tfrac{5}{54}\mathcal{H}^2\nonumber \\
    g_{\Omega,\pi}&=0,
  & g_{\Omega,K}&=\tfrac{10}{27}\mathcal{H}^2,
  & g_{\Omega,\eta_8} &=\tfrac{10}{27}\mathcal{H}^2.\label{eq:deccop}
\end{align}
This means that the NNLO fit to the decuplet baryon masses
has $10$ parameters: the six parameters describing the lattice spacing
dependence ($c_d$, $\bar c_d$, $\delta c_\Delta$, $\delta c_{\Sigma^*}$,
$\delta c_{\Xi^*}$, $\delta c_\Omega$),
the three parameters describing the linear NLO
chiral continuum limit behaviour
($\mathbbm{m}_{D0}$, $\overline{\mathbbm{t}}$, $\mathbbm{t}_D$)
and, in addition,
a parameter $\mathbbm{H}$ that is related to the 
LEC $\mathcal{H}$ in analogy to eq.~\eqref{eq:DFrescale}:
\begin{equation}
  \mathcal{H}^2 = \frac{{(4 \pi \mathbbm{F}_0)}^2}{\mathbbm{m}_{D0}^3}\mathbbm{H}^2.
  \label{eq:Hrescale}
\end{equation}
If fitting the decuplet baryons
alone, one could also have set $\mu=m_{D0}$, thereby removing one
of the arguments of the loop function eq.~\eqref{eq:decloopf}.
However, for the decuplet baryons octet baryon loops cannot be
neglected. This is discussed in the following subsection.
Note that including finite
volume effects in the parametrization does not involve additional LECs,
see appendix~\ref{sec:finitebaryon}.

\subsection{Octet-decuplet BChPT: including the small scale expansion}
\label{sec:fit3}
For most of our data points the gap between octet and decuplet
masses is of a similar size as the pion mass. Hence, we should
also consider that at NNLO the decuplet self-energies receive contributions
from octet baryon plus meson loops and
the octet energies receive contributions from decuplet loops.
In particular, this effect cannot be neglected when discussing
decuplet baryons since (with the exception of the $\Omega$)
these strongly decay into octet baryons
and mesons at the physical quark mass point. The small
scale expansion (SSE)~\cite{Hemmert:1997ye,Bernard:2005fy,Procura:2006bj}
is the theoretical framework to incorporate these baryon loop effects.
This was first worked out for the $\Delta$ baryon in refs.~\cite{Tiburzi:2005na}
and~\cite{Pascalutsa:2005nd} in
$\textmd{SU(2)}$ HBChPT and in $\textmd{SU(2)}$ BChPT, respectively,
and later generalized to $\textmd{SU(3)}$, employing
EOMS BChPT~\cite{MartinCamalich:2010fp}. It turns
out that these effects can be accounted for to NNLO with just one
additional LEC, $\mathcal{C}$:\footnote{For the LEC
  $\mathcal{C}$ we use the
  normalization of refs.~\cite{Jenkins:1991es,WalkerLoud:2004hf,Beane:2011pc},
  where $\mathcal{C}^2=g^2_{\Delta N}$~\cite{Beane:2004tw}.}
\begin{align}
  \label{eq:mixx}
  m_B\mapsto m_B+\mathcal{C}^2\frac{\delta^3}{{(4\pi F_0)}^2}
  \left[
    \xi_{B,\pi}h_B\left(\frac{M_{\pi}}{\delta}\right)
    +\xi_{B,K}h_B\left(\frac{M_{K}}{\delta}\right)
    +\xi_{B,\eta_8}h_B\left(\frac{M_{\eta_8}}{\delta}\right)
  \right],
\end{align}
where the coefficients $\xi_{B,P}$ are listed in table~\ref{tab:sse}
and in the usual SSE power counting $\delta=m_{D0}-m_0=\mathcal{O}(p)$ is an
additional small scale.

\begin{table}[thp]
  \caption{\label{tab:sse} The coefficients $\xi_{B,P}$ of
    eq.~\eqref{eq:mixx}.}
  \renewcommand{\arraystretch}{1.2}
  \begin{center}
\begin{tabular}{ccccccccc}
\toprule
        &$N$           &$\Lambda$    &$\Sigma$      &$\Xi$        &$\Delta$     &$\Sigma^*$    &$\Xi^*$       &$\Omega$\\\midrule
$\pi$   &$\frac{4}{3}$&$1$           &$\frac{2}{9}$ &$\frac{1}{3}$&$\frac{1}{3}$&$\frac{5}{18}$&$\frac{1}{6}$&$0$\\
$K$     &$\frac{1}{3}$ &$\frac{2}{3}$&$\frac{10}{9}$&$1$           &$\frac{1}{3}$&$\frac{2}{9}$ &$\frac{1}{3}$&$\frac{2}{3}$\\
$\eta_8$&$0$            &$0$           &$\frac{1}{3}$ &$\frac{1}{3}$&$0$           &$\frac{1}{6}$ &$\frac{1}{6}$&$0$\\\bottomrule
\end{tabular}
\end{center}
\end{table}

In general, the loop
functions~\cite{MartinCamalich:2010fp} will separately
depend on $M/m_0$ and on $\delta$. However, for simplicity,
for these transition terms we will only consider
the HBChPT limit~\cite{Jenkins:1991es,Beane:2004tw,WalkerLoud:2004hf},
in which the respective functions for octet and decuplet baryons read:
\begin{align}
  \label{eq:ssloop1}
  h_O(x)&=-\left(2-3x^2\right)
  \ln\left(\frac{x}{2}\right)-\frac{1}{2}x^2-2w(x),\\
  h_D(x)&=\left(2-3x^2\right)
  \ln\left(\frac{x}{2}\right)+\frac{1}{2}x^2-2w(-x),\\
  w(x)&=\left\{
  \begin{array}{ccc}
  {\left(x^2-1\right)}^{3/2}\arccos\left(x^{-1}\right)&,&|x|\geq 1\\
  {\left(1-x^2\right)}^{3/2}\ln\left(\left|x^{-1}+\sqrt{x^{-2}-1}\right|\right)&,&|x|< 1
  \end{array}\right..
  \label{eq:ssloop3}
\end{align}
For $w(x)$ we encounter both cases since we cover simulation
points with $M_\pi<300\,\textmd{MeV}\approx\delta$ as well as
with $M_\pi>\delta$. In the first case, the poles of the decuplet
baryon propagators will acquire imaginary parts in an infinite volume,
due to the possibility of a real $p$-wave decay into a pseudoscalar
meson and an octet baryon.
For a detailed discussion of the situation in a finite volume,
see, e.g., ref.~\cite{Bernard:2007cm}.
We take the real part of the logarithm --- hence the modulus in its argument ---
and we will only include stable decuplet baryons into our fits.
Again, our rescaled parameter $\mathbbm{C}$ is trivially related
to the LEC $\mathcal{C}$:
\begin{equation}
  \mathcal{C}^2=  \frac{{(4 \pi \mathbbm{F}_0)}^2}{(\mathbbm{m}_{D0}-\mathbbm{m}_0)^3}\mathbbm{C}^2.
  \label{eq:Crescale}
\end{equation}
The joint NNLO octet and decuplet fits have a total of 23 free parameters:
the 12 parameters of the octet parametrization, the 10 parameters of the
decuplet parametrization plus $\mathbbm{C}$. Twelve of these
are used to parameterize the lattice spacing dependence.
Like for the fits of section~\ref{sec:fit2}
without octet-decuplet transitions,
incorporating finite
volume effects does not involve additional LECs,
see appendix~\ref{sec:finitebaryon}.

\subsection{GMO: Taylor expansion about an SU(3) symmetric point}
\label{sec:taylor}
Instead of expanding about the chiral limit using BChPT,
one can also Taylor expand, e.g.,
about the symmetric point (where $M_K=M_{\pi}=M^*$), implementing
group theoretical constraints~\cite{Bietenholz:2011qq} that generalize
the Gell-Mann--Okubo relations~\cite{GellMann:1962xb,Okubo:1961jc}.
In this case the only assumptions we make are that
$\textmd{SU(3)}$ flavour symmetry is broken by the quark masses
and that to leading order these quark masses are proportional
to squared pseudoscalar meson masses.
This is less restrictive than ChPT as no additional assumption about
the symmetries of interactions mediated by Goldstone bosons is made.
Therefore, the Taylor expansion will involve a larger number of parameters.
In principle, one can expand, employing the same formulae, about any point
along the symmetric line, however, the domain of analyticity
cannot extend beyond $M_{\eta_8}\geq M_K\geq M_{\pi}\geq 0$.

At quadratic order in the quark masses we obtain
\begin{align}
  \label{eq:gmo1}
  \mathbbm{m}_O(\mathbbm{M}_\pi,\mathbbm{M}_K) &= \mathbbm{m}^*
  + \overline{\mathbbm{b}}\vphantom{b}^* \, \left(\overline{\mathbbm{M}}\vphantom{M}^2-\mathbbm{M}^{*2}\right) + \overline{\mathbbm{d}} \, \left(\overline{\mathbbm{M}}\vphantom{M}^2-\mathbbm{M}^{*2}\right)^2\nonumber\\
  &\quad + \delta \mathbbm{b}_O^* \, \delta\mathbbm{M}^2
  + \delta \mathbbm{d}_O \, \delta\mathbbm{M}^2\left(\overline{\mathbbm{M}}\vphantom{M}^2-\mathbbm{M}^{*2}\right)+\delta \mathbbm{e}_O\,\delta\mathbbm{M}^4,\\
    \mathbbm{m}_D(\mathbbm{M}_\pi,\mathbbm{M}_K) &= \mathbbm{m}^*_D
  + \overline{\mathbbm{t}}\vphantom{t}^* \, \left(\overline{\mathbbm{M}}\vphantom{M}^2-\mathbbm{M}^{*2}\right) + \overline{\mathbbm{u}} \, \left(\overline{\mathbbm{M}}\vphantom{M}^2-\mathbbm{M}^{*2}\right)^2\nonumber\\
  &\quad + \delta \mathbbm{t}_D^* \, \delta\mathbbm{M}^2
  + \delta \mathbbm{u}_D \, \delta\mathbbm{M}^2\left(\overline{\mathbbm{M}}\vphantom{M}^2-\mathbbm{M}^{*2}\right)+\delta \mathbbm{v}_D\,\delta\mathbbm{M}^4,
  \label{eq:gmo2}
\end{align}
for the octet and decuplet baryons, respectively, where
$\mathbbm{M}^{*2}=\tfrac{2}{3}\phi_4^*=0.740$ is chosen such that
it is close to $\overline{\mathbbm{M}}\vphantom{M}^2_{\text{ph}}$.
Since the above is a polynomial
in $\left(\overline{\mathbbm{M}}\vphantom{M}^2-\mathbbm{M}^{*2}\right)$,
we can absorb the dependence on $\mathbbm{M}^{*2}$ into
the coefficients of the expansion. At the above order this amounts to
the replacements
\begin{align}
  \mathbbm{m}^*&=\mathbbm{m}_0+\overline{\mathbbm{b}}\,\mathbbm{M}^{*2}+
  \overline{\mathbbm{d}}\,\mathbbm{M}^{*4}\quad
  &\overline{\mathbbm{b}}\vphantom{b}^*&=\overline{\mathbbm{b}}
  +2\overline{\mathbbm{d}}\,\mathbbm{M}^{*2},\quad
  &\delta\mathbbm{b}_O^*&=\delta\mathbbm{b}_O+\delta\mathbbm{d}_O\mathbbm{M}^{*2},\\
  \mathbbm{m}_D^*&=\mathbbm{m}_{D0}+\overline{\mathbbm{t}}\,\mathbbm{M}^{*2}+
\overline{\mathbbm{u}}\,\mathbbm{M}^{*4},\quad
  &\overline{\mathbbm{t}}\vphantom{t}^*&=\overline{\mathbbm{t}}
  +2\overline{\mathbbm{u}}\,\mathbbm{M}^{*2},\quad
  &\delta\mathbbm{t}_D^*&=\delta\mathbbm{t}_D+\delta\mathbbm{u}_D\mathbbm{M}^{*2}.
\end{align}
In terms of the new parameters, eqs.~\eqref{eq:gmo1} and~\eqref{eq:gmo2}
read
\begin{align}
  \label{eq:gmo1n}
  \mathbbm{m}_O(\mathbbm{M}_\pi,\mathbbm{M}_K) &= \mathbbm{m}_0
  + \overline{\mathbbm{b}}\,\overline{\mathbbm{M}}\vphantom{M}^2
  + \delta \mathbbm{b}_O\,\delta\mathbbm{M}^2
  + \overline{\mathbbm{d}}\,\overline{\mathbbm{M}}\vphantom{M}^4
  + \delta\mathbbm{d}_O\,\delta\mathbbm{M}^2\overline{\mathbbm{M}}\vphantom{M}^2+\delta \mathbbm{e}_O\,\delta\mathbbm{M}^4,\\
  \mathbbm{m}_D(\mathbbm{M}_\pi,\mathbbm{M}_K) &= \mathbbm{m}_{D0}
  + \overline{\mathbbm{t}}\,\overline{\mathbbm{M}}\vphantom{M}^2
  + \delta \mathbbm{t}_D\,\delta\mathbbm{M}^2
  + \overline{\mathbbm{u}}\,\overline{\mathbbm{M}}\vphantom{M}^4
  + \delta\mathbbm{u}_D\,\delta\mathbbm{M}^2\overline{\mathbbm{M}}\vphantom{M}^2+\delta \mathbbm{v}_D\,\delta\mathbbm{M}^4,
  \label{eq:gmo2n}
\end{align}
respectively, where
at linear order in the quark masses, one recovers
eqs.~\eqref{eq:linear1}--\eqref{su3cons}, reducing the
dependence to four and three $\textmd{SU(3)}$ BChPT LECs
for the octet and decuplet baryons, respectively.
While the expansions above
are valid near any point along the symmetric line, they can
only converge for positive pion masses, i.e.\
for $\delta \mathbbm{M}^2< 3\overline{\mathbbm{M}}\vphantom{M}^2$.
This explicitly excludes the chiral limit and also explains the
absence of terms that are non-analytic functions of the quark masses;
beyond the linear order, the above expansions are incompatible
with BChPT. However,
all our points, including the physical point, are far away from the
chiral limit so that the above parametrizations may still accurately
represent the data, as long as the ratio
$\delta \mathbbm{M}^2/\overline{\mathbbm{M}}\vphantom{M}^2$
is sufficiently small.

The parameters $\delta\mathbbm{b}_O$ and $\delta\mathbbm{t}_D$
satisfy the $\textmd{SU(3)}$ constraints~\eqref{su3cons1} and~\eqref{su3cons}
and analogous relations also apply to
$\delta\mathbbm{d}_O$ and $\delta\mathbbm{u}_D$:
\begin{align}
  \label{eq:constrain1}
\delta \mathbbm{d}_N&=\tfrac23(3\mathbbm{d}_F-\mathbbm{d}_D),\quad 
\delta \mathbbm{d}_\Lambda=-\tfrac43 \mathbbm{d}_D,\quad 
\delta \mathbbm{d}_\Sigma=\tfrac43 \mathbbm{d}_D,\quad 
\delta \mathbbm{d}_\Xi=-\tfrac23(3\mathbbm{d}_F+\mathbbm{d}_D),\\
\delta \mathbbm{u}_{\Delta}&=-\mathbbm{u}_D,\quad 
\delta \mathbbm{u}_{\Sigma^*}=0,\quad 
\delta \mathbbm{u}_{\Xi^*}=\mathbbm{u}_D,\quad 
\delta \mathbbm{u}_{\Omega}=2\mathbbm{u}_D.
\end{align}
There are
no $\textmd{SU(3)}$ constraints~\cite{Bietenholz:2011qq}
regarding the octet baryon parameters $\delta\mathbbm{e}_O$. 
In this case we end up with a total of 11 parameters
($\mathbbm{m}_0$, $\overline{\mathbbm{b}}$, $\mathbbm{b}_D$,
$\mathbbm{b}_F$, $\overline{\mathbbm{d}}$,
$\mathbbm{d}_D$, $\mathbbm{d}_F$,
$\delta\mathbbm{e}_N$,
$\delta\mathbbm{e}_\Lambda$,
$\delta\mathbbm{e}_\Sigma$,
$\delta\mathbbm{e}_\Xi$),
in addition
to the 6 parameters of eq.~\eqref{eq:fit2} parameterizing the lattice spacing
dependence ($c_o$, $\bar c_o$, $\delta c_N$, $\delta c_\Lambda$, $\delta c_\Sigma$, $\delta c_\Xi$).

For the decuplet baryons
one of the parameters accompanying the $\mathcal{O}(\delta \mathbbm{M}^4)$
terms
can be eliminated~\cite{Bietenholz:2011qq}: we rewrite the
$\delta\mathbbm{v}_D$ as
\begin{equation}
  \label{eq:constrain3}
  \delta\mathbbm{v}_{\Delta}=\mathbbm{v}_F-\mathbbm{v}_D,\quad
  \delta\mathbbm{v}_{\Sigma^*}=\mathbbm{v}_G,\quad
  \delta\mathbbm{v}_{\Xi^*}=\mathbbm{v}_G+\mathbbm{v}_D,\quad
  \delta\mathbbm{v}_{\Omega}=\mathbbm{v}_F+2\mathbbm{v}_D,
\end{equation}
such that we count $8+6$ parameters in total
($\mathbbm{m}_{D0}$, $\overline{\mathbbm{t}}$, $\mathbbm{t}_D$,
$\overline{\mathbbm{u}}$, $\mathbbm{u}_D$,
$\mathbbm{v}_D$,
$\mathbbm{v}_F$,
$\mathbbm{v}_G$ as well as
$c_d$, $\bar c_d$, $\delta c_\Delta$, $\delta c_{\Sigma^*}$,
$\delta c_{\Xi^*}$, $\delta c_\Omega$).

\section{Results for the scale $\boldsymbol{t_{0,\text{ph}}}$, the spectrum and the low energy constants}
\label{sec:results}
The methods used to compute baryonic two-point functions and to
obtain the masses from these are detailed in appendix~\ref{sec:spectrum}.
The statistical analysis methods, taking into account
autocorrelations in Monte Carlo time and correlations between different masses
obtained on the same ensemble, are described in
appendix~\ref{sec:statistical}. We carry out fully correlated
fits according to the parametrizations introduced in sections~\ref{sec:ex}
and~\ref{sec:physical}. In section~\ref{sec:naming} we introduce
the naming scheme of the fit functions that we employ.
We continue with explaining the fit strategy and cuts on the data in
section~\ref{sec:strat}. We then illustrate this for
the example of one parametrization in section~\ref{sec:illustrate},
before we determine the scale parameter $\sqrt{8t_{0,\text{ph}}}$
from the mass of the $\Xi$ baryon in section~\ref{sec:scale},
where we also detail our model averaging procedure and investigate
the systematics. In section~\ref{sec:barspec} we determine
the spectrum of octet and decuplet baryons and compare this to
expectations. Then, in section~\ref{sec:sigma}, we compute the
strange and light quark $\sigma$~terms of the octet and the
$\Omega$ baryons. In section~\ref{sec:lecres} we
determine the $\textmd{SU(3)}$ (H)BChPT LECs and
for the nucleon also the LO $\textmd{SU(2)}$ BChPT LECs.
Finally, we compare
the results for the scale parameter, the baryon spectrum
and the pion-nucleon
$\sigma$~term to literature
values in section~\ref{sec:resultsummary}.

\subsection{Fits carried out and the naming conventions used}
\label{sec:naming}
We find that linear fits (section~\ref{sec:fit1}) poorly describe
the data and do not consider these further.
When employing the (H)BChPT parametrizations (sections~\ref{sec:fit2}
and~\ref{sec:fit3}), we only fit the decuplet baryon masses together
with the octet masses (SSE, section~\ref{sec:fit3}). Fitting the
decuplet masses alone would amount to ignoring loops involving
an octet baryon and a meson.
In contrast, the GMO Taylor expansions (section~\ref{sec:taylor}) for octet
and decuplet baryons are independent of one another.

We label fits
generically as ``\textbf{multiplet(s) parametrization (FV) (SC$^{L|\infty}$)}'', where
\begin{itemize}
\item ``\textbf{octet (H)BChPT}'' refers to eq.~\eqref{eq:rescalem} with the
  loop-functions eq.~\eqref{eq:loopf} and eq.~\eqref{eq:loophb} for
  BChPT and HBChPT, respectively. The order $p^2$ coefficients
  can be expressed in terms of $\overline{\mathbbm{b}}$, $\mathbbm{b_D}$
  and $\mathbbm{b_F}$, see eq.~\eqref{su3cons2},
  whereas the $p^3$ coefficients
  can be written in terms of $\mathbbm{D}$ and $\mathbbm{F}$
  via eq.~\eqref{eq:octcop}. In addition, the
  (rescaled) mass parameter $\mathbbm{m}_0$ represents the mass
  of the octet in the $\textmd{SU(3)}$ chiral limit.
\item ``\textbf{octet-decuplet (H)BChPT}'' refers to simultaneous (SSE)
  fits utilizing
  eqs.~\eqref{eq:rescalem}, \eqref{eq:bchptdec} and~\eqref{eq:mixx}
  (with rescaled variables) of sections~\ref{sec:fit2} and~\ref{sec:fit3}.
  For BChPT the loop-functions
  eqs.~\eqref{eq:loopf}, \eqref{eq:decloopf}
  and~\eqref{eq:ssloop1}--\eqref{eq:ssloop3} are used, while
  for HBChPT the former two are replaced by the one defined in
  eq.~\eqref{eq:loophb}. There are five independent order $p^2$ coefficients,
  see eqs.~\eqref{su3cons2} and~\eqref{su3cons3}, namely
  $\overline{\mathbbm{b}}$, $\mathbbm{b}_D$, $\mathbbm{b}_F$,
  $\overline{\mathbbm{t}}$ and $\mathbbm{t}_D$.
  The order $p^3$ coefficients
  depend on the rescaled parameters
  $\mathbbm{D}$, $\mathbbm{F}$, $\mathbbm{H}$ and $\mathbbm{C}$, see
  eqs.~\eqref{eq:octcop} and~\eqref{eq:deccop} and table~\ref{tab:sse}.
  The relations between the rescaled parameters and the corresponding
  LECs are given in eqs.~\eqref{eq:DFrescale},
  \eqref{eq:Hrescale} and~\eqref{eq:Crescale}.
  In addition to the nine above parameters, we have the
  chiral octet and decuplet baryon masses $\mathbbm{m}_0$ and $\mathbbm{m}_{D0}$.
\item ``\textbf{octet GMO}'' and ``\textbf{decuplet GMO}'' correspond to eqs.~\eqref{eq:gmo1n}
  and~\eqref{eq:gmo2n} of section~\ref{sec:taylor}, respectively,
  with the $\textmd{SU(3)}$ constraints
  eqs.~\eqref{su3cons2}--\eqref{su3cons3} and
  \eqref{eq:constrain1}--\eqref{eq:constrain3}. This amounts to
  11 and 8 free parameters for the continuum limit
  parametrizations of the octet and decuplet masses, respectively.
\end{itemize}
An appended ``\textbf{FV}'' means that the (H)BChPT
finite volume effects of appendix~\ref{sec:finitebaryon} are included.
For ``octet BChPT~FV'' and ``octet HBChPT~FV'' fits
these are eqs.~\eqref{eq:baryonfse}
and~\eqref{eq:hbfse}, respectively. Regarding ``octet-decuplet
BChPT~FV'' and ``octet-decuplet
HBChPT~FV'', also eqs.~\eqref{eq:ssefv1}--\eqref{eq:ssefv2} and
\eqref{eq:heavy}--\eqref{eq:heavy2} are added, respectively.
These expressions do not include any additional parameters. For
``octet GMO~FV'' fits, the octet baryon masses are first
corrected for finite volume effects using estimates from the
octet BChPT~FV fits. We do not carry out any
``decuplet GMO~FV'' fits as the finite volume effects are very small.

We either append
``\textbf{SC$^{\infty}$}'' or ``\textbf{SC$^L$}'' to the fits involving
the decuplet masses, referring to stability cuts.
In the former case all decuplet baryons that will become unstable
at the given pseudoscalar masses in an infinite volume are excluded
from the fit, while
in the latter case only the decuplet baryons that are unstable
in the box of volume $L^3$ are disregarded.

For the lattice spacing dependence of the baryon masses,
we follow eq.~\eqref{eq:fit2}, which amounts to the six fit parameters
each  for the octet ($c_o$, $\bar{c}_o$, $\delta c_N$, $\delta c_{\Lambda}$, $\delta c_{\Sigma}$,
$\delta c_{\Xi}$) and decuplet baryons
($c_d$, $\bar{c}_d$ $\delta c_{\Delta}$, $\delta c_{\Sigma^*}$,
$\delta c_{\Xi^*}$, $\delta c_{\Omega}$).
This implies a total of 12, 23, 17 and 14
free parameters, respectively, for the octet (H)BChPT, octet-decuplet
(H)BChPT, octet GMO and decuplet GMO fits.
We also investigate the significance of the individual discretization
effects, adding or removing terms. Further details
are given in section~\ref{sec:scale}.
\subsection{Fit strategy and data cuts}
\label{sec:strat}
We carry out a number of fits to the pseudoscalar meson mass,
volume and lattice spacing dependence of the octet and decuplet
baryon masses. The continuum limit parametrizations are
introduced in sections~\ref{sec:fit1}--\ref{sec:taylor} and
the finite volume expressions are collected in
appendix~\ref{sec:finitebaryon}. The lattice spacing
dependence is parameterized according to eq.~\eqref{eq:fit2}.
The naming scheme for different fit ansätze is explained in
section~\ref{sec:naming} above.
We exclude ensembles with $M_{\pi}>450\,\textmd{MeV}$
or $L<2.3\,\textmd{fm}$ from all fits. We will use
the mass of the $\Xi$ baryon to fix the scale. The
masses of the pion and the kaon are then employed to determine the physical
point in the quark mass plane.

\begin{figure}[htp]
  \centering
  \resizebox{0.95\textwidth}{!}{\includegraphics[width=\textwidth]{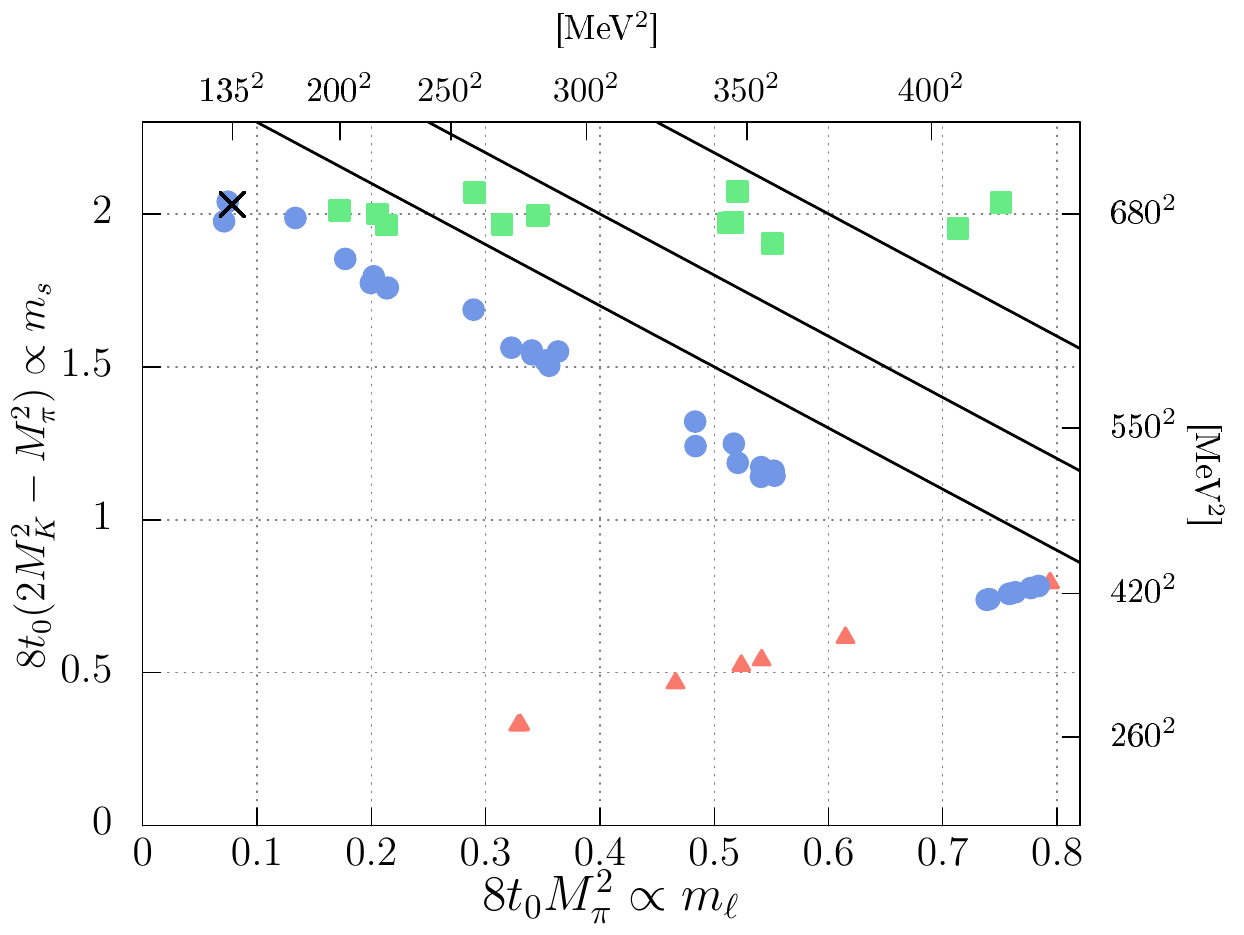}}
  \caption{The ensembles used for the six different lattice spacings
    shown in the quark mass plane. The green squares indicate
    the ensembles on the trajectory where the strange quark mass
    is kept constant, the blue
    circles indicate ensembles with a constant sum of the $N_f=2+1$
    quark masses and the red
    triangles are ensembles on the symmetric $m_s=m_{\ell}$ line.
    The latter two sets of points intersect
    near $8t_0M_\pi^2=2\phi_4^*/3=0.74$.
    The cross indicates the physical point and the black lines
    correspond to mass cuts at $\phi_4=12t_0\overline{M}\vphantom{M}^2=1.25$,
    $1.4$ and $1.6$, respectively. The physical point is at
    $\phi_4\approx 1.093$.
    \label{fig:masscut}}
\end{figure}

In order to explore the systematics, we vary the
continuum limit parametrization and, in addition, we explore the lattice
spacing dependence by including additional
lattice spacing dependent terms and/or by setting some
of the discretization terms to zero.
Moreover, we vary the fit ranges, excluding small volumes,
excluding the coarsest lattice spacing or ---
for the (H)BChPT fits --- excluding ensembles
with large average pion masses.
The latter mass cuts are illustrated in figure~\ref{fig:masscut},
which also provides an overview of the ensembles used.
The mass cuts are not applied when carrying out the GMO fits.
As explained in section~\eqref{sec:taylor},
in this case $\delta M^2$ is the parameter that governs the convergence
properties.
Since this is maximal for our two physical point ensembles and
we wish to obtain the spectrum for physical quark masses, it does
not make sense to impose a cut on $\delta M^2$.
Regarding the (H)BChPT fits, the two ensembles in the top right
corner of the figure are always excluded, i.e.\ we only
include ensembles with $\phi_4<1.6$. These two ensembles are, however,
always included in the GMO fits.

We start with one fit to illustrate the scale setting procedure, the
physical point determination and how the data that depend on
$M_K$, $M_{\pi}$, $L$ and $a$ are shifted to enable their visualization
in two-dimensional figures. We then investigate
the systematics and we present final results on the scale parameter,
the octet and the decuplet baryon masses at the physical point.
\subsection{Illustration of the scale setting procedure}
\label{sec:illustrate}
We start with an order $p^3$ (NNLO)
BChPT fit of the octet baryon masses obtained on the 44 ensembles with
$\phi_4<1.6$, $M_{\pi}<450\,\textmd{MeV}$ and $L>2.3\,\textmd{fm}$,
including the finite volume corrections
eq.~\eqref{eq:baryonfse} (the octet BChPT~FV fit, see section~\ref{sec:naming}).
Fourteen of these ensembles
are on the $m_s=m_{\ell}$ line. For the remaining 30 ensembles,
the four octet baryon masses are non-degenerate. In total
this amounts to 134 data points. These are fitted by 12 parameters,
which correspond to
$c_o$, $\bar{c}_o$, $\delta c_N$, $\delta c_\Lambda$, $\delta c_\Sigma$ and $\delta c_{\Xi}$, accounting
for lattice spacing effects, and to the LECs $m_0$, $b_0$,
$b_D$, $b_F$, $F$ and $D$, describing the quark mass and
the volume dependence.\footnote{We have already corrected
the meson masses for volume effects, see section~\ref{sec:finite}.}
We obtain the value
$\chi^2/N_{\text{DF}}=1.16$, taking into account all correlations
between the four baryon and the two meson masses.
Note that the $6\times 6$ covariance matrices reduce to
$2\times 2$ matrices on the symmetric line $m_s=m_{\ell}$.
The quality of this fit suggests that
our data cannot resolve the additional LECs that would appear
at order $p^4$.

\begin{figure}[htp]
  \centering
  \resizebox{0.95\textwidth}{!}{\includegraphics[width=\textwidth]{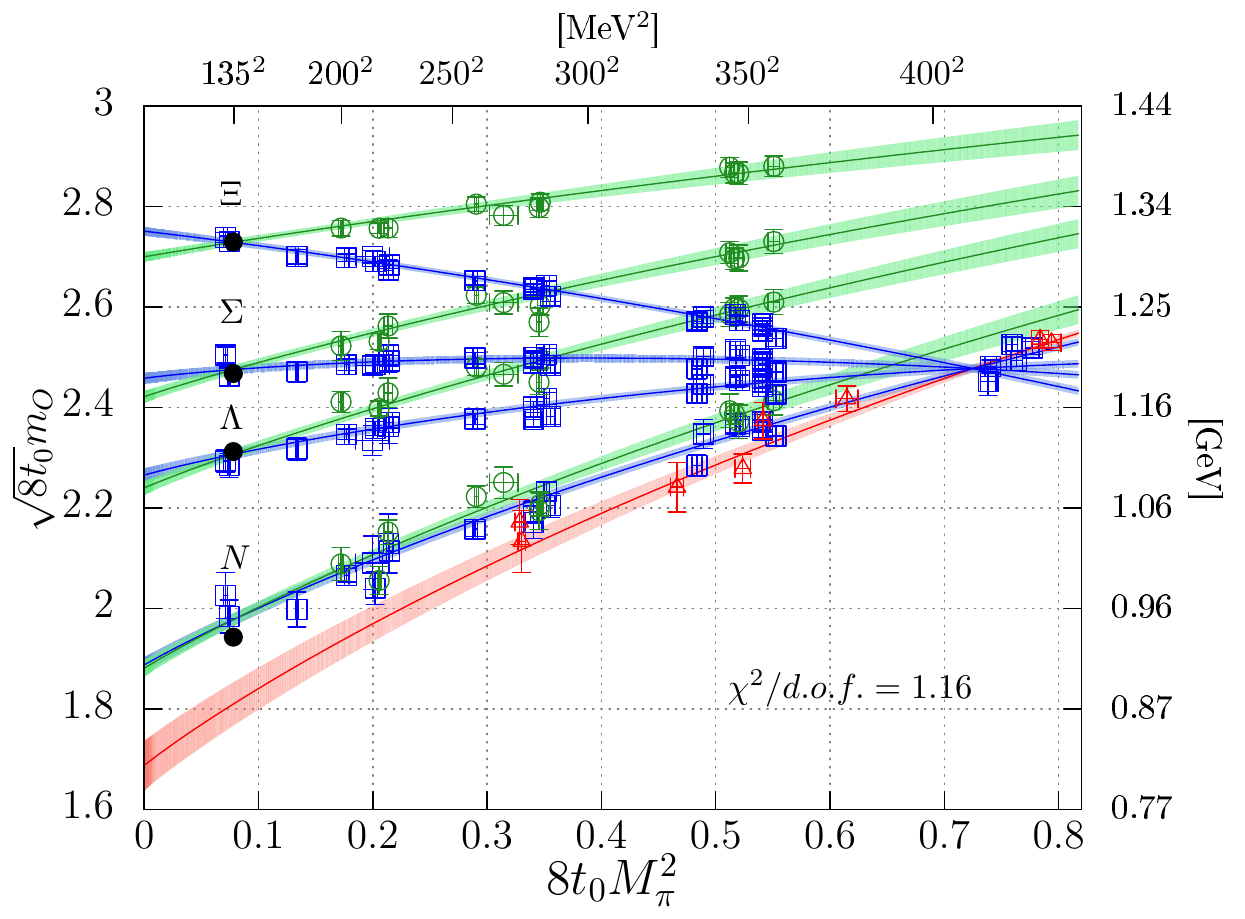}}
  \caption{Octet BChPT~FV fit (see section~\ref{sec:naming} for the naming
    convention and references to the equations used) for $\phi_4<1.6$.
    The data points are
    corrected for the fitted lattice spacing and finite volume effects.
    The blue squares ($\Tr M=\text{const}$)
    and the green circles ($\widehat{m}_s\approx\text{const}$)
    are in addition shifted according to the
    fit to kaon masses that correspond to constant values of
    $\phi_4=12t_0\overline{M}\vphantom{M}^2$ and
    $8t_0(2M_K^2-M_\pi^2)$, respectively, which coincide
    with those at the physical point.
    The red triangles denote the symmetric $m_s=m_{\ell}$ line, i.e.\
    $M_K=M_{\pi}$.
    The curves with error bands are the projections of the continuum limit fit
    onto these trajectories. The black circles are the experimental
    masses, corrected for QCD and QED isospin breaking effects, where
    $m_{\Xi}$ was used to set the scale, see
    eqs.~\eqref{eq:physical} and~\eqref{eq:physpoint}. The red band approaches
    the $\textmd{SU(3)}$ chiral limit.
    \label{fig:bchptall}}
\end{figure}
The resulting dependence on the pion mass, with the data points shifted
to the infinite volume and continuum limits,
is shown in figure~\ref{fig:bchptall}.
The scale $t_{0,\text{ph}}$ is then computed from the continuum
limit parametrization of the $\Xi$ baryon mass as a function
of the squared pion and kaon masses (all in units of $8t_0$),
according to eqs.~\eqref{eq:physical} and~\eqref{eq:physpoint}
via an iterative procedure. The uncertainties of the experimental
hadron masses that have been corrected regarding the up and down quark mass
splitting and electrical charge effects (see appendix~\ref{sec:isospin})
are implemented via pseudo-bootstrap distributions in the error
analysis. Once $t_{0,\text{ph}}$ has been determined, the physical
point on the $8t_0M_{\pi}^2$ axis is known too. For comparison,
we also show the (pseudo\nobreakdash-)experimental values, with isospin breaking
effects removed, of table~\ref{tab:contmasses} as
solid black circles in the figure. While the agreement with the experimental
value of $m_{\Xi}$ --- that has been used to set the scale --- is trivial,
the other three masses are predicted. The red curve with
$m_s=m_{\ell}$ approaches the value $\sqrt{8t_{0,\text{ch}}}m_0$
in the chiral limit. The kaon masses along the blue curves
are such that
$8t_0(2M_K^2+M_\pi^2)=8t_{0,\text{ph}}(2M^2_{K,\text{ph}}+M^2_{\pi,\text{ph}})$,
i.e.\ $\overline{M}^2$ is kept constant in units of $t_0$, while
the green curves correspond to
$8t_0(2M_K^2-M_\pi^2)=8t_{0,\text{ph}}(2M^2_{K,\text{ph}}-M^2_{\pi,\text{ph}})$,
i.e.\ the strange quark mass is kept approximately fixed.
These two sets of curves intersect at the physical point.
The original data are not exactly aligned along these two sets,
therefore, the blue squares and green circles depicted in the figure have
been shifted somewhat, according to the fit, to the respective kaon masses.

\begin{figure}[htp]
  \centering
  \resizebox{0.49\textwidth}{!}{\includegraphics[width=\textwidth]{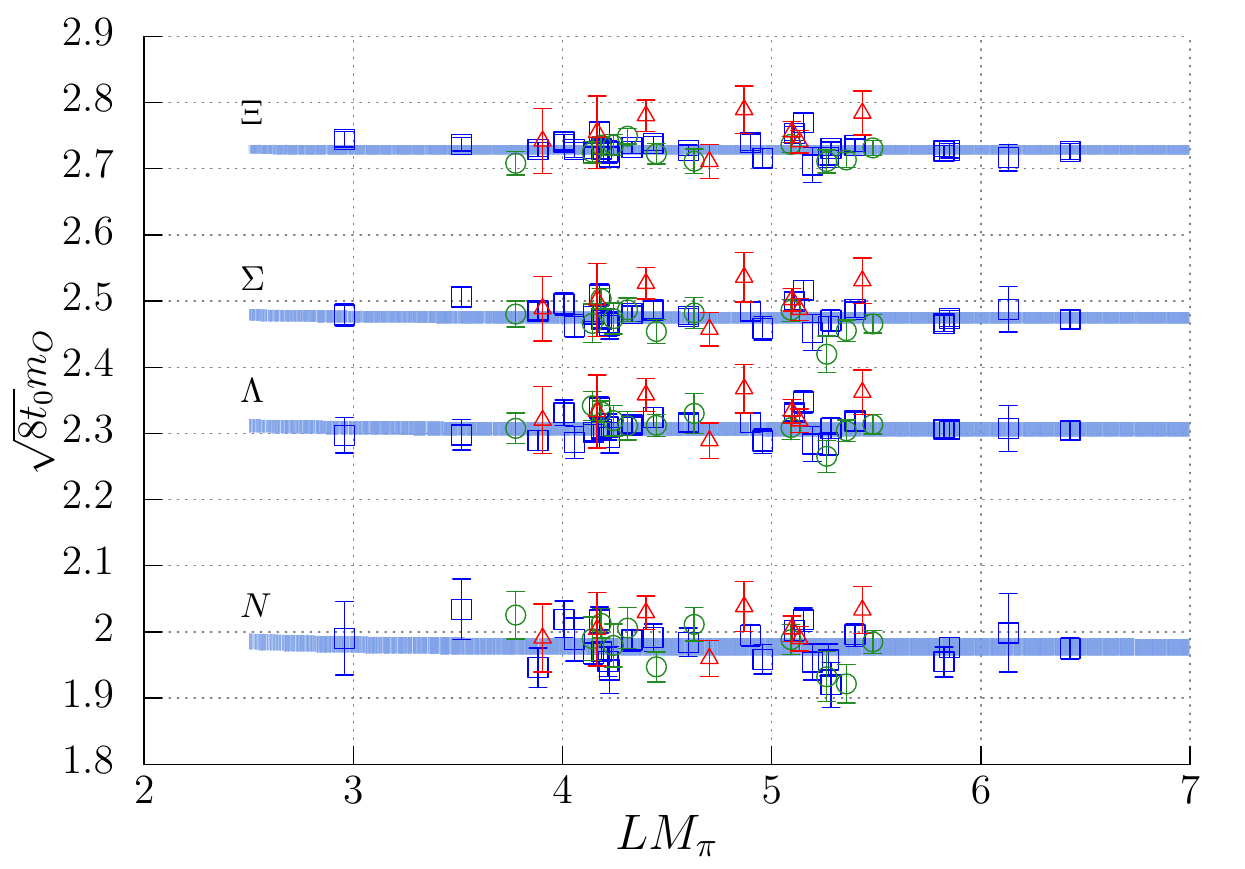}}
  \resizebox{0.49\textwidth}{!}{\includegraphics[width=\textwidth]{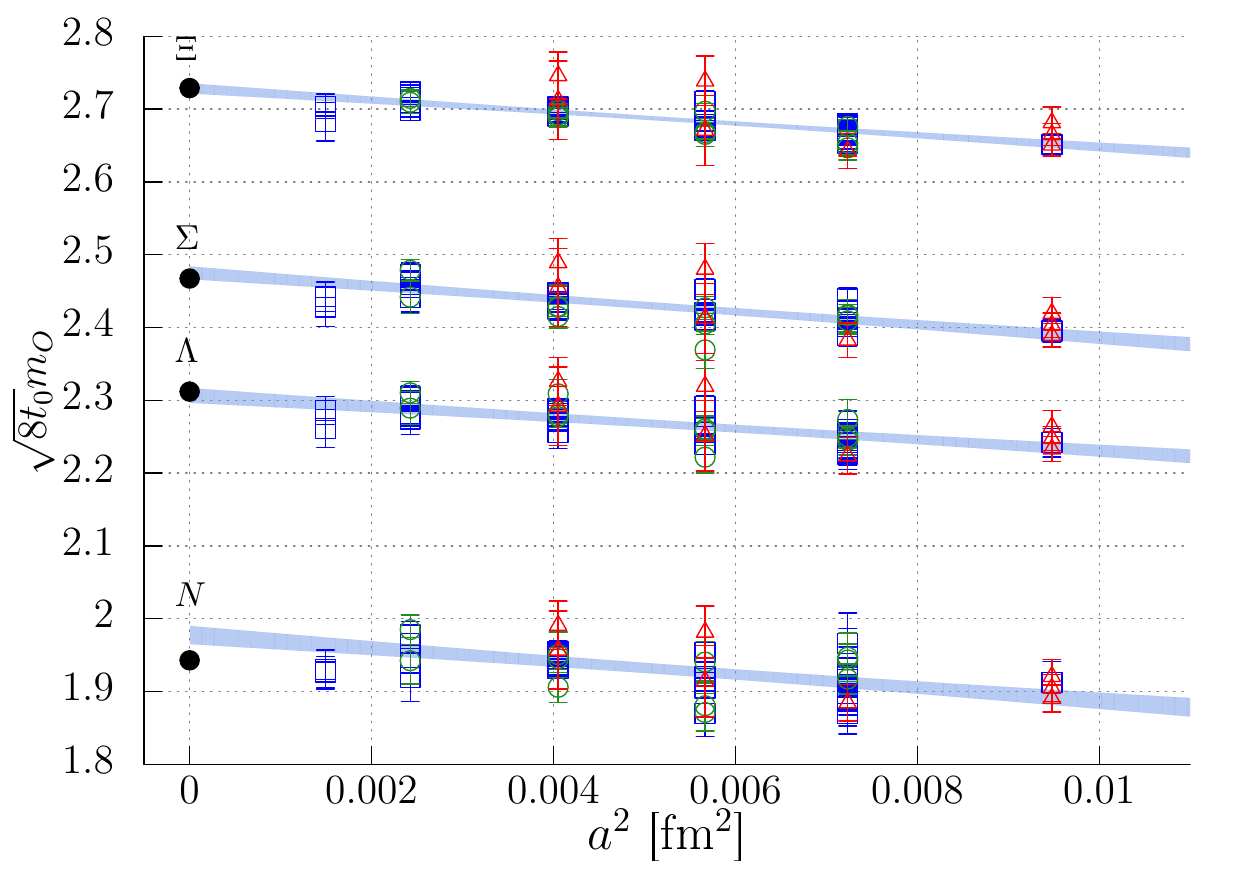}}
  \caption{The volume dependence (left) and the lattice spacing dependence
    (right) of the four octet baryon masses.
    The underlying octet BChPT~FV fit (see section~\ref{sec:naming})
    is identical to the one shown in
    figure~\ref{fig:bchptall}. Again, red triangles, blue squares and
    green circles correspond to the $m_s=m_{\ell}$, $\Tr M=\text{const}$
    and $\hat{m}_s\approx\text{const}$ quark mass trajectories, respectively.
    \label{fig:volaeff}}
\end{figure}
We show other projections of the same fit in figure~\ref{fig:volaeff}.
In this case all the data are projected to the physical point in terms of the
pion and kaon masses. On the left, additionally, these are also
projected to the continuum limit and shown as a function of the
linear spatial lattice dimension in units of the inverse pion mass.
The blue bands correspond to the fit. The data look rather flat
and also the volume dependence suggested by the BChPT~FV fit is
mild. Nevertheless, we will see that omitting finite volume effects
deteriorates the fit quality, without reducing the number of parameters.
In the right panel of figure~\ref{fig:volaeff} the dependence of the
data (projected onto the physical quark masses and infinite volume) on
the squared lattice spacing is shown. This is significant,
however, the slope of $\sqrt{8t_0}m_O$ as a function of $a^2$
is similar for all four baryons.
It is also obvious from this figure that there is some tension
between the nucleon mass determined by the fit and the experimental
point on the level of two to three standard deviations. However,
for the fit discussed above,
the $\chi^2$-value is larger than the number of degrees of freedom.
We will vary the fit ranges and parametrization to explore the
systematics. We start by determining the scale.
\subsection{Determination of the scale parameter $\boldsymbol{\sqrt{8t_{0,\text{ph}}}}$}
\label{sec:scale}
In order to explore the systematics, we first impose different cuts on the data
for the octet BChPT~FV fit (defined in section~\ref{sec:naming})
and then we consider different parametrizations.
Regarding the fit ranges,
\begin{itemize}
  \item we include or exclude our coarsest {\bf lattice spacing},
  \item we impose the {\bf volume} cuts $LM_\pi>4$, $LM_\pi>3.5$ or
    we use all our data with $L>2.3\,\textmd{fm}$ and
  \item we impose the cuts on the average {\bf meson mass}
    $12t_0\overline{M}\vphantom{M}^2<1.6$,
    $12t_0\overline{M}\vphantom{M}^2<1.4$ or
    $12t_0\overline{M}\vphantom{M}^2<1.25$, see figure~\ref{fig:masscut},
    where $12t_0\overline{M}\vphantom{M}^2\approx 1.09$ at the physical
    point.
\end{itemize}
This gives a total of 18 different fits with
values $\chi^2/N_{\text{DF}}$ ranging from 0.94 (with the cuts
$LM_\pi>4$ and $12t_0\overline{M}\vphantom{M}^2<1.25$) to
$\chi^2/N_{\text{DF}}=1.20$ (with the cuts $a<0.09\,\textmd{fm}$ and
$12t_0\overline{M}\vphantom{M}^2<1.6$ but no cut on the volume,
except for $L>2.3\,\textmd{fm}$).

We use the Akaike Information Criterion (AIC)~\cite{Akaike:1998zah} to
assign a weight
\begin{align}
  w_j= A \exp\left[-\frac{1}{2}\left(\chi^2_j-N_{\text{DF},j}+k_j\right)\right]
\end{align}
to the result of each fit $j\in\{1,\ldots,N_M\}$,
see, e.g.,  eq.~(161) of the e-print version of ref.~\cite{Borsanyi:2020mff}
and references therein.\footnote{Recently, instead of
subtracting $N_\text{DF}-k$ from the $\chi^2$-value in the exponent,
in ref.~\cite{Jay:2020jkz} it has been suggested to subtract
$-2n_{\text{cut}}-2k=\text{const}+2N_{\text{DF}}$ instead, where $n_{\text{cut}}$
is the number of removed data points. This seems counter-intuitive:
for a good fit $\chi^2\sim N_{\text{DF}}$, therefore, this change
results in a very strong preference for fits that include as many
data points as possible, even if the corresponding $\chi^2/N_{\text{DF}}$-values
are significantly larger.}
The normalization $A$ is such that $\sum_i^{N_M} w_i = 1$.
$\chi^2_j$ denotes the $\chi^2$-value of the fit $j$, $k_j$
the number of fit parameters and $N_{\text{DF},j} = n_j - k_j$
the number of degrees of freedom. The above equation extends the AIC
to also include fits where the number of data points $n_j$ is varied
and not only the fit function. It assumes no correlations between
the removed and the remaining data points. Moreover, the parametrization
should not vary with the included data. Both criteria are satisfied
since we exclude whole ensembles from the analysis, while maintaining the same
parametrizations. One concern with this procedure may be the preference
for fits with low $\chi^2$-values, even if these are much smaller than
$N_{\text{DF}}$. However, this is not a problem for our current
analysis where for the ``best'' fit that enters the averaging we obtain
$\chi^2/N_{\text{DF}}\approx 0.94$, which, given the error on the error,
is not significantly different from one.

For each parameter $a$ that we are interested in, we generate for
each fit $j$ a bootstrap distribution of $a_j$, $a_j(b)$, with $N_b$ bootstrap
samples $b$. We then determine the one-$\sigma$ confidence interval
by sorting the values $a_j(b)$ in ascending order and then determining
the 15.9\% and 84.1\% percentiles, counting each entry $a_j(b)$
according to its weight $w_j/N_b$. The central values that we quote
correspond to the median of this distribution. Usually, this is very
close to the naive weighted average $\sum_jw_ja_j$.

\begin{figure}[htp]
  \centering
  \resizebox{0.95\textwidth}{!}{\includegraphics[width=\textwidth]{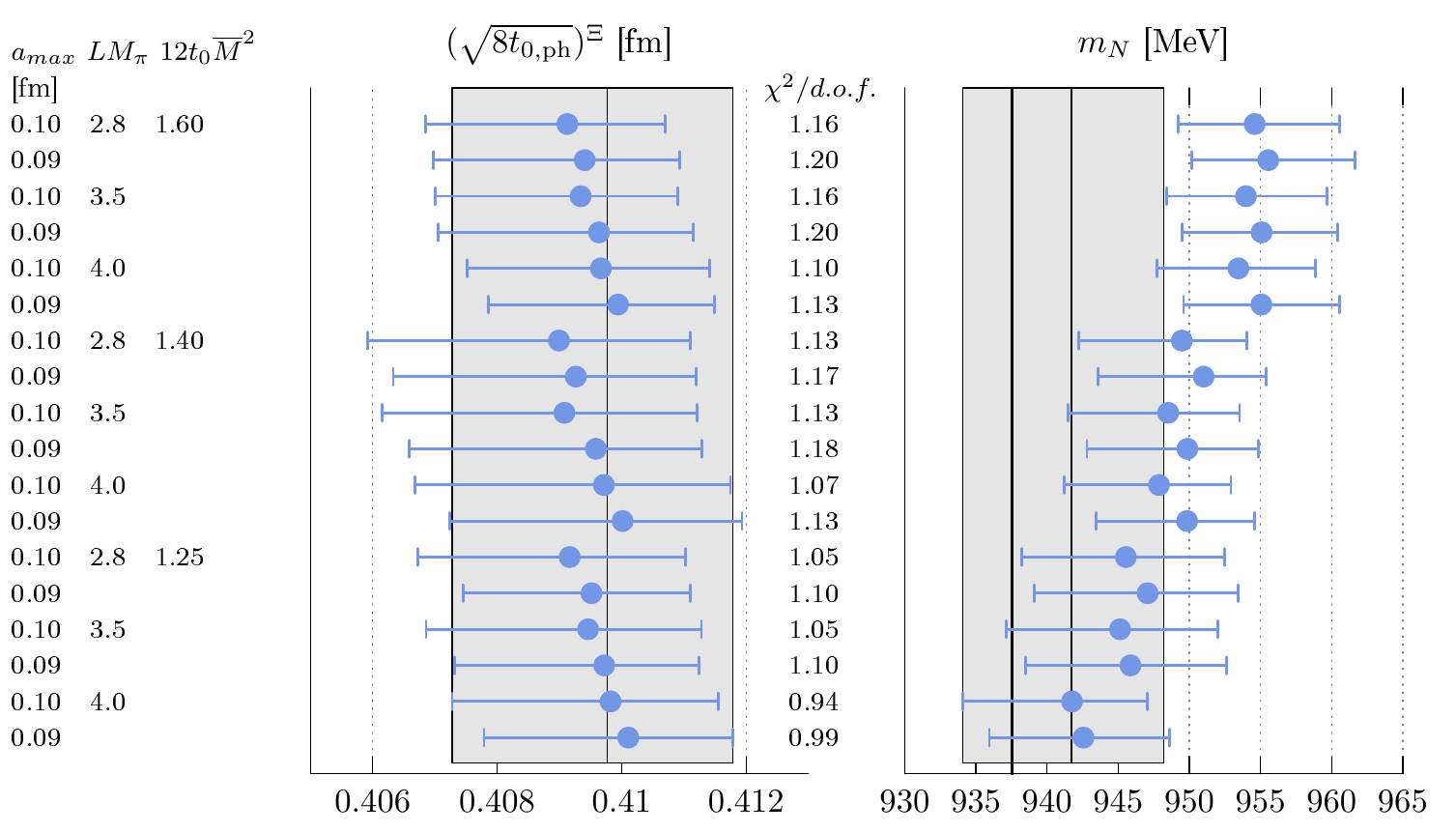}}
  \caption{Results for $\sqrt{8t_{0,\text{ph}}}$ determined from the mass
    of the $\Xi$ baryon, together with results for
    the nucleon mass $m_N$ from the
    BChPT~FV fits (see section~\ref{sec:naming}).
    On the left the cuts on the lattice spacing, the
    spatial lattice extent in units of the pion mass and
    the average pseudoscalar meson mass are indicated. In the centre
    the corresponding $\chi^2/N_{\text{DF}}$-values are shown. The grey
    boxes correspond to the 68\% confidence intervals
    of the AIC averaging procedure. Also indicated as a thick
    vertical line is the experimental nucleon mass, corrected for
    isospin breaking effects.
    \label{fig:t0average}}
\end{figure}

The procedure is illustrated for $\sqrt{8t_{0,\text{ph}}}$, obtained from
$m_{\Xi}$, and the nucleon mass $m_N$
in figure~\ref{fig:t0average}, where the grey boxes with thin vertical
lines (indicating the central values) correspond to the outcome of the
AIC averaging procedure. Also shown as a thick vertical line is the
experimental nucleon mass, corrected for QCD and QED isospin breaking effects.
Within our uncertainty of about $7\,\textmd{MeV}$, we are able to reproduce
this mass. Note that the fit discussed in section~\ref{sec:illustrate}
above corresponds to the first fit shown in figure~\ref{fig:t0average}
with $\chi^2/N_{\text{DF}}=1.16$, which overestimates the nucleon mass
by almost three standard deviations. While $t_{0,\text{ph}}$ is
quite independent of the data cuts, the nucleon mass decreases
systematically with the cut on $12t_0\overline{M}\vphantom{M}^2$.
This can easily be understood from figure~\ref{fig:bchptall}: the
$\Xi$ baryon data are quite precise and its physical point value is
well constrained by the intersection of the two
trajectories in the quark mass plane. Therefore, it depends only
mildly on the parametrization and the cuts. In contrast, not only are
the statistical errors of the nucleon mass data larger but also
the two trajectories that intersect at the physical point are not
very different from each other, resulting in a less precise
determination.

\begin{figure}[htp]
  \centering
  \resizebox{0.95\textwidth}{!}{\includegraphics[width=\textwidth]{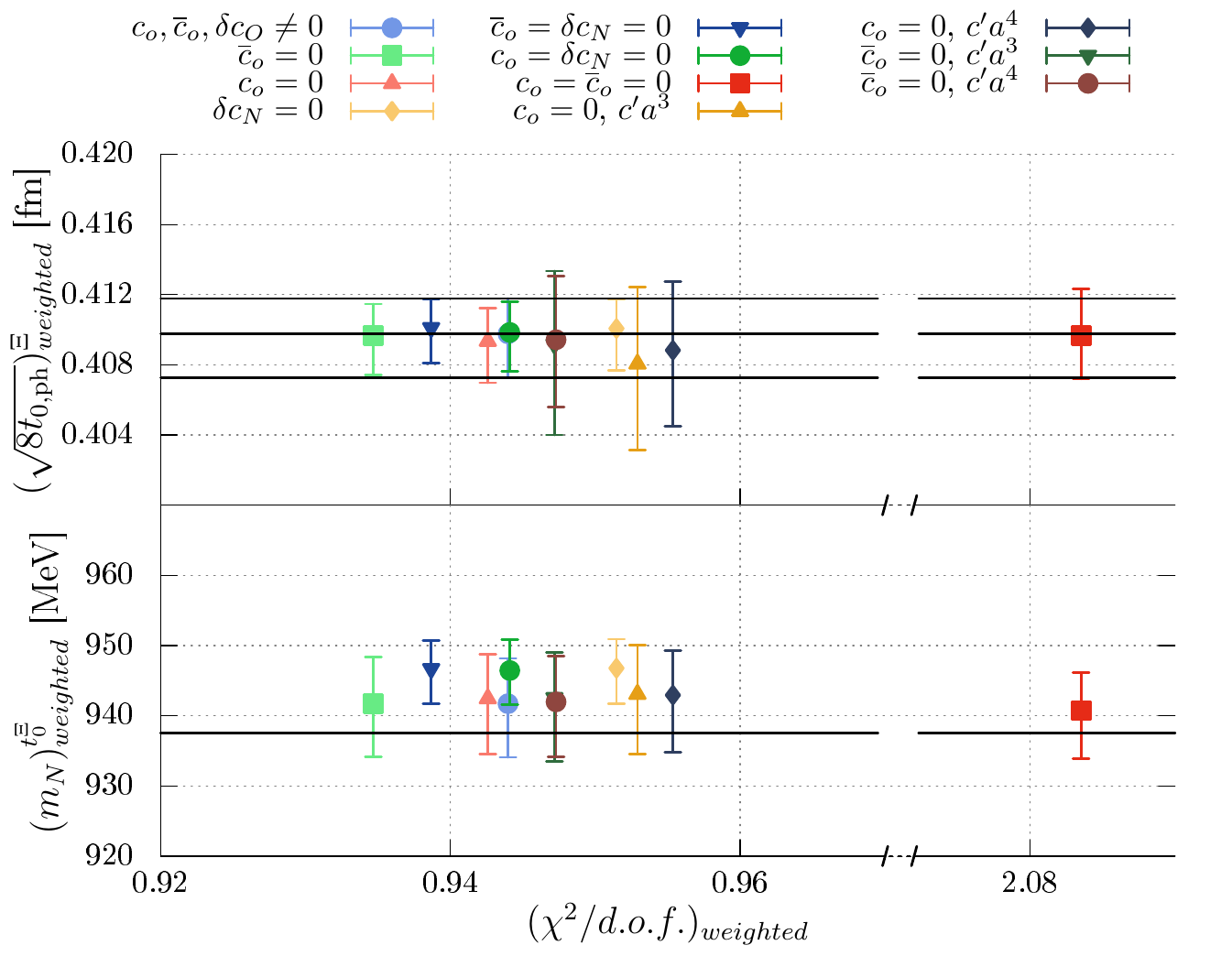}}
  \caption{Results for $\sqrt{8t_{0,\text{ph}}}$ and the nucleon mass from the
    octet BChPT~FV fit (see section~\ref{sec:naming}),
    when varying the lattice spacing corrections, setting different
    discretization terms within eq.~\eqref{eq:fit2} to zero
    and/or adding $a^3$ or $a^4$ terms as
    indicated in the labelling.
    For each fit form the result from the AIC averaging over the 18
    different ensemble cuts is plotted against the corresponding
    reduced $\chi^2$-value, also AIC averaged over the 18 cuts.
    The horizontal lines displayed
    for $\sqrt{8t_{0,\text{ph}}}$ correspond to our final result, obtained
    with $c_o,\bar{c}_o,\delta c_N,\delta c_\Lambda \delta c_\Sigma,\delta c_\Xi \neq 0$, also indicated by the blue circle.
    The isospin corrected experimental value of the nucleon mass is also
    shown as a black horizontal line.
  \label{fig:t0aeff}}
\end{figure}
We investigate different parametrizations of the lattice spacing
dependence. For $O\in\{N,\Lambda,\Sigma,\Xi\}$,
we assumed the generic form $(a^2/8t_0^*)(c_o+\bar{c}_o\overline{M}\vphantom{M}^28t_0
+\delta c_O\delta\overline{M}\vphantom{M}^28t_0)$, see eq.~\eqref{eq:fit2}.
We carry out our AIC procedure separately for 10 additional functional
forms for these effects. The results are shown in figure~\ref{fig:t0aeff}.
Within our standard octet BChPT~FV fit ($c_o,\bar{c}_o,\delta c_O\neq 0$),
$\delta c_N$ is statistically compatible with zero. This motivates us
to carry out an additional 11-parameter fit ($\delta c_N=0$) and
two 10-parameter fits ($\delta c_N=c_o=0$ and $\delta c_N=\bar{c}_o=0$).
All these fits give
somewhat larger nucleon masses but have little impact on
$t_{0,\text{ph}}$. The latter two choices result in
slightly smaller AIC weighted $\chi^2/N_{\text{DF}}$-values than the
original ansatz, due to the reduced number of parameters.
Setting either $c_o=0$ or $\bar{c}_o=0$ alone gives even smaller
$\chi^2/N_{\text{DF}}$-values.
In contrast, setting $c_o=\bar{c}_o=0$ gives a very large
$\chi^2$ (red square): an $a^2$ correction term that does not
depend on $\delta M$ is required, however, the fit cannot
discriminate between $c_o$ and $\bar{c}_o$ since, except for some
less precise $m_s=m_{\ell}$  data points, $8t_0\overline{M}\vphantom{M}^2$
varies only very little. This also means that it is impossible to
achieve stable fits, when adding $a^3$ or $a^4$ lattice corrections
without either setting $c_o=0$ or $\bar{c}_o=0$. Therefore, we replace either
the $c_o$-term or the $\bar{c}_o$-term by
$c'a^3(t_0^*)\vphantom{t_0}^{-3/2}$ or
$c'a^4(t_0^*)\vphantom{t_0}^{-2}$. The two $c_o=0$ choices somewhat
decrease the quality of the fit while the two $\bar{c}_o=0$
choices do not significantly affect $\chi^2/N_{\text{DF}}$.

Carrying out the AIC procedure
over all the 198 results that are obtained when combining the
18 cuts with the 11 different parametrizations gives similar values
for $t_{0,\text{ph}}$ and $m_N$ with
slightly smaller errors than the original fit. All the
fit forms that decrease the $\chi^2/N_{\text{DF}}$ are based on
reducing the number of parameters in a way that is physically not well
motivated. It must be stressed, however, that (with the exception of
the $c_o=\bar{c}_o=0$ fit) the variations in terms of the
reduced $\chi^2$-value are minimal.
We conclude that the systematics regarding
the continuum limit extrapolation are already covered within the
AIC error bar of our original fit.

\begin{figure}[htb]
  \centering
  \resizebox{0.95\textwidth}{!}{\includegraphics[width=\textwidth]{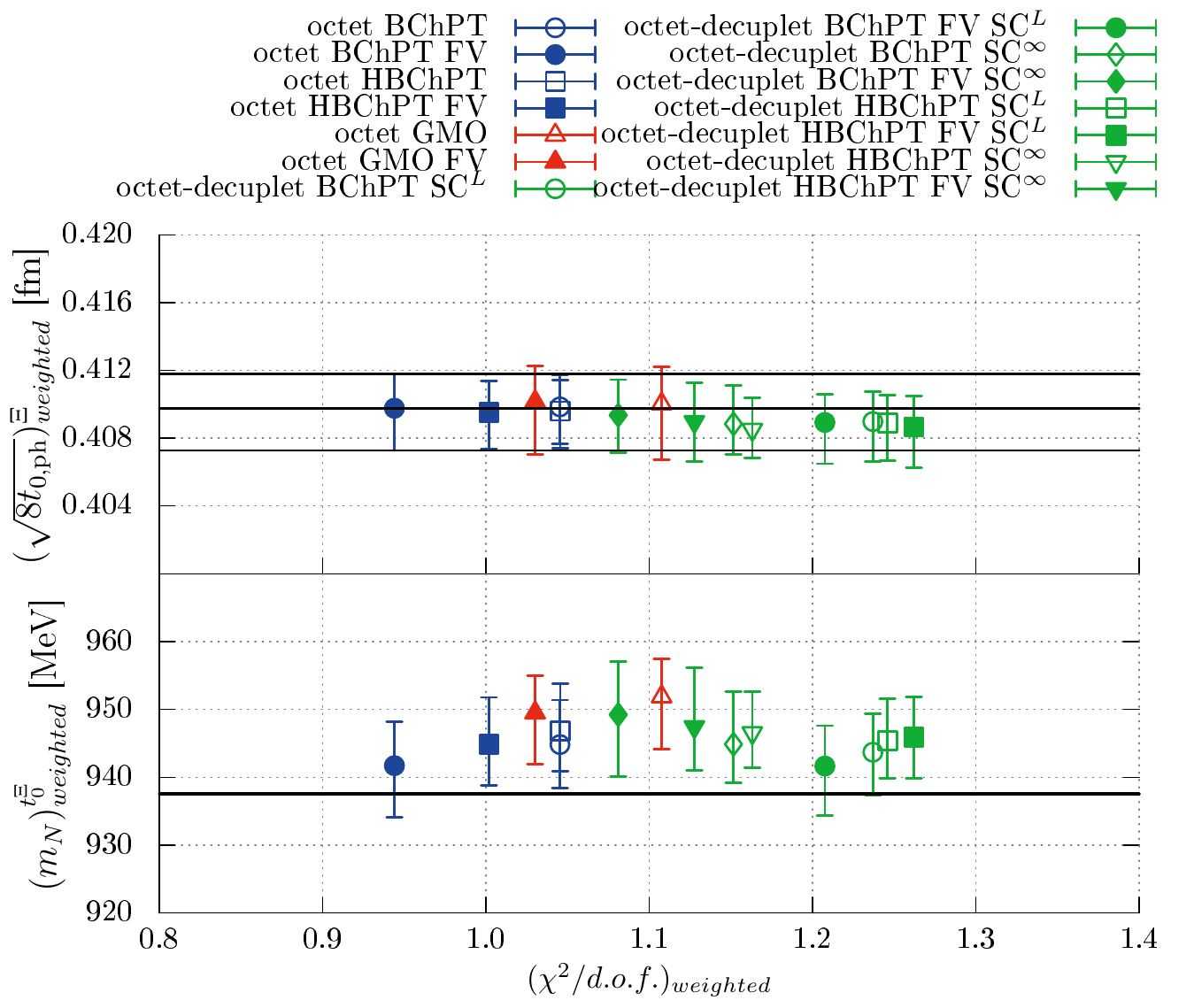}}
  \caption{Results for $\sqrt{8t_{0,\text{ph}}}$ and the nucleon mass obtained
    from the AIC procedure for different continuum parametrizations, see
    section~\ref{sec:naming}.
    Blue symbols correspond to the
    octet (H)BChPT parametrization, red symbols to the octet GMO
    fits and green symbols to simultaneous octet-decuplet (H)BChPT fits.
    Solid symbols include finite
    volume effects. Circles and diamonds correspond to BChPT,
    squares and downward triangles to HBChPT. Regarding the joint
    octet-decuplet fits, only data for stable decuplet baryons were
    included. We adopt two stability cuts: ``SC$^L$'' stands for
    stability in the finite volume, while ``SC$^\infty$'' denotes stability
    even for an infinite volume.
  \label{fig:t0fitfo}}
\end{figure}

Finally, in figure~\ref{fig:t0fitfo} we investigate the impact of
different parametrizations of the continuum dependence of the
baryon masses on the meson masses and the volume.
The naming conventions and references to the explicit parametrizations
can be found in section~\ref{sec:naming}.
Solid symbols include finite volume (FV) effects, open symbols do not.
The best fit (solid blue circle, octet BChPT~FV) is the one discussed above.
Its HBChPT equivalent is the next best fit. The GMO parametrization
has a larger number of parameters (17 instead of 12) but also a larger
$\chi^2/N_{\text{DF}}$. This polynomial expansion does not include finite
volume effects. Therefore, for the GMO~FV fit we correct the data ``by hand'',
using the finite volume corrections from the octet BChPT~FV fit discussed in
section~\ref{sec:illustrate}. Additionally, we display results from simultaneous
octet-decuplet (H)BChPT fits,
where we carry out two stability cuts regarding the $\Delta$,
$\Sigma^*$ and $\Xi^*$ baryons which at the physical point can strongly decay
into $N\pi$, $\Lambda\pi$ or $\Sigma\pi$ and $\Xi\pi$, respectively,
namely requiring stability in the infinite volume limit (SC$^\infty$) or
stability in the finite volume (SC$^L$). These fits contain 23 free parameters,
12 accounting for lattice artefacts, and 11 LECs. However,
the number of data points is much larger since eight instead of four
baryon masses are fitted simultaneously.

With one exception, the quality of the fits improves when finite volume
effects are included. In principle, one could AIC average the HBChPT~FV and the
BChPT~FV results, however, due to the better $\chi^2$-values, the
result would be
dominated by the EOMS BChPT parametrizations.
Note that no AIC averaging can be carried out across the three
classes of parametrizations, i.e.\ (H)BChPT, octet-decuplet (H)BChPT and
GMO, because the data sets are not the same. Only in the
octet-decuplet (H)BChPT fits do the decuplet masses contribute while within the
GMO fits no cut on $\phi_4=12t_0\overline{M}\vphantom{M}^2$ is imposed and data
with $\phi_4>1.6$ are included. Similarly, it is also not possible
to average over the two classes of simultaneous octet-decuplet fits
(SC$^L$ and SC$^\infty$) as different sets of data points are excluded
per ensemble in each case. Finally, we remark that
for each of the continuum parametrizations
investigated, we also varied the form of the lattice spacing dependence
as discussed above and reached similar conclusions as for the
octet BChPT~FV fits.

All methods give very consistent results and errors. Therefore, as
our final result we quote the AIC average of the BChPT~FV
fit to the octet baryon masses:
\begin{equation}
  \label{eq:t0result}
  \sqrt{8t_{0,\text{ph}}}=0.4098^{(20)}_{(25)}\,\textmd{fm}.
\end{equation}
We find $\sqrt{8t_0^*}$ to be smaller by $10^{-4}\,\textmd{fm}$ than this
value, see eq.~\eqref{eq:t0starp}.
\subsection{The baryon spectrum}
\label{sec:barspec}
\begin{figure}[htp]
  \centering
  \resizebox{0.49\textwidth}{!}{\includegraphics[width=\textwidth]{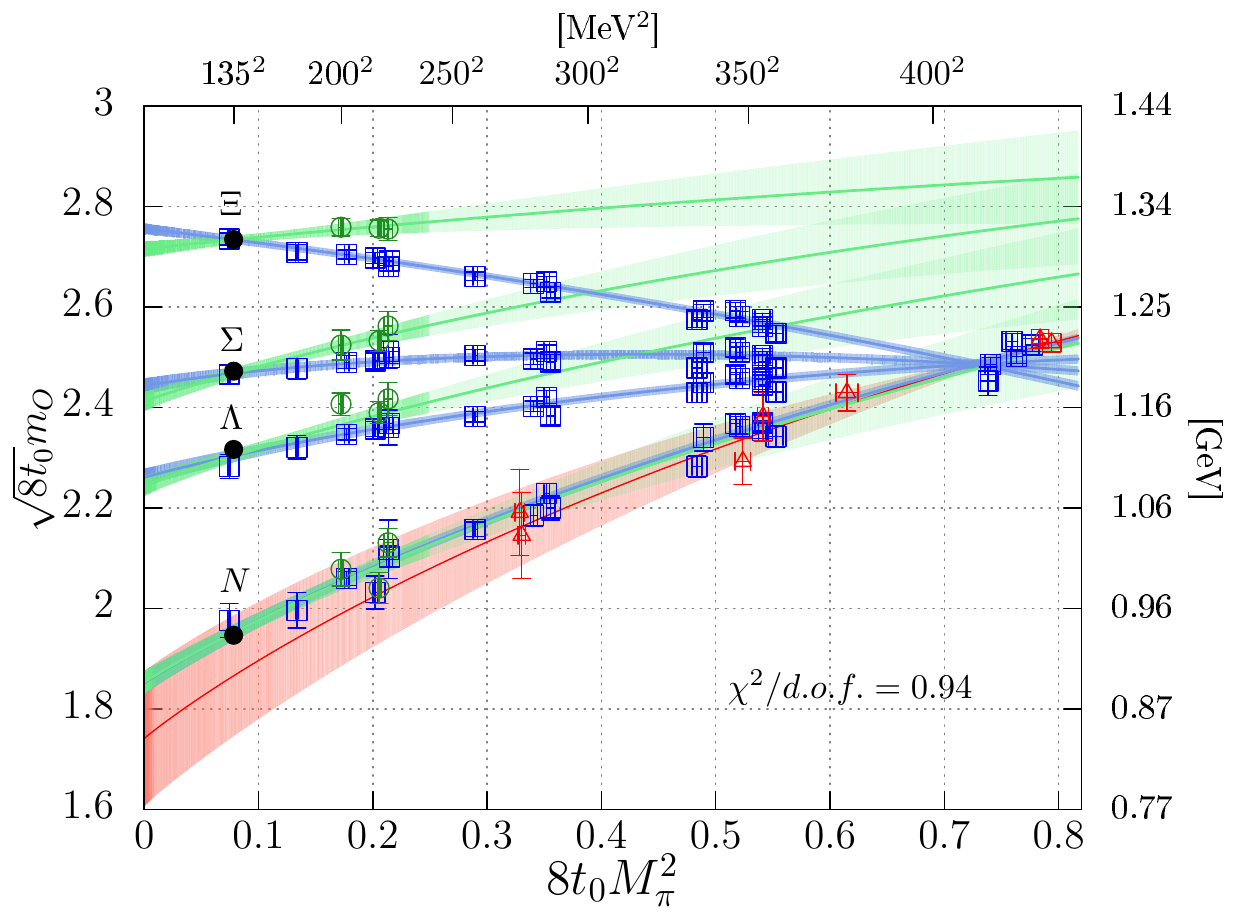}}
  \resizebox{0.49\textwidth}{!}{\includegraphics[width=\textwidth]{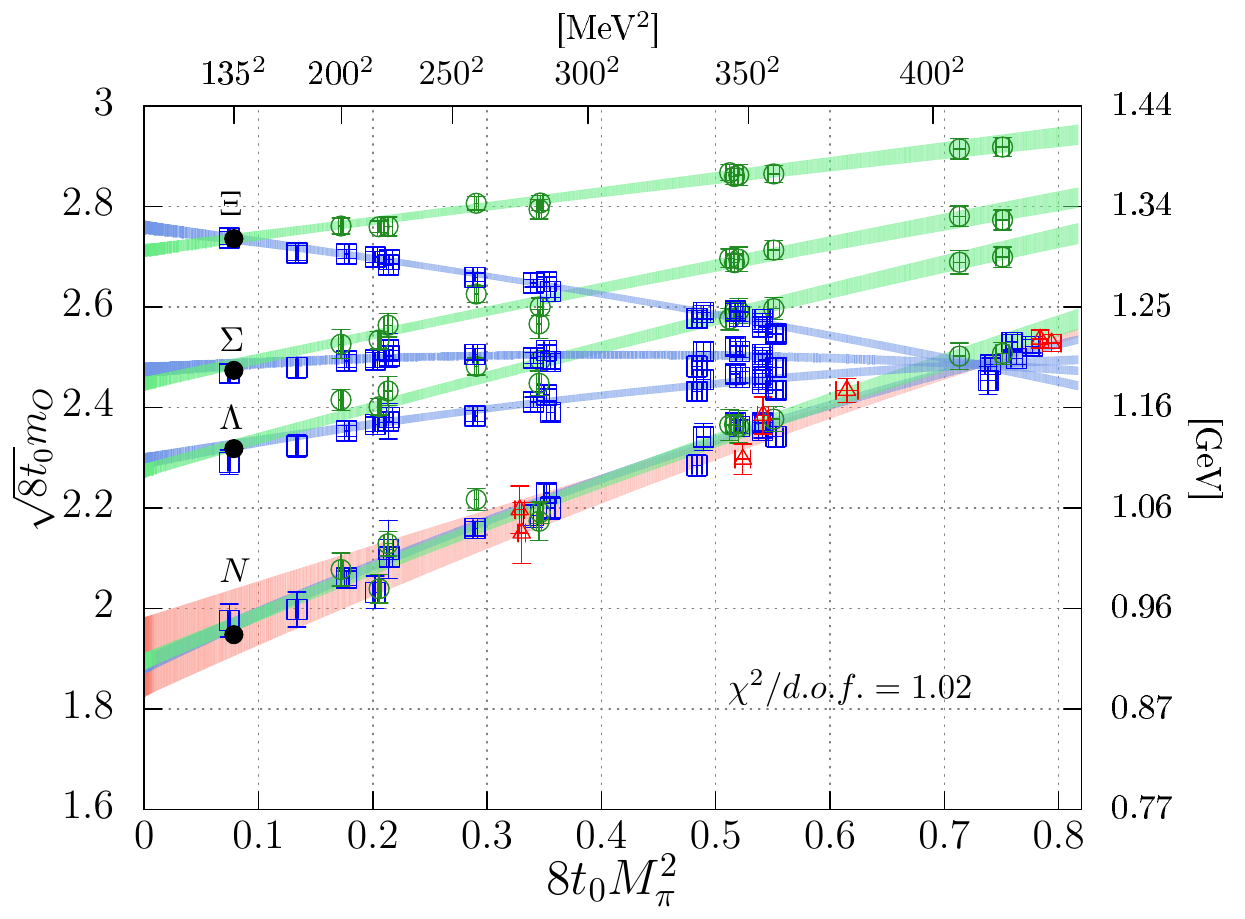}}
  \caption{Left: the same octet BChPT~FV fit as in figure~\ref{fig:bchptall}
    but with the cuts $\phi_4>1.25$ and $LM_{\pi}>4$ that resulted in
    the smallest $\chi^2/N_{\text{DF}}$.
    The lighter parts of the green error bands indicate the region that has
    been discarded from the fit. Right: the best polynomial 
    fit (GMO~FV)
    to the octet baryon spectrum. As in figure~\ref{fig:bchptall} the
    data points have been shifted to the continuum, infinite volume limit
    and to the kaon masses that correspond
    to the respective trajectories in the quark mass plane.
    References to the formulae used can be found in section~\ref{sec:naming}.
    \label{fig:bestoct}}
\end{figure}
As a by-product of the scale setting procedure outlined above
we also determine the baryon spectrum. The fits to the octet
baryon masses and our AIC model averaging procedure
have already been explained and discussed in sections~\ref{sec:illustrate}
and~\ref{sec:scale}. Examples for the nucleon mass were shown
next to $\sqrt{8t_{0,\text{ph}}}$ in
figures~\ref{fig:t0average}, \ref{fig:t0aeff} and~\ref{fig:t0fitfo}.
A fit to the whole octet was illustrated in figure~\ref{fig:bchptall}.
In the left panel of figure~\ref{fig:bestoct} we show the corresponding
octet BChPT~FV fit that carried the highest weight in the
AIC averaging procedure,
with the data shifted in the same way to $a=0$, $L=\infty$ and
to kaon masses corresponding to the correct physical point trajectories
as in section~\ref{sec:illustrate}. The data cuts for this fit
with $\chi^2/N_{\text{DF}}=0.94$
were $LM_{\pi}>4$ and $\phi_4>1.25$. For comparison, in the right panel
we show the continuum limit projection of our best polynomial fit
(GMO~FV, $\chi^2/N_{\text{DF}}=1.02$). In this case, the volume
was restricted to $LM_{\pi}>4$ and all data with $8t_0M_{\pi}^2<0.8$ were
included. No further cut in the quark mass plane was imposed since
restricting the parameter governing the convergence of the expansion,
$\delta M^2$, would have excluded the physical point.
In this fit form only group theory constraints
are implemented but no assumption is made regarding the interactions.
Therefore, the number of parameters entering the continuum limit
parametrization is 11 instead of 6, however, more data points
are included in the fit. Both fit forms
give adequate descriptions of the data and result in similar
predictions regarding the scale parameter and the baryon masses.
However, in terms of the fit quality, the BChPT parametrization
seems to be preferred by the data.

\begin{figure}[htp]
  \centering
  \resizebox{0.49\textwidth}{!}{\includegraphics[width=\textwidth]{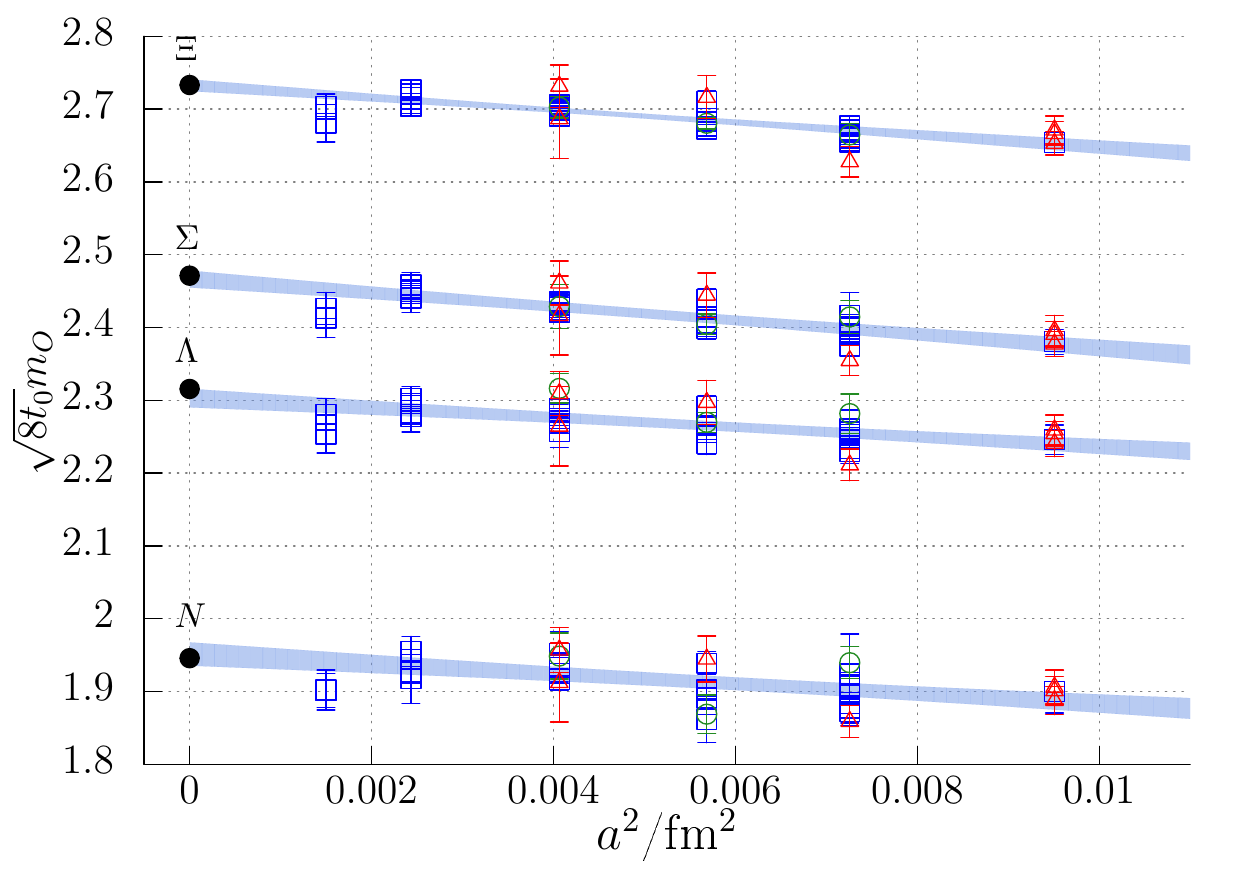}}
  \resizebox{0.49\textwidth}{!}{\includegraphics[width=\textwidth]{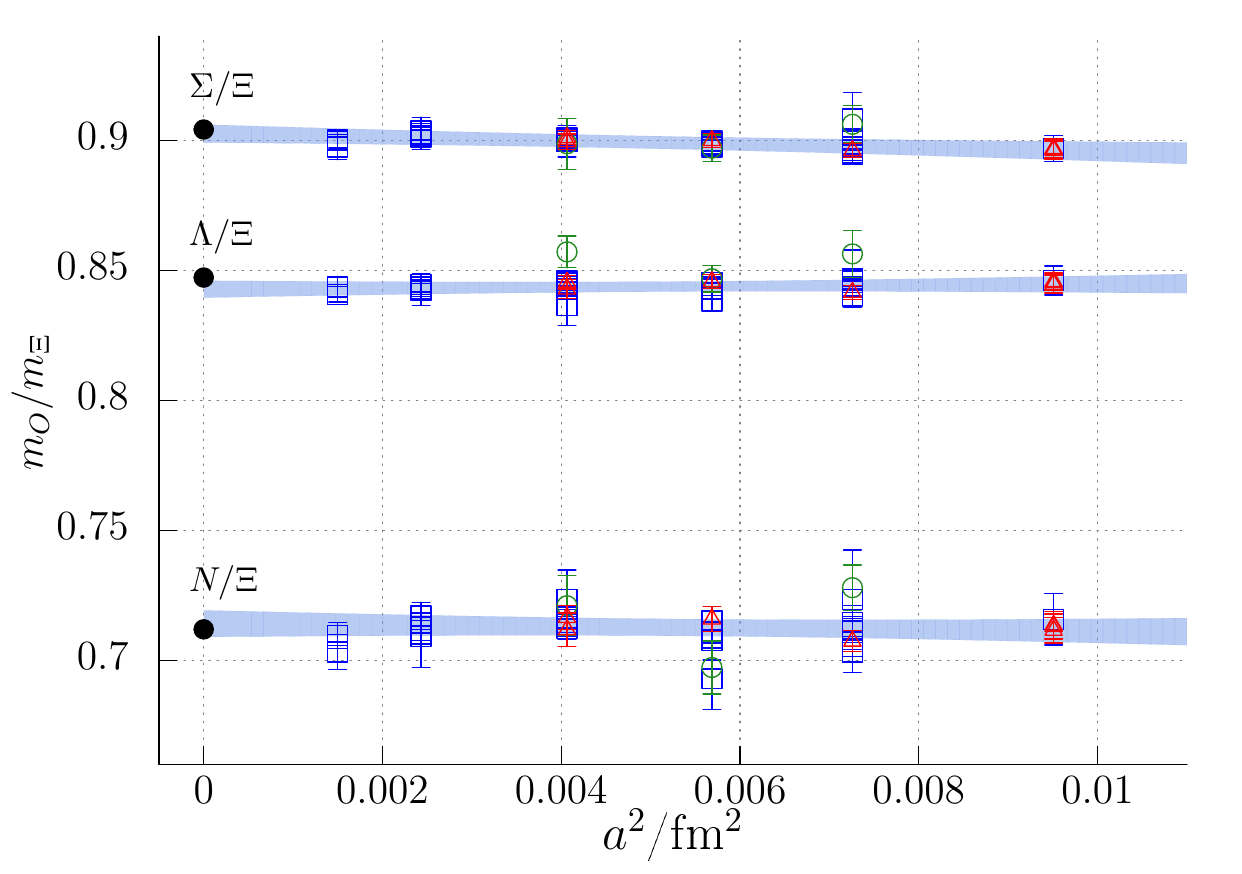}}
  \caption{Continuum limit extrapolations for the octet baryon masses.
    The data are shifted to the infinite
    volume and the physical point in the quark mass plane according to
    the octet BChPT~FV parametrization (see section~\ref{sec:naming}),
    using the fit with the best quality
    ($\chi^2/N_{\text{DF}}=0.94$).
    Red triangles, blue squares and
    green circles correspond to the $m_s=m_{\ell}$, $\Tr M=\text{const}$
    and $\hat{m}_s\approx\text{const}$ quark mass trajectories, respectively.
    Left: the lattice spacing dependence of the masses in units of
    $\sqrt{8t_{0,\text{ph}}}$. Right: the lattice spacing dependence
    of ratios of the baryon masses over the mass of the $\Xi$.
    \label{fig:bestaeff}}
\end{figure}
In the left panel of figure~\ref{fig:bestaeff} we show for our best fit
the dependence of the combinations $\sqrt{8t_0}m_O$ at the physical point
in the infinite volume limit
on the squared lattice spacing: there is a difference of about 3\% between
the values obtained on our coarsest lattice ($a\approx 0.098\,\textmd{fm}$)
and the continuum limit. Note that in this representation we
are only sensitive to four combinations of the six independent order
$a^2$ terms since the pion and kaon masses are set to the physical ones.
In the right panel we display ratios of the other octet baryon masses over
the mass of the $\Xi$. We find that
ratios of octet baryon masses have a much smaller dependence
on $a$, which we cannot resolve within our present errors,
than the combinations $\sqrt{8t_0}m_O$.
Note that unlike in the fit shown in the right panel of
figure~\ref{fig:volaeff}, that carries a very small weight in the
AIC averaging procedure, the nucleon mass agrees with the physical one within
errors.

Finite volume effects have been discussed in sections~\ref{sec:finite}
and illustrated for the baryons in
figures~\ref{fig:fsebaryon} and~\ref{fig:volaeff}. The corresponding
analytical expression eq.~\eqref{eq:baryonfse}
can be found in appendix~\ref{sec:finitebaryon}.
These are generally mild for our lattice sizes, however, they can have an
impact on the fit quality even for $LM_{\pi}>4$ as can be seen
in figure~\ref{fig:t0fitfo} above; with one exception the fits including the
finite volume terms (full symbols) give smaller reduced $\chi^2$-values.
Note that the number
of fit parameters is not affected by this. In the approach of polynomially
expanding in $\delta M^2$ and $\overline{M}\vphantom{M}^2$ (GMO),
no statement can be made about finite volume effects. In this case, we
just adjust the data for the effects predicted by BChPT, which results in
smaller $\chi^2$-values for the GMO fits.

\begin{figure}[htp]
  \centering
  \resizebox{0.49\textwidth}{!}{\includegraphics[width=\textwidth]{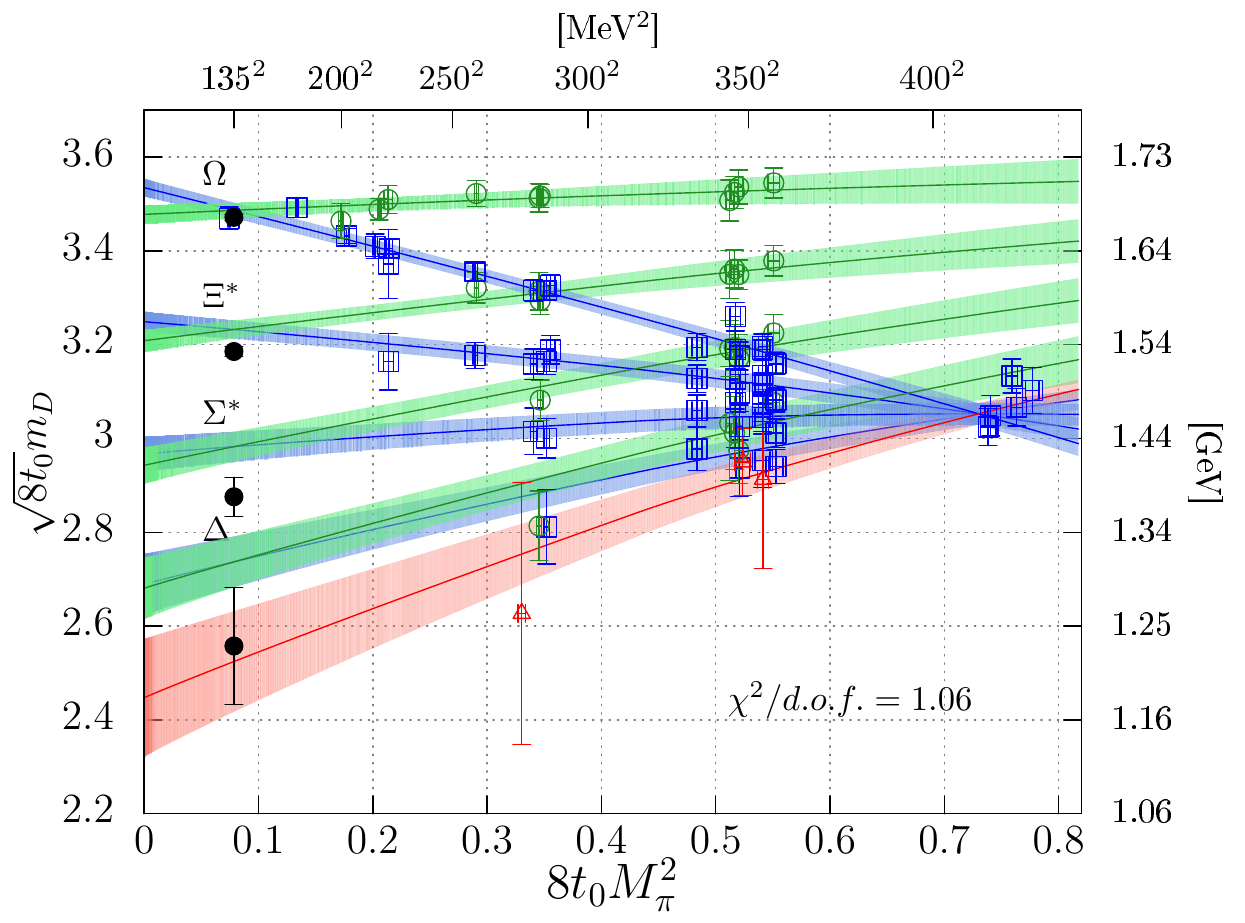}}
  \resizebox{0.49\textwidth}{!}{\includegraphics[width=\textwidth]{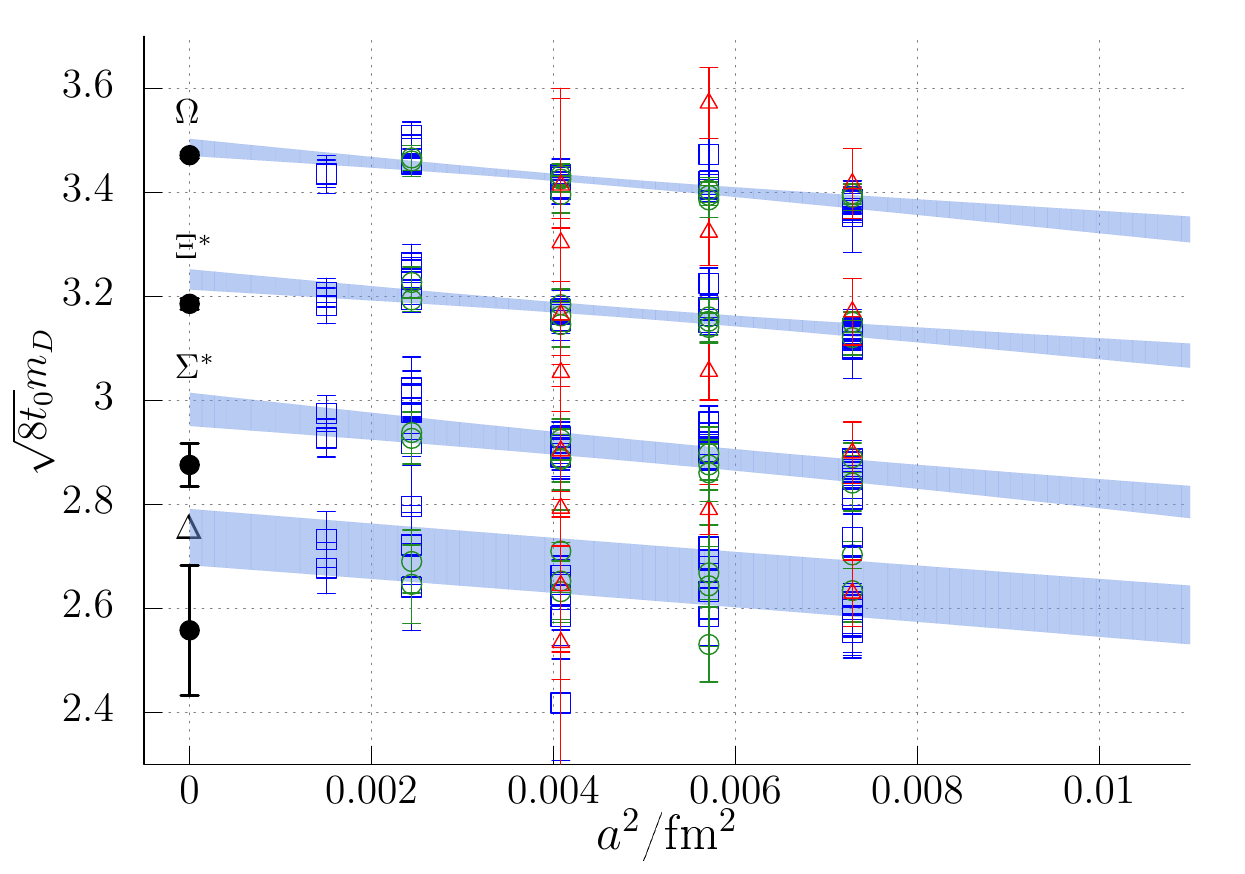}}
  \caption{Results in the decuplet sector for a simultaneous octet-decuplet
    BChPT~FV~SC$^{\infty}$ fit (see section~\ref{sec:naming}).
    Shown is the fit with highest
    weight that enters the AIC procedure. The data cuts are $\phi_4<1.6$,
    $a<0.09\,\textmd{fm}$ and $LM_{\pi}>4$. Decuplet masses that can strongly
    decay into a pion and an octet baryon for an infinite volume
    (SC$^{\infty}$) are ignored and not displayed. The experimental results
    (black circles) correspond to the Breit--Wigner masses plus and minus
    half the Breit--Wigner widths. Left: the continuum limit dependence on
    $M_{\pi}^2$. Right: the dependence on $a^2$.
    \label{fig:decup}}
\end{figure}
In figure~\ref{fig:decup} we show the results of a simultaneous
octet-decuplet BChPT~FV~SC$^\infty$ fit for the decuplet sector
(for details, see section~\ref{sec:naming}).
The corresponding results for the octet baryons are very similar to those
shown in previous figures. The scale is set self-consistently
by requiring that $m_{\Xi}$ assumes the physical mass at the physical point,
see eqs.~\eqref{eq:physical} and~\eqref{eq:physpoint}.
Not shown or fitted are masses of decuplet baryons that can
strongly decay to a pion and an octet baryon in an infinite volume
since this goes beyond the formalism that we apply here.
The $\Omega$ baryon mass agrees with the expectation. In principle,
we could also have used this to set the scale. However, the
error is larger than for the $\Xi$ baryon. The results for the
other three baryons are slightly above
the sums of the Breit--Wigner masses and the half-widths: in the regime
in which we discard the data also the prediction has to be taken
with a grain of salt since we neglect that the poles of the resonances
acquire imaginary parts.

In the right panel of figure~\ref{fig:decup} we show the $a^2$
dependence of $\sqrt{8t_0}m_D$ at the physical point. Also in this case
the difference between the results at our coarsest lattice spacing
(not included into this particular fit) and in the continuum limit
are about 3\%. Again, ratios comprised of a decuplet baryon mass
divided by an octet baryon mass (not shown), are independent
of $a$ within our present errors.

\begin{figure}[thp]
  \centering
  \resizebox{0.95\textwidth}{!}{\includegraphics[width=\textwidth]{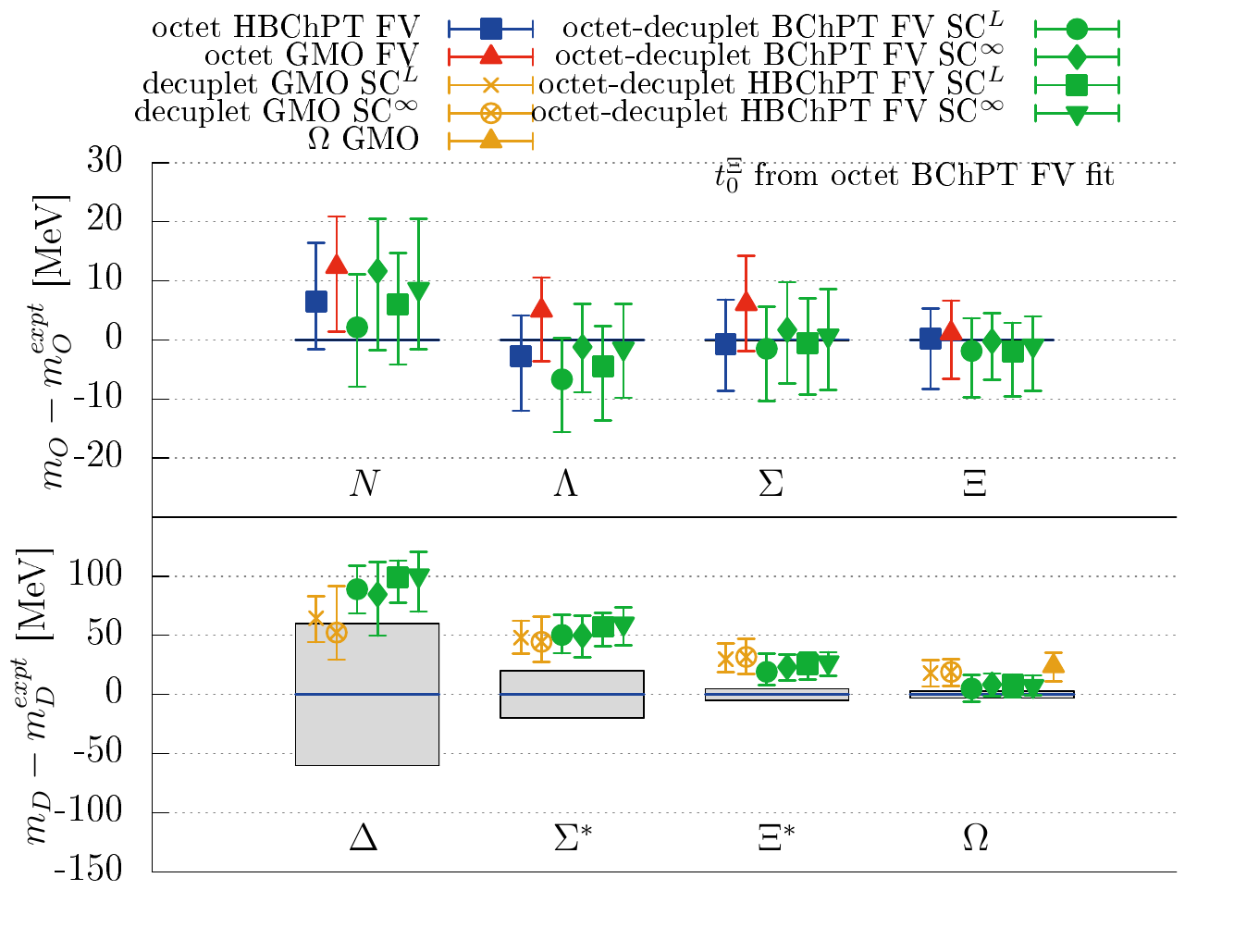}}
  \caption{Comparison of the various fit results for the light baryon masses
    with the experimental values. The scale is set using the
    $m_{\Xi}$ mass from the octet BChPT~FV fit. For the fit forms
    already covered, the symbols are the same as those used in
    figure~\ref{fig:t0fitfo} and the labelling corresponds to that
    of section~\ref{sec:naming}. For the $\Omega$ baryon alone we
    carry out an additional nine parameter fit ($\Omega$ GMO) with terms
    proportional to
    $\overline{\mathbbm{M}}\vphantom{M}^2$,
    $\delta \mathbbm{M}^2$, $\overline{\mathbbm{M}}\vphantom{M}^2\delta \mathbbm{M}^2$,
    $\overline{\mathbbm{M}}\vphantom{M}^4$, $\delta \mathbbm{M}^4$,
    $\mathbbm{a}^2$, $\mathbbm a^2\mathbbm{M}\vphantom{M}^2$
    and $\mathbbm{a}^2\delta \mathbbm{M}^2$.
    \label{fig:specover}}
\end{figure}
\begin{table}[htb]
  \caption{\label{tab:specresults}
    Comparison of our physical point continuum limit predictions for
    the lower lying light baryon masses (with all
    systematics included in the error) with the ``experimental'' expectations of
    iso-symmetric QCD (table~\ref{tab:contmasses}). For orientation,
    for the unstable baryons $\Delta$, $\Sigma^*$ and $\Xi^*$, we give
    the Breit--Wigner masses, with the respective half-widths as errors.
    The $\Xi$ baryon mass of the octet BChPT~FV fits
    was used to set the scale. The fit functions are referenced
    in section~\ref{sec:naming}.
}
\begin{center}
    \begin{tabular}{ccccc}
      \toprule
fit& $m_N$ & $m_\Lambda$ &  $m_\Sigma$ & $m_\Xi$ \\\midrule
octet BChPT FV &  941.7$^{(6.5)}_{(7.6)}$ &  1110.0$^{(4.2)}_{(4.2)}$ & 1188.4$^{(4.8)}_{(3.5)}$ & ---\\      
octet-decuplet BChPT FV~SC$^\infty$ & 949.2$^{~(8.8)}_{(13.4)}$ &1114.5$^{(7.3)}_{(7.7)}$  & 1192.4$^{(8.1)}_{(9.1)}$ & 1316.7$^{(4.7)}_{(6.5)}$  \\
expt. & 937.54(6)  & 1115.68(1) & 1190.66(12) & 1316.9(3)\\\midrule
fit& $m_\Delta$ & $m_{\Sigma^*}$ &  $m_{\Xi^*}$ & $m_\Omega$ \\\midrule
octet-decuplet BChPT FV SC$^\infty$ & 1314.6$^{(27.2)}_{(34.9)}$ & 1432.9$^{(16.8)}_{(18.5)}$ &1555.2$^{(10.3)}_{(11.6)}$ & 1678.0$^{~(9.3)}_{(10.0)}$ \\
expt. & 1230(60)  & 1383(20) & 1532(5) & 1669.5(3.0)\\\bottomrule
  \end{tabular}
\end{center}
\end{table}
In figure~\ref{fig:specover} we compare our results for the baryon masses
from various fits
to our expectations, subtracting the (pseudo)-experimental values
$m_B^{\text{expt}}$. For the (H)BChPT parametrizations, we always include
finite volume effects. The symbols are the same as those used in 
figure~\ref{fig:t0fitfo}. Again, for the decuplet baryons we implement
two cuts (in addition to those on the maximal $\phi_4$):
only considering baryons that are stable in the finite volume
(SC$^L$) and baryons that are stable in the infinite volume (SC$^\infty$).
In addition, we carry out GMO fits to the decuplet baryon masses
(orange symbols). We also show the result of such a fit, using
the $\Omega$ baryon mass alone (orange triangle). The results of the
octet BChPT~FV fits
and of the octet-decuplet BChPT~FV~SC$^\infty$
fits are displayed in
table~\ref{tab:specresults} and shown in the summary
figure~\ref{fig:baryonspectrum} below. For all the results
shown in figure~\ref{fig:specover},
the scale $\sqrt{8t_{0,\text{ph}}}$ was set
from the mass of the $\Xi$ baryon, obtained via the octet BChPT~FV fits.
In general we see good agreement between the different
results and for the octet baryons also with the experimental masses.
Regarding the $\Omega$ baryon, with the decuplet GMO fits there is a tension on
the level of 1.5 to 2 standard deviations with experiment, while the
octet-decuplet BChPT
and HBChPT fits agree. The values of
table~\ref{tab:specresults} indicate that within our errors
of $4\,\textmd{MeV}$ (for the $\Lambda$ and $\Sigma$),
$7\,\textmd{MeV}$ (for the nucleon) and $10\,\textmd{MeV}$
(for the $\Omega$), we reproduce the
experimental masses. The differences regarding the unstable
decuplet baryons have already been addressed above.

\begin{figure}[thp]
  \centering
  \resizebox{0.95\textwidth}{!}{\includegraphics[width=\textwidth]{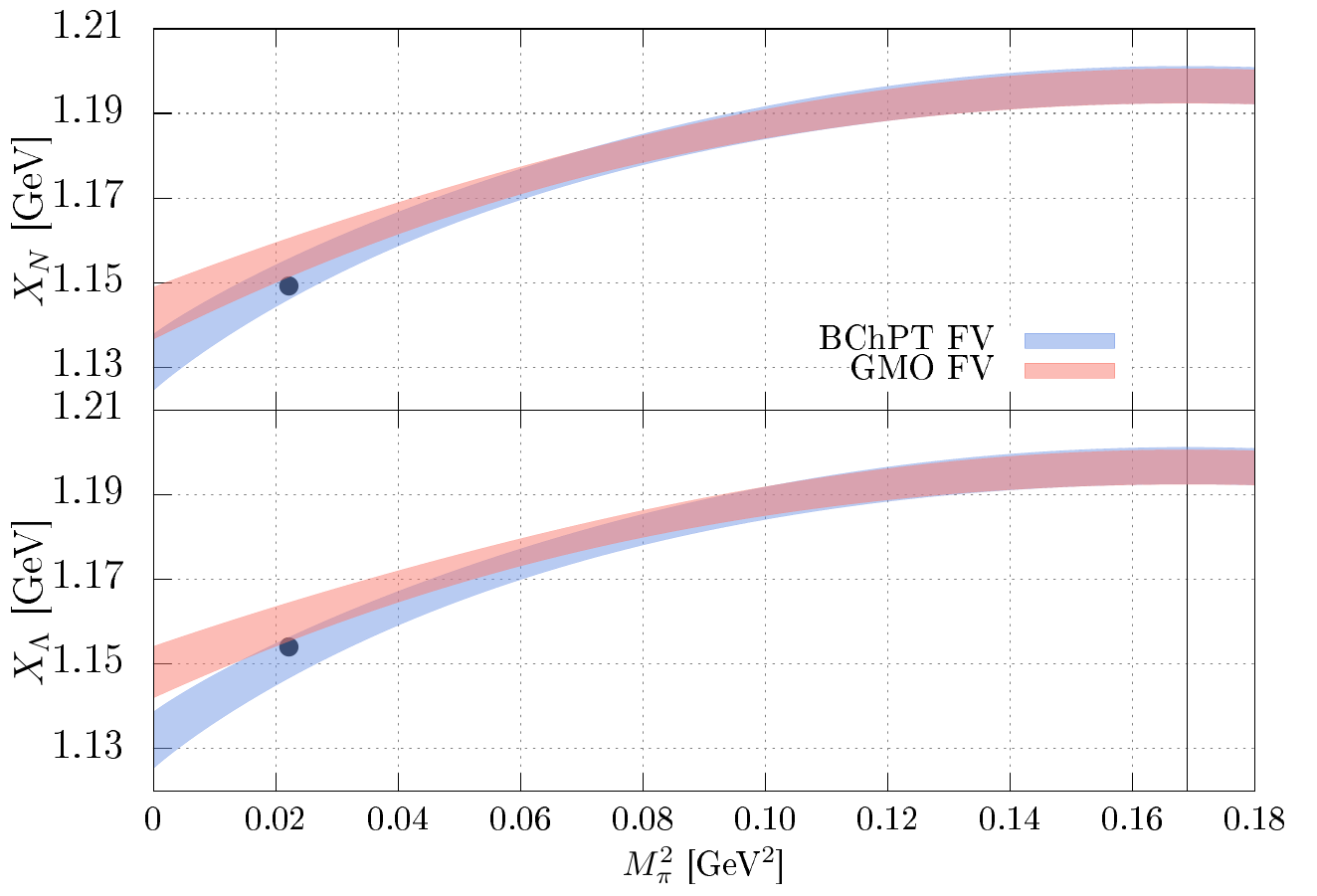}}
  \caption{The mass averages $X_N$ (top) and $X_\Lambda$ (bottom)
    defined in eq.~\eqref{eq:singmas} as functions of $M_\pi^2$, keeping
    $3X_{\pi}^2=3\overline{M}\vphantom{M}^2=2M_K^2+M_\pi^2=2M_{K,\text{ph}}^2+M_{\pi,\text{ph}}^2$ fixed, in the continuum
    limit. Shown are the results of the best BChPT~FV (blue bands) and
    GMO~FV (red bands) fits to the baryon octet. The vertical line denotes the
    flavour symmetric point $M_\pi=M_K=411\,\textmd{MeV}$ where all baryon
    masses are equal. The black points are the experimental values.
    The parametrizations used are referenced in section~\ref{sec:naming}.
    \label{fig:average}}
\end{figure}
Finally, we investigate the breaking of $\textmd{SU(3)}$ flavour symmetry
in the baryon masses. Following ref.~\cite{Bietenholz:2011qq}, we
define the averages
\begin{align}
  \label{eq:singmas}
  X_N=\frac{1}{3}\left(m_N+m_\Sigma+m_\Xi\right),\quad
  X_\Lambda=\frac12\left(m_{\Lambda}+m_\Sigma\right),\quad
  X_\pi^2=\frac13\left(M_\pi^2+2M_K^2\right)=\overline{M}\vphantom{M}^2
\end{align}
that do not receive any linear contributions in the
flavour breaking parameter $\delta m=m_s-m_{\ell}$.
In figure~\ref{fig:average}
we show continuum limit results for $X_N$ and $X_\Lambda$ as functions of
$M_\pi^2$ along the line where the flavour averaged pseudoscalar mass
is fixed to its experimental value $X_\pi=411\,\textmd{MeV}$.
This approximately corresponds to our $\Tr M=\text{const}$ trajectory
in the quark mass plane. 
Shown are the results of the best BChPT~FV fit (blue bands) and the
best polynomial fit (GMO~FV, red bands) for $X_N$ (top) and $X_\Lambda$ (bottom).
Since $b_D$ and $b_F$ cancel from these combinations and
$b_0$ is fixed for $X_\pi=\text{const}$,
any dependence on $M_\pi^2$ is due to $F$ and $D$
in the BChPT case and the two combinations
$\delta e_N+\delta e_\Sigma+\delta e_\Xi$ and
$\delta e_\Lambda+\delta e_\Sigma$ for the GMO parametrization
eq.~\eqref{eq:gmo1}.
The scale was set from the mass of the $\Xi$ baryon, using the
octet BChPT~FV fit, which explains why the experimental
values (black circles) are
not in the centre of the error bands. The vertical line denotes the
symmetric point $M_\pi=M_K=X_\pi$, where all the octet baryon masses agree.
As we have not carried out the AIC averaging but just selected the
fit ranges that gave the smallest $\chi^2/N_{\text{DF}}$, the error bands do not
include all systematics. The difference, however, is not big; as
an example one may compare the grey band to the entry with
$\chi^2/N_{\text{DF}}=0.94$ in the right panel of figure~\ref{fig:t0fitfo}.

The BChPT parametrizations of $X_N$ and $X_\Lambda$ are not identical
but they are extremely close to each other, whereas differences between the
GMO curves are visible. As for the individual baryon masses, the curvature
suggested by BChPT is somewhat larger than that of the polynomial
expansion. However, in the range $M_\pi>127\,\text{MeV}$ where we
have data, the parametrizations agree within errors.
The difference between $X_N$ and $X_\Lambda$ at the
physical point cannot be resolved within our present
accuracy but both predictions agree with experiment
(solid circles). The experimental value for this difference
reads $X_N-X_{\Lambda}\approx -5\,\textmd{MeV}$. This is
related to the Gell-Mann--Okubo relation~\cite{GellMann:1962xb,Okubo:1961jc}
\begin{equation}
  \frac{m_N+m_\Xi}{2}-\frac{3\Lambda+\Sigma}{4}=\frac{3}{2}(X_N-X_\Lambda)
  \approx 0
\end{equation}
and expected to be small since the average masses only differ by
quadratic and higher
order terms in the $\textmd{SU(3)}$ symmetry breaking parameter
$\delta m$ from their values at the symmetric point, where these must agree.
However, we see that both averages at the physical point
differ by over $40\,\textmd{MeV}$ from the value at $m_s=m_{\ell}$
(vertical line). There is no obvious (group theoretical or other)
reason why this effect has the same sign or a similar size for
both averages, $X_N$ and $X_{\Lambda}$.
In view of the size of the observed higher than linear order corrections,
the accuracy of the GMO relation is somewhat surprising.
It could easily have been violated by as much as $50\,\textmd{MeV}$
rather than $5\,\textmd{MeV}$.
One may speculate that the variation of $X_N$ and $X_{\Lambda}$
as a function of $M_{\pi}^2$ when $\overline{M}=X_{\pi}$ is kept constant
may dominantly be due to $\textmd{SU(3)}$ breaking
effects in the mesonic sector. However, this is hard to reconcile with
section~\ref{sec:t02}, where we found $\phi_{4,\text{opt}}\approx
\phi_{4,\text{ph}}$ ($\phi_4=12t_0X^2_\pi$).

\subsection{The $\boldsymbol{\sigma}$ terms of the octet baryons and the
  $\boldsymbol{\Omega}$ baryon}
\label{sec:sigma}
We define the baryon sigma terms as
\begin{equation}
  \sigma_{qB}=m_q\frac{\partial m_B}{\partial m_q},\quad
  \sigma_{\pi B}=\sigma_{uB}+\sigma_{dB}.
\end{equation}
Following appendix~\ref{sec:sigmadef}, we extract these
terms from the dependence of the
mass of baryons $B$ on the pion and kaon masses, using the
GMOR relations. We carry this out for all
the octet baryons as well as for the $\Omega$ baryon. We distinguish between
$\tilde{\sigma}_{\pi B}$ and $\sigma_{\pi B}$ as well as between
$\tilde{\sigma}_{sB}$ and $\sigma_{sB}$, where the $\tilde{\sigma}$ terms
refer to the logarithmic derivatives with respect to squared pion masses, see
their definition eq.~\eqref{eq:sigmagmor}, while the usual $\sigma$~terms
are the derivatives with respect to the quark masses as defined above
(and in eq.~\eqref{eq:sigmadef}). The difference is due to the corrections
to the GMOR relations that are worked out in
appendix~\ref{sec:gmor2},\footnote{We take into account
the leading dependence of $t_0$ on the meson masses
when determining $\tilde{\sigma}$ from the rescaled
variables (see eq.~\eqref{eq:correctt0}).
For $\sigma$ we also include the higher order terms
eq.~\eqref{eq:correctt1} (as well as
the NLO corrections to the GMOR relations).}
where we also determine the required combinations of
mesonic NLO LECs, see eq.~\eqref{eq:l4569}.
As can be seen from the results in table~\ref{tab:sigmaresults},
the differences between assuming the GMOR relations and accounting
for the violations are negligible for the pion $\sigma$~terms.
While the differences between $\sigma_{sB}$ and $\tilde{\sigma}_{sB}$
are sizeable for the less-well constrained strange $\sigma$~terms,
also in this case these are much smaller than the respective
errors.

\begin{table}[htb]
  \caption{\label{tab:sigmaresults}
    The $\sigma$~terms for the octet baryons and for the $\Omega$ baryon. The $\tilde{\sigma}_{\pi B}$
    and $\tilde{\sigma}_{sB}$ are obtained from the pseudoscalar mass dependence of the
    baryon masses, assuming the GMOR relations to be valid. The usual $\sigma$~terms take violations
    of the GMOR relations into account. The results are obtained
    using the octet~BChPT~FV and, for the $\Omega$ baryon, the
    octet-decuplet~BChPT~FV parametrizations (see section~\ref{sec:naming}),
    including AIC averaging analogous to
    figure~\ref{fig:t0average}.
  }
  \begin{center}
    \begin{tabular}{cccccccc}
      \toprule
      $B$&$\tilde{\sigma}_{\pi B}$/$\textmd{MeV}$&$\sigma_{\pi B}$/$\textmd{MeV}$&$\tilde{\sigma}_{sB}$/$\textmd{MeV}$&$\sigma_{sB}$/$\textmd{MeV}$\\
      \midrule
      $N$      &44.0$^{(4.4)}_{(4.7)}$&43.9$^{(4.7)}_{(4.7)}$&~~~4$^{(59)}_{(61)}$&~16$^{(58)}_{(68)}$\\
      $\Lambda$&27.6$^{(4.3)}_{(4.9)}$&28.2$^{(4.3)}_{(5.4)}$&113$^{(63)}_{(60)}$&144$^{(58)}_{(76)}$\\
      $\Sigma$ &24.9$^{(4.6)}_{(5.0)}$&25.9$^{(3.8)}_{(6.1)}$&194$^{(68)}_{(61)}$&229$^{(65)}_{(70)}$\\
      $\Xi$    &10.1$^{(4.4)}_{(5.4)}$&11.2$^{(4.5)}_{(6.4)}$&267$^{(70)}_{(68)}$&311$^{(72)}_{(83)}$\\
      $\Omega$ &~5.8$^{(5.5)}_{(3.8)}$&~6.9$^{(5.3)}_{(4.3)}$&391$^{(92)}_{(56)}$&421$^{(89)}_{(59)}$
  \\\bottomrule
  \end{tabular}
\end{center}
\end{table}

To the order of BChPT implemented here, it is consistent to ignore
corrections to the GMOR relations. Nevertheless, we choose
to remove this small source of systematic uncertainty
(see appendix~\ref{sec:gmor2}), in particular,
since these corrections are visible in the squared meson mass data as
functions of the quark masses.
Regarding the baryon mass dependence on the pseudoscalar masses,
within the present statistical errors 
we do not need to go beyond $\mathcal{O}(p^3)$, as we have seen above.
The results shown in
table~\ref{tab:sigmaresults} are consistent with the naive expectation that
(a) the sum of the $\sigma$~terms becomes bigger with the baryon mass
and (b) that baryons with a larger number of strange valence quarks
have bigger strange $\sigma$~terms but smaller pion $\sigma$~terms.
Within errors the $\sigma$~terms for the $\Lambda$, the $\Sigma$ and the $\Xi$
baryons agree with previous lattice
determinations~\cite{Shanahan:2012wh,Durr:2015dna}.
We refrain from predicting the $\sigma$~terms of the unstable decuplet
baryons and we are unaware of previous lattice calculations of the
$\sigma$~terms for the $\Omega$ baryon. However, predictions from
fits to collections of
lattice data on octet and decuplet baryon masses can also be found,
e.g., in refs.~\cite{MartinCamalich:2010fp,Ren:2012aj,Ren:2013oaa,Lutz:2014oxa,Lutz:2018cqo}.
We discuss $\sigma_{\pi N}$ in more detail in section~\ref{sec:resultsummary}
below. It is evident from our distribution of ensembles, see, e.g.,
figure~\ref{fig:masscut}, that the strange quark mass is varied
very little close to the physical point. Therefore, the strange
$\sigma$~terms at present are not very well constrained. This can be
improved upon by direct determinations of the respective matrix
elements~\cite{Petrak:2021aqf}.

\begin{figure}[htp]
  \centering
  \resizebox{0.95\textwidth}{!}{\includegraphics[width=\textwidth]{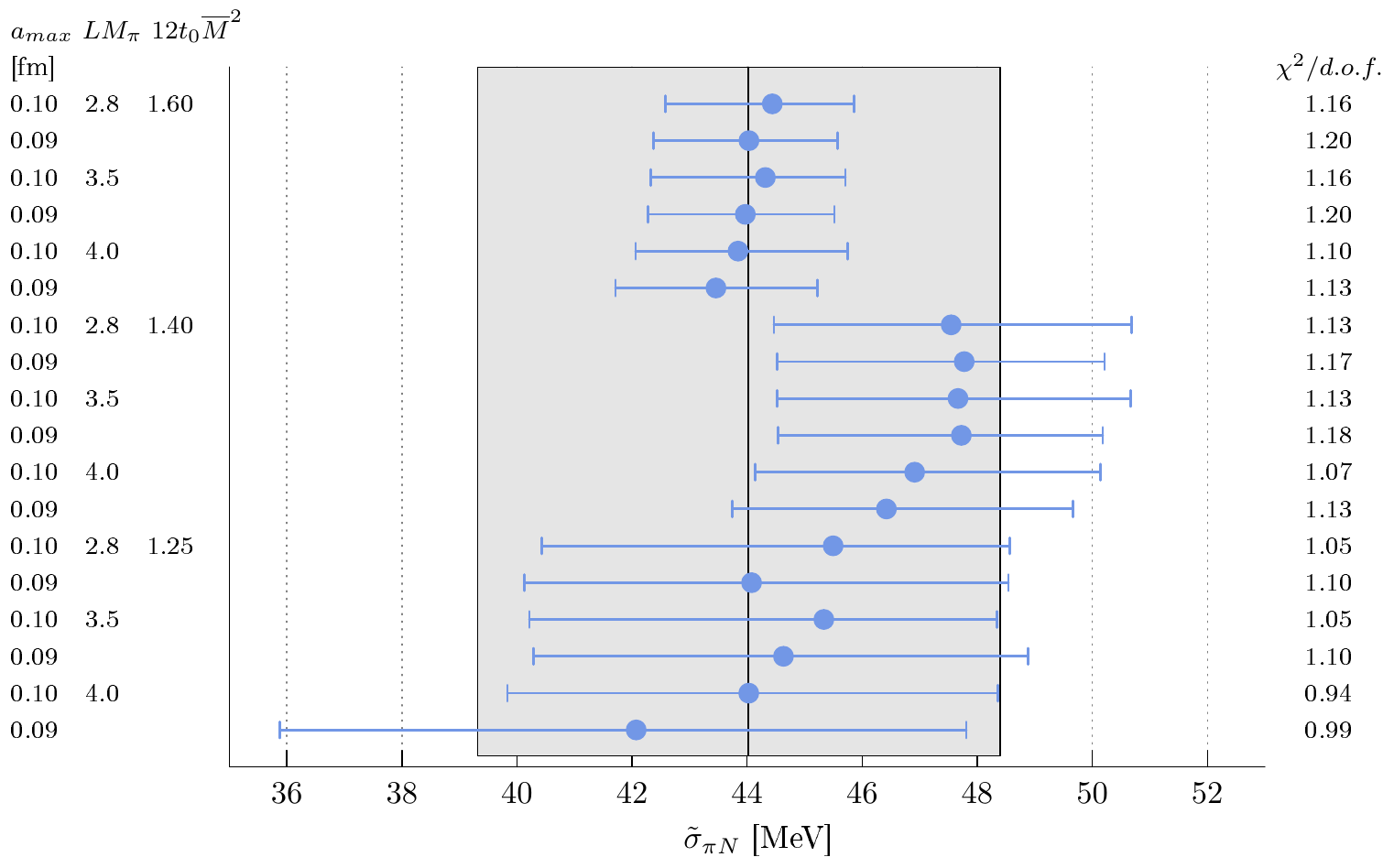}}
  \caption{The same as figure~\ref{fig:t0average} for $\tilde{\sigma}_{\pi N}$.
    The grey band represents the final result obtained
    via the AIC averaging procedure.
    The underlying BChPT~FV parametrization is referenced in
    section~\ref{sec:naming} and the data cuts are explained
    in section~\ref{sec:scale}.\label{fig:sigmafinal}}
\end{figure}

\begin{figure}[htp]
  \centering
  \resizebox{0.95\textwidth}{!}{\includegraphics[width=\textwidth]{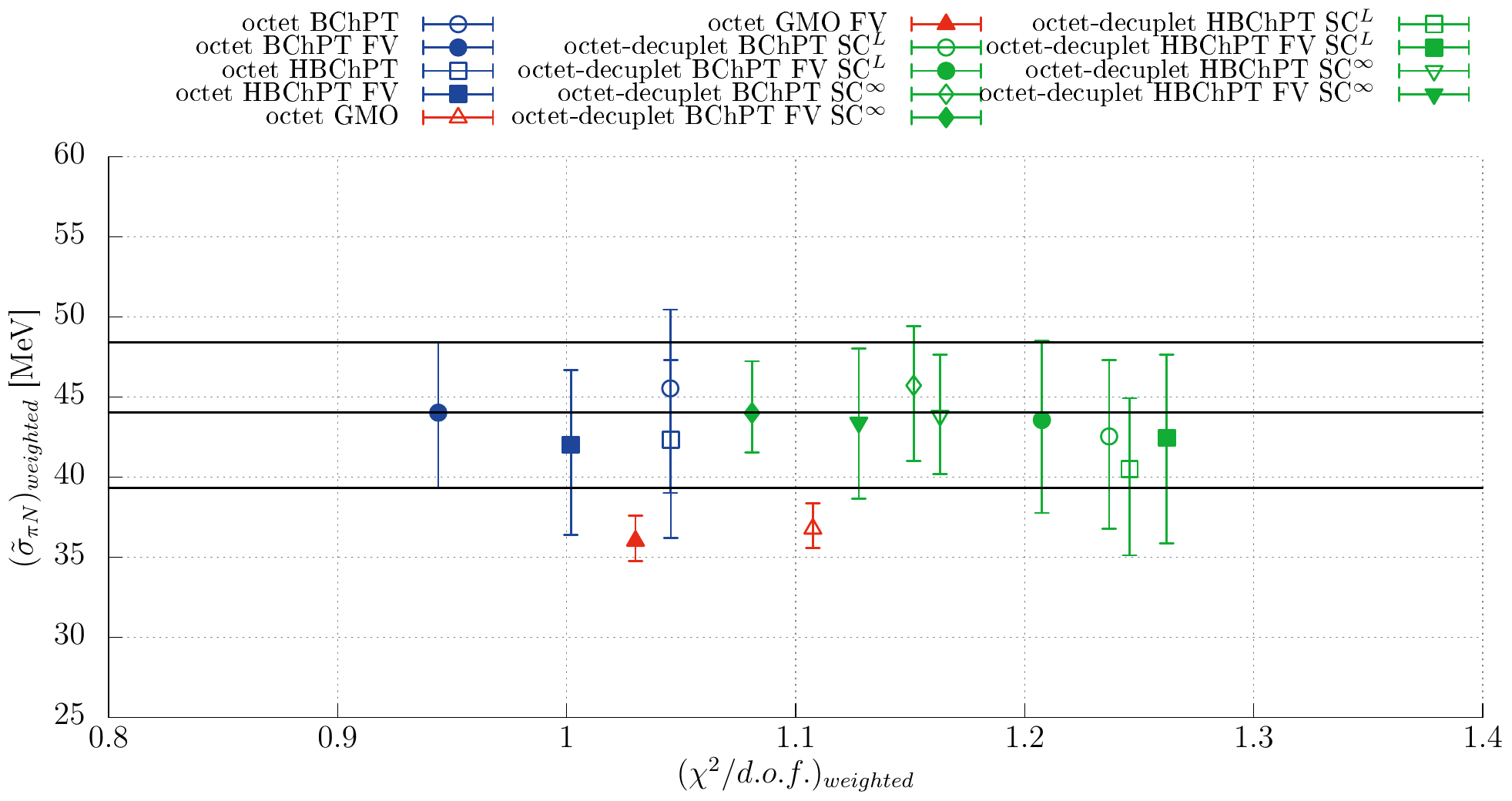}}
  \caption{The same as
    figure~\ref{fig:t0fitfo} for $\tilde{\sigma}_{\pi N}$.
    The parametrizations are referenced in
    section~\ref{sec:naming}.\label{fig:sigmafinal2}}
\end{figure}

In the statistical analysis we follow section~\ref{sec:scale}, also regarding the
cuts and model variation. We illustrate the dependence on different cuts with
respect to the lattice spacing,
the volume and the average squared pseudoscalar mass for the example of the
$\tilde{\sigma}_{\pi N}$ in figure~\ref{fig:sigmafinal}.
As mentioned above, the difference between $\tilde{\sigma}_{\pi N}$
and the pion-nucleon $\sigma$~term $\sigma_{\pi N}$ is negligible, see
table~\ref{tab:sigmaresults}.
Unlike the nucleon mass shown in figure~\ref{fig:t0average},
$\tilde{\sigma}_{\pi N}$ shows very little dependence on the
data cuts, apart from an increase in the statistical error as
the set of ensembles is reduced. We also investigated whether setting the scale
using the nucleon mass rather than the cascade mass had
any impact on the $\sigma$~terms, and, in particular, on $\sigma_{\pi N}$.
However, the resulting shifts only amounted to negligible fractions of
the errors.

In figure~\ref{fig:sigmafinal2} we investigate the
dependence on the parametrization: all fits give
very consistent results, with the exception of the GMO ansätze
that carry much smaller errors and give lower central values.
This is related to a reduced curvature of these
baryon mass parametrizations near the physical point, relative
to the (H)BChPT fits, see figure~\ref{fig:bestoct}.
A similarly consistent picture arises for the remaining
octet baryon $\sigma$~terms. The $\sigma$~terms of the
$\Omega$ baryon in
table~\ref{tab:sigmaresults} are determined via the
simultaneous octet-decuplet BChPT~FV~SC$^{\infty}$ fit. The GMO fits give
very similar results for $\sigma_{s\Omega}$ but with somewhat
smaller errors.

\subsection{SU(3) and SU(2) (H)BChPT low energy constants}
\label{sec:lecres}
\begin{table}[htb]
  \caption{\label{tab:lecresults}
    The LO $\textmd{SU(3)}$ octet and decuplet LECs. BChPT is
    employed as well as HBChPT. For the decuplet baryons we only carry out
    (H)BChPT fits including octet baryon loops, while the octet baryons are
    fitted with and without the decuplet baryons. We also include
    $m_0$ and $m_{D0}$ determined via the $\mathcal{O}(M^4)$ GMO expansions.
    The relevant equations are referenced in section~\ref{sec:naming}.
}
\begin{center}
    {
    \begin{tabular}{cccccc}
      \toprule
 fit& $\chi^2/N_{\text{DF}}$ & $m_0$ & $F/(4\pi F_0)$& $D/(4\pi F_0)$ & $F/D$\\
&    (weighted)     & [$\textmd{MeV}$]  &  [$\textmd{GeV}^{-1}$] &  [$\textmd{GeV}^{-1}$] & \\\midrule
 octet &0.94&821$^{(71)}_{(53)}$ & 0.383$^{(39)}_{(51)}$ & 0.638$^{(52)}_{(51)}$ & 0.606$^{(83)~}_{(97)}$\\
 BChPT FV & & & & &  \\
 octet &1.00&836$^{(86)}_{(61)}$ & 0.287$^{(33)}_{(30)}$ & 0.460$^{(41)}_{(42)}$ & 0.617$^{(105)}_{~(77)}$\\ 
 HBChPT FV & & & & & \\
octet-decuplet &1.08 &809$^{(33)}_{(45)}$ &0.324$^{(59)}_{(42)}$ & 0.509$^{(109)}_{~(91)}$ & 0.641$^{(155)}_{(127)}$\\         
 BChPT FV SC$^{\infty}$& & & & & \\
octet-decuplet  &1.13 &818$^{(60)}_{(69)}$ & 0.255$^{(48)}_{(51)}$ & 0.378$^{(99)}_{(84)}$ & 0.677$^{(166)}_{(138)}$\\
HBChPT FV SC$^{\infty}$& & & & & \\
 octet & 1.03 & 896$^{(34)}_{(31)}$ & & &\\
   GMO FV & & & & & \\
\midrule
&         & $m_{D0}$ & $|\mathcal{H}|/(4\pi F_0)$& $|\mathcal{C}|/(4\pi F_0)$ & $\delta$\\
 &        & [$\textmd{MeV}$]   &  [$\textmd{GeV}^{-1}$] &  [$\textmd{GeV}^{-1}$] & [$\textmd{MeV}$]\\\midrule
 octet-decuplet & 1.08 & 1147$^{(74)~}_{(91)}$ & 0.66$^{(59)}_{(66)}$  &0.45$^{(14)}_{(15)}$ & 333$^{(79)~}_{(84)}$\\         
 BChPT FV SC$^{\infty}$& & & & & \\
 octet-decuplet  &1.13 &1128$^{(118)}_{(289)}$ & 0.02$^{(74)}_{~(2)}$ &  0.47$^{(15)}_{(18)}$ & 305$^{(113)}_{(304)}$\\         
HBChPT FV SC$^{\infty}$& & & & & \\
 decuplet &0.88 &  1249$^{(112)}_{(118)}$ & &  & \\         
 GMO SC$^{\infty}$& & & & & \\
 \bottomrule
  \end{tabular}}
\end{center}
\end{table}
\begin{table}[htb]
  \caption{\label{tab:lecresults2}
    The NLO $\textmd{SU(3)}$ octet and decuplet LECs. BChPT is
    employed as well as HBChPT. For the decuplet baryons we only carry out
    (H)BChPT fits including octet baryon loops, while the octet baryons are
    fitted with and without the decuplet baryons. 
    In addition, we
    list the corresponding linear coefficients of the GMO expansion,
    truncated at
    $\mathcal{O}(M^4)$.
    The relevant equations are referenced in section~\ref{sec:naming}.
}
\begin{center}
    {
    \begin{tabular}{ccccc}
      \toprule
 fit& $\chi^2/N_{\text{DF}}$ & $b_0$ &  $b_D$ & $b_F$\\
&    (weighted)     & [$\textmd{GeV}^{-1}$] &  [$\textmd{GeV}^{-1}$] &  [$\textmd{GeV}^{-1}$]\\\midrule
 octet & 0.94 & $-0.739^{(70)}_{(84)}$ & 0.056$^{(43)}_{(39)}$ & $-0.440^{(40)}_{(26)}$\\
 BChPT FV & & & &\\
 octet &1.00 &  $-0.649^{(80)}_{(75)}$ &  0.052$^{(30)}_{(34)}$ &  $-0.399^{(35)}_{(24)}$\\ 
 HBChPT FV & & & &\\
octet-decuplet &1.08 & $-0.706^{(56)}_{(69)}$ & 0.083$^{(33)}_{(35)}$ & $-0.384^{(28)}_{(44)}$\\
 BChPT FV SC$^{\infty}$& & & &\\
octet-decuplet  &1.13 &$-0.662^{(73)}_{(78)}$ &0.080$^{(27)}_{(37)}$   &$-0.377^{(43)}_{(28)}$\\
HBChPT FV SC$^{\infty}$ & & & &\\
 octet & 1.03&$-0.389^{(49)}_{(53)}$ &0.092$^{(9)~}_{(7)}$ &$-0.243^{(6)~}_{(9)}$\\
   GMO FV & & & &\\
\midrule
&         & $t_{D0}$ &  $t_D$ & \\
 &        & [$\textmd{GeV}^{-1}$] &  [$\textmd{GeV}^{-1}$] & \\\midrule
 octet-decuplet & 1.08 &0.42$^{(22)}_{(17)}$ & 0.33$^{(12)}_{(64)}$  & \\         
 BChPT FV SC$^{\infty}$& & & & \\
 octet-decuplet  &1.13 & 0.44$^{(57)}_{(22)}$  & 0.28$^{(9)~}_{(3)}$  & \\         
HBChPT FV SC$^{\infty}$& & & &\\
 decuplet &0.88 &  0.10$^{(36)}_{(35)}$ &  0.32$^{(3)~}_{(2)}$ & \\         
 GMO SC$^{\infty}$ & & & &\\
 \bottomrule
  \end{tabular}}
\end{center}
\end{table}
Our continuum limit extrapolated results include some of the
$\textmd{SU(3)}$ and $\textmd{SU(2)}$
(H)BChPT LECs. To determine these and to estimate their uncertainties, we
employ the methods explained in the previous subsections.
For the $\textmd{SU(2)}$ LECs see also appendix~\ref{sec:su2lec}. We carry out
the analysis using the octet(-decuplet) BChPT~FV (SC$^{\infty}$)
as well as the octet-(decuplet) HBChPT~FV (SC$^{\infty}$) parametrizations
(see section~\ref{sec:naming}).
For comparison we also give the leading terms $m_0$ and $m_{D0}$
and the linear coefficients of the octet GMO~FV and the
decuplet GMO~SC$^{\infty}$ fits.
Note that for the linear coefficients,
we have taken into account the mass dependence of
$\sqrt{8t_0}$, see eqs.~\eqref{eq:bbmshift0}--\eqref{eq:bbmshift2}.

All results for the LO and NLO LECs
are shown in tables~\ref{tab:lecresults} and~\ref{tab:lecresults2}.
Differences between the different (H)BChPT parametrizations are indicative of
higher order effects. We find consistent values for the
LO LEC $m_0$. The LECs $F$ and $D$,
while formally also of leading order, only appear within
the loop corrections when considering the baryon self-energies.
Therefore, these are subject to uncertainties larger than 10\%.
Both have the tendency to be smaller when using the HBChPT~FV
compared to the BChPT~FV parametrization. Including
decuplet loops reduces their values too. The ratio $F/D$, however,
is quite stable, albeit with large errors. Regarding the
additional LECs related to the decuplet baryons, $m_{D0}$ is
basically the same when using HBChPT and BChPT while $\mathcal{H}$
is compatible with zero within errors that are as large as the
central value of $D$.
We find $\mathcal{C}$ to be small too
but to differ from zero by three standard deviations.
The GMO fits give larger central values for
the baryon masses in the chiral limit, $m_0$ and $m_{D0}$,
but the results are compatible with the (H)BChPT results
within one standard deviation. The larger
values are due to the smaller curvatures of the corresponding fits.

Our parametrizations only depend on the ratios
$D/F_0$, $F/F_0$, $\mathcal{C}^2/F_0^2$ and $\mathcal{H}^2/F_0^2$.
In ref.~\cite{Bali:2022qja} some of us determined
the combination $\sqrt{8t_{0,\text{ch}}}F_0=
0.1502^{(56)}_{(29)}$ by analysing the quark mass
dependence of the pseudoscalar meson mass and
its decay constant along the symmetric line, as well as
$\sqrt{8t_{0,\text{ch}}}m_0=1.57^{(5)}_{(6)}$, when carrying out an
analogous analysis for the octet baryon mass and the axial
charges. Combining this with our result
$\sqrt{8t_{0,\text{ch}}}=0.4170^{(22)}_{(27)}\,\textmd{fm}$
(see eqs.~\eqref{eq:t0ratio} and~\eqref{eq:t0starp}) gives
\begin{equation}
  \label{eq:oldval}
  4\pi F_0=893^{(34)}_{(18)}\,\textmd{MeV}
\end{equation}
and $m_0=743^{(24)}_{(29)}\,\textmd{MeV}$.
This value of $m_0$, which is compared to literature values
in figure~11 of ref.~\cite{Bali:2022qja}, is by about 1.4 standard
deviations smaller than the BChPT value of table~\ref{tab:lecresults},
\begin{equation}
  \label{eq:m0bchp}
  m_0=821^{(71)}_{(53)}\,\textmd{MeV}.
\end{equation}
Note that while in the present study we have included a larger number
of ensembles with a better coverage of the quark mass plane,
the work~\cite{Bali:2022qja} is based on a joint analysis of
the octet baryon axial charges and its mass, which better constrains
$F$ and $D$. In particular, the result $F/D=0.612^{(14)}_{(12)}$ was
obtained which compares well with the values shown for this ratio in
table~\ref{tab:lecresults}.
Ideally, one would repeat such a simultaneous analysis
incorporating all available ensembles.

Using the value eq.~\eqref{eq:oldval} for the combination $4\pi F_0$
gives
\begin{equation}
  \label{eq:lolecs}
F=0.34^{(4)}_{(5)},\quad
D=0.57(5),\quad
|\mathcal{C}|=0.40(13),\quad
|\mathcal{H}|=0.59^{(53)}_{(59)},
\end{equation}
where for $F$ and $D$ we quote the result of the octet BChPT~FV fits
while for $\mathcal{C}$ and $\mathcal{H}$
we have chosen the octet-decuplet BChPT~FV~SC$^\infty$ result.
The $\textmd{SU(6)}$ quark model expectation reads
$6D= 9F= -2\mathcal{H}= -3 \mathcal{C}$ (see, e.g.,
ref.~\cite{Jenkins:1991es}), which is consistent with the large-$N_c$
limit~\cite{Dashen:1993ac}.
Our results satisfy the first equality within errors.
However, both $|\mathcal{H}|$ and $|\mathcal{C}|$ are smaller
than expected. The axial charge in the chiral limit
is obtained as $\mathring{g}_A=F+D\approx 0.8$, which is quite
small too, given that $g_A\approx 1.27$~\cite{Workman:2022ynf}
at the physical point. Using the axial charges, albeit
only along the $m_s=m_{\ell}$ line, in ref.~\cite{Bali:2022qja}
some of us obtained the larger values $F=0.447(7)$ and $D=0.730(11)$.
Using these values, within a preliminary analysis of directly determined
$\sigma$~terms~\cite{Petrak:2021aqf}, it was only possible to describe
the data if a larger value for the chiral decay constant $F_0$
was admitted. The same appears to hold in the present analysis of
the baryon spectrum. We conclude that the data are
well described by the BChPT parametrization at
$\mathcal{O}(p^3)$, however, there appears to be some tension with
the axial charges on the level of 20\%, regarding the LECs.
This hints at contributions from higher order corrections
that can only be resolved by simultaneously analysing a larger set
of observables. Values for $F$ and $D$ from other determinations can be
found in figure~12 of ref.~\cite{Bali:2022qja}.

From our preferred simultaneous fits to the octet and decuplet baryon masses
(octet-decuplet BChPT~FV~SC$^\infty$),
we obtain
\begin{equation}
  \label{eq:mobchp2}
  m_{D0}=1147^{(74)}_{(91)}\,\textmd{MeV},\quad
  \delta=333^{(79)}_{(84)}\,\textmd{MeV}
\end{equation}
and $m_0=818^{(60)}_{(69)}\,\textmd{MeV}$, which
agrees well with the value extracted from the octet baryons
alone, see eq.~\eqref{eq:m0bchp}.
The central value of $\delta$ differs slightly from that of $m_{D0}-m_0$
since we quote the median of the respective sum of AIC weighted bootstrap
histograms.

Regarding the linear NLO coefficients of table~\ref{tab:lecresults2},
the differences between BChPT and
HBChPT are rather small. The central values of $b_D$ and $b_F$ are somewhat
larger when including the decuplet loops. However,
in the GMO fits $b_0$ comes out by almost
a factor of two smaller, while $b_F$ is systematically
larger than in the (H)BChPT fits. We remark that the data are not
well described by the linear $\mathcal{O}(p^2)$ parametrization.
To get the physics right, the inclusion of $\mathcal{O}(p^3)$ terms is
necessary within (H)BChPT, which increases $b_0$
by a factor of two. From our preferred fits (octet BChPT~FV
for $b_0$, $b_D$ and $b_F$ and octet-decuplet BChPT~FV~SC$^{\infty}$
for $t_{D0}$,\footnote{
We renamed this LEC to avoid confusion with the scale parameter $t_0$.}
and $t_D$) we obtain:
\begin{align}
  \label{eq:nlolecres}
  b_0&=-0.739^{(70)}_{(84)},&
  b_D&= 0.056^{(43)}_{(39)},&
  b_F=-0.440^{(40)}_{(26)},\\
  t_{D0}&=0.42^{(22)}_{(17)},&
  t_D&=0.33^{(12)}_{(64)}.&\label{eq:nlolecres2}
\end{align}

\begin{table}[htb]
  \caption{\label{tab:su2lecresults}
    The $\textmd{SU(2)}$ LECs $m^0_N$ and $c_1$ for the nucleon,
    using BChPT and HBChPT both with and without the inclusion
    of decuplet loops.
    The relevant parametrizations are referenced in section~\ref{sec:naming}.
    The relation to the $\textmd{SU(3)}$ LECs is worked out in
    appendix~\ref{sec:su2lec} and the LECs are defined in
    eq.~\eqref{eq:suscpt}.
}
\begin{center}
    \begin{tabular}{ccccccccc}
      \toprule
 fit& $\chi^2/N_{\text{DF}}$ & $m^{0}_N$ & $c_1$ \\
&    (weighted)     & [$\textmd{MeV}$]   & [$\textmd{GeV}^{-1}$] \\\midrule
 octet  BChPT FV& 0.94 & 893.2$^{(9.3)~}_{(8.6)~}$ & $-0.920^{(59)}_{(96)}$ \\
 octet  HBChPT FV &1.00& 898.3$^{(11.7)}_{(12.0)}$ & $-0.817^{(59)}_{(75)}$ \\
 octet-decuplet  BChPT FV~SC$^{\infty}$ &1.08 &  888.1$^{(12.1)}_{(10.4)}$ & $-0.924^{(66)}_{(84)}$ \\
octet-decuplet  HBChPT FV SC$^{\infty}$ &1.13 & 894.9$^{(12.1)}_{(10.6)}$ & $-0.823^{(55)}_{(74)}$ \\
 \bottomrule
  \end{tabular}
\end{center}
\end{table}

We now address the $\textmd{SU(2)}$ LECs.
Following appendix~\ref{sec:su2lec}, we determine
$m_0^N$, i.e.\ the nucleon mass in the $\textmd{SU(2)}$
chiral limit, as well as
$c_1$, the coefficient of the linear contribution to $m_N(M_{\pi}^2)$
at a fixed physical strange quark mass, $-4c_1 M_{\pi}^2$, see
equation~\eqref{eq:suscpt}. The results
from the (H)BChPT fits with and without decuplet loops are
displayed in table~\ref{tab:su2lecresults}. While $m_N^0$ is very stable
across the four classes of fits, $c_1$ comes out systematically smaller
in the HBChPT parametrizations. However, the decuplet loops
appear to have little impact, as was the case for $b_0$ which is the dominant
contribution to $c_1$, see eq.~\eqref{eq:calcc1}. The results from
our preferred octet BChPT~FV fits read
\begin{equation}
  \label{eq:ressu2lecs}
  m_N^0=893.2^{(9.3)}_{(8.6)}\,\textmd{MeV},\quad
  c_1=-0.920^{(59)}_{(96)}\,\textmd{GeV}^{-1}.
\end{equation}
The nucleon mass in the $\textmd{SU(2)}$ chiral limit
$m_N^0$ is much better constrained than the nucleon mass
in the $\textmd{SU(3)}$ chiral limit,
$m_0=821^{(71)}_{(53)}\,\textmd{MeV}$, since we have many ensembles
close to the former point in the quark mass plane.
Note that Hoferichter et al.~\cite{Hoferichter:2015hva} predict
$c_1\sim -1.1\,\textmd{GeV}^{-1}$ (see table~7 of that reference), which is
somewhat larger in magnitude than our result. Consistent with this, they also
obtain a larger $\sigma_{\pi N}$-value~\cite{Hoferichter:2015dsa} (see below).

Assuming that $m_N(M_{\pi}^2)$ at a fixed value of the strange quark mass
is a concave function, one can
easily derive the inequalities
\begin{align}
  898\,\textmd{MeV}&\approx m_N-\tilde{\sigma}_{\pi N}>m_N^0=893.2^{(9.3)}_{(8.6)}
  \textmd{MeV},\\
  44\,\textmd{MeV}&\approx\tilde{\sigma}_{\pi N}<-4c_1 M_{\pi,\text{ph}}^2=67^{(4)}_{(7)}\,\textmd{MeV},
\end{align}
which are both satisfied. On the one hand, increasing $\sigma_{\pi N}$ would
violate the first inequality, unless the nucleon mass in the
$\textmd{SU(2)}$ chiral limit was smaller. On the other hand,
$\sigma_{\pi N}/M_{\pi}^2$ must be larger in the $\textmd{SU(2)}$ chiral
limit than at the physical point, which results in another,
less stringent upper limit for $\sigma_{\pi N}$.

\subsection{Discussion of $\boldsymbol{t_{0,\text{ph}}}$,
  the baryon spectrum and $\boldsymbol{\sigma_{\pi N}}$}
\label{sec:resultsummary}

Above we determined the scale parameter $t_{0,\text{ph}}$ from the mass
of the $\Xi$ baryon, see eq.~\eqref{eq:t0result}.
In section~\ref{sec:scale} we have demonstrated that this result is
robust against variations of the parametrizations of the mass dependence,
finite volume effects and continuum limit extrapolation as well
as regarding different cuts imposed on the data. The
input quantities used to fix the physical
point and to set the scale, i.e.\ $M_{\pi}$, $M_K$ and $m_{\Xi}$, are also very
precisely known, even after correcting these for isospin breaking effects.
We found the lattice spacing dependence of $\sqrt{8t_{0,\text{ph}}}m_{\Xi}$ to
be less pronounced than that observed using the same lattice action
when determining the pseudoscalar decay constants in units of
$\sqrt{8t_0}$~\cite{Bruno:2016plf,Strassberger:2021tsu}. However,
it is still significant: 
on our coarsest lattice, the combination
$\sqrt{8t_{0,\text{ph}}}m_\Xi$ differs by a bit more than 3\%
from the continuum limit
value, see the left panel of figure~\ref{fig:bestaeff}. Nevertheless, with
six lattice spacings at our disposal, this extrapolation is well controlled
so that the most dominant uncertainty of our determination of $t_0$
is by far the statistical error, in particular, of the mass of the $\Xi$ baryon.

\begin{figure}[thp]
  \centering
  \resizebox{0.95\textwidth}{!}{\includegraphics[width=\textwidth]{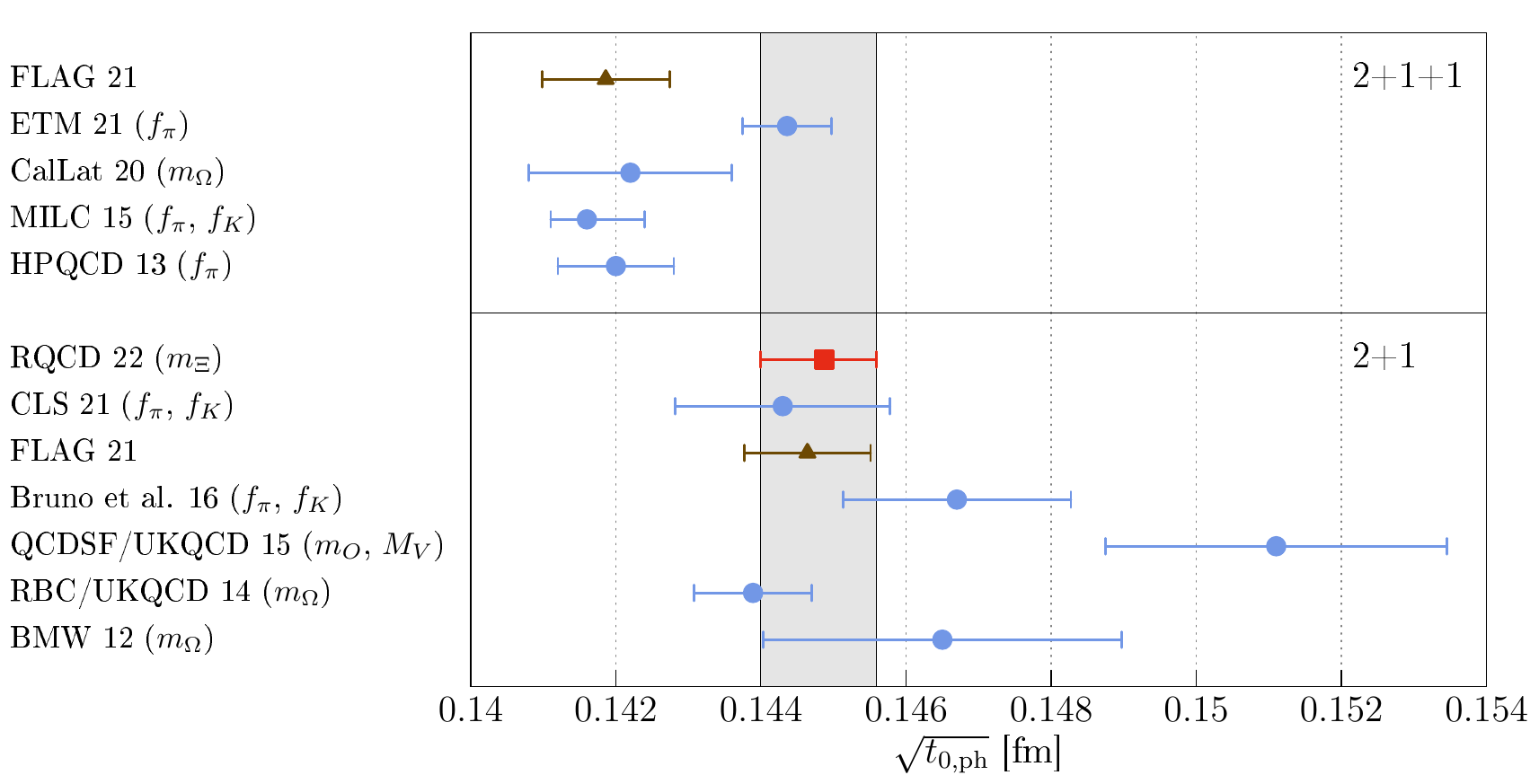}}
  \caption{Comparison of our determination of $\sqrt{t_{0,\text{ph}}}$
    from the $\Xi$ baryon mass (red square and shaded grey region) to
    other determinations of this quantity with $N_f=2+1$ flavours
    (CLS~21~\cite{Strassberger:2021tsu}, Bruno~et~al.~16~\cite{Bruno:2016plf},
    QCDSF/UKQCD~15~\cite{Bornyakov:2015eaa}, RBC/UKQCD~14~\cite{Blum:2014tka}
    and BMW~12~\cite{Borsanyi:2012zs}) and
    $N_f=2+1+1$ flavours (ETM~21~\cite{ExtendedTwistedMass:2021qui},
    CalLat~20~\cite{Miller:2020evg}, MILC~15~\cite{MILC:2015tqx}
    and HPQCD~13~\cite{Dowdall:2013rya}). The experimental
    quantities used as an input to set the scale are indicated in brackets. 
    Also shown as green triangles are the
    averages of the FLAG Review 2021~\cite{Aoki:2021kgd}.
  \label{fig:t0compare}}
\end{figure}

The scale parameter $t_{0,\text{ph}}$ has been determined previously in
$N_f=2+1$ QCD~\cite{Borsanyi:2012zs,Blum:2014tka,Bornyakov:2015eaa,Bruno:2016plf,Strassberger:2021tsu} and
$N_f=2+1+1$ QCD~\cite{Dowdall:2013rya,MILC:2015tqx,Miller:2020evg,ExtendedTwistedMass:2021qui}, using different input quantities and
different extrapolation strategies.
In figure~\ref{fig:t0compare} we compare our result (red square and shaded
grey region) with previous determinations as well as with the 2021 FLAG
averages~\cite{Aoki:2021kgd}. 
HPQCD~\cite{Dowdall:2013rya} and ETM~\cite{ExtendedTwistedMass:2021qui}
used the pion decay constant to set the scale while MILC~\cite{MILC:2015tqx},
Bruno et al.~\cite{Bruno:2016plf} as well as CLS~\cite{Strassberger:2021tsu}
(both on CLS ensembles) used combinations of
the pion and kaon decay constants. MILC carried out an interpolation
to a decay constant of a hypothetical meson with two quarks of mass
$0.4\,m_s$ and used this to set the scale while
the two CLS-based determinations
extrapolated/interpolated the flavour averaged combination
$\sqrt{t_0}f_{\pi K}\coloneqq\sqrt{t_0}(2f_K+f_\pi)/3$ to the physical point
in the continuum limit. In contrast BMW~\cite{Borsanyi:2012zs},
RBC/UKQCD~\cite{Blum:2014tka} and CalLat~\cite{Miller:2020evg}
used the mass of the $\Omega$ baryon to set the scale.
Finally, QCDSF/UKQCD~\cite{Bornyakov:2015eaa} averaged a result
from the average octet baryon mass with a result from the
average vector meson mass. This is not entirely unproblematic since
the $\rho$ meson at the physical point has quite a substantial decay width.
All in all, with the exception of this last result, there is agreement
within 1.6\,$\sigma$ or less between any pair of $N_f=2+1$ determinations,
regardless of whether the $\Omega$ mass, the $\Xi$ mass or pseudoscalar decay
constants were used. Also the $N_f=2+1+1$ determination with
twisted mass fermions (ETM~21~\cite{ExtendedTwistedMass:2021qui})
is consistent with our result. However, the three $N_f=2+1+1$ determinations
that were obtained using the highly improved staggered quark (HISQ)
sea quark action (CalLat~20~\cite{Miller:2020evg}, MILC~15~\cite{MILC:2015tqx}
and HPQCD~13~\cite{Dowdall:2013rya}) appear to suggest a somewhat lower value.

\begin{figure}[htp]
  \centering
  \resizebox{0.95\textwidth}{!}{\includegraphics[width=\textwidth]{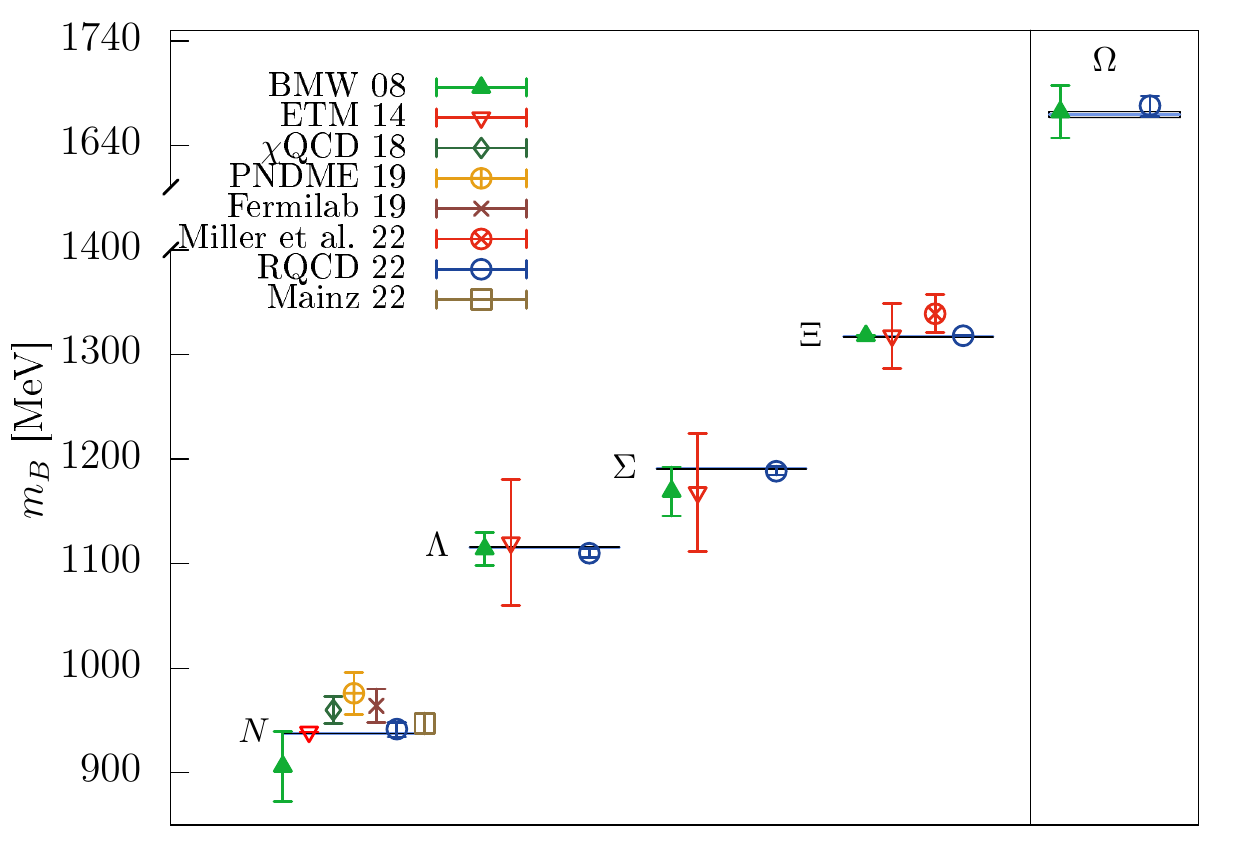}}
  \caption{Comparison of our spectrum of strongly stable baryons (RQCD~22, blue circles) with
    other $N_f=2+1$ and
    $N_f=2+1+1$ continuum limit extrapolated results. BMW~08~\cite{Durr:2008zz}
    ($N_f=2+1$) also used the $\Xi$ baryon mass to set the scale, whereas
    ETM~14~\cite{Alexandrou:2014sha} ($N_f=2+1+1$) employed the nucleon mass and
    no prediction
    of the $\Omega$ baryon mass was made since this was used to fix the strange
    quark mass. Other determinations of either the nucleon
    or the cascade mass include $\chi$QCD~18~\cite{Yang:2018nqn}
    ($N_f=2+1$,
    scale from the $\Omega$ baryon mass~\cite{Blum:2014tka}),
    PNDME~19~\cite{Jang:2019jkn} ($N_f=2+1+1$, scale
    from ref.~\cite{MILC:2012znn}, indirectly using the pion and kaon
    decay constants),
    Fermilab~19~\cite{Lin:2019pia} ($N_f=2+1+1$, scale from
    ref.~\cite{FermilabLattice:2019ugu} using the value
    of the parameter $w_0$~\cite{Borsanyi:2012zs}, determined in
    HPQCD~13~\cite{Dowdall:2013rya} from the pion decay constant),
    Miller~et~al.~22~\cite{Miller:2022vcm}
    ($N_f=2+1+1$, scale set from the $\Omega$ mass of
    CalLat~20~\cite{Miller:2020evg}),
    Mainz~22~\cite{Ottnad:2022axz}
    ($N_f=2+1$, scale set via $t_0$ from the pion and kaon decay
    constants by Bruno~et~al.~16~\cite{Bruno:2016plf}).
  \label{fig:specoth}}
\end{figure}

In figure~\ref{fig:specoth} we compare our spectrum of strongly stable
baryons with previous determinations in $N_f=2+1(+1)$ QCD where a
continuum limit extrapolation was attempted. The most comprehensive
study so far was carried out for $2+1$ fermions by BMW~08~\cite{Durr:2008zz},
who also set the scale using the mass of the $\Xi$ baryon.
In that case 18 ensembles were employed across three lattice spacings
$a\approx 0.125\,\textmd{fm}$, 
$a\approx 0.085\,\textmd{fm}$ and 
$a\approx 0.065\,\textmd{fm}$ with pion masses
$M_{\pi}>190\,\textmd{MeV}$ and statistics typically between
1000 and 2000~MDUs. Apart from having longer HMC runs,
employing almost 50 ensembles (see table~\ref{tab:parameters}),
we significantly vary the strange quark mass too,
cover six lattice spacings $0.039\,\textmd{fm}\lesssim a\lesssim
0.098\,\textmd{fm}$ and go down to the physical pion mass
(see table~\ref{tab:physparams} and
figures~\ref{fig:ensembles}--\ref{fig:volumes}).
While BMW~08 also employed Wilson quarks, our action is
non-perturbatively order $a$ improved.
All of this, together with the improved quark smearing
used (see appendix~\ref{sec:numerics}), enables
very significant reductions of the errors, see
figure~\ref{fig:specoth}. Another comprehensive
study was carried out by ETM~14~\cite{Alexandrou:2014sha}
using $2+1+1$ twisted mass fermions. In that case
the nucleon mass was used to set the scale and the $\Omega$ mass
was employed to match the strange quark mass and could therefore
not be predicted. Ten ensembles across three
lattice spacings $0.065\,\textmd{fm}\lesssim
a\lesssim 0.094\,\textmd{fm}$ were realized with pion
masses $M_{\pi}>210\,\textmd{MeV}$. Additional continuum limit
results on either the nucleon
mass
($\chi$QCD~18~\cite{Yang:2018nqn},
PNDME~19~\cite{Jang:2019jkn},
MILC~19~\cite{Lin:2019pia} and
Mainz~22~\cite{Ottnad:2022axz})
or the cascade mass (Miller et al.~22~\cite{Miller:2020evg}) alone
are included too. For details on the scale input, see
the caption of figure~\ref{fig:specoth}. These determinations employed
overlap fermions on top of $N_f=2+1$ domain wall fermions~\cite{Yang:2018nqn},
Wilson-clover fermions on top of $N_f=2+1+1$ HISQ~\cite{Jang:2019jkn},
$N_f=2+1+1$ HISQ~\cite{Lin:2019pia} and a subset of the $N_f=2+1$ CLS
ensembles that we use~\cite{Ottnad:2022axz}.

\begin{figure}[thp]
  \centering
  \resizebox{0.95\textwidth}{!}{\includegraphics[width=\textwidth]{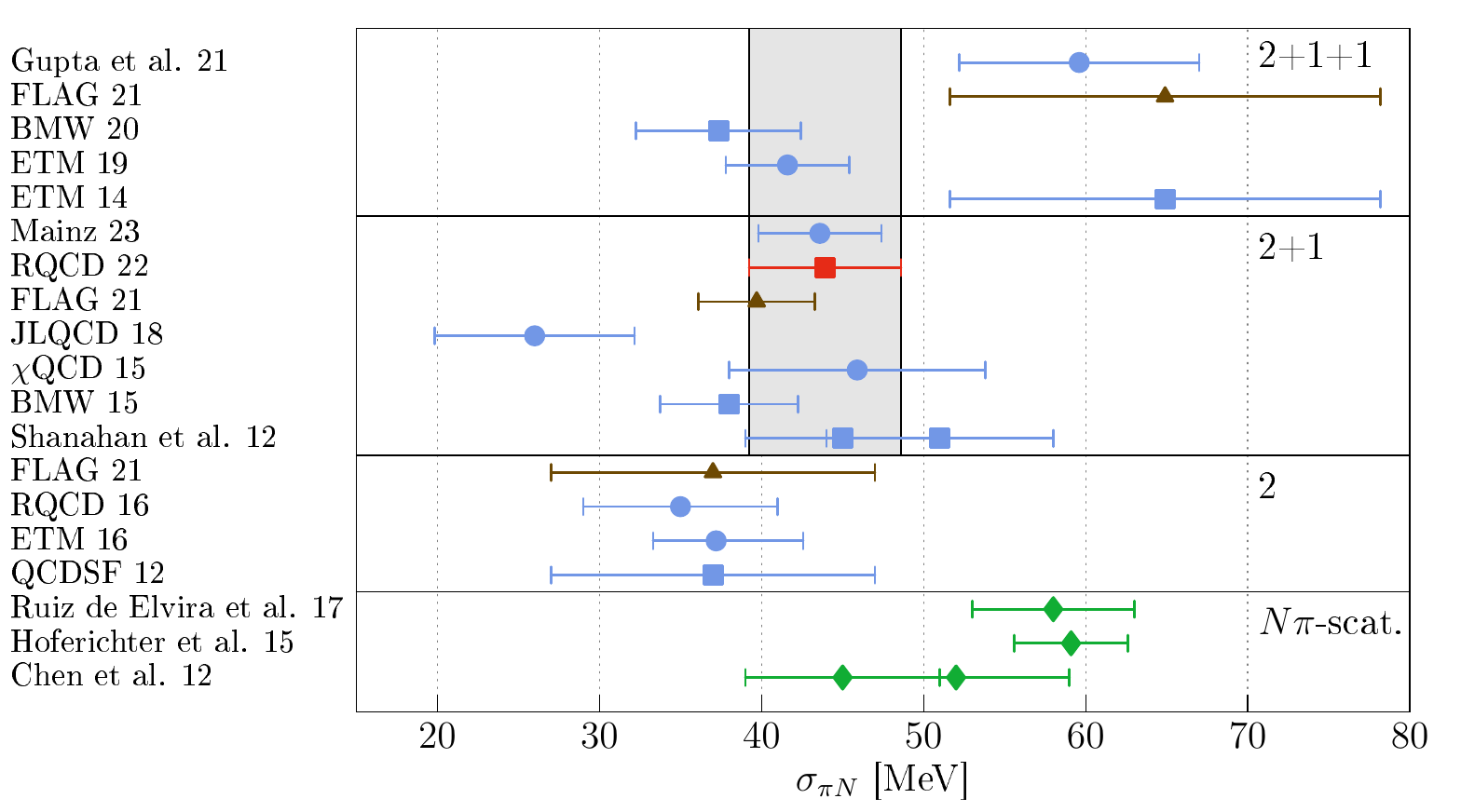}}
  \caption{Comparison of our determination of $\sigma_{\pi N}$
    (red square and shaded grey region) to
    recent determinations of this quantity from pion-nucleon scattering
    data (Chen et al.~12~\cite{Chen:2012nx},
    Hoferichter et al.~15~\cite{Hoferichter:2015dsa},
    Ruiz de Elvira et al.~17~\cite{RuizdeElvira:2017stg}),
    lattice simulations with $N_f=2$ flavours
    (QCDSF~12~\cite{Bali:2012qs}, ETM~16~\cite{Abdel-Rehim:2016won},
    RQCD~16~\cite{Bali:2016lvx}), $N_f=2+1$ flavours
    (Shanahan et al.~12~\cite{Shanahan:2012wh}, BMW~15~\cite{Durr:2015dna},
    $\chi$QCD~15~\cite{Yang:2015uis}, JLQCD~18~\cite{Yamanaka:2018uud}),
    Mainz~23~\cite{Agadjanov:2023jha}
    and $N_f=2+1+1$ flavours
    (ETM~14~\cite{Alexandrou:2014sha}, ETM~19~\cite{Alexandrou:2019brg},
    BMW~20~\cite{Borsanyi:2020bpd}, Gupta et al.~21~\cite{Gupta:2021ahb}).
    Also shown as brown triangles are the
    averages of the FLAG Review 2021~\cite{Aoki:2021kgd}. Circles correspond
    to direct determinations, squares to fits to the quark
    mass dependence of the nucleon mass.
  \label{fig:sigma}}
\end{figure}

In figure~\ref{fig:sigma} we compare our result on $\sigma_{\pi N}$
(RQCD~22, red square and the shaded grey region)
with other determinations.\footnote{Further efforts to determine
the $\sigma$~terms from
fits to collections of lattice baryon mass data can be found, e.g., in
refs.~\cite{MartinCamalich:2010fp,Ren:2012aj,Ren:2013oaa,Lutz:2014oxa,Lutz:2018cqo}. Particularly puzzling results (also for the LECs) were  obtained in a
recent fit~\cite{Lutz:2023xpi} to the same baryon mass data that we present here: not only the value
found for $\sigma_{\pi N}$ is much larger than ours but also the error given
is three and a half-fold 
smaller, while the strangeness $\sigma$~term
is reported to be significantly negative
($\sigma_{sN}\approx -316(76)\,\textmd{MeV}$)!}
The lattice determinations are all statistically
consistent with our result.
Only JLQCD~18~\cite{Yamanaka:2018uud} suggest a much smaller central value,
while ETM~14~\cite{Alexandrou:2014sha} (however, superseded in part by
ETM~19~\cite{Alexandrou:2019brg}) and, more recently,
Gupta et al.~\cite{Gupta:2021ahb}
suggest quite large values for this parameter.
We refer the reader to the FLAG~2021 Review~\cite{Aoki:2021kgd} for a more
detailed comparison of the different lattice results.
Also shown in the figure are the FLAG~21 averages as well as determinations
using $N\pi$ scattering data. Note that
refs.~\cite{Hoferichter:2015dsa,RuizdeElvira:2017stg}
take account of isospin breaking effects, while all the lattice results refer to
iso-symmetric QCD. In ref.~\cite{Hoferichter:2015dsa} isospin effects are
estimated to increase the pion-nucleon $\sigma$~term by about $3\,\textmd{MeV}$.
The magnitude of this effect is not unexpected since to leading non-trivial
order the $\sigma$~term is proportional to the squared pion mass and, for
instance, $M_{\pi^+}^2/M_{\pi^0}^2\approx 1.07$. Nevertheless, even
when adding $3\,\textmd{MeV}$ to our iso-symmetric QCD prediction,
\begin{equation}
  \sigma_{\pi N}=(43.9\pm 4.7)\,\textmd{MeV},
\end{equation}
this is still by more than 1.6 standard deviations
smaller than the latest determination from pion-nucleon scattering data,
$\sigma_{\pi N}=58(5)\,\textmd{MeV}$~\cite{RuizdeElvira:2017stg}.
An earlier determination by the same authors gave
$\sigma_{\pi N}=59.1(3.5)\,\textmd{MeV}$~\cite{Hoferichter:2015dsa}.

\section{Summary of the main results}
\label{sec:summary}
We summarize our findings, starting with results that are specific to
our lattice action, continuing with the $\sigma$~terms, the LECs
and the scale setting parameter, and concluding with the light
baryon spectrum.
\subsection{Results that are specific to the lattice action}
\label{sec:lattb}
We employed non-perturbatively order $a$ improved Wilson
fermions~\cite{Sheikholeslami:1985ij,Bulava:2013cta}
and the tree-level Symanzik improved gauge action~\cite{Weisz:1982zw}.
For details on the action and the implementation
for $N_f=2+1$ fermions, see ref.~\cite{Bruno:2014jqa}.
We determined the scale parameter $t_0^*$ in lattice units
as a function of the inverse lattice coupling
$\beta=6/g^2$. This parameter is defined at the point in the
quark mass plane where $m_u=m_d=m_s$ and $\phi_4^*= 12t_0^*M_{\pi}^2= 1.11$.
Within the range $3.34\leq\beta\leq 3.85$, $(t_0^*/a^2)(g^2)$
is given by the interpolating formula~\eqref{eq:t0interpol},
which approaches the perturbative two-loop expectation at large $\beta$.
The dependence on $\beta$ is visualized in the right panel
of figure~\ref{fig:ambarratio}, where in addition we show $t_0/a^2$ at the
$\textmd{SU(3)}$ chiral point. The results for the $\beta$-values at which
we carried out the simulations are displayed in table~\ref{tab:t0star},
along with the lattice spacings that are obtained using
$\sqrt{8t_0^*}=0.4097^{(20)}_{(25)}\,\textmd{fm}$.

An interpolating formula for the $N_f=3$ critical hopping parameter
value $\kappa_{\text{crit}}(g^2)$ is given in eq.~\eqref{eq:kappainter} and
shown in the right panel of figure~\ref{fig:rmkappa}.
This interpolation will approach the two-loop result at large
$\beta$-values. The corresponding values, as well as $\kappa^*$ and
$am^*=\tfrac12(1/\kappa^*-1/\kappa_{\text{crit}})$, are displayed in
table~\ref{tab:kappa}. In table~\ref{tab:kappas} we list
the starting points on the $m_s=m_{\ell}$ line
for the different $\beta$-values
of trajectories of constant $m_u+m_d+m_s$ that will touch the
physical point in terms of $\kappa_{\text{opt}}$ and
$\phi_{4,\text{opt}}$. The physical point is
defined by $M_{\pi,\text{ph}}=134.8(3)\,\textmd{MeV}$
and $M_{K,\text{ph}}=494.2(3)\,\textmd{MeV}$, where the scale is
set using $\sqrt{8t_{0,\text{ph}}}=0.4098^{(20)}_{(25)}\,\textmd{fm}$.

We give interpolating formulae for the combination of
renormalization factors $Z(g^2)=(Z_mZ_P/Z_A)(g^2)$ and
the ratio of singlet and octet mass renormalization constants
$r_m(g^2)=(Z_m^s/Z_m^{ns})(g^2)$ in eqs.~\eqref{eq:zinter}
and~\eqref{eq:rminter}, respectively. The former is compared to
results from the literature in the right panel of figure~\ref{fig:ZBC}
and the latter in the left panel of figure~\ref{fig:rmkappa}.
The interpolating formulae are such that $Z(g^2)$ and $r_m(g^2)$
will approach the perturbative one- and two-loop expectations,
respectively, at small values of $g^2$.
The numerical values at our simulation points are tabulated
in table~\ref{tab:fitresults}. We include the combinations of
$\mathcal{O}(a)$ improvement coefficients $\mathcal{A}$, $\mathcal{B}_0$,
$\mathcal{C}_0$ and $\mathcal{D}_0$ that are defined
in eqs.~\eqref{eq:adef}--\eqref{eq:ddef} (see also
eqs.~\eqref{eq:fitdiff} and~\eqref{eq:fitsum}) in the same table.
The corresponding interpolations for $\mathcal{B}_0$ and
$\mathcal{C}_0$ are given in eq.~\eqref{eq:pade} (with
the coefficients of table~\ref{tab:fitresults2}) and plotted in
the left panel of figure~\ref{fig:ZBC}. These interpolations also
approach the respective one-loop expectations at small values of $g^2$.
Finally, estimates of integrated autocorrelation times for
$t_0$ can be found in the last column of table~\ref{tab:parameters}.

\subsection{The $\boldsymbol{\sigma}$~terms and the BChPT low energy constants}
\label{sec:sigmab}
All the pion-baryon and strange-baryon $\sigma$~terms for the nucleon,
the $\Lambda$, the $\Sigma$, the $\Xi$ and the $\Omega$ can be found in
the third and fifth columns of table~\ref{tab:sigmaresults} with all
systematics included in the error estimates. We refrain from
determining the $\sigma$~terms for the unstable resonances $\Delta$,
$\Sigma^*$ and $\Xi^*$. The results, that are the first continuum limit
determinations of $\sigma$~terms other than for the nucleon,
refer to iso-symmetric QCD. We compare $\sigma_{\pi N}$ (RQCD~22)
to previous determinations in figure~\ref{fig:sigma} and
discuss this in section~\ref{sec:resultsummary}.
The $\sigma$~terms are computed from the dependence of
the baryon masses on the pseudoscalar masses, also
taking into account the dependence of the pseudoscalar masses on the quark
masses. In particular, the strange $\sigma$~terms are not very well
constrained since we do not sufficiently vary the strange quark mass
near the physical point. This will be improved upon in the
near future by incorporating direct determinations of the
scalar matrix elements~\cite{Petrak:2021aqf}.

In tables~\ref{tab:lecresults} and~\ref{tab:lecresults2} we list the
leading order and the NLO $\textmd{SU(3)}$ LECs, employing BChPT in the EOMS
regularization as well as HBChPT, both including and not including
decuplet loops. In table~\ref{tab:su2lecresults} we also display
the corresponding results for the nucleon mass in the $\textmd{SU(2)}$
chiral limit and the $\textmd{SU(2)}$ slope parameter $c_1$.
While the errors given also reflect the systematics due to
different cuts regarding the lattice spacing, the volume and
also the maximum average squared pseudoscalar mass, 
the results depend on the ChPT parametrization due to different
higher order contributions. In terms of the reduced $\chi^2$-values,
our data are better described by the BChPT expressions than by HBChPT.

We find $m_0=821^{(71)}_{(53)}\,\textmd{MeV}$ and
$m_{D0}=1147^{(74)}_{(91)}\,\textmd{MeV}$ from the preferred
BChPT fits to the octet baryons and to the octet and
decuplet baryons, respectively. This then gives
$\delta=m_{D0}-m_0=333^{(79)}_{(84)}\,\textmd{MeV}$.
We display the corresponding values for $F$, $D$, $|\mathcal{C}|$
and $|\mathcal{H}|$ in eq.~\eqref{eq:lolecs}. From the
baryon masses alone one can only determine ratios of these LECs
with respect to the pseudoscalar decay constant in the
$N_f=3$ chiral limit, $F_0$, for which we used the result
of ref.~\cite{Bali:2022qja}. In the same reference, where
the baryon mass and the axial charges were
analysed simultaneously, albeit only for the $m_s=m_{\ell}$ ensembles,
$m_0$ comes out somewhat smaller and $F$ and $D$ somewhat larger
than here. These ambiguities are most likely
due to higher order terms in the chiral expansion, which we are
unable to resolve within the present statistical uncertainties
using the baryon spectrum alone. To investigate this
more systematically, a simultaneous analysis of the baryon
spectrum and the charges is planned.

The NLO LECs $b_0$, $b_D$, $b_F$, $t_{D0}$ (renamed from $t_0$
to avoid confusion with the scale parameter) and $t_D$ are listed
in table~\ref{tab:lecresults2} and the values from the preferred
fits displayed in eqs.~\eqref{eq:nlolecres} and~\eqref{eq:nlolecres2}.
Due to the larger statistical errors of the decuplet masses and the smaller
number of available data points for which these are stable, $t_{D0}$
and $t_D$ are subject to substantial errors.
Regarding $\textmd{SU(2)}$ BChPT, our preferred results for
$m_N^0$ and $c_1$ are shown in eq.~\eqref{eq:ressu2lecs}.
Since we have many ensembles that are close to the $N_f=2$ chiral
limit, the $\textmd{SU(2)}$ mass parameter
$m_N^0=893.2^{(9.3)}_{(8.6)}\,\textmd{MeV}$
is well determined and robust against changes of the parametrization.
Finally, we obtained the preliminary results in eq.~\eqref{eq:l4569}
for the combinations of the mesonic $\textmd{SU(3)}$ LECs
$L_{85}=2L_8-L_5$ and $L_{64}=2L_6-L_4$.

\subsection{The scale parameter $\boldsymbol{t_0}$}
We determined the scale parameter $t_0$ in $N_f=2+1$ QCD
using the mass of the $\Xi$
baryon, along with $M_\pi$ and $M_K$ to define the physical
quark mass point. Using this input, we observe that
$t_0$ is quite robust against cuts in the data and changes
of the parametrization with respect to the lattice spacing,
the meson mass and the volume dependence, see
figures~\ref{fig:t0average}, \ref{fig:t0aeff} and~\ref{fig:t0fitfo}.
To summarize our results,
here we give values of $t_0$ at four different points in the quark
mass plane,
\begin{itemize}
\item $t_{0,\text{ph}}$, the value at the physical point 
  (where $\phi_4=\phi_{4,\text{ph}}=1.093^{(11)}_{(13)}$),
\item $t_0^*$, the value at the point where $\phi_4=\phi_4^*=1.11$ and
  $m_s=m_{\ell}$,
\item $t_{0,\text{ch}}$, the value in the $N_f=3$ chiral limit (where $\phi_4=0$) and
\item $t_{0,\text{ch2}}$, the value in the $\textmd{SU(2)}$ chiral limit
  (where $\phi_4=\phi_{4,\text{ch2}}=1.014^{(11)}_{(13)}$):
\end{itemize}
\begin{align}
  \sqrt{8t_{0,\text{ph}}}&=0.4098^{(20)}_{(25)}\,\textmd{fm},&
  \sqrt{8t_0^*}&=0.4097^{(20)}_{(25)}\,\textmd{fm},\label{eq:resultt0}\\
  \sqrt{8t_{0,\text{ch}}}&=0.4170^{(22)}_{(27)}\,\textmd{fm},&
  \sqrt{8t_{0,\text{ch2}}}&=0.4108^{(22)}_{(25)}\,\textmd{fm}.
\end{align}
All systematics are included in the error estimates.
A comparison of $\sqrt{8t_{0,\text{ph}}}$
with other determinations of this parameter
is shown in figure~\ref{fig:t0compare}
and discussed in section~\ref{sec:resultsummary}.
The leading meson mass dependence of $t_0$ in the continuum limit is given
in eq.~\eqref{eq:t0chi} and the respective
LEC $k_1$~\cite{Bar:2013ora} is displayed in eq.~\eqref{eq:k1result}. Within
the present accuracy, we are unable to resolve any higher
order corrections.

\subsection{The light baryon spectrum}
\label{sec:lightb}
\begin{figure}[htp]
  \centering
  \resizebox{0.95\textwidth}{!}{\includegraphics[width=\textwidth]{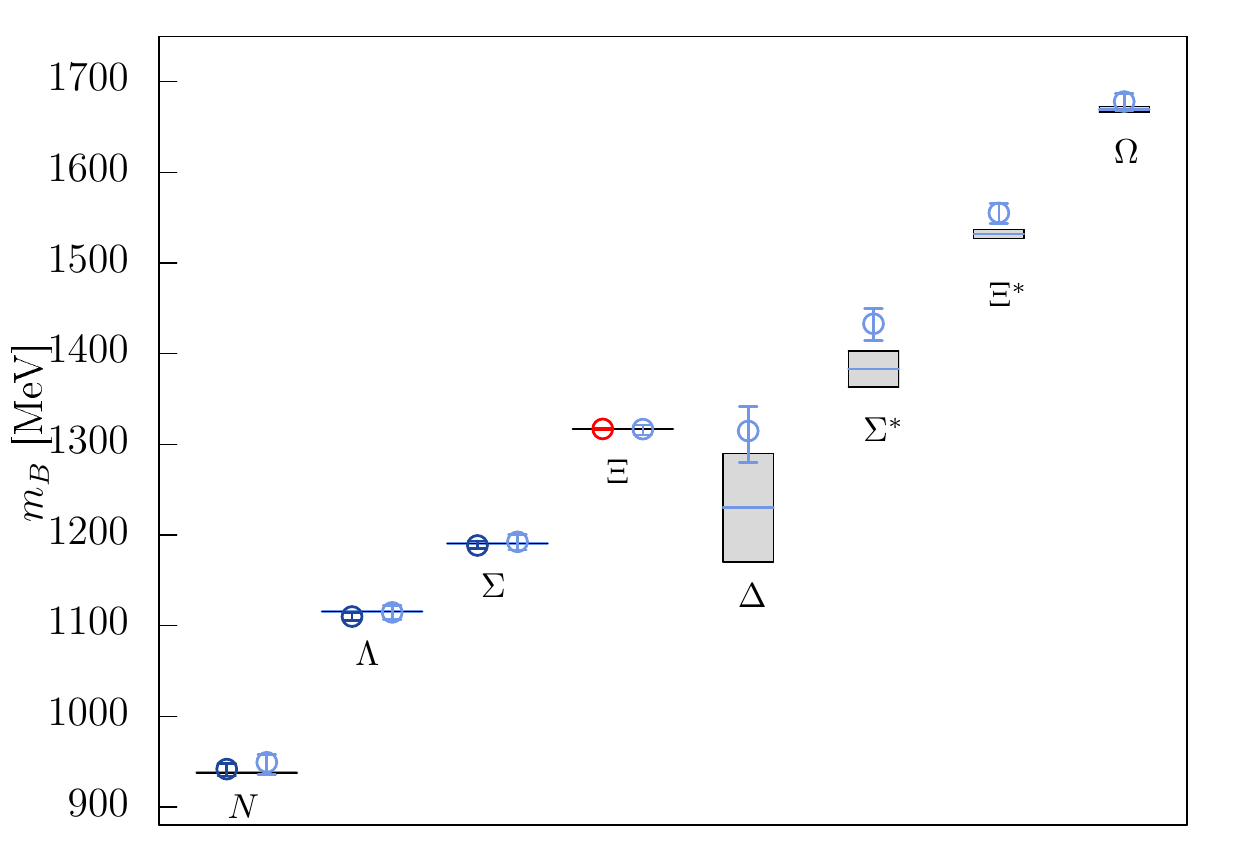}}
  \caption{The continuum limit baryon spectrum at the physical point,
    obtained via the AIC averaging procedure. The dark blue symbols
    correspond to results obtained from the octet baryons. The
    mass of the $\Xi$ baryon (red circle) was used to
    set the scale. The light symbols correspond to simultaneous BChPT fits
    to the octet and decuplet baryon spectrum. The lines and grey
    boxes depict the isospin violation corrected experimental
    values of table~\ref{tab:contmasses}, including the uncertainties.
    The $\Delta$, $\Sigma^*$ and $\Xi^*$ masses correspond to the maxima of the
    respective Breit--Wigner distributions and their errors reflect the
    widths of these distributions.
  \label{fig:baryonspectrum}}
\end{figure}
In figure~\ref{fig:specover} we investigate the impact of the
parametrization of the continuum meson mass dependence on the
baryon spectrum. The results from the preferred parametrization
(BChPT including finite volume effects) are listed in
table~\ref{tab:specresults}. In particular, we obtain for the
nucleon mass $m_N=941.7^{(6.5)}_{(7.6)}\,\textmd{MeV}$, which should be
compared to the value $937.5\,\textmd{MeV}$ that is
obtained, when correcting the experimental proton and neutron masses
for the isospin breaking QED and QCD effects.
The other stable baryons also agree within errors with the expectations.
Note that the uncertainty on the nucleon mass is just $\sim 7\,\textmd{MeV}$,
i.e.\ less than 0.8\%, while the uncertainties regarding the $\Lambda$ and
$\Sigma$ baryon masses are $\sim 4\,\textmd{MeV}$, i.e.\ less than
0.4\%. Our data driven approach with a very small residual dependence
on the parametrization was possible due to the excellent coverage
of the quark mass plane as well as by the availability of six lattice spacings
with $a^2$ varying by a factor of more than six.
In particular, the physical quark mass point is tightly constrained due to
the intersection of the $\Tr M=\text{const}$ and $\widehat{m}_s\approx\text{const}$
trajectories, see, e.g.,
figures~\ref{fig:bchptall} and~\ref{fig:bestoct} for the octet baryon masses.
The error reduction in comparison to previous continuum
limit results, where the errors stated did not always include all
sources of systematics, is evident from figure~\ref{fig:specoth}.

In figure~\ref{fig:baryonspectrum} we visualize the results
for the light octet and decuplet baryon masses given in
table~\ref{tab:specresults}. The lines and boxes correspond
to the experimental values, corrected for isospin breaking effects,
of table~\ref{tab:contmasses}, including the uncertainties of these
estimates. For the $\Delta$, the $\Sigma^*$ and the $\Xi^*$ we show the
Breit--Wigner masses rather than the real parts of their poles, together with
half the respective widths of the Breit--Wigner distributions as errors.
The three dark circles are our predictions from
the averaged fits to the octet baryon masses alone. The red circle is the
corresponding $\Xi$ baryon mass, which was used to set the scale
for the other predictions. As mentioned above,
we find agreement with experiment within sub-percent level errors.
The light circles are the results of
AIC averaged joint BChPT fits to the octet and decuplet baryons. Also
here we find agreement with the expectation for the octet baryons, as
well as for the $\Omega$ baryon.

Regarding the remaining decuplet
resonances, namely the $\Delta$, the $\Sigma^*$ and the $\Xi^*$, our
predictions from fits where only stable baryons are considered
lie above the positions of the experimental resonances by slightly
more than one decay width.
This illustrates the limitations of determining ``masses'' and
other properties of baryon resonances in the absence of
a dedicated scattering study relating the spectrum of QCD in a
finite volume to the resonance parameters~\cite{Gockeler:2012yj},
e.g., within the
framework of (H)BChPT~\cite{Detmold:2015qwf}. Such an
investigation~\cite{Andersen:2017una,Silvi:2021uya}
would require additional interpolators that specifically
couple to octet baryons plus pions as well as additional volumes and/or
non-zero momentum frames.

The Gell-Mann--Okubo mass relation for octet baryons
implies that the flavour averaged mass combinations
$X_N$ and $X_\Lambda$ defined in eq.~\eqref{eq:singmas} agree
at linear order in the $\textmd{SU(3)}$ symmetry breaking parameter
$\delta m$ if the average squared pseudoscalar mass is kept
constant. Using the experimental masses, this relation is
only violated by about 0.5\%, which is consistent with our lattice data.
This suggests that non-linear
$\textmd{SU(3)}$ symmetry breaking effects are
similarly small at the physical point.
One interesting observation is summarized in figure~\ref{fig:average}:
the flavour breaking effects for these two mass
combinations between the symmetric and the physical points turn out
to be about 4\%, an order of magnitude larger than the violation of
the GMO relation. Since the correction is very
similar for both averages shown, most of this effect cancels
in the GMO relation.

\section{Conclusions and outlook}
\label{sec:conclusions}
This article is based on Coordinated Lattice
Simulations (CLS) ensembles~\cite{Bruno:2014jqa,Bali:2016umi,Mohler:2017wnb}
that have been generated using the {\sc openQCD}~\cite{Luscher:2012av} code,
employing the non-perturbatively $\mathcal{O}(a)$ improved Wilson action on
top of the tree-level improved Symanzik gauge action. This study
utilizes the unique combination of a large variation in the
lattice spacing and good coverage of the quark mass plane: for
the simultaneous continuum, infinite volume and physical mass
point extrapolation we have determined two pseudoscalar and
eight baryon masses (as well as $t_0/a^2$ and the quark masses) on
almost 50 distinct gauge ensembles, encompassing six lattice spacings
(covering a factor of more than six in terms of $a^2$) that are
scattered around three distinct trajectories in the quark mass plane,
including the physical point. The main results on the $\sigma$~terms,
$\textmd{SU(3)}$ and $\textmd{SU(3)}$ ChPT
low energy constants, the scale parameter $t_{0,\text{ph}}$ and the
light baryon spectrum are summarized in
sections~\ref{sec:sigmab}--\ref{sec:lightb} above.

Since Wilson fermions provide an excellent compromise between
theoretical rigour and computational affordability
within the landscape of fermion formulations available in lattice QCD,
further simulation points will be realized in the future.
We determined interpolations for a number of parameters, e.g., the
critical hopping parameter $\kappa_{\text{crit}}$, the
ratio of the singlet over the non-singlet quark mass renormalization
constants $r_m$, various combinations of order~$a$ improvement
coefficients as well as the scale $t_0^*/a^2$
as functions of the inverse coupling parameter $\beta$,
see section~\ref{sec:lattb} for references to the relevant equations,
figures and tables. Having mapped
out the parameter space for our action will enable a very efficient
planning of new simulation points.
Another important result is the determination of the lattice scales
$t_{0,\text{ph}}$ and $t_0^*$, see eq.~\eqref{eq:resultt0}, from the
experimentally very well known $\Xi$ baryon mass.

Following up on this work, we plan investigations of the quark masses,
pseudoscalar decay constants and nucleon matrix elements.
Of particular interest in this context is $\textmd{SU(3)}$ ChPT.
On the one hand, due to the heavier strange quark mass, its convergence
properties at the physical point must be inferior to those of
$\textmd{SU(2)}$ ChPT. On the other hand, the increase of the
number of LECs when going from $\textmd{SU(2)}$
to $\textmd{SU(3)}$ is much smaller than the increase of the number
of independent observables that can be used to constrain these.
Here we have demonstrated that within sub-percent level
accuracy the mass and volume dependence
of the whole octet baryon spectrum in the iso-symmetric
continuum limit can be parameterized in terms of just six LECs,
at least for
$\overline{M}\vphantom{M}^2=(2M_{K}^2+M_{\pi}^2)/3<(440\,\textmd{MeV})^2$
and $M_{\pi}>130\,\textmd{MeV}$. 
In the future we plan a simultaneous baryon ChPT analysis of the axial
charges~\cite{Bali:2019svt,RQCD:2019jai,Bali:2022qja},
directly determined $\sigma$~terms~\cite{Petrak:2021aqf}
and the baryon spectrum to establish the universality of the
LECs and to arrive at increased precision regarding
these observables and $t_{0,\text{ph}}$.

\acknowledgments
GB, SC and DJ acknowledge support through the European Union’s Horizon 2020 research and innovation programme under the Marie Skłodowska\nobreakdash-Curie grant agreement no.~813942 (ITN EuroPLEx). SC and SW received support through the German Research Foundation (DFG) grant CO~758/1-1. Part of the work of GB and SW was funded by the German Federal Ministry of Education and Research (BMBF) grant no.~05P18WRFP1. Additional support from the EU through grant agreement no.~824093 (STRONG~2020) is gratefully acknowledged, as well as initial stage funding through the DFG collaborative research centre SFB/TRR\nobreakdash-55. This work was also supported by DFG grant NFDI 39/1 (PUNCH4NFDI).\par
We thank all our CLS colleagues for discussions and the joint production of the gauge ensembles. Moreover, we thank Benjamin Gläßle, Fabian Kaiser, Andreas Rabenstein, Philipp Wein and Thomas Wurm for their contributions at the beginning of this project regarding code development, the generation of some of the two-point functions and cross checks.\par
The authors gratefully acknowledge the \href{https://www.gauss-centre.eu}{Gauss Centre for Supercomputing (GCS)} for providing computing time through the \href{http://www.john-von-neumann-institut.de}{John von Neumann Institute for Computing (NIC)} on the supercomputer JUWELS~\cite{juwels} and, in particular, on the Booster partition of the supercomputer JURECA~\cite{jureca} at \href{http://www.fz-juelich.de/ias/jsc/}{J\"ulich Supercomputing Centre (JSC)}. GCS is the alliance of the three national supercomputing centres HLRS (Universität Stuttgart), JSC (Forschungszentrum Jülich), and LRZ (Bayerische Akademie der Wissenschaften), funded by the BMBF and the German State Ministries for Research of Baden\nobreakdash-Württemberg (MWK), Bayern (StMWFK) and Nordrhein\nobreakdash-Westfalen (MIWF).  The authors also acknowledge the In\-ter\-dis\-ci\-plin\-ary Centre for Mathematical and Computational Modelling (ICM) of the University of Warsaw for computer time on Okeanos (grant Nos.\ GA67\nobreakdash-12, GA69\nobreakdash-20 and GA72\nobreakdash-26) and AGH Cyfronet Computing Center under Grant ID pionda, nspt, hadronspectrum. Additional computations were carried out on the SFB/TRR\nobreakdash-55 QPACE~2~\cite{Arts:2015jia} Xeon~Phi installation and the iDataCool cluster in Regensburg, on the QPACE~3 Xeon~Phi cluster of SFB/TRR\nobreakdash-55 hosted at JSC and the Regensburg Athene~2 cluster. The authors also thank the JSC for their support and for providing services and computing time on the HDF Cloud cluster~\cite{hdfcloud} at JSC, funded via the Helmholtz Data Federation (HDF) programme.\par
Most of the ensembles were generated using \href{https://luscher.web.cern.ch/luscher/openQCD/}{\sc openQCD}~\cite{Luscher:2012av} within the \href{https://wiki-zeuthen.desy.de/CLS/}{Coordinated Lattice Simulations (CLS)} effort. A few additional ensembles were generated employing the {\sc BQCD}\nobreakdash-code~\cite{Nakamura:2010qh} on the QPACE supercomputer of SFB/TRR\nobreakdash-55. For the computation of the correlation functions we used a modified version of the {\sc Chroma}~\cite{Edwards:2004sx} software package along with the {\sc Lib\-Hadron\-Analysis} library and the multigrid solver implementation of Refs.~\cite{Heybrock:2015kpy,Georg:2017zua} (see also ref.~\cite{Frommer:2013fsa}) as well as the IDFLS solver~\cite{Luscher:2007es} of {\sc openQCD}.

\appendix
\section{Correcting for electrical and mass isospin breaking effects}
\label{sec:isospin}
In our simulations we neglect isospin breaking effects.
Isospin breaking is controlled by two small parameters, the quark
mass difference $(m_d-m_u)/\Lambda$ and the fine structure constant
$\alpha_{\rm fs}$. Both parameters are of the size $10^{-2}$. The factorization
of isospin breaking into QCD and QED effects is not unique, see,
e.g., refs.~\cite{Bijnens:1993ae,Moussallam:1997xx,Gasser:2003hk}.
A popular way is to choose the $\MS$ scheme at the scale
$\mu=2\,\textmd{GeV}$~\cite{Gasser:2003hk} in order to disentangle
these effects. Regarding the baryons we will deviate from this prescription
as the ambiguity between different procedures is smaller
than the precision required here.

We match our simulation parameters to a world where isospin
breaking effects are subtracted.
The starting point is the splitting between neutral and charged
pions. Unlike the charged pion, the neutral pion two-point function
receives contributions from a disconnected quark line diagram.
Formally, this is of second order in the QCD and QED isospin breaking
parameters. Restricting ourselves to first order corrections,
the mass difference is entirely electromagnetic in nature and
the isospin corrected pion mass can be obtained, e.g.,
following the conventions outlined in ref.~\cite{Gasser:2003hk}.
We use the so-obtained pion and kaon mass
values of the FLAG~2016 Review~\cite{Aoki:2016frl} and
define the physical point in the quark mass plane at each
lattice spacing as the position where
\begin{equation}
\label{eq:pik}
M_{\pi,\text{ph}}=134.8(3)\,\textmd{MeV}\quad\text{and}\quad
M_{K,\text{ph}}=494.2(3)\,\textmd{MeV}.
\end{equation}
These values correspond to electrically neutral isospin-averaged
pions and kaons, see the discussion in Sections 3.1.1 
and 3.1.2 of ref.~\cite{Aoki:2016frl}.

Also in the case of a baryon $B$, isospin breaking effects can be
parameterized in terms of QCD- and QED-related
baryon mass shifts, $\Delta m^{\text{QED}}_B$
and $\Delta m^{\text{QCD}}_B$: we start by decomposing
the on-shell matrix element for the isovector vector current
between baryons $B^{Q+1}$ and $B^Q$ (that differ by $\Delta I_3=1$
in their isospin)
for $p'\approx p$ into formfactors:
\begin{equation}\partial^{\mu}\langle B^{Q+1}(p')|\bar{d}\gamma_{\mu}u|B^Q(p)\rangle=\partial^{\mu}
\bar{u}_{B^{Q+1}}(p')\left[g_{V,B}\gamma_\mu+\ldots\right]u_{B^Q}(p),\end{equation}
where $g_{V,B}=1$ to leading order in the symmetry breaking parameters
and the ellipses denote terms that vanish for $p'=p$.
Combining this with the vector Ward
identity $i\partial^{\mu}\bar{d}\gamma_{\mu}u=(m_u-m_d)\bar{d}u$,
one obtains
\begin{equation}
  \Delta m^{\text{QCD}}_B=g_{S,B}(m_u -m_d)
  \label{eq:masssh}
\end{equation}
as the QCD contribution to the
mass difference, where $g_{S,B}$ denotes the isovector scalar
charge in the iso-symmetric limit. Since we neither precisely
know the isovector scalar charges $g_{S,B}$ nor the quark
mass difference $m_u-m_d$, we resort to current algebra arguments ---
which is sufficient for the precision that is required in the present
context. We assume that the splittings can be explained
by two parameters, $\delta m^{\text{QED}}>0$ and $\delta m^{\text{QCD}}>0$,
that are the same across an $\textmd{SU(3)}$ multiplet,
the former multiplying the difference of the squared electrical
charges and the latter the isospin difference.
This assumption implies the Coleman--Glashow theorem~\cite{Coleman:1961jn}
$\Delta m_N-2\Delta m_{\Sigma}+\Delta m_{\Xi}=0$.
It also corresponds to $\Delta m_B^{\text{QCD}}=-\delta m^{\text{QCD}}$,
i.e.\  the scalar couplings $g_{S,B}$ are assumed to be independent
of the baryon $B$. The proportionality of $\delta m^{\text{QED}}$ to
the square of the electric charge is consistent
with the Dashen theorem~\cite{Dashen:1969eg}.
In other words, to leading order
in the isospin breaking effects, the masses of neutral particles do not receive
QED contributions. We obtain
\begin{align}
  \label{eq:iso1}
\Delta m_N &= m_p - m_n \approx -\delta m^{\text{QCD}}+\delta m^{\text{QED}},\\
2\Delta m_{\Sigma} &= m_{\Sigma^+} - m_{\Sigma^-} \approx -2\delta m^{\text{QCD}},\\
\Delta m_{\Xi} &= m_{\Xi^0} - m_{\Xi^-} \approx -\delta m^{\text{QCD}}-\delta m^{\text{QED}}.
  \label{eq:iso3}
\end{align}
The resulting Coleman--Glashow theorem was confirmed to hold within an accuracy
of $0.13\,\textmd{MeV}$ in ref.~\cite{Borsanyi:2014jba}, while the experimental
value reads $0.06(23)\,\textmd{MeV}$~\cite{Workman:2022ynf}.

Plugging in the experimental masses gives
\begin{align}
\delta m^{\text{QED}}&=\frac12
\left(\Delta m_N-\Delta m_{\Xi}\right)=2.78(11)\,\textmd{MeV},\nonumber\\
&=\Delta m_N-\Delta m_{\Sigma}=2.75(4)\,\textmd{MeV},\label{eq:mqed}
\end{align}
where
we will use the second, more precise difference as our central value.
However, as the Coleman--Glashow relation does not necessarily hold
with the same accuracy, we will apply
the larger error of $0.11\,\textmd{MeV}$.

We can now proceed to
compute the mass of a hypothetical uncharged isospin symmetric
$\Sigma$ baryon:
\begin{equation}
\label{eq:sigma}
m_{\Sigma}=\frac12\left(m_{\Sigma^+}+m_{\Sigma^-}\right)-\delta m^{\text{QED}}
=1190.66(12)\,\textmd{MeV}.
\end{equation}
However, the charge-neutral $\Sigma^0$ baryon has the mass
$m_{\Sigma^0}=1192.64(2)\,\textmd{MeV}$. In other words:
computing $\tfrac12(m_{\Sigma^+}+m_{\Sigma^-})-m_{\Sigma^0}
\approx 0.8\,\textmd{MeV}$ gives an estimate of
$\delta m^{\text{QED}}$ that is much smaller than the
$2.75\,\textmd{MeV}$ of eq.~\eqref{eq:mqed}.
Isospin breaking enables mixing between the two $I_3=0$ baryons
$\Sigma^0$ and $\Lambda$. However, the impact
of this on the baryon masses should be quadratic
in the isospin breaking parameters and also there are
numerical indications that the mixing angle
is small~\cite{Horsley:2014koa,Horsley:2015lha}.
Therefore, the difference between the $\Sigma^0$ mass and eq.~\eqref{eq:sigma}
strongly suggests that more than two parameters are needed to
parameterize the isospin violations of the octet baryon masses,
in spite of the experimental accuracy of
the Coleman--Glashow relation: the spatial extension of a baryon and therefore 
its QED mass shift may depend on its strange quark content.
Moreover, as we discussed above (see eq.~\eqref{eq:masssh}),
the QCD mass shifts are proportional to the isovector scalar
couplings $g_{S,B}$, which in general will depend on the strangeness of
the baryon too. Indeed, in ref.~\cite{Borsanyi:2014jba} the Coleman--Glashow
theorem was confirmed to hold, however,
$\Delta m_B^{\text{QCD}}$ was reported to differ between the octet
baryons $N$, $\Sigma$ and $\Xi$. Nevertheless, here we will use the
value eq.~\eqref{eq:sigma} but we have to
keep in mind that its uncertainty may be bigger than the suggested
error, i.e.\ in the worst case it could be as large as $2\,\textmd{MeV}$.

\begin{table}[t]
  \caption{(Pseudo-)experimental masses, corrected for the isospin breaking
    effects due to the light quark mass difference and electric charges.
    Note that for the baryons we have chosen one particular prescription
    so that an additional uncertainty of up to 2\,\textmd{MeV} may exist.
    Regarding the three unstable decuplet baryons, we display their
    Breit--Wigner masses with the Breit--Wigner half-widths as errors.
    \label{tab:contmasses}}
  \begin{center}
  \begin{tabular}{cc}\toprule
    mass&value$/\textmd{MeV}$\\\midrule
    $M_{\pi}$&134.8(3)\\
    $M_K$&494.2(3)\\
    $m_N$&937.54(6)\\
    $m_{\Lambda}$&1115.68(1)\\
    $m_{\Sigma}$&1190.66(12)\\
    $m_{\Xi}$&1316.9(3)\\
    $m_{\Delta}$&1230(60)\\
    $m_{\Sigma^*}$&1383(20)\\
    $m_{\Xi^*}$&1532(5)\\
    $m_{\Omega}$&1669.5(3.0)\\\bottomrule
  \end{tabular}
  \end{center}
\end{table}

In this article we use the cascade mass for the scale
setting. We compute
\begin{equation}
m_{\Xi}\coloneqq \frac12\left(m_{\Xi^0}+m_{\Xi^-}-\delta m^{\text{QED}}\right)
=1316.9(3)\,\textmd{MeV},
\end{equation}
where we carry out the isospin average and correct the $\Xi^-$ mass
for the charge effect. Similarly, the isospin symmetric nucleon mass can be
obtained as
\begin{equation}
m_{N}\coloneqq \frac12\left(m_{n}+m_{p}-\delta m^{\text{QED}}\right)
=937.54(6)\,\textmd{MeV}.
\end{equation}
Again, the errors displayed above may be
underestimated: considering the deviation of the $\Sigma^0$ mass
from the expectation, the real uncertainty of removing isospin
breaking effects could be as large as 2\textperthousand, which, however,
is much smaller than the uncertainty of 5\textperthousand\ of the
scale determination that we carry out here.

Regarding the decuplet, in this case only the $\Omega$ is a
narrow resonance, while $\Xi^*(1530)$, $\Sigma^*(1385)$ and
$\Delta(1232)$
have decay widths of about $10\,\textmd{MeV}$, $40\,\textmd{MeV}$ and
$120\,\textmd{MeV}$, respectively. For the real parts of the
$\Xi^*$ poles experiment gives
\begin{equation}
  m_{\Xi^{*-}}-m_{\Xi^{*0}}=3.2(6)\,\textmd{MeV}\approx \delta m^{\text{QCD}}+\delta m^{\text{QED}}.
\end{equation}
For the octet we found $\delta m^{\text{QCD}}\approx 4\,\textmd{MeV}$.
So the above difference is already saturated by the difference of the quark
masses. This means that due to the presence of strong decay channels,
we cannot reliably determine $\delta m^{\text{QED}}$ for the decuplet
baryons.
Regarding the $\Omega$, it is conceivable that this is spatially
more compact than octet baryons, due to the heavier strange quark, so
the octet value $\delta m^{\text{QED}}\approx 2.75(11)\,\textmd{MeV}$
may be an underestimate. In view of the above, we subtract $3\,\textmd{MeV}$
from the $\Omega^-$ baryon mass and add this as our systematic uncertainty,
arriving at
\begin{equation}
  m_{\Omega}=1669.5(3.0)\,\textmd{MeV}.
\end{equation}
Determining the other decuplet baryon poles reliably
will require a dedicated scattering study. Since
this is not carried out in the present article, we will
compare our results to the experimental Breit--Wigner mass
values of the $\Xi^*$,
the $\Sigma^*$ and the $\Delta$ decuplet baryons, with the
errors given by half of the Breit--Wigner widths.

We summarize the masses of isospin
symmetric QCD discussed above in table~\ref{tab:contmasses}.

\section{Finite size effects on baryon masses}
\label{sec:finitebaryon}
The expected dependence of the pseudoscalar masses on the spatial
lattice extent $L$ is presented in section~\ref{sec:finite}.
Here we provide the corresponding expressions regarding the
octet and decuplet baryon masses.
The finite volume effects in a cubic box of size $L^3$
have been computed in $\textmd{SU(2)}$ and in $\textmd{SU(3)}$
HBChPT and covariant BChPT, first at order $p^3$ and then at
order $p^4$~\cite{AliKhan:2003ack,Beane:2004tw,WalkerLoud:2004hf,Procura:2006bj,WalkerLoud:2008bp,Geng:2011wq,Ren:2012aj}, also including
baryon loop effects~\cite{Beane:2004tw,Procura:2006bj,Geng:2011wq,Lutz:2014oxa,Lutz:2018cqo}.
To be consistent
with the dependence on the pseudoscalar masses, here we will also
only make use of the order $p^3$ results. Allowing for pseudoscalar
loops only, the volume dependence in BChPT
is given as~\cite{AliKhan:2003ack,Procura:2006bj,Geng:2011wq,Ren:2012aj}
\begin{align}
  \label{eq:baryonfse}
  m_B(L)=m_B+\frac{1}{(4\pi F_0)^2}\!\!\sum_{P\in\{\pi,K,\eta_8\}}\!\!\!\!\!g_{B,P}M_P^3\,2
  \int_0^{\infty}\!\!\!\dd{x}\sum_{\mathbf{n}\neq\mathbf{0}}K_0\left(\lambda_{M}
  |\mathbf{n}|\sqrt{1+x^2-\frac{M_P}{m_{0|D0}}x}\right),
\end{align}
where the couplings $g_{B,P}$ are expressed in terms of the LECs
$D$, $F$ and $\mathcal{H}$ in eqs.~\eqref{eq:octcop}
and~\eqref{eq:deccop} and $\lambda_P=LM_P$.
For the octet and decuplet baryons,
$m_0$ and $m_{D0}=m_0+\delta$ should be used, respectively, in the argument of
the square root. The finite volume corrections
are independent of the covariant
ultraviolet regulator (e.g., the EOMS or the IR scheme),
however, the HBChPT formulae differ somewhat since terms proportional to
$M_P/m_0$ are consistently neglected in this case. In the heavy baryon
limit the integration can be carried out
analytically~\cite{Beane:2004tw,WalkerLoud:2004hf,Detmold:2005pt,WalkerLoud:2008bp}:
\begin{align}
  \label{eq:hbfse}
  2\int_0^{\infty}\!\!\!\dd{x}\,K_0\left(\lambda_{M}
  |\mathbf{n}|\sqrt{1+x^2-\frac{M_P}{m}x}\right)\stackrel{m\rightarrow\infty}{\longrightarrow}
  \pi\frac{e^{-\lambda_P|\mathbf{n}|}}{\lambda_P|\mathbf{n}|}.
\end{align}

In addition, one encounters corrections from loops involving transitions
between octet and decuplet baryons~\cite{Procura:2006bj,Geng:2011wq}:
\begin{align}
  m_O(L)&\mapsto
  m_O(L)+
  \frac{\mathcal{C}^2}{(4\pi F_0)^2}
  \frac{m_{0}^2}{m_{D0}^2}\!\!\sum_{P\in\{\pi,K,\eta_8\}}\!\!\!\!\!\xi_{O,P}M_P^2
  \,\delta\!\int_0^{\infty}\!\!\!\dd{y}\,\Biggl\{\!\left[2+\frac{\delta}{m_0}(1-y)\right]g_O(y)
  \nonumber\\
  &\qquad\qquad\qquad\times\sum_{\mathbf{n}\neq\mathbf{0}}\left[g_O(y)K_0(\lambda_P|\mathbf{n}|g_O(y))
    -\frac{K_1(\lambda_P|\mathbf{n}|g_O(y))}{\lambda_P|\mathbf{n}|}\right]\Biggr\}
  \label{eq:ssefv1}
\end{align}
with
\begin{align}
  g_O(y)&=\sqrt{\left(1-\frac{\delta^2}{M^2_P}\right)\left(1-\frac{\delta}{m_0}y\right)
    +\frac{\delta^2}{M_P^2}(1+y)^2}.
\end{align}
The coefficients $\xi_{B,P}$ can be found in table~\ref{tab:sse}.
The decuplet masses receive similar octet loop corrections:
\begin{align}
  m_D(L)&\mapsto
  m_D(L)+
  \frac{\mathcal{C}^2}{(4\pi F_0)^2}
  \frac{m_{D0}^2}{m_0^2}\!\!\sum_{P\in\{\pi,K,\eta_8\}}\!\!\!\!\!\xi_{O,P}M_P^2
  \,\delta\!\int_0^{\infty}\!\!\!\dd{y}\,\Biggl\{\!\left[2-\frac{\delta}{m_{D0}}(1+y)\right]g_D(y)
  \nonumber\\
  &\qquad\qquad\qquad\times\sum_{\mathbf{n}\neq\mathbf{0}}\left[g_D(y)K_0(\lambda_P|\mathbf{n}|g_D(y))
    -\frac{K_1(\lambda_P|\mathbf{n}|g_D(y))}{\lambda_P|\mathbf{n}|}\right]\Biggr\}
\end{align}
with
\begin{align}
  g_D(y)&=\sqrt{\left(1-\frac{\delta^2}{M^2_P}\right)\left(1-\frac{\delta}{m_{D0}}y\right)
    +\frac{\delta^2}{M_P^2}(1-y)^2}.\label{eq:ssefv2}
\end{align}
The decuplet baryons may become unstable for $M_P<\delta$ and the infinite
volume results as well as the finite volume corrections acquire
imaginary parts. As for the infinite volume result,
we take the real part and restrict the fit to only include
decuplet baryons that are stable.
For imaginary $g_D(y)$,
$K_0$ becomes imaginary while $K_1$ remains real, which means that
the integrand is purely imaginary. Taking the real part, therefore,
corresponds to omitting the part of the integration range where
the argument of the square root in $g_D(y)$ is negative. The corresponding
$y$-interval is symmetric
about $y=1-\delta/(2m_{D0})+M_P^2/(2 m_{D0}\delta)$.

In the heavy baryon limit, the above integrals simplify to~\cite{Beane:2004tw,WalkerLoud:2008bp}
\begin{align}
  \delta\int_0^{\infty}\cdots
  \stackrel{m_0\rightarrow\infty}{\longrightarrow}
  2\delta\int_0^{\infty}\!\!\!\dd{y}\,\Biggl\{f_{\pm}(y)\sum_{\mathbf{n}\neq\mathbf{0}}
  \left[f_{\pm}(y)K_0(\lambda_P|\mathbf{n}|f_{\pm}(y))
    -\frac{K_1(\lambda_P|\mathbf{n}|f_{\pm}(y))}{\lambda_P|\mathbf{n}|}\right]\Biggr\}
\label{eq:heavy}
\end{align}
with
\begin{equation}
  f_{\pm}(y)
  =\sqrt{1-\frac{\delta^2}{M_P^2}+\frac{\delta^2}{M_P^2}(y\pm 1)^2},
  \label{eq:heavy2}
\end{equation}
where the plus (minus) sign is for the octet (decuplet)
baryons. For $M_P<\delta$, the integrand becomes imaginary
in the range
\begin{equation}
  y\in\left( 1-\sqrt{1-\frac{M_P^2}{\delta^2}},
  1+\sqrt{1-\frac{M_P^2}{\delta^2}}
  \right),
\end{equation}
which we omit from the integration region.\footnote{Note that in this case
the integral from zero
to the lower limit is equal to that from the upper limit to 2.}
In the octet case, for $\lambda_P\gg 1$ the
integral~\eqref{eq:heavy} has the limiting behaviour
\begin{equation}
  \sqrt{2\pi}\,\delta\, \frac{M_P^2}{\delta^2}
  \sum_{\mathbf{n}\neq\mathbf{0}}
  \frac{e^{-\lambda_P|\mathbf{n}|}}{(\lambda_P|\mathbf{n}|)^{3/2}},
\end{equation}
which one may further approximate, setting $|\mathbf{n}|=1$ and
replacing the sum by a factor of six.
For small values of the ratio $\delta/M_P$,
this limit is only approached slowly. Here we use the full
(H)BChPT expressions eqs.~\eqref{eq:heavy} and~\eqref{eq:heavy2}.

\section{The $\boldsymbol{\sigma}$ terms}
\label{sec:sigmadef}
We define the $\sigma$~terms
\begin{align}
  \label{eq:sigmadef}
  \sigma_{qB}
  =m_q\left[\frac{\left\langle B\left|\bar{q}\mathds{1}q\right|B\right\rangle}
    {2m_B}-
    V_3\left\langle\Omega\left|\bar{q}\mathds{1}q\right|\Omega\right\rangle\right]
    =m_q\frac{\partial m_B}{\partial m_q},
\end{align}
where $|\Omega\rangle$ denotes the vacuum, $V_3$ the spatial volume
and the states at zero momentum
are normalized as $\langle B|B\rangle=2m_BV_3$.
The $\sigma$~terms can approximately be obtained from the dependence of
the baryon masses $m_B$ on the squared pseudoscalar meson masses $M^2_P$.
We discuss this below, as well as the corrections we encounter due
to the mass rescaling $\mathbbm{m}_B=\sqrt{8t_0m}_B$ and
$\mathbbm{M}_P^2=8t_0M_P^2$. We then further correct the formulae,
taking into account violations of the GMOR relations, in
appendix~\ref{sec:gmor2}, where we also determine
the combinations of the mesonic $\textmd{SU(3)}$ LECs
$2L_8-L_5$ and $2L_6-L_4$.

\subsection{The $\boldsymbol{\sigma}$ terms from the meson mass dependence}
In addition to the baryon masses and the LECs, it is also possible to
obtain the $\sigma$~terms:
by relating the derivatives with respect to a quark mass $m_q$
to derivatives with respect to squared pseudoscalar masses via
the GMOR relations, it is easy to see that
\begin{equation}
  \tilde{\sigma}_{\pi B}=M_{\pi}^2\left(\frac23\frac{\partial m_B}{\partial\overline{M}\vphantom{M}^2}-\frac{\partial m_B}{\partial\delta M^2}\right),\quad
  \tilde{\sigma}_{sB}=\left(2M_K^2-M_{\pi}^2\right)\left(\frac{1}{3}\frac{\partial m_B}{\partial\overline{M}
    \vphantom{M}^2}+
  \frac{\partial m_B}{\partial\delta M^2}\right),\label{eq:sigmagmor}
\end{equation}
where $\sigma_{\pi B}=\sigma_{uB}+\sigma_{dB}$ and
$\sigma_{qB}=\tilde{\sigma}_{qB}\left[1+\mathcal{O}(M^2)\right]$.
The order $M^2$ corrections~\cite{Gasser:1984gg} to the GMOR relations
will depend on the combinations $2L_8-L_5$ and $2L_6-L_4$
of the mesonic $\textmd{SU(3)}$ LECs, in addition to $F_0$ and $B_0$.
This is discussed in appendix~\ref{sec:gmor2} below.
These NLO mesonic ChPT corrections should
be taken into account when determining the derivatives
using the GMO Taylor expansion, which includes order $M^4$ terms.
Within the chiral power counting, formally these terms only
need to be included in BChPT at NNNLO, whereas we only employ
the NNLO expressions (which is sufficient to describe the baryon mass data).
However, when fitting our quark masses as functions of the pion masses,
it turns out that corrections to the GMOR relations need to be taken
into account. Hence, for determining the $\sigma$~terms, we will
include them. We will distinguish between $\tilde{\sigma}_{qB}$,
the $\sigma$~terms determined assuming linear dependencies of
the quark masses on the squared pseudoscalar masses, and
$\sigma_{qB}$, the results that take into account the leading
violations of the GMOR relations. 

We remark that the LECs are defined in the
$\textmd{SU(3)}$ chiral limit while the $\sigma$~terms are given
at the physical point in the quark mass plane. Our fits are carried out
for masses that are rescaled in units of $\sqrt{8t_0}$, which in
turn depends on the
meson masses. The leading dependence of $t_0$ on the pseudoscalar masses,
eq.~\eqref{eq:t0chi}, results in the relations
\begin{align}
  \left.\frac{\partial\mathbbm{m}_{O|D}}{\partial{\overline{\mathbbm{M}}\vphantom{M}^2}}\right|_{\text{ph}}
  &=\left.\frac{1}{\sqrt{8t_0}}\
  \frac{\partial m_{O|D}}{\partial{\overline{M}\vphantom{M}^2}}\right|_{\text{ph}}+
  \frac{\tilde{k}_1}{2}\sqrt{8t_0}m_{0|D0}
  +\mathcal{O}\left(\overline{M}\vphantom{M}^2\right),\\
  \left.\frac{\partial\mathbbm{m}_B}{\partial{\delta\mathbbm{M}^2}}\right|_{\text{ph}}
  &= \left.\frac{1}{\sqrt{8t_0}}\frac{\partial m_B}{\partial{\delta M^2}}\right|_{\text{ph}}+\mathcal{O}\left(\overline{M}\vphantom{M}^2\right)
\end{align}
between the derivatives of the fit formulae and those that
we are interested in.
Up to order $\overline{M}\vphantom{M}^2$ corrections, this amounts to
\begin{align}
  \label{eq:correctt0}
  M_P^2\frac{\partial m_{O|D}}{\partial\overline{M}^2}
  &=
  \frac{\mathbbm{M}_P^2}{\sqrt{8t_{0,\text{ph}}}}
  \left(\frac{\partial\mathbbm{m}_{O|D}}{\partial\overline{\mathbbm{M}}\vphantom{M}^2}-\frac{\tilde{k}_1}{2}\mathbbm{m}_{0|D0}\right),\\
  M_P^2\frac{\partial m_B}{\partial\delta M^2}
 &=\frac{\mathbbm{M}_P^2}{\sqrt{8t_{0,\text{ph}}}}
  \frac{\partial\mathbbm{m}_B}
       {\partial\delta\mathbbm{M}^2},
       \label{eq:correctt2}
\end{align}
which is consistent with the relation~\eqref{eq:bbmshift} between
$\overline{\mathbbm{b}}$ and $\overline{b}$
(or between $\overline{\mathbbm{t}}$ and $\overline{t}$ for
the decuplet baryons) as well
as with eq.~\eqref{eq:bbmshift0}. Within our statistical accuracy
we cannot detect any higher order corrections to the dependence
of $t_0$ on the pseudoscalar masses eq.~\eqref{eq:t0chi}. Since we
will also include corrections to the GMOR relations,
we expand eq.~\eqref{eq:correctt0} one order higher (neglecting
higher order contributions to $t_0$) and obtain
for the case of the octet baryons
\begin{align}
  \label{eq:correctt1}
  M_P^2\frac{\partial m_O}{\partial\overline{M}^2}
  &=
  \frac{\mathbbm{M}_P^2}{\sqrt{8t_{0,\text{ph}}}}
  \left[\frac{\partial\mathbbm{m}_O}{\partial\overline{\mathbbm{M}}\vphantom{M}^2}-\frac{\tilde{k}_1}{2}\left(\mathbbm{m}_0+\overline{\mathbbm{b}}\,
      \overline{\mathbbm{M}}{\vphantom{M}}^2+\delta \mathbbm{b}_O\delta \mathbbm{M}^2\right)
    +\frac{\tilde{k}_1^2}{2}\overline{\mathbbm{M}}{\vphantom{M}}^2\mathbbm{m}_0\right].
\end{align}
The expression for decuplet baryons is analogous, replacing
$\mathbbm{m}_0\mapsto\mathbbm{m}_{D0}$,
$\overline{\mathbbm{b}}\mapsto\overline{\mathbbm{t}}$ and
$\delta\mathbbm{b}_O\mapsto\delta\mathbbm{t}_D$.
We will include this effect into our determinations of both
$\tilde{\sigma}_{qB}$ and $\sigma_{qB}$.

\subsection{Impact of corrections to the GMOR relations on the $\boldsymbol{\sigma}$~terms}
\label{sec:gmor2}
The general strategy of determining the $\sigma$~terms was outlined above,
including the impact of rescaling all quantities in
units of $\sqrt{8t_0}$, see eq.~\eqref{eq:correctt1}.
Here we consider higher order corrections
to eq.~\eqref{eq:sigmagmor} due to violations of the GMOR relations.

The $\sigma$~terms are defined in eq.~\eqref{eq:sigmadef}:
\begin{equation}
  \sigma_{qB}=m_q\frac{\partial m_B}{\partial m_q}=\sum_P
  m_q\frac{\partial M_P^2}{\partial m_q}\frac{\partial m_B}{\partial M_P^2}.\label{eq:pslo}
\end{equation}
For a linear dependence of the meson masses $M_P^2$ on the quark
masses $m_q$ this results in $\tilde{\sigma}_{qB}=\sigma_{qB}$,
where $\tilde{\sigma}_{qB}$ is defined in eq.~\eqref{eq:sigmagmor}.
At our level of precision, we find order $p^4$ terms to be necessary to
describe the dependence of the meson masses on the quark masses, whereas
the order $p^3$ BChPT expansion suffices to parameterize the
baryon masses. Therefore, we wish to expand the right hand side of the
above equation to order
$p^4$, even though this is not required in the BChPT power counting.
For the GMO expansion, where we include terms
proportional to $M_P^4$, expanding the meson masses
one order higher is necessary also for consistency.

Expanding eq.~\eqref{eq:pslo} to order $p^4$,
meson masses that appear
within order $p^3$ terms (or order $M_P^4$ terms
in the GMO expansion) can be substituted with the
leading order expressions
\begin{equation}
M_{\pi}^2=2B_0m_{\ell},\quad
M_{K}^2=B_0\left(m_{\ell}+m_s\right),\quad
M_{\eta_8}^2=B_0\left(\frac{2}{3}m_{\ell}+\frac{4}{3}m_s\right).\label{eq:single}
\end{equation}
Using
\begin{equation}
M^2_{\eta_8}=\frac43M_K^2-\frac13M_{\pi}^2,
\end{equation}
the leading order derivatives with respect to the logarithms of the
quark masses read
\begin{align}
  \label{eq:line1}
  m_{\ell}\frac{\partial M_{\pi}^2}{\partial m_{\ell}}&=M_{\pi}^2,\quad
  &m_{\ell}\frac{\partial M_{K}^2}{\partial m_{\ell}}&=\frac12M_{\pi}^2,\quad
  &m_{\ell}\frac{\partial M_{\eta_8}^2}{\partial m_{\ell}}&=\frac13M_{\pi}^2,\\
  m_s\frac{\partial M_{\pi}^2}{\partial m_s}&=0,\quad
  &m_s\frac{\partial M_{K}^2}{\partial m_s}&=\frac12(2M_K^2-M_{\pi}^2),\quad
  &m_s\frac{\partial M_{\eta_8}^2}{\partial m_s}&=\frac23(2M_K^2-M_{\pi}^2).
\label{eq:line2}
\end{align}
Plugging this into eq.~\eqref{eq:pslo} gives eq.~\eqref{eq:sigmagmor},
where we replace the derivatives by eqs.~\eqref{eq:correctt0}
and~\eqref{eq:correctt2}. At order $p^4$ the $\sigma$~terms will
differ from the $\tilde{\sigma}$~terms, defined in eq.~\eqref{eq:sigmagmor}.
We will consistently correct for this difference to order $p^4$.
This amounts to substituting eq.~\eqref{eq:correctt0} with
eq.~\eqref{eq:correctt1} for the mass-dependence
of $t_0$ and expanding eq.~\eqref{eq:single}
to the next order, substituting eqs.~\eqref{eq:line1} and~\eqref{eq:line2}
with the corresponding derivatives of the higher order expression.
These will depend on mesonic LECs that we shall also determine
from the dependence of the quark masses on the pseudoscalar meson masses.
Below we explain how the result is obtained.

The order $p^4$ corrections that we address below
only affect the terms proportional to
$\bar{b}$ and $\delta b_B$
(see, e.g., eq.~\eqref{eq:octetc})
and are therefore not required for $M^2_{\eta_8}$.
We define
\begin{equation}
\mu_P^2=\frac12\frac{M_P^2}{(4\pi F_0)^2}\ln\left(\frac{M_P^2}{\mu^2}\right),
\end{equation}
which we will need for $P=\pi$ and $P=\eta_8$. Up to higher orders,
we can carry out the replacements
\begin{equation}
\mu_{\pi}^2=\frac{B_0}{(4\pi F_0)^2}m_{\ell}\ln\left(\frac{2B_0m_{\ell}}{\mu^2}\right),
\quad
\mu_{\eta_8}^2=\frac{B_0}{(4\pi F_0)^2}\frac13(m_{\ell}+2m_s)\ln\left[
\frac{2}{3}\frac{B_0(m_{\ell}+2m_s)}{\mu^2}\right].
\end{equation}
The order $p^4$ relations read~\cite{Gasser:1984gg}
\begin{align}
  \label{eq:mpi}
  M_{\pi}^2&=2B_0m_{\ell}\left\{1+\mu^2_{\pi}-\frac13\mu^2_{\eta_8}+
    \frac{16B_0}{F_0^2}\left[m_{\ell}L_{85}+(m_s+2m_{\ell})L_{64}\right]\right\},\\
  M_K^2&=B_0(m_s+m_{\ell})\left\{1+\frac23\mu_{\eta_8}^2
    +\frac{8B_0}{F_0^2}\left[(m_s+m_{\ell})L_{85}+2(m_s+2m_{\ell})L_{64}\right]\right\},\label{eq:mK}
\end{align}
where
\begin{equation}
\label{eq:scalerun}
L_{85}(\mu)=2L_8(\mu)-L_5(\mu),\quad
L_{64}(\mu)=2L_6(\mu)-L_4(\mu)
\end{equation}
are  combinations of scale-dependent LECs. It is easy to see that
\begin{equation}
L_{85}(\mu')=L_{85}(\mu)-\frac{1}{12}\frac{1}{16\pi^2}\ln\left(\frac{\mu^2}{\mu^{\prime\,2}}\right),\quad
L_{64}(\mu')=L_{85}(\mu)+\frac{1}{72}\frac{1}{16\pi^2}\ln\left(\frac{\mu^2}{\mu^{\prime\,2}}\right).
\end{equation}
Inverting eqs.~\eqref{eq:mpi} and~\eqref{eq:mK} gives
\begin{align}
  \label{eq:ml}
2B_0m_{\ell}&=M_{\pi}^2\left\{1-\mu_{\pi}^2+\frac13\mu_{\eta_8}^2-\frac{8}{F_0^2}
\left[M_{\pi}^2L_{85}+\left(2M_K^2+M_{\pi}^2\right)L_{64}\right]\right\},\\
2B_0m_s&=\left(2M_K^2-M_{\pi}^2\right)\left[1-\frac{8}{F_0^2}\left(2M_K^2+M_{\pi}^2\right)L_{64}\right]+M_{\pi}^2\mu_{\pi}^2-\frac13\left(4M_K^2+M_{\pi}^2\right)\mu_{\eta_8}^2\nonumber\\
&\quad -\frac{8}{F_0^2}\left(2M_K^4-M_{\pi}^4\right)L_{85}.
\label{eq:ms}
\end{align}

In order to compute the contributions to the $\sigma$~terms according to
eq.~\eqref{eq:pslo} consistently at
order $p^4$, we need to multiply the derivatives with respect to
the quark mass of the squared pion masses eqs.~\eqref{eq:mpi}
and~\eqref{eq:mK} that accompany the
coefficients $\bar{b}$ and $\delta b_B$ within eq.~\eqref{eq:octetc}
by the quark masses eqs.~\eqref{eq:ml} and~\eqref{eq:ms}.
As discussed above, for the higher order corrections to
the baryon masses it is sufficient to truncate according to
eqs.~\eqref{eq:line1}--\eqref{eq:line2}. The respective
expressions are generated automatically, using {\sc SymPy}~\cite{10.7717/peerj-cs.103}.

In our fits we rescale $m_q$ and $M_P^2$ into units of $\sqrt{8t_0}$.
The LECs are all defined in the $\textmd{SU(3)}$ chiral limit,
in units of $\sqrt{8t_{0,\text{ch}}}$. This difference implies
a shift of the LEC combination $L_{64}$, due to
$t_{0}=t_{0,\text{ch}}(1+\tilde{k}_18t_0\overline{M}\vphantom{M}^2)$. We define
$\widetilde{L}_{64}$ as the fitted LEC for the dependence of
the quark masses on the squared meson masses given in units of
$\sqrt{8t_0}$ and, starting from
eq.~\eqref{eq:ml}, derive the relation
\begin{align}
2m_{\ell}B_0&=\frac{\sqrt{t_0}}{\sqrt{t_{0,\text{ch}}}}M_{\pi}^2
\left\{1-\mu_{\pi}^2+\frac13\mu_{\eta_8}^2-\frac{8}{F_0^2}
\left[M_{\pi}^2L_{85}+\left(2M_K^2+M_{\pi}^2\right)\widetilde{L}_{64}\right]\right\}\nonumber\\
&=M_{\pi}^2\left\{1+\cdots-\left(2M_K^2+M_{\pi}^2\right)\left(\frac{8}{F_0^2}\widetilde{L}_{64}-\frac{1}{6}8t_0\tilde{k}_1\right)\right\}.
\end{align}
Comparison with eq.~\eqref{eq:ml} gives
\begin{equation}
L_{64}=\widetilde{L}_{64}-\frac{\tilde{k}_1}{48}\,8t_0F_0^2.
\end{equation}
Within eq.~\eqref{eq:pslo} we need the derivatives of the
pseudoscalar masses with respect to the quark mass. Therefore,
within the analytic expression, $L_{64}$ should be used
instead of $\widetilde{L}_{64}$. The difference is
small but of a similar magnitude as $L_{64}$ itself.

The combinations of LECs $L_{85}$ and $L_{64}$ turn out to be numerically small,
however, they are accompanied by $1/F_0^2$,
rather than by $1/(4\pi F_0)^2$ and $(4\pi)^2\approx 158$.
MILC~\cite{MILC:2010hzw} and HPQCD~\cite{Dowdall:2013rya} give values
at the scale $\mu=M_{\eta_8}\approx 576\,$MeV. Converting these
results to the standard scale $\mu=M_{\rho}\approx 770\,$MeV
(see eq.~\eqref{eq:scalerun}) results in
\begin{equation}
L_{85}=-0.20(11)^{(45)}_{(19)}\cdot 10^{-3},\quad
L_{64}=0.04(24)^{(32)}_{(27)}\cdot 10^{-3}
\end{equation}
for MILC and
\begin{equation}
L_{85}=-0.15(20)\cdot 10^{-3},\quad
L_{64}=0.23(17)\cdot 10^{-3}
\end{equation}
for HPQCD.
$L_6$ and $L_4$ are suppressed
in $1/N_c$ relative to $L_8$ and $L_5$, which is not obvious
from the above combinations which are all consistent with zero,
due to cancellations.
Some of us determined a value $L_{85}=0.50(34)\cdot 10^{-3}$
in a large $N_c$ NLO $\textmd{U(3)}$ ChPT analysis of the
$\eta/\eta'$ meson system~\cite{Bali:2021qem}. In this approach,
at this order $L_6=L_4=0$ and there
is no scale dependence of $L_{85}$.
Finally, Bijnens and Ecker~\cite{Bijnens:2014lea} obtain the values
\begin{equation}
L_{85}=-0.12(21)\cdot 10^{-3},\quad
L_{64}=-0.02(10)\cdot 10^{-3}
\end{equation}
from a phenomenological fit (column BE14 of table~3 in their article,
with a fixed value of $L_4$). Again, only upper limits could be set.

In the absence of precise and reliable literature values, we determined
these parameters from our quark mass data. We obtain at the
scale $\mu=770\,\textmd{MeV}\approx 1.6/\sqrt{8t_0}$
\begin{equation}
  \label{eq:l4569}
  \frac{L_{85}}{F_0^2}=7.8(4.8)\cdot 10^{-9}\,\textmd{MeV}^{-2},\quad
  \frac{L_{64}}{F_0^2}=-1.5(3.2)\cdot 10^{-9}\,\textmd{MeV}^{-2}.
\end{equation}
Setting, for instance, $F_0=71\,\textmd{MeV}$, the central values
would correspond to
$L_{85}\approx 0.04\cdot 10^{-3}$ and $L_{64}\approx -0.008\cdot 10^{-3}$.
The smallness of these NLO LECs at this scale does not mean that
the impact of the higher order is completely negligible since also
the logarithmic terms $\mu_\pi^2$ and $\mu_{\eta_8}^2$ enter the expressions.

\section{SU(2) BChPT low energy constants}
\label{sec:su2lec}
The $\textmd{SU(2)}$ (H)BChPT LECs can easily be derived from their
$\textmd{SU(3)}$
counterparts. Regarding the decuplet sector, the $\Delta$ appears to
be the most interesting particle, however, due to its unstable nature,
its self-energy acquires an imaginary part, whose inclusion is
beyond the scope of the present study. Hence, we will restrict ourselves
to the LECs that are governing the dependence of the nucleon mass
on the pion mass, neglecting
decuplet loops. In principle, similar LECs for the pion
mass dependence of the $\Lambda$, the $\Sigma$ and the $\Xi$ can be
obtained, however, we refrain from doing this since
$\textmd{SU(3)}$ BChPT is more adequate as
a framework to study processes involving
the different octet baryons. In $\textmd{SU(2)}$ (H)BChPT the $\mathcal{O}(p^3)$
dependence is given as (see, e.g., ref.~\cite{AliKhan:2003ack})
\begin{equation}
  m_N(M_{\pi})=m_N^0
  -4c_1M_{\pi}^2+\frac{3}{2}\frac{{g_A^0}^{\!2}}{\left(4\pi F_{\pi}^0\right)^2}
  {m_N^0}^{\!\!3}f_O\left(\frac{M_{\pi}}{m_N^0}\right),
  \label{eq:suscpt}
\end{equation}
where $m_N^0$, $g_A^0$ and $F_{\pi}^0$ denote the nucleon mass,
the axial charge and the pion decay constant
in the $\textmd{SU(2)}$ chiral limit, respectively,
at the physical strange quark mass.
The loop function $f_O$ for the EOMS regularization
is defined in eq.~\eqref{eq:loopf}. In the
heavy baryon limit, $f_O(x)=-\pi x^3$.
To our order in ChPT, keeping the strange quark mass fixed corresponds to
varying the kaon mass as a function of the pion mass, according to
\begin{equation}
  2M_K^2-M_\pi^2\eqqcolon M_{\bar{s}s}^2=M_{\bar{s}s,\text{ph}}^2\approx \left(686\,\textmd{MeV}\right)^2.
\end{equation}
Since $\overline{M}\vphantom{M}^2=(M_{\bar{s}s}^2+2M_{\pi}^2)/3$, this means that
$\overline{M}\vphantom{M}_{\text{ph}}^2-
\overline{M}\vphantom{M}_{\text{ch2}}^2=
\tfrac23 M_{\pi,\text{ph}}^2$, where the subscript ``ch2'' indicates the
$\textmd{SU(2)}$ chiral limit.
This then relates the $t_0$ parameters between the
two points: using eqs.~\eqref{eq:t0chi} and~\eqref{eq:k1result}, we obtain
\begin{equation}
  t_{0,\text{ph}}=\left[1+\tilde{k}_18t_0\left(\overline{M}_{\text{ph}}^2-
    \overline{M}_{\text{ch2}}^2\right)\right]t_{0,\text{ch2}}=0.9976(3)\,t_{0,\text{ch2}}.
  \label{eq:t0ch2}
\end{equation}
Evaluating the parametrization~\eqref{eq:rescalem} for $O=N$ at
$\mathbbm{M}_{\pi}=0$, $\mathbbm{M}_K=\sqrt{8t_{0,\text{ch2}}}\,M_{K,\text{ch2}}$
and $\mathbbm{M}_{\eta_8}=\sqrt{8t_{0,\text{ch2}}}\,M_{\eta_8,\text{ch2}}$,
where $M_{K,\text{ch2}}=\sqrt{1/2}\,M_{\bar{s}{s},\text{ph}}$
and $M_{\eta_8,\text{ch2}}=\sqrt{2/3}\,M_{\bar{s}s,\text{ph}}$,
gives $\mathbbm{m}_N^0=\sqrt{8t_{0,\text{ch2}}}\,m_N^0$, i.e.\ the nucleon
mass in the $\textmd{SU(2)}$ chiral limit in units of $\sqrt{8t_{0,\text{ch2}}}$.
The conversion between $\sqrt{8t_{0,\text{ch2}}}$ and
$\sqrt{8t_{0,\text{ph}}}=0.4098^{(20)}_{(25)}\,\textmd{fm}$ is given
in eq.~\eqref{eq:t0ch2} above. Note that
computing the expression~\eqref{eq:octetc} --- that is given in
physical units --- for the pseudoscalar
masses in the $\textmd{SU(2)}$ chiral limit at order $p^3$ is equivalent
to the above procedure.

Regarding the LEC $c_1$, a comparison between eq.~\eqref{eq:suscpt} and
eq.~\eqref{eq:octetc} gives
\begin{align}
  \label{eq:calcc1}
  c_1=-\frac{1}{6}\bar{b}+\frac{1}{4}\delta b_N=b_0+\frac{b_D}{2}+\frac{b_F}{2}.
\end{align}
This is a relation between LECs that are all given in physical units.
Therefore, in this case there exist no subtleties related to changes of $t_0$
between different points in the quark mass plane.

Note that to leading non-trivial order $\sigma_{\pi N}=-4c_1M_{\pi,\text{ph}}^2$.
Since the LECs $F$ and $D$ are not overly well-determined in the
present study, we refrain from predicting $g_A^0$, i.e.\
$g_A$ in the $\textmd{SU(2)}$
chiral limit. This is best left to simulations of the axial nucleon matrix
element in the forward limit.

\section{Details of the spectrum and quark mass determinations}
\label{sec:spectrum}
We define the interpolators that are used to create and destroy the
pseudoscalar mesons and baryons. We then detail the quark smearing
and the source positions employed. The latter are relevant regarding
ensembles with open boundary conditions in time. After explaining the
extraction of masses and their correlations from the resulting two-point
functions, in appendix~\ref{sec:masstables} we tabulate
the resulting hadron and AWI quark masses in lattice units as well as $t_0/a^2$
for all the ensembles.

\subsection{Hadron interpolators and smearing}
\label{sec:numerics}
In our simulations the pseudoscalar mesons, the octet and the decuplet baryons
are destroyed, respectively, using the relativistic interpolators
\begin{align}
  I_{\pi} &= \overline{d} \gamma_5 u,\quad
   I_K = \overline{s} \gamma_5 u,\\\nonumber
I_N         &= \epsilon_{abc} u_a \left[ u_b^{\intercal} C \gamma_5 d_c \right],\quad
I_{\Lambda} = \epsilon_{abc} s_a \left[ u_b^{\intercal} C \gamma_5 d_c \right],\quad
I_\Sigma    = \epsilon_{abc} u_a \left[ u_b^{\intercal} C \gamma_5 s_c \right],\\
I_\Xi       &= \epsilon_{abc} s_a \left[ s_b^{\intercal} C \gamma_5 u_c \right],\\\nonumber
I_\Delta     &= \epsilon_{abc} \left( 2u_a \left[u_b^{\intercal} C\gamma_- d_c\right] + d_a \left[u_b^{\intercal} C\gamma_- u_c\right] \right),\\\nonumber
I_{\Sigma^*} &= \epsilon_{abc} \left( 2u_a \left[u_b^{\intercal} C\gamma_- s_c\right] + s_a \left[u_b^{\intercal} C\gamma_- u_c\right] \right),\\
I_{\Xi^*}    &= \epsilon_{abc} \left( 2s_a \left[s_b^{\intercal} C\gamma_- u_c\right] + u_a \left[s_b^{\intercal} C\gamma_- s_c\right] \right),
\quad I_\Omega     = \epsilon_{abc} s_a \left[ s_b^{\intercal} C\gamma_- s_c \right].
\end{align}
The superscript ``$^\intercal$'' indicates the
transpose in Dirac spinor space, $C = \gamma_2\gamma_0$ is the
charge conjugation matrix and $\gamma_- = \tfrac12(\gamma_2 + i\gamma_1)$.
The quark fields $u(x)$ and $d(x)$ need to be distinguished when
carrying out the Wick contractions for the hadronic two-point functions
but they share the same mass. The indices $a, b, c\in\{1,2,3\}$ run over
fundamental colour. The spinor index of the baryonic interpolators
is suppressed. Note that the above naive implementation
of $I_{\Lambda}$ also has overlap with the $\Sigma^0$
baryon. However, since $m_{\Lambda}<m_{\Sigma}$ this does not inhibit us from
extracting the correct ground state signal. In addition we
compute two-point functions, destroying the pseudoscalar mesons
with the local axial currents $\bar{q}\gamma_0\gamma_5 u$, where
$q\in\{d,s\}$, in order to determine the AWI quark masses.

Within these interpolators we employ either local or smeared quark
field operators. For the baryons we only consider smeared-smeared
two-point functions while for the mesons we analyse
smeared-smeared as well as smeared-local two-point functions,
the latter to determine the quark masses. We project all interpolators
to zero momentum:
\begin{equation}
  I_X(t)\coloneqq\sum_{\vec{x}}I_X(\vec{x},t).
\end{equation}
Exploiting spatial translational invariance of two-point functions,
in practice we only explicitly carry out the momentum projection at
the sink (i.e.\ we utilize
point-to-all propagators), however, some additional measurements
are carried out using stochastic timeslice-to-all propagators.

The two-point function for a pseudoscalar meson $P$ is given as
\begin{equation}
  C_P(t)=\langle \Omega|I_P(t+t_0)I_P^{\dagger}(t_0)|\Omega\rangle,
\end{equation}
where the vacuum expectation value on the right hand side is
obtained as an ensemble average of the Wick-contracted two-point function
and $t_0$ denotes the source position in Euclidean time.
Defining the parity and spin projectors
$\Gamma_{\text{unpol}}=\tfrac12(\mathds{1}+\gamma_0)$ and
$\Gamma_{\text{pol}}=\tfrac12\Gamma_{\text{unpol}}(\mathds{1}+i\gamma_3\gamma_5)$,
we compute the two-point functions
\begin{align}
  \label{eq:bar2pt}
  C_O(t)&=\Tr \left(\Gamma_{\text{unpol}}\langle\Omega| I_O(t+t_0)\overline{I}_O(t_0)|\Omega\rangle\right),\\
  C_D(t)&=\Tr \left(\Gamma_{\text{pol}}\langle\Omega| I_D(t+t_0)\overline{I}_D(t_0)|\Omega\rangle\right)
  \label{eq:bar2pt2}
\end{align}
for the octet and decuplet baryons, respectively.
The trace is over spinor space, the time separation $t$ is positive
and the conjugation has the usual
meaning: $\overline{I}=I^{\dagger}\gamma_0$. The phases are set
such that all the two-point functions are positive.

The above interpolators are either local or Wuppertal
smeared~\cite{Gusken:1989ad,Gusken:1989qx}, employing spatially
APE smeared gauge links~\cite{Falcioni:1984ei} as parallel transporters
that are iteratively constructed via
\begin{equation}
\label{ape}
U_i^{(n+1)}(\vec{x})= P_{\textmd{SU}(3)}\!\!\left(U_i^{(n)}(\vec{x})+\delta\sum_{|j|\neq i}
U_j^{(n)}(\vec{x})U_i^{(n)}(\vec{x}+\unitj)U^{\dagger(n)}_j(\vec{x}+\uniti)\right),
\end{equation}
where $i\in\{1,2,3\}$, $j\in\{\pm 1,\pm 2,\pm 3\}$ and
$\unitj$ denotes a vector of length $a$
pointing into the direction $j$, i.e.\
the sum is over the four spatial ``staples'' surrounding
the link that connects $\vec{x}$ with $\vec{x}+\uniti$,
$U_i(\vec{x})$. Note that $U_{-i}(\vec{x})=U^{\dagger}_i(\vec{x}-\uniti)$
and $\{U^{(0)}_i(\vec{x})\}$ is the original gauge field.
Since the smearing is local in time, the
time index is suppressed. $P_{\textmd{SU}(3)}$ is a gauge covariant
projector onto the $\textmd{SU}(3)$
group, defined by maximizing $\mathrm{Re}\Tr[A^{\dagger} P_{\textmd{SU}(3)}(A)]$.
We iterate eq.~\eqref{ape} 25 times, using the weight factor $\delta=0.4$.
Using these smeared gauge transporters
$\overline{U}_i(\vec{x})=U^{(25)}_i(\vec{x})$, we Wuppertal smear the
quark fields $q$, successively applying the smearing operator $\Phi$ that is
defined as
\begin{equation}
  (\Phi q)(\vec{x}) = \frac{1}{1+6\varepsilon} \left(q(\vec{x})+\varepsilon \sum_{j=\pm 1}^{\pm 3} \overline{U}_j(\vec{x}) q(\vec{x}+\unitj)\right),
\end{equation}
either at the sink to propagators that have been obtained by solving the
discretized Wilson--Dirac equation or to
point sources
$q_{a\alpha}(\vec{x})=\delta_{\vec{x}\vec{0}}\delta_{aa_0}\delta_{\alpha\alpha_0}$
for $a_0=1,2,3$, where the smearing only needs to be carried out
for one value of the spin index $\alpha$ since $\Phi$ commutes with the
spin structure. The normalization $(1+6\varepsilon)^{-1}$
is arbitrary and is introduced to avoid numerical overflow for
high iteration counts.

On a free field configuration,
i.e.\ $\overline{U}_i(\vec{x})=\mathds{1}$, the root mean squared
(rms) smearing radius for the gauge invariant combination
$||q(\vec{x})||\coloneqq
\sqrt{q^{\dagger}(\vec{x})q(\vec{x})}$ for a large iteration count $n$ in an
infinite volume reads~\cite{Bali:2016lva}
\begin{equation}
  \label{eq:radius}
  r=\sqrt{\frac{6\varepsilon}{1+6\varepsilon}}\,a\sqrt{n},
\end{equation}
where $\varepsilon$ should be positive. Here we set $\varepsilon=0.25$.
As can be seen from the above equation, to maintain a constant
radius in physical units, the number of iterations $n$ needs to
be scaled in proportion to $a^{-2}$. Moreover, it turns out that
an optimal ground state overlap requires
$r$ to be increased with decreasing quark mass.

\setlength\LTcapwidth{\columnwidth}
\begin{longtable}{ccccccccc}
  \caption{\label{tab:smearing}Numbers of smearing iterations $n_\ell$
    and $n_s$ for the light
    and strange quarks, respectively, as well as the corresponding root
    mean squared smearing radii $r_\ell$ and $r_s$. We also include the number
    of sources $N_{\text{src}}$ employed for point-to-all propagators.
}\\\toprule
trajectory & id
& ${M_{\pi}}/{\textmd{MeV}}$
& ${M_K}/{\textmd{MeV}}$
& $n_\ell$
& $n_s$
& ${r_\ell}/{\textmd{fm}}$
& ${r_s}/{\textmd{fm}}$
& $N_{\text{src}}$
\\
\hline\endfirsthead
\caption{\small List of smearing parameters (continued).}\\\toprule
trajectory & id
& ${M_{\pi}}/{\textmd{MeV}}$
& ${M_K}/{\textmd{MeV}}$
& $n_\ell$
& $n_s$
& ${r_\ell}/{\textmd{fm}}$
& ${r_s}/{\textmd{fm}}$
& $N_{\text{src}}$
\\
\hline\endhead
\hline
\multicolumn{9}{r}{\textit{Continued on next page}} \\
\endfoot
\hline
\endlastfoot
\multicolumn{9}{c}{$\beta=3.34, a = 0.098\,\mathrm{fm}$} \\\hline
\multirow{3}{*}{$m_s=m_{\ell}$}
           & A651
           & 556
           & 556
           & 120
           & 120
           & 0.55
           & 0.55
           & 3
           \\
           & A652
           & 432
           & 432
           & 150
           & 150
           & 0.62
           & 0.62
           & 3
           \\
           & A650
           & 371
           & 371
           & 160
           & 160
           & 0.64
           & 0.64
           & 3
           \\
           \hline
\multirow{2}{*}{$\overline{m}=m_{\text{symm}}$}
           & A653
           & 429
           & 429
           & 150
           & 150
           & 0.61
           & 0.61
           & 3
           \\
           & A654
           & 338
           & 459
           & 185
           & 165
           & 0.68
           & 0.64
           & 3
           \\
           \hline
\multicolumn{9}{c}{$\beta=3.4, a = 0.085\,\mathrm{fm}$} \\\hline
\multirow{3}{*}{$m_s=m_{\ell}$}
           & rqcd019
           & 608
           & 608
           & 150
           & 150
           & 0.54
           & 0.54
           & 3
           \\
           & rqcd021
           & 340
           & 340
           & 250
           & 250
           & 0.67
           & 0.67
           & 3
           \\
           & rqcd017
           & 236
           & 236
           & 320
           & 320
           & 0.73
           & 0.73
           & 3
           \\
           \hline
\multirow{12}{*}{$\overline{m}=m_{\text{symm}}$}
           & U103
           & 420
           & 420
           & 220
           & 220
           & 0.63
           & 0.63
           & 8
           \\
           & H101
           & 423
           & 423
           & 220
           & 220
           & 0.64
           & 0.64
           & 20
           \\
           & U102
           & 357
           & 445
           & 250
           & 210
           & 0.66
           & 0.62
           & 36
           \\
           & H102a
           & 359
           & 444
           & 250
           & 210
           & 0.66
           & 0.62
           & 20
           \\
           & H102b
           & 354
           & 442
           & 250
           & 210
           & 0.66
           & 0.62
           & 20
           \\
           & U101
           & 271
           & 464
           & 300
           & 200
           & 0.69
           & 0.60
           & 36
           \\
           & H105
           & 281
           & 468
           & 300
           & 200
           & 0.73
           & 0.61
           & 20
           \\
           & N101
           & 281
           & 467
           & 300
           & 200
           & 0.72
           & 0.61
           & 33
           \\
           & S100
           & 214
           & 476
           & 350
           & 170
           & 0.77
           & 0.57
           & 33
           \\
           & C101
           & 222
           & 476
           & 350
           & 170
           & 0.77
           & 0.57
           & 20
           \\
           & D101
           & 222
           & 476
           & 350
           & 170
           & 0.79
           & 0.58
           & 3
           \\
           & D150
           & 127
           & 482
           & 440
           & 140
           & 0.84
           & 0.52
           & 32
           \\
           \hline
\multirow{3}{*}{$\widetilde{m}_s=\widetilde{m}_{s,\text{ph}}$}
           & H107
           & 368
           & 550
           & 250
           & 160
           & 0.67
           & 0.56
           & 3
           \\
           & H106
           & 273
           & 520
           & 250
           & 160
           & 0.67
           & 0.56
           & 3
           \\
           & C102
           & 223
           & 504
           & 350
           & 160
           & 0.77
           & 0.56
           & 3
           \\
           \hline
\multicolumn{9}{c}{$\beta=3.46, a = 0.075\,\mathrm{fm}$} \\\hline
\multirow{3}{*}{$m_s=m_{\ell}$}
           & rqcd029
           & 713
           & 713
           & 180
           & 180
           & 0.52
           & 0.52
           & 3
           \\
           & rqcd030
           & 319
           & 319
           & 355
           & 355
           & 0.68
           & 0.68
           & 3
           \\
           & X450
           & 265
           & 265
           & 400
           & 400
           & 0.73
           & 0.73
           & 3
           \\
           \hline
\multirow{4}{*}{$\overline{m}=m_{\text{symm}}$}
           & B450
           & 421
           & 421
           & 270
           & 270
           & 0.61
           & 0.61
           & 16
           \\
           & S400
           & 354
           & 445
           & 310
           & 260
           & 0.66
           & 0.61
           & 20
           \\
           & N401
           & 287
           & 464
           & 375
           & 250
           & 0.72
           & 0.60
           & 20
           \\
           & D450
           & 216
           & 480
           & 480
           & 200
           & 0.78
           & 0.55
           & 32
           \\
           \hline
\multirow{4}{*}{$\widetilde{m}_s=\widetilde{m}_{s,\text{ph}}$}
           & B451
           & 422
           & 577
           & 270
           & 200
           & 0.62
           & 0.54
           & 3
           \\
           & B452
           & 352
           & 548
           & 310
           & 200
           & 0.65
           & 0.54
           & 3
           \\
           & N450
           & 287
           & 528
           & 375
           & 200
           & 0.70
           & 0.54
           & 3
           \\
           & D451
           & 219
           & 507
           & 480
           & 200
           & 0.78
           & 0.54
           & 32
           \\
           \hline
\multicolumn{9}{c}{$\beta=3.55, a = 0.064\,\mathrm{fm}$} \\\hline
\multirow{3}{*}{$m_s=m_{\ell}$}
           & B250
           & 713
           & 713
           & 250
           & 250
           & 0.44
           & 0.44
           & 3
           \\
           & X250
           & 350
           & 350
           & 445
           & 445
           & 0.65
           & 0.65
           & 3
           \\
           & X251
           & 268
           & 268
           & 540
           & 540
           & 0.71
           & 0.71
           & 3
           \\
           \hline
\multirow{7}{*}{$\overline{m}=m_{\text{symm}}$}
           & H200
           & 422
           & 422
           & 390
           & 390
           & 0.61
           & 0.61
           & 3
           \\
           & N202
           & 414
           & 414
           & 390
           & 390
           & 0.61
           & 0.61
           & 20
           \\
           & N203
           & 348
           & 445
           & 445
           & 375
           & 0.65
           & 0.61
           & 20
           \\
           & S201
           & 290
           & 471
           & 540
           & 360
           & 0.71
           & 0.61
           & 3
           \\
           & N200
           & 286
           & 466
           & 540
           & 360
           & 0.69
           & 0.58
           & 20
           \\
           & D200
           & 202
           & 484
           & 660
           & 290
           & 0.78
           & 0.55
           & 20
           \\
           & E250
           & 131
           & 493
           & 795
           & 285
           & 0.82
           & 0.54
           & 32
           \\
           \hline
\multirow{3}{*}{$\widetilde{m}_s=\widetilde{m}_{s,\text{ph}}$}
           & N204
           & 353
           & 549
           & 445
           & 285
           & 0.66
           & 0.54
           & 3
           \\
           & N201
           & 287
           & 527
           & 540
           & 285
           & 0.72
           & 0.54
           & 3
           \\
           & D201
           & 200
           & 504
           & 660
           & 285
           & 0.77
           & 0.54
           & 3
           \\
           \hline
\multicolumn{9}{c}{$\beta=3.7, a = 0.049\,\mathrm{fm}$} \\\hline
\multirow{1}{*}{$m_s=m_{\ell}$}
           & N303
           & 646
           & 646
           & 440
           & 440
           & 0.51
           & 0.51
           & 3
           \\
           \hline
\multirow{4}{*}{$\overline{m}=m_{\text{symm}}$}
           & N300
           & 425
           & 425
           & 640
           & 640
           & 0.59
           & 0.59
           & 18
           \\
           & N302
           & 348
           & 455
           & 750
           & 620
           & 0.64
           & 0.59
           & 20
           \\
           & J303
           & 259
           & 479
           & 950
           & 525
           & 0.70
           & 0.55
           & 3
           \\
           & E300
           & 176
           & 496
           & 800
           & 310
           & 0.67
           & 0.44
           & 16
           \\
           \hline
\multirow{3}{*}{$\widetilde{m}_s=\widetilde{m}_{s,\text{ph}}$}
           & N305
           & 428
           & 584
           & 640
           & 465
           & 0.60
           & 0.52
           & 3
           \\
           & N304
           & 353
           & 558
           & 750
           & 465
           & 0.62
           & 0.51
           & 3
           \\
           & J304
           & 261
           & 527
           & 950
           & 465
           & 0.70
           & 0.52
           & 3
           \\
           \hline
\multicolumn{9}{c}{$\beta=3.85, a = 0.039\,\mathrm{fm}$} \\\hline
\multirow{1}{*}{$m_s=m_{\ell}$}
           & N500
           & 604
           & 604
           & 650
           & 650
           & 0.48
           & 0.48
           & 3
           \\
           \hline
\multirow{2}{*}{$\overline{m}=m_{\text{symm}}$}
           & J500
           & 413
           & 413
           & 1000
           & 1000
           & 0.57
           & 0.57
           & 3
           \\
           & J501
           & 336
           & 448
           & 1225
           & 1025
           & 0.61
           & 0.57
           & 3
           \\
           \hline
\end{longtable}

In table~\ref{tab:smearing} we list our approximate pion and kaon masses as
well as the number of smearing iterations $n_{\ell}$ and $n_s$ for
the light and strange quarks and the resulting rms smearing
radii $r_{\ell}$ and $r_s$. These have been calculated numerically
on time slices of a few gauge configurations via
\begin{equation}
  r^2=\frac{\left(\prod_{i=1}^3\sum_{x_i/a=-N_s/2}^{N_s/2-1}\right)\,||q(\vec{x})||\,\vec{x}^2}
  {\sum_{\vec{x}}\,||q(\vec{x})||}.
\end{equation}
The results are slightly smaller
than eq.~\eqref{eq:radius} would suggest due to volume effects
and also since the APE smeared
gauge transporters are somewhat rougher than unit gauge fields.
We kept the radii approximately
constant in physical units across the lattice spacings.
The light quark radius $r_{\ell}$ depends on $M_\pi^2$ while
the strange quark radius $r_s$ depends on $2M_K^2-M_\pi^2$.

\subsection{Source positions and extraction of the masses}
\label{sec:positions}

We distinguish between open boundary conditions (obc) and
periodic boundary conditions (pbc) in time.
Regarding obc, we expect order $g^2 a$ lattice artefacts as well
as physical states of scalar quantum numbers (the lightest one having
a mass of approximately $2M_{\pi}$) to propagate from the
boundaries into the bulk. These contributions
will be exponentially suppressed with the distance from the
boundaries. Therefore, at some minimum separation
$t_{\text{bound}}$, translational invariance in time is
effectively restored for most of the two-point functions.
Below we explain how we estimate $t_{\text{bound}}$,
how two-point functions with different source and sink positions
are combined in the analysis, how the fit ranges are chosen
and how the pseudoscalar, AWI quark and baryon masses are extracted.

We employ several temporal source positions $t_0$
for point sources for each configuration as indicated in the last column of
table~\ref{tab:smearing}. With pbc one can naively
average the two-point functions over the source positions.
In this case, for the mesons we can naively average forward
and backward ($t\mapsto -t$) propagating two-point functions,
while for the baryons, in addition, we have to replace
$\gamma_0\mapsto -\gamma_0$
within the projectors in eqs.~\eqref{eq:bar2pt} and~\eqref{eq:bar2pt2}.
Also the backward propagating
two-point functions for pseudoscalar mesons that are destroyed by
local axial currents (which are used to compute the AWI quark masses,
see eq.~\eqref{eq:awimass})
acquire a minus sign.
We label the time slices as
\begin{equation}
  t'=0, a, \ldots, T-a,\quad\text{where}\quad T=N_ta.
\end{equation}
With pbc the temporal dimension of the lattice is $T$,
while with obc it is $T-a$. In the latter case, the two-point
function starting at $t_0$, propagating in the forward
direction, can be averaged with that starting at
$T-a-t_0$, propagating in the backward direction, as above.
Other than this, we keep the source positions
separate throughout the analysis, enabling us to detect
any violation of translational invariance. Moreover,
the source and sink positions $t'=t_0$ and $t'=t_0+t$ (with $t$
positive or negative), respectively, should be restricted to
$t_{\text{bound}}\leq t'\leq T-a-t_{\text{bound}}$ with
$t_{\text{bound}}$ yet to be determined (see below). This means that for large
values of $|t|$ a smaller number of sources contributes than indicated
in table~\ref{tab:smearing}.

In the case of pbc, the spatial and temporal source positions
are selected randomly for each configuration, while for
obc we employ fixed temporal (but random spatial) positions, where
for each source at $t'=t_0$, we also place a source at $t'=T-a-t_0$.

Regarding the pseudoscalar (and pseudoscalar-axial)
two-point functions, additional measurements have been carried out
employing the ``one-end-trick''~\cite{Sommer:1994gg,Foster:1998wu},
using spin-explicit~\cite{Bernardson:1993he,Viehoff:1997wi} stochastic
complex $\mathbb{Z}_2$-sources. These techniques were first
combined in ref.~\cite{McNeile:2006bz}. It turns out to be advantageous in
terms of the signal over noise ratio not to seed the stochastic source
on the whole time slice $t_0$ but just on a subset~\cite{Sommer:1994gg}.
We carry out this ``thinning'' in a random pattern (to exclude overlap with
non-zero momentum contributions), with the occupation ratio 0.02.
Each source consists of one random vector for each of the four spin components
with support on the thinned set of points.
In the case of obc, the temporal positions of the one-end sources usually are
$t_0=a$ and $t_0=T-2a$, one time slice away from the boundary,
at the positions of the first non-trivial spatial links. In some cases
the distance was chosen somewhat larger. We employ two
stochastic estimates for each of the two time slices.
For pbc on each configuration we use four different randomly
selected time slices. The sources and sinks
of the one-end two-point functions used in this analysis are
local, without quark smearing applied.
We separately analyse the mesonic two-point functions from the
conventional point sources
and from the one-end sources and find consistent results.
Whenever this leads to an error reduction, we carry out a combined fit
to the data obtained using the two methods.

\begin{table}
  \caption{Ranges for the separation $t_{\text{bound}}$ from the boundaries
    of obc ensembles.\label{tab:boundary}}
\begin{center}
  \begin{tabular}{cccccc}
    \toprule
    $\beta$&3.4&3.46&3.55&3.7&3.85\\
    \midrule
    $t_{\text{bound}}/a$&24--29&31--32&31--40&31--43&45--55\\
    $t_{\text{bound}}/\textmd{fm}$&2--2.5&2.3--2.4&2--2.5&1.5--2.1&1.7--2.1\\
    \bottomrule
  \end{tabular}
\end{center}
\end{table}

When determining the boundary region on obc ensembles, we
observe by varying the source position, $t_0$, that
the pion two-point function is more sensitive with respect to
the proximity of the source or the sink to the boundaries than any
other two-point function.
Moreover, if the pion source is placed in the centre of the lattice,
ground state dominance is achieved at smaller
source-sink separations $t$ than for the one-end sources near
the boundaries. Therefore, we define $t_{\text{bound}}$ as the minimum
separation of the sink from the boundary for
the local-local pion two-point function with the source placed at
$t_0=a$, that is needed for the boundary/excited state
contributions to become smaller than one quarter of the statistical error.
In this determination we follow the strategy
that we also employ for the ground state
mass extractions that is described below.
All the point source positions that we use are outside of the
respective boundary regions, defined in terms of $t_{\text{bound}}$.
An overview of the resulting ranges is provided in table~\ref{tab:boundary}.
Note that none of the $\beta=3.34$ ensembles have obc, while
at $\beta=3.46$ only two ensembles with obc exist, where
one-end source measurements were taken (S400 and N401).

Our mass determinations follow a two step procedure: first we
determine a time range of ground state dominance and then we
carry out one-state fits for the hadron and AWI quark masses.
We carry out both uncorrelated and correlated fits. The latter fits
are utilized to monitor the fit quality in terms of the
$\chi^2/N_{\text{DF}}$-values.
However, in many cases the statistics is insufficient for
a reliable determination of the large covariance matrices
between the values of the two-point functions at different times.
Therefore, the final results are obtained from uncorrelated fits
and the errors computed via the bootstrap procedure,
also taking autocorrelations into account, see appendix~\ref{sec:statistical}.
We remark that the mean values for correlated and uncorrelated fits
are in good agreement in almost all of the cases.

\begin{figure}[thp]
  \centering
  \resizebox{0.95\textwidth}{!}{\includegraphics[width=\textwidth]{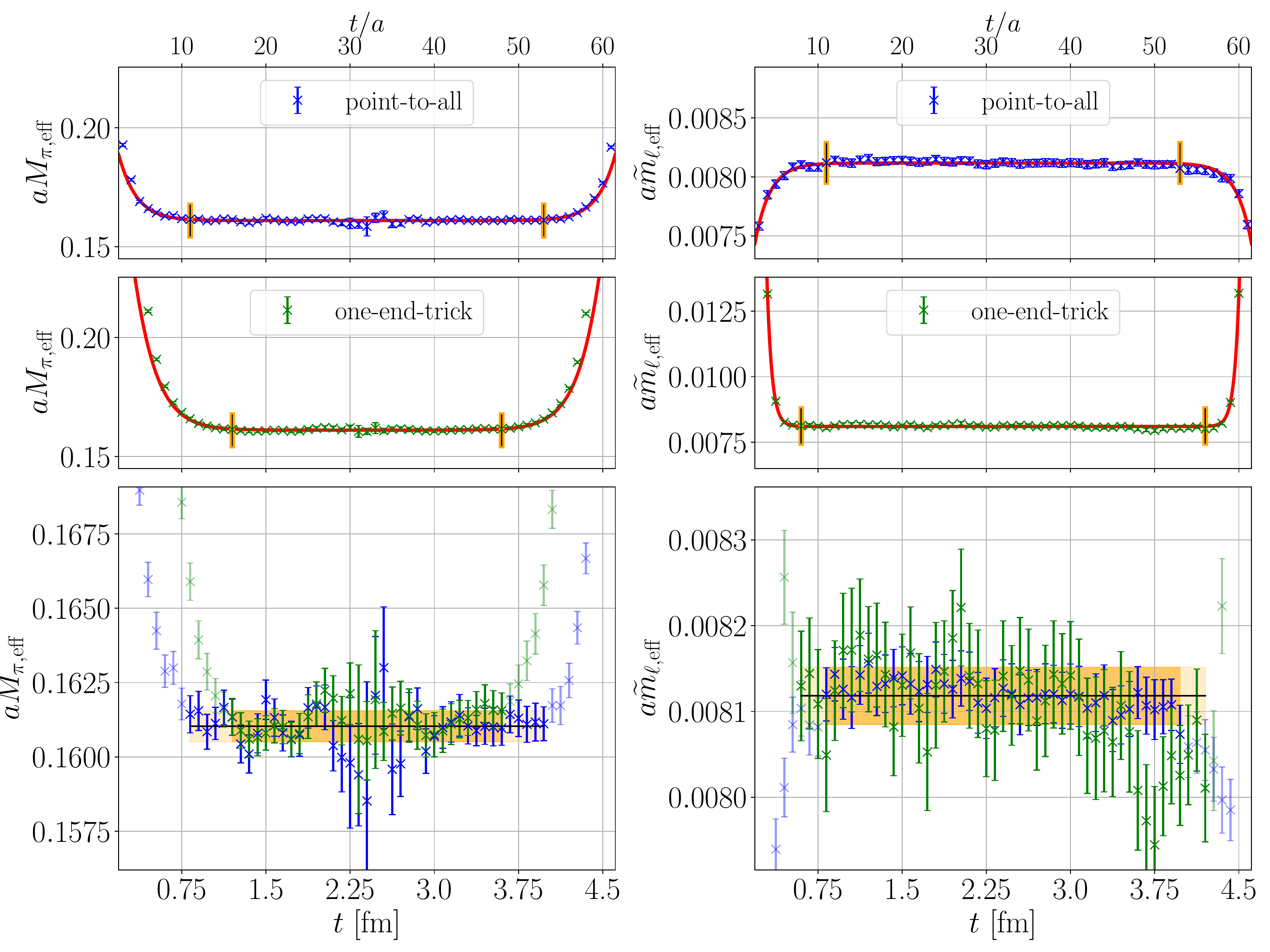}}
  \caption{\label{fig:effm_B450} Left: the effective mass $a M_{\pi,\text{eff}}$
    of the pion for ensemble B450 (pbc, $a\approx 0.075\,\textmd{fm}$,
    $M_\pi\approx 420\,\textmd{MeV}$, see table~\ref{tab:physparams}).
    The upper plot shows the ``point-to-all'' smeared-smeared
    data and the middle plot
    the ``one-end-trick'' local-local data.
    The two-point functions have been averaged
    over sixteen and four source positions, respectively.
    The red line corresponds to the fit
    result of the two-state ansatz given in eq.~\eqref{eq:2pt_excited}.
    The vertical orange bars indicate the resulting fit ranges for
    one-state fits (setting $B_P=0$),
    see the main text for details. In the last row
    the effective masses for both data sets are shown, together with
    the final (one-state) fit result, where the solid black
    line with orange error band
    indicates the result. The data that are not included in the
    fit are shown as light symbols. Right: the corresponding
    figures for the AWI quark mass, using the parametrization
    eq.~\eqref{eq:2pt_AWI} and setting $B_m=0$ for the ground state
    fit. In this case the point-to-all two-point functions are smeared-local
    and the one-end ones local-local.}
\end{figure}

Starting from the mesonic two-point functions with pbc,
we carry out fits according to the ansatz
\begin{equation}
  C_P(t)=A_P\cosh\left[M_P(t-T/2)\right]+B_P\cosh\left[M_P'(t-T/2)\right]
    \label{eq:2pt_excited}
\end{equation}
with the fit parameters $A_P$, $B_P$, $M_P$ and $M'_P$,
increasing the minimum $t$-value until we find an acceptable
representation of the data (by means of the correlated
$\chi^2/N_{\text{DF}}$-value). In the spirit of ref.~\cite{Blossier:2010vz},
we determine the minimum value $t_{\min}$ as the value
from which onwards the (effective) contribution of the
excited state term (proportional
to $B_P$) to the two-point function $C_P(t)$
becomes smaller than one quarter of the statistical error
of the data. Note that whenever more than one two-state fit was
found to adequately represent the data, the largest resulting
value for $t_{\min}$ is chosen.
To obtain the ground state mass, we subsequently fit $C_P(t)$ within the
range $t_{\min} \le  t \le t_{\max} = T-t_{\min}$ to the ground
state only, i.e.\ we set $B_P=0$ within eq.~\eqref{eq:2pt_excited}.
The analogous procedure is implemented for the ratio eq.~\eqref{eq:awimass}
of smeared pseudoscalar-local axial over smeared
pseudoscalar-local pseudoscalar two-point functions to obtain
AWI quark masses,
\begin{equation}
  \label{eq:2pt_AWI}
  \widetilde{m}_{\text{eff}}(t)=\widetilde{m}+B_m\cosh\left[m'(t-T/2)\right],
\end{equation}
where the second term is a lattice artefact ($B_m=\mathcal{O}(a^2)$)
and $m'\approx M_P'-M_P$.
Once the region of ground state dominance has been determined,
$\widetilde{m}(t)$ is averaged over this region. 

The above procedure is illustrated for the pbc ensemble
B450 in figure~\ref{fig:effm_B450}, where we show data both
for point sources (smeared-smeared for the pion, smeared-local
for the AWI quark mass) and for stochastic one-end sources
(local-local in both cases, hence the larger values of
$t_{\min}$ for the pion). In the left of the figure,
the effective mass
\begin{equation}
  M_{P,\text{eff}}=\frac{1}{2a}
  \left|\arccosh\left(\frac{C_P(t-a)}{C_P(T/2)}\right)
    -\arccosh\left(\frac{C_P(t+a)}{C_P(T/2)}\right)\right|
\end{equation}
is compared to $|(\dd/\dd{}t)\arccosh[C_P(t)/C_P(T/2)]|$.
Both expressions will approach $M_P$ in the limit $T\gg t\gg 0$.

\begin{figure}[t]
  \centering
  \resizebox{0.95\textwidth}{!}{\includegraphics[width=\textwidth]{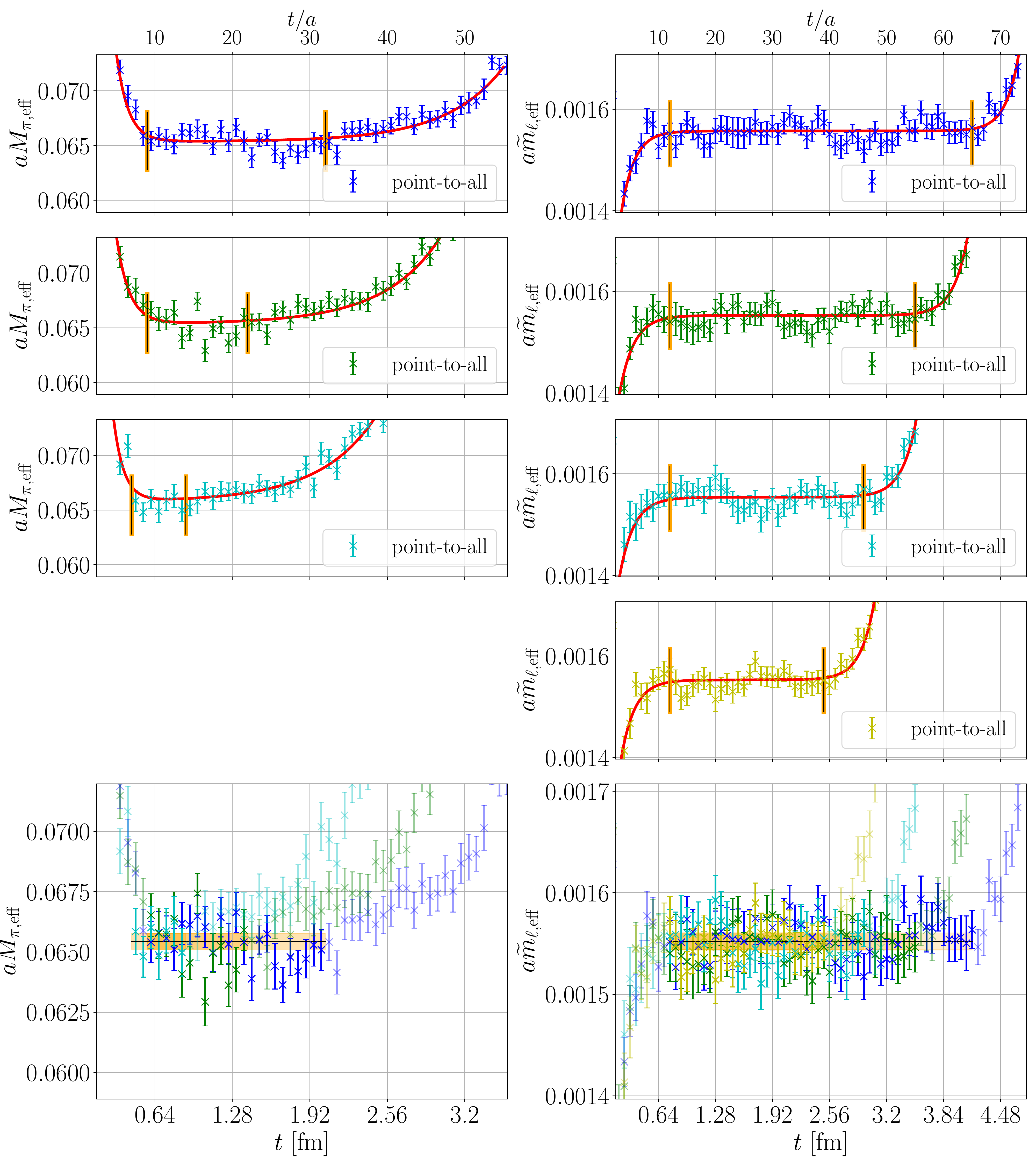}}
  \caption{\label{fig:effm_D200}Left: the effective mass of the pion for
    ensemble D200 (obc, $a\approx 0.064\,\textmd{fm}$,
    $M_\pi\approx 200\,\textmd{MeV}$, see table~\ref{tab:physparams}).
    Data from eighteen point source positions are shown, which are pairwise
    forward/backward averaged between $t_0$ and $T-a-t_0$.
    Within each panel, in addition we average data from
    three nearby source positions. The red lines, orange bars, black line
    and orange error band
    in the four panels have the same meaning as in figure~\ref{fig:effm_B450},
    however, the fit form
    defined in eq.~\eqref{eq:2pt_excited_obc} is used for the two-state
    fit and eq.~\eqref{eq:2pt_ground_obc} for the one-state fit.
    Right: the same for the light AWI quark mass, where two additional
    source positions (shown in the fourth row) are within the
    range that we average over. The fit range is
    determined using eq.~\eqref{eq:pcac_excited_obc}.}
\end{figure}

Regarding obc, the above expression is replaced by
\begin{equation}
  C_P(t)=A_P\left[\exp\left(-M_Pt\right)-\exp\left(-M_P(2T'-t)\right)\right]+B_P\exp\left(-M_P't\right)
    \label{eq:2pt_excited_obc}
\end{equation}
with an additional fit parameter $T'\approx T$.
The forward and backward
propagating two-point functions are symmetrized as explained above
so that $t>0$.
The ground state mass is then extracted via a fit to
\begin{equation}
  C_P(t)=A_P \exp(-M_P t),
    \label{eq:2pt_ground_obc}
\end{equation}
for the fit range $t_{\min} \le t \le t_{\max}$. The start time
$t_{\min}$ is determined in the same way as described above as
the time where the excited state contribution, proportional to $B_P$,
becomes smaller than one quarter of the statistical error of $C_P(t)$.
The upper limit of the fit range, $t_{\max}<T-a-t_{\text{bound}}$, corresponds
to the time at which $A_Pe^{M_P(t-2T')}$
becomes larger than one quarter of the statistical error.
Since the resulting values of $t_{\min}$ and $t_{\max}$
will depend on the statistical error of the two-point function,
one danger of this procedure is that for each individual source
position a value for $t_{\min}$ may be
suggested that is smaller than if the correlators for all the
different sources were averaged.
For pbc we average the two-point functions over all the source
positions in any case. The same is carried out for obc data with
one-end sources since these are always placed at the same
distance from the boundaries. Regarding obc two-point functions
with point sources, we first reconfirm that translational invariance
is effectively restored within the region
$t_{\text{bound}}\leq t+t_0\leq T-a-t_{\text{bound}}$ for the different
source positions $t_0$, before we average two-point functions obtained
for nearby values of $t_0$ into up to four groups. We then
determine the fit range for each of this smaller number of groups.
We implement the same procedure for the baryonic two-point functions.

In the left of figure~\ref{fig:effm_D200}, the determination
of the fit ranges is shown for the pion mass on
ensemble D200. The one-state fit to
determine the ground state mass (shown in the bottom left of the
figure) is carried out simultaneously for the different source positions,
using the respective fit ranges. The effective mass
\begin{equation}
  M_{P,\text{eff}}(t)=\frac{1}{a}\ln\left(\frac{C_P(t-a/2)}{C(t+a/2}\right)
\end{equation}
is compared to the derivative of the parametrization
$-(\dd/\dd{t})\ln C_P(t)$. Both expressions will
approach $M_P$ at large values of $t$.

Turning to the AWI quark mass with obc, we encounter,
in addition to the usual lattice artefacts, also contributions from
the boundary. This motivates the ansatz
\begin{equation}
  \widetilde{m}_{\text{eff}}(t)= \widetilde{m} + B_m\exp\left(-m' t\right)
  + D_m\exp\left(m'' t\right)
    \label{eq:pcac_excited_obc}
\end{equation}
with the fit parameters $\widetilde{m}$, $B_m$, $D_m$,
$m'\approx M_P'-M_P$ and $m''\approx 2M_{\pi}$,
where $\widetilde{m}$ is the AWI quark mass of interest, the second
term describes the usual lattice spacing effects or, if one-end
sources are used, also boundary effects and the last term encapsulates
the boundary effects at large times.
Using this ansatz, we determine $t_{\min}$ and $t_{\max}$ in the
same way as above
and carry out the fit to a constant for the resulting range.
This is shown for the example of point sources for ensemble D200
in the right of figure~\ref{fig:effm_D200}. As with pbc,
smeared-local and local-local two-point functions are used
to obtain the AWI quark masses with
point sources and with one-end sources (not shown), respectively.

\begin{figure}[t]
  \centering
  \resizebox{0.49\textwidth}{!}{\includegraphics[width=\textwidth]{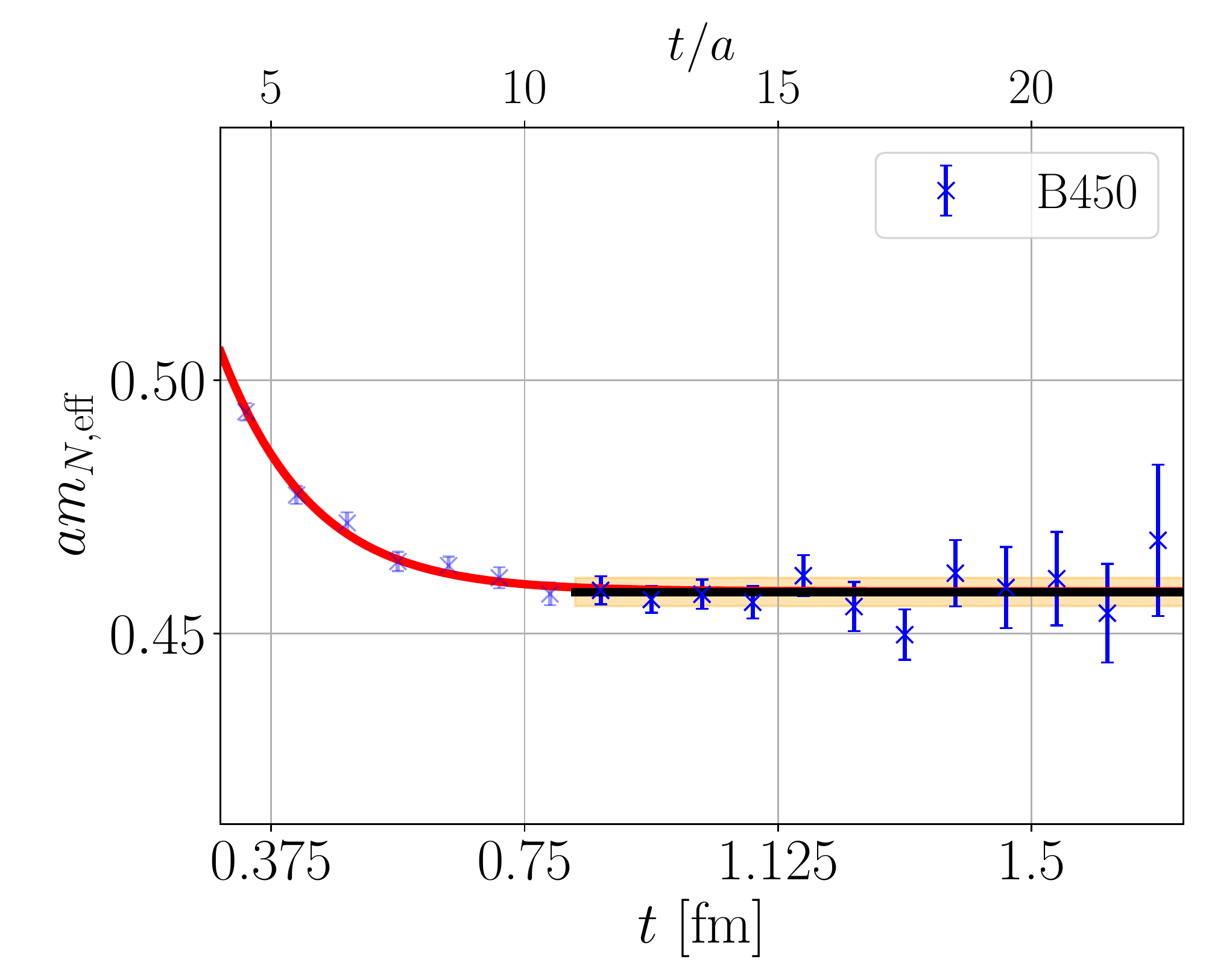}}
  \resizebox{0.49\textwidth}{!}{\includegraphics[width=\textwidth]{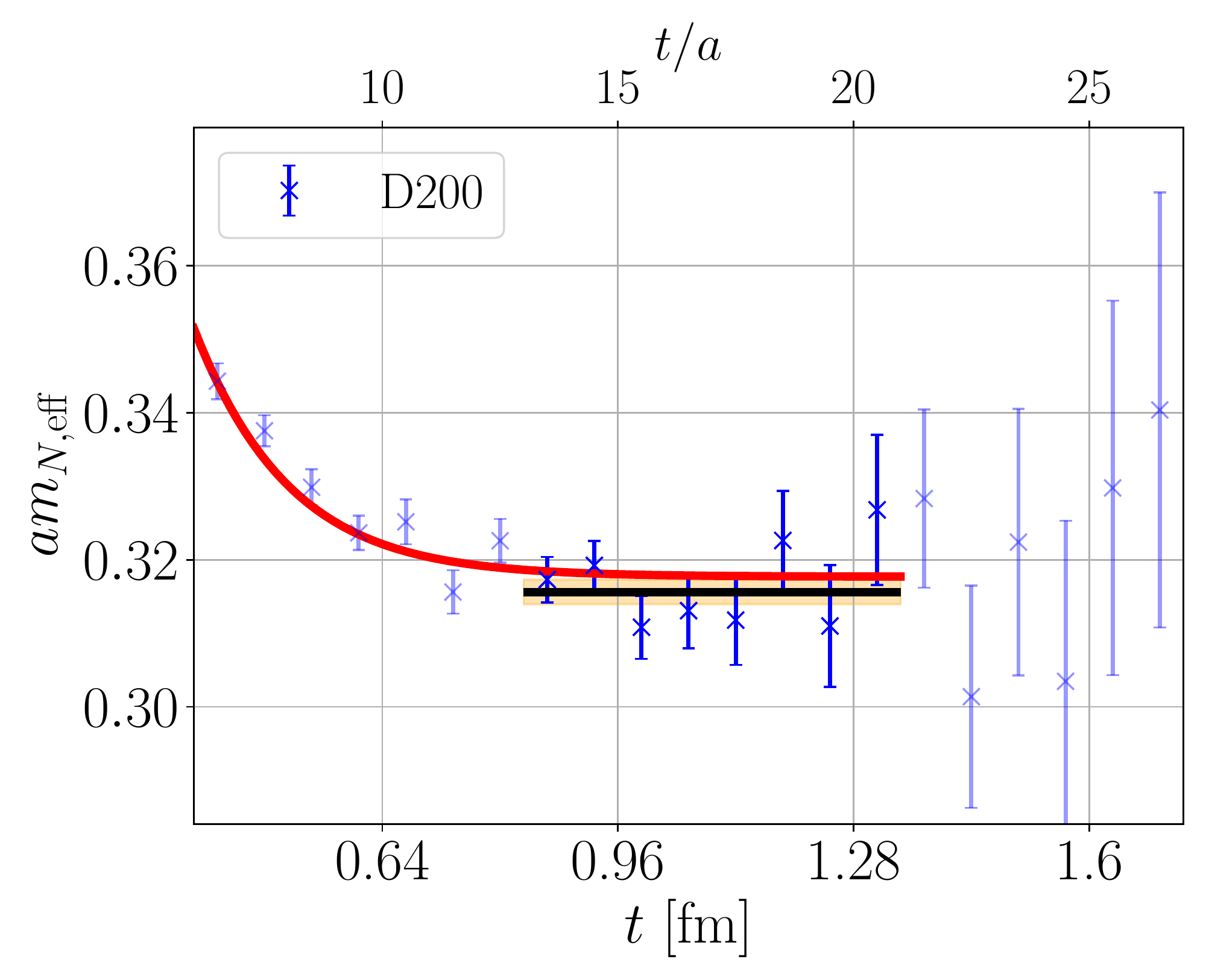}}
  \resizebox{0.49\textwidth}{!}{\includegraphics[width=\textwidth]{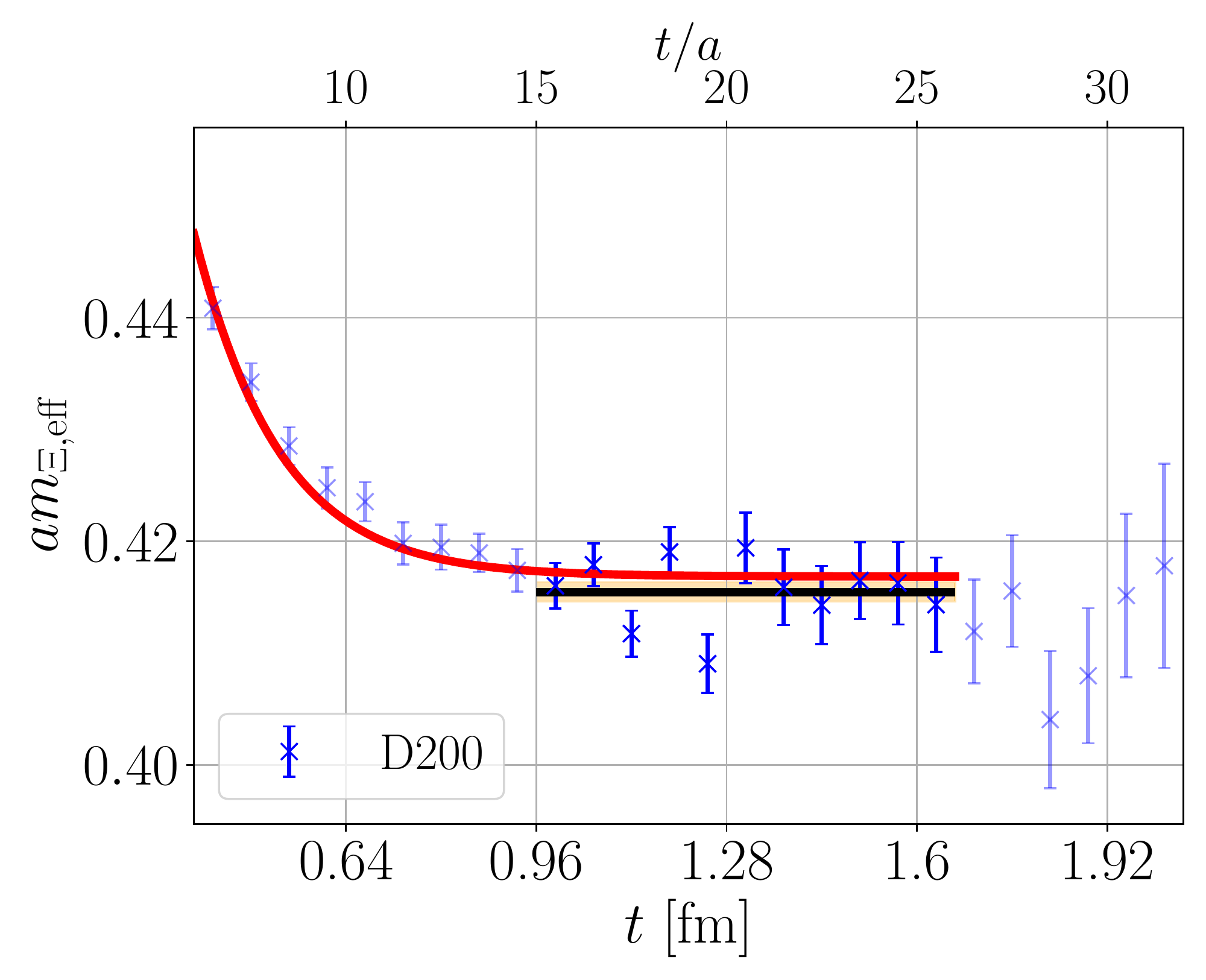}}
  \resizebox{0.49\textwidth}{!}{\includegraphics[width=\textwidth]{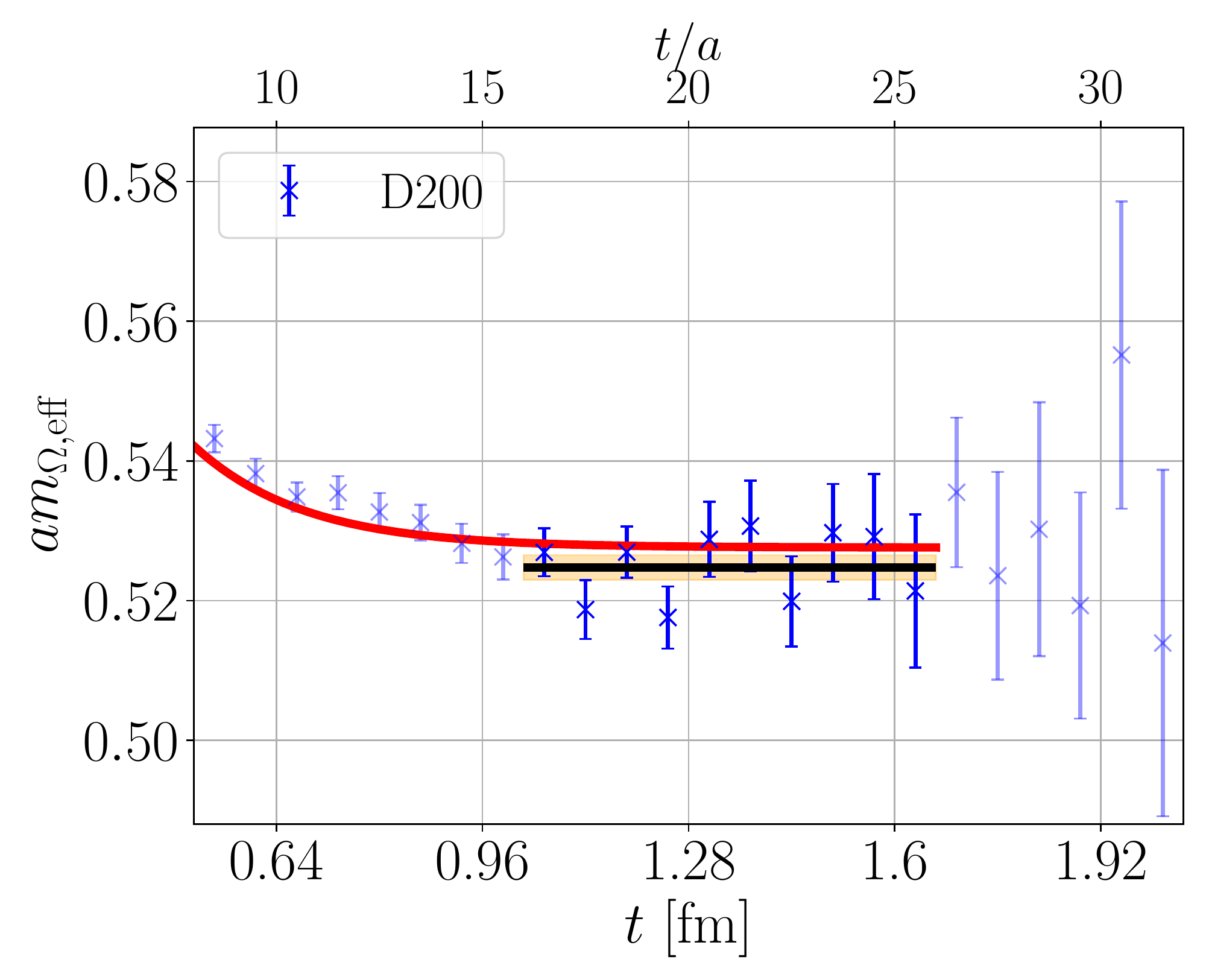}}
  \caption{\label{fig:effm_baryon} Effective masses for the nucleon
    for ensembles B450 (pbc, $a\approx 0.075\,\textmd{fm}$,
    $M_\pi\approx 420\,\textmd{MeV}$)
    and D200 (obc, $a\approx 0.064\,\textmd{fm}$,
    $M_\pi\approx 200\,\textmd{MeV}$) are shown (top left and top right
    panel, respectively)
    as well as for the $\Xi$ and the $\Omega$ baryon for D200
    (lower left and lower right panel, respectively).
    For B450 the two-point function has been averaged over sixteen sources.
    For D200 twenty sources have been used within the range
    $37a\leq t_0\leq 90a$, where for the average over the two-point
    functions also the sink positions
    $t_0+t$ are restricted to the same range.
    The red lines and the orange bands
    have the same meaning as in figure~\ref{fig:effm_B450},
    however, the fit form
    defined in eq.~\eqref{eq:2pt_excited_obc_bar} is used for the two-state
    fit and $B_B$ is set to zero within this equation for the one-state fit.}
\end{figure}

As was already noticed and explained in appendix~B of ref.~\cite{Hamber:1983vu}
(see also ref.~\cite{Lepage:1989hd}),
the relative error of two-point functions other than for the
pseudoscalar mesons increases exponentially
with $t$. In view of this, we use the ansatz
\begin{equation}
  C_B(t)=A_B\exp(-m_Bt)+B_B\exp(-m_B't)
    \label{eq:2pt_excited_obc_bar}  
\end{equation}
for the baryons, both for obc and pbc, to determine the value of
$t_{\min}$ for the subsequent one-state fit, where
$t_{\text{bound}} \le t_{\min}$ and
$t_{\max} \le T- a-t_{\text{bound}}$. In addition, we restrict
$t_{\max}$ to values where the noise over signal ratio of the two-point
function satisfies the condition
$\Delta C_B(t)/C_B(t)<0.25$.
In figure~\ref{fig:effm_baryon}, we show the effective masses
for the nucleon on ensemble B450 as well as for the nucleon,
the $\Xi$ and the $\Omega$ baryons on ensemble D200.
Also shown are the two-state fits, used to determine $t_{\min}$,
and the results of the subsequent ground state fits. The data
shown are averaged over the different source positions.
For the baryon spectroscopy only smeared-smeared
two-point functions with point sources are employed.

Note that $t_{\min}$ is determined on binned data, where a suitable bin
size is chosen in order to account for autocorrelations.
The data shown in figures~\ref{fig:effm_B450}--\ref{fig:effm_baryon}
are binned accordingly.
The final statistical errors of the masses and derived quantities
are computed from bootstrap distributions, as detailed in
appendix~\ref{sec:boot}.
The widths of these distributions are rescaled, taking autocorrelations
into account by means of a binning analysis, see appendices~\ref{sec:bin}
and~\ref{sec:autocorr}.
Note that the mean values are always extracted from unbinned data,
while the error bands shown for the masses resulting from the
one-state fits correspond to the infinite bin size.
In addition, for each ensemble a normalized covariance matrix between all the
meson, baryon and AWI quark masses is computed as explained in
appendix~\ref{sec:autocorr}, which also incorporates the
autocorrelation effects. For $m_s= m_{\ell}$ the dimension of
this matrix is four (meson mass, quark mass, octet baryon mass and
decuplet baryon mass), while for $m_s\neq m_{\ell}$ it is
twelve (two meson and two quark masses each, as well as
four octet and four decuplet baryon masses). For fits of
the baryon masses as functions of the meson masses we take
correlations between the two mesons, the four octet baryons and
the four decuplet baryons into account as well as between the
baryons and the mesons, see appendix~\ref{sec:leastsquares}.

\begin{figure}[htp]
  \centering
  \resizebox{0.49\textwidth}{!}{\includegraphics[width=\textwidth]{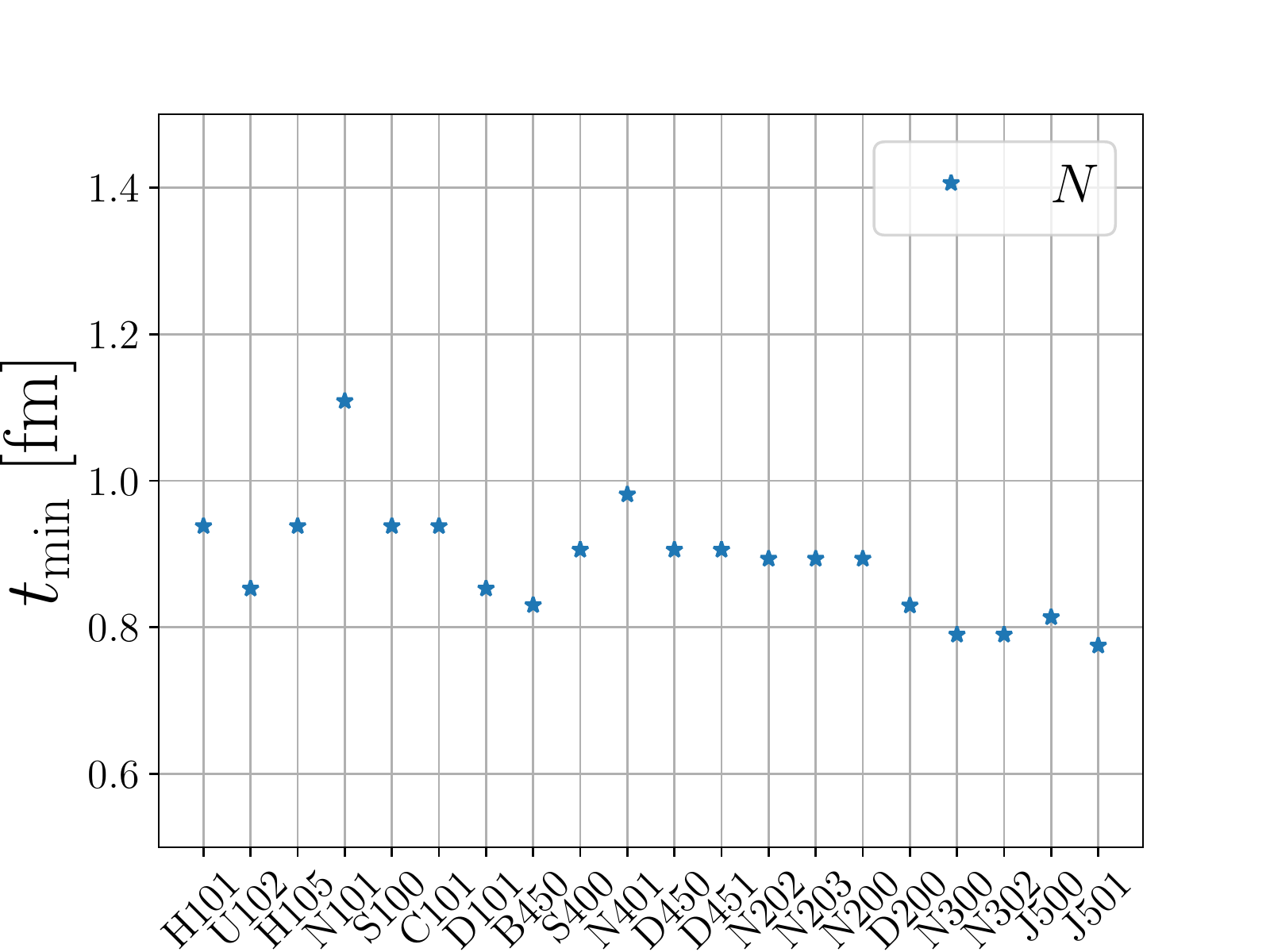}}
  \resizebox{0.49\textwidth}{!}{\includegraphics[width=\textwidth]{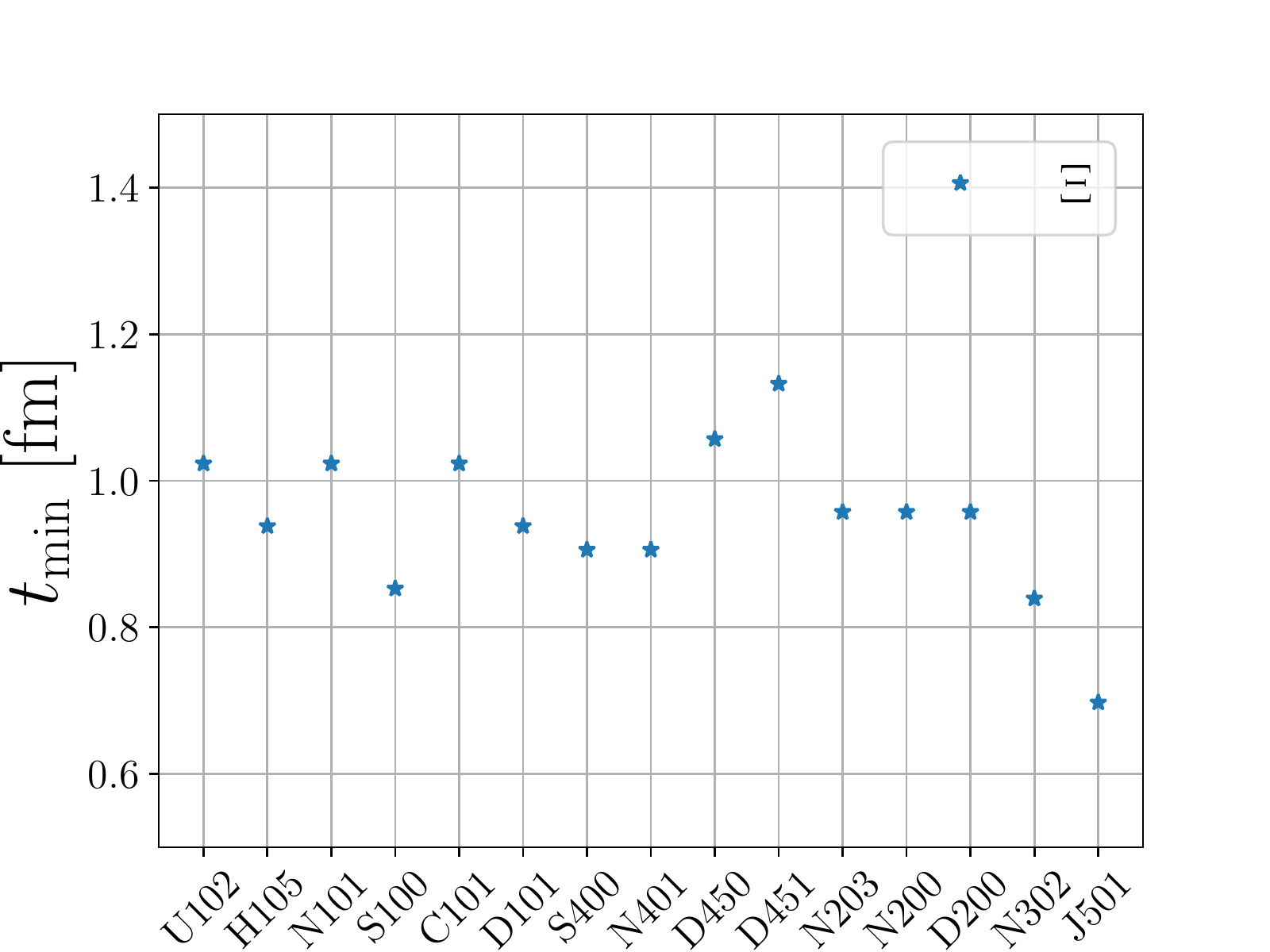}}
  \caption{\label{fig:fitrange_baryon}Starting times $t_{\min}$
    of the one-state fits for the nucleon $N$ (left) and the $\Xi$ baryon
    (right) for the subset of ensembles with high statistics.}
\end{figure}

We have kept the smearing radii almost constant across the different
lattice spacings, however, the fit ranges will also depend on the
quark masses and, in particular, on the statistics. In addition,
at least for the AWI masses, $t_{\min}$ will decrease with the lattice spacing.
This complicates a systematic comparison of fit ranges across the different
ensembles. In figure~\ref{fig:fitrange_baryon} we show the $t_{\min}$
values for the nucleon and the cascade baryon for ensembles where
we have particularly good statistics. These are sorted in the order
of a decreasing lattice spacing. Within each lattice spacing group
(distinguishable
by the first numerical digit of the ensemble~id, see
table~\ref{tab:physparams}), the squared pion mass
$M_\pi^2\propto 3 m_{\ell}$  decreases from left to right for the nucleon,
while the squared $\eta_8$ mass $M_{\eta_8}^2\propto 2m_s+m_{\ell}$, which is
more relevant for the $\Xi$ baryon, increases. Regarding the cascade, we
omit ensembles with $m_s=m_{\ell}$,
since in this case the two-point functions for the two baryons are the same.
With our smearing and statistics, $t_{\min}$ typically varies between
$0.8\,\textmd{fm}$ and $1\,\textmd{fm}$ for the nucleon and
on average it is somewhat larger for the $\Xi$ baryon.

\subsection{Tables of the masses and $\boldsymbol{t_0/a^2}$}
\label{sec:masstables}

In table~\ref{tab:mass_sym} we list for the $m_s=m_{\ell}$ ensembles
the pseudoscalar meson mass $M_P=M_\pi=M_K$, the AWI quark mass
$\widetilde{m}_q=\widetilde{m}_{\ell}=\widetilde{m}_s$,
the octet baryon mass $m_O=m_N=m_{\Lambda}=m_{\Sigma}=m_{\Xi}$
and the decuplet baryon mass $m_D=m_{\Delta}=m_{\Sigma^*}=m_{\Xi^*}=m_{\Omega}$
as well as the scale parameter $t_0$, all in lattice units.
In table~\ref{tab:mass_mes}
the pseudoscalar meson masses $M_{\pi}$ and
$M_K$, the AWI quark masses $\widetilde{m}_{\ell}$ and
$\widetilde{m}_s$ as well as the scale parameter $t_0$
are displayed for the ensembles not listed in
table~\ref{tab:mass_sym} (i.e.\ for
$m_s\neq m_{\ell}$).
In table~\ref{tab:mass_oct} the octet baryon masses
$m_N$, $m_{\Lambda}$, $m_{\Sigma}$ and $m_\Xi$ are shown
for the ensembles not already listed in table~\ref{tab:mass_sym}.
In table~\ref{tab:mass_dec} the
decuplet baryon masses $m_{\Delta}$, $m_{\Sigma^*}$, $m_{\Xi^*}$ and
$m_{\Omega}$ are collected for the ensembles with $m_s\neq m_\ell$.
We regard two of the $\Delta$ and two of the
$\Sigma^*$ entries as unreliable since the respective mass values
are larger than those of non-interacting pairs of a nucleon or a
$\Lambda$ and a pion in a $P$-wave
for these volumes. The corresponding entries are
displayed in {\it Italics}. A much larger number of decuplet baryon entries
will become unstable at the given quark masses in an infinite volume.
We indicate these cases, where one would expect large volume effects,
in \underline{\smash{\it underlined Italics}}.

\setlength\LTcapwidth{\columnwidth}
\begin{small}
\begin{longtable}{ccccccc}
  \caption{\label{tab:mass_sym}The pseudoscalar mass $aM_P$, the
    AWI quark mass
    $a\widetilde{m}_q$, the octet baryon mass $am_O$ and the
    decuplet baryon mass $am_D$ as
    well as the scale parameter $t_0/a^2$ in lattice units for the
    ensembles with equal light and strange quark masses $m_s=m_{\ell}$. The
    decuplet baryon will become unstable in
    the infinite volume for the two entries indicated in
    \underline{\smash{\it underlined Italics}}.}\\\toprule
trajectory & id
& $aM_P$
& $a\widetilde{m}_q$
& $am_O$
& $am_D$
& $t_0/a^2$
\\
\hline\endfirsthead
\caption{\small List of masses for $m_s=m_{\ell}$ (continued).}\\\toprule
trajectory & id
& $aM_P$
& $a\widetilde{m}_q$
& $am_O$
& $am_D$
& $t_0/a^2$
\\
\hline\endhead
\hline
\multicolumn{7}{r}{\textit{Continued on next page}} \\
\endfoot
\hline
\endlastfoot
\multicolumn{7}{c}{$\beta=3.34, a = 0.098\,\mathrm{fm}$} \\\hline
\multirow{3}{*}{$m_s=m_{\ell}$}
           & A651
           & 0.27507(87)
           & 0.017314(93)
           & 0.6715(44)
           & 0.8033(63)
           & 1.9200(47)
           \\
           & A652
           & 0.2140(10)
           & 0.010782(97)
           & 0.5842(41)
           & 0.689(13)
           & 2.1697(56)
           \\
           & A650
           & 0.1835(13)
           & 0.00803(10)
           & 0.5469(54)
           & 0.663(13)
           & 2.2878(72)
           \\
           \hline
\multirow{1}{*}{$\overline{m}=m_{\text{symm}}$}
           & A653
           & 0.21245(93)
           & 0.010663(88)
           & 0.5855(37)
           & 0.7168(73)
           & 2.1729(50)
           \\
           \hline
\multicolumn{7}{c}{$\beta=3.4, a = 0.085\,\mathrm{fm}$} \\\hline
\multirow{3}{*}{$m_s=m_{\ell}$}
           & rqcd019
           & 0.26281(66)
           & 0.018095(73)
           & 0.6268(23)
           & 0.7353(54)
           & 2.4795(81)
           \\
           & rqcd021
           & 0.14702(88)
           & 0.005983(59)
           & 0.4508(47)
           & 0.566(12)
           & 3.032(15)
           \\
           & rqcd017
           & 0.1022(15)
           & 0.002793(89)
           & 0.388(13)
           & \underline{\smash{\it 0.514(20)}}
           & 3.251(13)
           \\
           \hline
\multirow{2}{*}{$\overline{m}=m_{\text{symm}}$}
           & U103
           & 0.18158(60)
           & 0.008936(42)
           & 0.5193(30)
           & 0.638(10)
           & 2.8815(57)
           \\
           & H101
           & 0.18286(57)
           & 0.009197(39)
           & 0.5074(18)
           & 0.6178(71)
           & 2.8545(81)
           \\
           \hline
\multicolumn{7}{c}{$\beta=3.46, a = 0.075\,\mathrm{fm}$} \\\hline
\multirow{3}{*}{$m_s=m_{\ell}$}
           & rqcd029
           & 0.27253(52)
           & 0.021714(49)
           & 0.6169(22)
           & 0.7162(42)
           & 2.976(11)
           \\
           & rqcd030
           & 0.12221(68)
           & 0.004750(40)
           & 0.3957(90)
           & 0.466(22)
           & 3.914(15)
           \\
           & X450
           & 0.10144(62)
           & 0.003300(28)
           & 0.3764(61)
           & \underline{\smash{\it 0.4902(68)}}
           & 3.9935(92)
           \\
           \hline
\multirow{1}{*}{$\overline{m}=m_{\text{symm}}$}
           & B450
           & 0.16103(49)
           & 0.008118(33)
           & 0.4582(24)
           & 0.5619(47)
           & 3.663(11)
           \\
           \hline
\multicolumn{7}{c}{$\beta=3.55, a = 0.064\,\mathrm{fm}$} \\\hline
\multirow{3}{*}{$m_s=m_{\ell}$}
           & B250
           & 0.23052(70)
           & 0.018769(37)
           & 0.5237(29)
           & 0.6104(35)
           & 4.312(18)
           \\
           & X250
           & 0.11321(39)
           & 0.004899(21)
           & 0.3597(51)
           & 0.435(28)
           & 5.283(28)
           \\
           & X251
           & 0.08684(40)
           & 0.002877(23)
           & 0.3185(85)
           & 0.382(42)
           & 5.483(26)
           \\
           \hline
\multirow{2}{*}{$\overline{m}=m_{\text{symm}}$}
           & H200
           & 0.13653(53)
           & 0.006865(23)
           & 0.3968(30)
           & 0.4792(67)
           & 5.150(16)
           \\
           & N202
           & 0.13389(35)
           & 0.006856(15)
           & 0.3799(18)
           & 0.4637(61)
           & 5.165(14)
           \\
           \hline
\multicolumn{7}{c}{$\beta=3.7, a = 0.049\,\mathrm{fm}$} \\\hline
\multirow{1}{*}{$m_s=m_{\ell}$}
           & N303
           & 0.16153(30)
           & 0.012570(12)
           & 0.3742(21)
           & 0.4391(40)
           & 7.743(23)
           \\
           \hline
\multirow{1}{*}{$\overline{m}=m_{\text{symm}}$}
           & N300
           & 0.10647(38)
           & 0.0055137(68)
           & 0.3035(13)
           & 0.3711(54)
           & 8.576(21)
           \\
           \hline
\multicolumn{7}{c}{$\beta=3.85, a = 0.039\,\mathrm{fm}$} \\\hline
\multirow{1}{*}{$m_s=m_{\ell}$}
           & N500
           & 0.11862(67)
           & 0.0084940(80)
           & 0.2878(22)
           & 0.3333(55)
           & 12.912(72)
           \\
           \hline
\multirow{1}{*}{$\overline{m}=m_{\text{symm}}$}
           & J500
           & 0.08119(34)
           & 0.0042100(37)
           & 0.2313(26)
           & 0.2834(35)
           & 14.013(34)
           \\
           \hline
\end{longtable}
\end{small}

\setlength\LTcapwidth{\columnwidth}
\begin{small}
\begin{longtable}{ccccccc}
  \caption{\label{tab:mass_mes}The pion mass $a M_\pi$, the kaon mass $a M_K$, the light AWI quark mass $a\widetilde{m}_\ell$, the strange AWI quark mass $a\widetilde{m}_s$ and the scale parameter $t_0/a^2$ in lattice units for the ensembles with $m_s\neq m_{\ell}$. Results for the other ensembles are shown in table~\ref{tab:mass_sym}.}\\\toprule
trajectory & id
& $aM_\pi$
& $aM_K$
& $a\widetilde{m}_\ell$
& $a\widetilde{m}_s$
& $t_0/a^2$
\\
\hline\endfirsthead
\caption{\small List of pseudoscalar and AWI quark masses (continued).}\\\toprule
trajectory & id
& $aM_\pi$
& $aM_K$
& $a\widetilde{m}_\ell$
& $a\widetilde{m}_s$
& $t_0/a^2$
\\
\hline\endhead
\hline
\multicolumn{7}{r}{\textit{Continued on next page}} \\
\endfoot
\hline
\endlastfoot
\multicolumn{7}{c}{$\beta=3.34, a = 0.098\,\mathrm{fm}$} \\\hline
\multirow{1}{*}{$\overline{m}=m_{\text{symm}}$}
           & A654
           & 0.1673(11)
           & 0.22719(91)
           & 0.006501(83)
           & 0.018108(90)
           & 2.1950(77)\\
           \hline
\multicolumn{7}{c}{$\beta=3.4, a = 0.085\,\mathrm{fm}$} \\\hline
\multirow{10}{*}{$\overline{m}=m_{\text{symm}}$}
           & U102
           & 0.15498(84)
           & 0.19251(61)
           & 0.006380(59)
           & 0.013791(58)
           & 2.8932(63)
           \\
           & H102a
           & 0.15321(98)
           & 0.19091(78)
           & 0.006436(67)
           & 0.013789(61)
           & 2.8840(89)
           \\
           & H102b
           & 0.15499(92)
           & 0.19194(77)
           & 0.006554(60)
           & 0.013866(54)
           & 2.8792(90)
           \\
           & U101
           & 0.1184(24)
           & 0.2005(13)
           & 0.00362(11)
           & 0.018393(95)
           & 2.934(11)
           \\
           & H105
           & 0.1215(13)
           & 0.20234(64)
           & 0.003976(68)
           & 0.018714(58)
           & 2.8917(65)
           \\
           & N101
           & 0.12133(58)
           & 0.20156(30)
           & 0.003972(33)
           & 0.018685(23)
           & 2.8948(39)
           \\
           & S100
           & 0.0929(31)
           & 0.20551(57)
           & 0.00229(11)
           & 0.02117(10)
           & 2.9212(91)
           \\
           & C101
           & 0.09589(63)
           & 0.20561(33)
           & 0.002427(27)
           & 0.021210(33)
           & 2.9176(38)
           \\
           & D101
           & 0.0958(11)
           & 0.20572(45)
           & 0.002500(35)
           & 0.021280(33)
           & 2.910(10)
           \\
           & D150
           & 0.05500(79)
           & 0.20834(17)
           & 0.000799(20)
           & 0.023643(24)
           & 2.9476(30)
           \\
           \hline
\multirow{3}{*}{$\widetilde{m}_s=\widetilde{m}_{s,\text{ph}}$}
           & H107
           & 0.15921(73)
           & 0.23746(53)
           & 0.006656(47)
           & 0.023976(52)
           & 2.7193(76)
           \\
           & H106
           & 0.1182(20)
           & 0.22472(67)
           & 0.003799(62)
           & 0.024031(68)
           & 2.8227(68)
           \\
           & C102
           & 0.09647(77)
           & 0.21783(36)
           & 0.002468(39)
           & 0.023958(37)
           & 2.8682(47)
           \\
           \hline
\multicolumn{7}{c}{$\beta=3.46, a = 0.075\,\mathrm{fm}$} \\\hline
\multirow{3}{*}{$\overline{m}=m_{\text{symm}}$}
           & S400
           & 0.13554(42)
           & 0.17035(38)
           & 0.005679(28)
           & 0.012605(28)
           & 3.6919(74)
           \\
           & N401
           & 0.10987(56)
           & 0.17759(37)
           & 0.003795(28)
           & 0.016460(35)
           & 3.6844(52)
           \\
           & D450
           & 0.08256(41)
           & 0.18354(12)
           & 0.002077(18)
           & 0.019429(19)
           & 3.7076(75)
           \\
           \hline
\multirow{4}{*}{$\widetilde{m}_s=\widetilde{m}_{s,\text{ph}}$}
           & B451
           & 0.16141(57)
           & 0.22047(32)
           & 0.007893(32)
           & 0.022061(32)
           & 3.4265(72)
           \\
           & B452
           & 0.13492(47)
           & 0.20973(34)
           & 0.005525(30)
           & 0.022001(26)
           & 3.5286(66)
           \\
           & N450
           & 0.10967(31)
           & 0.20176(18)
           & 0.003659(18)
           & 0.022044(19)
           & 3.5920(42)
           \\
           & D451
           & 0.08371(31)
           & 0.19385(15)
           & 0.002120(16)
           & 0.021843(19)
           & 3.6684(36)
           \\
           \hline
\multicolumn{7}{c}{$\beta=3.55, a = 0.064\,\mathrm{fm}$} \\\hline
\multirow{5}{*}{$\overline{m}=m_{\text{symm}}$}
           & N203
           & 0.11249(30)
           & 0.14399(24)
           & 0.004739(16)
           & 0.011051(13)
           & 5.1465(63)
           \\
           & S201
           & 0.09453(47)
           & 0.15228(37)
           & 0.003135(19)
           & 0.014142(17)
           & 5.1638(91)
           \\
           & N200
           & 0.09244(29)
           & 0.15061(24)
           & 0.003156(12)
           & 0.014146(12)
           & 5.1600(71)
           \\
           & D200
           & 0.06544(33)
           & 0.15652(15)
           & 0.0015521(84)
           & 0.0172194(86)
           & 5.1793(39)
           \\
           & E250
           & 0.04228(23)
           & 0.159370(61)
           & 0.0006446(73)
           & 0.0188850(74)
           & 5.2027(41)
           \\
           \hline
\multirow{3}{*}{$\widetilde{m}_s=\widetilde{m}_{s,\text{ph}}$}
           & N204
           & 0.11427(33)
           & 0.17734(29)
           & 0.004792(13)
           & 0.018906(12)
           & 4.9473(79)
           \\
           & N201
           & 0.09276(31)
           & 0.17040(22)
           & 0.003149(14)
           & 0.018853(14)
           & 5.0427(75)
           \\
           & D201
           & 0.06476(42)
           & 0.16302(18)
           & 0.001551(15)
           & 0.018872(15)
           & 5.1378(66)
           \\
           \hline
\multicolumn{7}{c}{$\beta=3.7, a = 0.049\,\mathrm{fm}$} \\\hline
\multirow{3}{*}{$\overline{m}=m_{\text{symm}}$}
           & N302
           & 0.08716(41)
           & 0.11373(36)
           & 0.0037228(78)
           & 0.0090864(78)
           & 8.539(19)
           \\
           & J303
           & 0.06488(19)
           & 0.11975(16)
           & 0.0020511(61)
           & 0.0123411(47)
           & 8.615(14)
           \\
           & E300
           & 0.04403(20)
           & 0.12397(15)
           & 0.0009277(55)
           & 0.0145418(80)
           & 8.6241(74)
           \\
           \hline
\multirow{3}{*}{$\widetilde{m}_s=\widetilde{m}_{s,\text{ph}}$}
           & N305
           & 0.10719(35)
           & 0.14598(34)
           & 0.0054860(88)
           & 0.0152409(83)
           & 8.181(19)
           \\
           & N304
           & 0.08855(33)
           & 0.13961(31)
           & 0.0037156(93)
           & 0.0152961(80)
           & 8.322(20)
           \\
           & J304
           & 0.06545(18)
           & 0.13181(14)
           & 0.0020500(61)
           & 0.0152663(44)
           & 8.497(12)
           \\
           \hline
\multicolumn{7}{c}{$\beta=3.85, a = 0.039\,\mathrm{fm}$} \\\hline
\multirow{1}{*}{$\overline{m}=m_{\text{symm}}$}
           & J501
           & 0.06599(26)
           & 0.08799(23)
           & 0.0027380(36)
           & 0.0071717(31)
           & 13.928(39)
           \\
           \hline
\end{longtable}
\end{small}

\setlength\LTcapwidth{\columnwidth}
\begin{longtable}{cccccc}
  \caption{\label{tab:mass_oct}\small The octet baryon masses $am_O$, $O\in\{N,\Lambda,\Sigma,\Xi\}$, in lattice units for the ensembles with $m_s\neq m_{\ell}$. Results for the other ensembles are shown in table~\ref{tab:mass_sym}.}\\\toprule
trajectory & id
& $am_N$
& $am_\Lambda$
& $am_\Sigma$
& $am_\Xi$
\\
\hline\endfirsthead
\caption{\small List of octet baryon masses (continued).}\\\toprule
trajectory & id
& $am_N$
& $am_\Lambda$
& $am_\Sigma$
& $am_\Xi$
\\
\hline\endhead
\hline
\multicolumn{6}{r}{\textit{Continued on next page}} \\
\endfoot
\hline
\endlastfoot
\multicolumn{6}{c}{$\beta=3.34, a = 0.098\,\mathrm{fm}$} \\\hline
\multirow{1}{*}{$\overline{m}=m_{\text{symm}}$}
           & A654
           & 0.5423(65)
           & 0.5711(48)
           & 0.5850(40)
           & 0.6070(30)
           \\\pagebreak[4]
           \hline
           \multicolumn{6}{c}{$\beta=3.4, a = 0.085\,\mathrm{fm}$} \\\hline
\multirow{10}{*}{$\overline{m}=m_{\text{symm}}$}
           & U102
           & 0.4899(43)
           & 0.5059(49)
           & 0.5150(41)
           & 0.5287(31)
           \\*
           & H102a
           & 0.4797(28)
           & 0.4986(26)
           & 0.5079(26)
           & 0.5235(23)
           \\*
           & H102b
           & 0.4747(39)
           & 0.4948(34)
           & 0.5046(32)
           & 0.5201(27)
           \\*
           & U101
           & 0.456(11)
           & 0.4888(80)
           & 0.5108(64)
           & 0.5432(53)
           \\*
           & H105
           & 0.4397(66)
           & 0.4845(33)
           & 0.5096(22)
           & 0.5392(17)
           \\
           & N101
           & 0.4412(38)
           & 0.4873(18)
           & 0.5050(18)
           & 0.5376(11)
           \\
           & S100
           & 0.4219(91)
           & 0.4731(60)
           & 0.5019(33)
           & 0.5465(21)
           \\
           & C101
           & 0.4237(35)
           & 0.4788(17)
           & 0.5026(19)
           & 0.5433(11)
           \\
           & D101
           & 0.427(10)
           & 0.4775(66)
           & 0.5053(59)
           & 0.5416(42)
           \\
           & D150
           & 0.4033(88)
           & 0.4594(40)
           & 0.4997(25)
           & 0.5482(14)
           \\
           \hline
\multirow{3}{*}{$\widetilde{m}_s=\widetilde{m}_{s,\text{ph}}$}
           & H107
           & 0.4997(45)
           & 0.5408(36)
           & 0.5632(35)
           & 0.5941(28)
           \\
           & H106
           & 0.4594(54)
           & 0.5053(44)
           & 0.5325(37)
           & 0.5687(33)
           \\
           & C102
           & 0.4341(48)
           & 0.4927(58)
           & 0.5179(48)
           & 0.5588(33)
           \\
           \hline
\multicolumn{6}{c}{$\beta=3.46, a = 0.075\,\mathrm{fm}$} \\\hline
\multirow{3}{*}{$\overline{m}=m_{\text{symm}}$}
           & S400
           & 0.4261(22)
           & 0.4441(17)
           & 0.4530(17)
           & 0.4665(14)
           \\
           & N401
           & 0.3965(38)
           & 0.4309(26)
           & 0.4492(25)
           & 0.4758(17)
           \\
           & D450
           & 0.3653(50)
           & 0.4252(21)
           & 0.4479(24)
           & 0.4865(13)
           \\
           \hline
\multirow{4}{*}{$\widetilde{m}_s=\widetilde{m}_{s,\text{ph}}$}
           & B451
           & 0.4721(42)
           & 0.5043(31)
           & 0.5203(26)
           & 0.5452(22)
           \\
           & B452
           & 0.4403(52)
           & 0.4764(39)
           & 0.4975(29)
           & 0.5290(20)
           \\
           & N450
           & 0.3985(68)
           & 0.4471(43)
           & 0.4678(48)
           & 0.5105(29)
           \\
           & D451
           & 0.3694(49)
           & 0.4335(27)
           & 0.4566(28)
           & 0.4986(16)
           \\
           \hline
\multicolumn{6}{c}{$\beta=3.55, a = 0.064\,\mathrm{fm}$} \\\hline
\multirow{5}{*}{$\overline{m}=m_{\text{symm}}$}
           & N203
           & 0.3624(18)
           & 0.3782(15)
           & 0.3863(15)
           & 0.3980(13)
           \\
           & S201
           & 0.3549(69)
           & 0.3799(56)
           & 0.3942(49)
           & 0.4151(31)
           \\
           & N200
           & 0.3423(22)
           & 0.3727(14)
           & 0.3863(13)
           & 0.4084(10)
           \\
           & D200
           & 0.3156(17)
           & 0.3605(13)
           & 0.3816(14)
           & 0.41547(84)
           \\
           & E250
           & 0.3020(48)
           & 0.3497(33)
           & 0.3764(30)
           & 0.4188(14)
           \\
           \hline
\multirow{3}{*}{$\widetilde{m}_s=\widetilde{m}_{s,\text{ph}}$}
           & N204
           & 0.3721(24)
           & 0.4056(19)
           & 0.4201(21)
           & 0.4469(16)
           \\
           & N201
           & 0.3415(32)
           & 0.3870(22)
           & 0.4029(27)
           & 0.4352(21)
           \\
           & D201
           & 0.3200(50)
           & 0.3709(36)
           & 0.3875(48)
           & 0.4243(22)
           \\
           \hline
\multicolumn{6}{c}{$\beta=3.7, a = 0.049\,\mathrm{fm}$} \\\hline
\multirow{3}{*}{$\overline{m}=m_{\text{symm}}$}
           & N302
           & 0.2862(19)
           & 0.2992(16)
           & 0.3065(16)
           & 0.3161(14)
           \\
           & J303
           & 0.2583(16)
           & 0.2862(11)
           & 0.3012(12)
           & 0.32079(82)
           \\
           & E300
           & 0.2384(40)
           & 0.2784(23)
           & 0.2972(25)
           & 0.3258(14)
           \\
           \hline
\multirow{3}{*}{$\widetilde{m}_s=\widetilde{m}_{s,\text{ph}}$}
           & N305
           & 0.3092(18)
           & 0.3325(16)
           & 0.3414(17)
           & 0.3595(14)
           \\
           & N304
           & 0.2892(36)
           & 0.3175(26)
           & 0.3299(27)
           & 0.3509(21)
           \\
           & J304
           & 0.2674(22)
           & 0.2996(17)
           & 0.3169(17)
           & 0.3395(13)
           \\
           \hline
\multicolumn{6}{c}{$\beta=3.85, a = 0.039\,\mathrm{fm}$} \\\hline
\multirow{1}{*}{$\overline{m}=m_{\text{symm}}$}
           & J501
           & 0.2165(17)
           & 0.2304(15)
           & 0.2351(17)
           & 0.2442(14)
           \\
           \hline
\end{longtable}

\setlength\LTcapwidth{\columnwidth}
\begin{longtable}{cccccc}
  \caption{\label{tab:mass_dec}\small The decuplet baryon masses $am_D$,
    $D\in\{\Delta, \Sigma^*, \Xi^*, \Omega\}$, in lattice units for the ensembles with $m_s\neq m_{\ell}$. Results for the other ensembles are shown in table~\ref{tab:mass_sym}. Masses for decuplet baryons that can strongly decay into
    pairs of octet baryons and pions in the finite volume
    are unreliable and displayed in {\it Italics}. Additional entries
    for baryons
    that will become unstable in the infinite volume are indicated in
    \underline{\smash{\it underlined Italics}}.}\\\toprule
trajectory & id
& $am_\Delta$
& $am_{\Sigma^*}$
& $am_{\Xi^*}$
& $am_\Omega$
\\
\hline\endfirsthead
\caption{\small List of decuplet baryon masses (continued).}\\\toprule
trajectory & id
& $am_\Delta$
& $am_{\Sigma^*}$
& $am_{\Xi^*}$
& $am_\Omega$
\\
\hline\endhead
\hline
\multicolumn{6}{r}{\textit{Continued on next page}} \\
\endfoot
\hline
\endlastfoot
\multicolumn{6}{c}{$\beta=3.34, a = 0.098\,\mathrm{fm}$} \\\hline
\multirow{1}{*}{$\overline{m}=m_{\text{symm}}$}
           & A654
           & 0.6954(74)
           & 0.7115(59)
           & 0.7289(69)
           & 0.7459(67)
           \\
           \hline
\multicolumn{6}{c}{$\beta=3.4, a = 0.085\,\mathrm{fm}$} \\\hline
\multirow{10}{*}{$\overline{m}=m_{\text{symm}}$}
           & U102
           & 0.6131(79)
           & 0.6276(63)
           & 0.6417(53)
           & 0.6563(56)
           \\
           & H102a
           & 0.588(11)
           & 0.6092(52)
           & 0.6258(40)
           & 0.6417(32)
           \\
           & H102b
           & 0.5866(69)
           & 0.6046(52)
           & 0.6220(41)
           & 0.6388(34)
           \\
           & U101
           & 0.586(13)
           & 0.600(21)
           & 0.633(11)
           & 0.6734(82)
           \\
           & H105
           & \underline{\smash{\it 0.576(13)}}
           & 0.6033(89)
           & 0.6360(79)
           & 0.6689(69)
           \\
           & N101
           & \underline{\smash{\it 0.577(14)}}
           & 0.6028(80)
           & 0.6359(46)
           & 0.6681(31)
           \\
           & S100
           & \underline{\smash{\it 0.567(11)}}
           & \underline{\smash{\it 0.597(11)}}
           & \underline{\smash{\it 0.6429(69)}}
           & 0.6838(40)
           \\
           & C101
           & \underline{\smash{\it 0.556(17)}}
           & \underline{\smash{\it 0.6020(66)}}
           & 0.6394(43)
           & 0.6817(27)
           \\
           & D101
           & {\it 0.5833(91)}
           & \underline{\smash{\it 0.587(19)}}
           & 0.634(12)
           & 0.676(14)
           \\
           & D150
           & \underline{\smash{\it 0.477(22)}}
           & \underline{\smash{\it 0.578(13)}}
           & \underline{\smash{\it 0.6402(51)}}
           & 0.6892(38)
           \\
           \hline
\multirow{3}{*}{$\widetilde{m}_s=\widetilde{m}_{s,\text{ph}}$}
           & H107
           & 0.637(10)
           & 0.6712(54)
           & 0.7055(41)
           & 0.7378(40)
           \\
           & H106
           & \underline{\smash{\it 0.581(21)}}
           & 0.613(12)
           & 0.6662(68)
           & 0.7066(53)
           \\
           & C102
           & \underline{\smash{\it 0.5846(44)}}
           & \underline{\smash{\it 0.606(10)}}
           & \underline{\smash{\it 0.6579(65)}}
           & 0.7078(47)
           \\
           \hline
\multicolumn{6}{c}{$\beta=3.46, a = 0.075\,\mathrm{fm}$} \\\hline
\multirow{3}{*}{$\overline{m}=m_{\text{symm}}$}
           & S400
           & 0.5473(56)
           & 0.5539(69)
           & 0.5635(51)
           & 0.5759(40)
           \\
           & N401
           & \underline{\smash{\it 0.5141(84)}}
           & \underline{\smash{\it 0.5428(53)}}
           & 0.5728(43)
           & 0.5979(32)
           \\
           & D450
           & \underline{\smash{\it 0.488(10)}}
           & \underline{\smash{\it 0.5362(60)}}
           & \underline{\smash{\it 0.5730(48)}}
           & 0.6111(44)
           \\
           \hline
\multirow{4}{*}{$\widetilde{m}_s=\widetilde{m}_{s,\text{ph}}$}
           & B451
           & 0.584(12)
           & 0.6142(51)
           & 0.6414(38)
           & 0.6683(30)
           \\
           & B452
           & 0.556(20)
           & 0.5879(91)
           & 0.6196(75)
           & 0.6476(59)
           \\
           & N450
           & 0.508(12)
           & \underline{\smash{\it 0.5600(94)}}
           & 0.6059(61)
           & 0.6417(41)
           \\
           & D451
           & \underline{\smash{\it 0.5096(96)}}
           & \underline{\smash{\it 0.5429(85)}}
           & \underline{\smash{\it 0.5866(50)}}
           & 0.6285(33)
           \\\pagebreak[4]
           \hline
\multicolumn{6}{c}{$\beta=3.55, a = 0.064\,\mathrm{fm}$} \\\hline
\multirow{5}{*}{$\overline{m}=m_{\text{symm}}$}
           & N203
           & 0.4467(89)
           & 0.4620(67)
           & 0.4743(60)
           & 0.4871(45)
           \\
           & S201
           & 0.467(14)
           & 0.467(14)
           & 0.477(12)
           & 0.5035(74)
           \\
           & N200
           & 0.426(13)
           & 0.4575(69)
           & 0.4845(39)
           & 0.5083(24)
           \\
           & D200
           & \underline{\smash{\it 0.4228(47)}}
           & \underline{\smash{\it 0.4581(41)}}
           & \underline{\smash{\it 0.4898(27)}}
           & 0.5247(18)
           \\
           & E250
           & \underline{\smash{\it 0.374(19)}}
           & {\it 0.4503(83)}
           & \underline{\smash{\it 0.4906(49)}}
           & 0.5287(29)
           \\
           \hline
\multirow{3}{*}{$\widetilde{m}_s=\widetilde{m}_{s,\text{ph}}$}
           & N204
           & 0.470(10)
           & 0.4996(55)
           & 0.5271(42)
           & 0.5519(34)
           \\
           & N201
           & \underline{\smash{\it 0.4441(81)}}
           & 0.4763(68)
           & 0.5109(48)
           & 0.5451(33)
           \\
           & D201
           & \underline{\smash{\it 0.433(11)}}
           & \underline{\smash{\it 0.4575(81)}}
           & \underline{\smash{\it 0.4948(62)}}
           & 0.5308(55)
           \\
           \hline
\multicolumn{6}{c}{$\beta=3.7, a = 0.049\,\mathrm{fm}$} \\\hline
\multirow{3}{*}{$\overline{m}=m_{\text{symm}}$}
           & N302
           & 0.368(11)
           & 0.3754(72)
           & 0.3846(44)
           & 0.3943(32)
           \\
           & J303
           & \underline{\smash{\it 0.3330(91)}}
           & \underline{\smash{\it 0.3572(50)}}
           & 0.3811(28)
           & 0.4034(16)
           \\
           & E300
           & {\it 0.332(10)}
           & {\it 0.3651(55)}
           & \underline{\smash{\it 0.3920(30)}}
           & 0.4191(23)
           \\
           \hline
\multirow{3}{*}{$\widetilde{m}_s=\widetilde{m}_{s,\text{ph}}$}
           & N305
           & 0.3901(42)
           & 0.4085(33)
           & 0.4258(28)
           & 0.4426(24)
           \\
           & N304
           & 0.3610(74)
           & 0.3865(55)
           & 0.4098(29)
           & 0.4331(36)
           \\
           & J304
           & \underline{\smash{\it 0.3445(71)}}
           & \underline{\smash{\it 0.3704(44)}}
           & 0.4010(31)
           & 0.4257(26)
           \\
           \hline
\multicolumn{6}{c}{$\beta=3.85, a = 0.039\,\mathrm{fm}$} \\\hline
\multirow{1}{*}{$\overline{m}=m_{\text{symm}}$}
           & J501
           & 0.2796(42)
           & 0.2877(32)
           & 0.2945(27)
           & 0.3010(27)
           \\
           \hline
\end{longtable}

\section{Statistical analysis}
\label{sec:statistical}
We detail our statistical methods, starting
with an exposition on jackknife~\cite{jackknife1,jackknife2} and
bootstrap~\cite{efron2,Efron} resampling techniques
that are used with respect to primary (correlation functions),
secondary (masses determined on individual ensembles) and
tertiary (extrapolated masses, LECs etc.) observables.
In appendix~\ref{sec:bin} we then explain how we take into account
autocorrelations in the Monte Carlo time series both for the
variances of observables and for the reduced correlation matrices
between observables by means of a binning and extrapolation
procedure. This is illustrated for some examples in appendix~\ref{sec:autocorr}.
Finally, in appendix~\ref{sec:leastsquares} we detail how we
take into account the errors of the arguments of our fit functions
as well as correlations between the fitted masses and the arguments,
also incorporating prior information on $t_0/a^2$ and the
$\Xi$ baryon mass into the fits to secondary data.
We also employ a model averaging procedure which is explained
in section~\ref{sec:scale} of the main text.

\subsection{Resampling: jackknife and bootstrap analysis}
\label{sec:boot}
We assume that a particular ensemble contains $N$ gauge configurations and
that $I$ different observables $A^i$, $i=1,\ldots, I$
are computed on these configurations,
with the set of results given as $\{A_1^i,A_2^i,\ldots,A_N^i\}$.
These could, e.g., be correlation functions $C_H(t)$ for a hadron
$H$ at a time $t$, where $H$ and $t$ are encoded as the
superscript $i$. For these observables we
define ensemble averages
\begin{equation}
\label{eq:ensemble}
\langle A^i\rangle=\frac{1}{N}\sum_{n=1}^{N}A^i_n.
\end{equation}
We can then move on to
compute secondary observables (also known as derived observables),
e.g., hadron masses, from these ensemble averages:
\begin{equation}
f_k=f_k(\{\langle A^i\rangle\})=
f_k(\langle A\rangle),\end{equation}
where the dependence on the arguments could be non-parametric,
e.g., the result of a least squares fit. Above, we introduced
the short-hand notation $\langle A\rangle=\{\langle A^i\rangle\}$.
Within the remainder of this subsection we will omit the superscripts
$i$ but keep in mind
that in general the secondary observables $f_k$
will depend on several primary observables $A$.

In addition to the original sample $\{A_1,\ldots,A_N\}$
we can also introduce a single elimination jackknife sample
$\{A^{(J)}_1,\ldots, A^{(J)}_N\}$, where
\begin{equation}
A^{(J)}_j = \frac{1}{N-1}\left(N\langle A\rangle - A_j\right).
\end{equation}
In this case the number of jackknifes equals the number of
configurations: $J=N$. Note that
\[
\frac{1}{N}\sum_{j=1}^N A^{(J)}_j = \langle A\rangle,
\]
however, for secondary quantities in general
\[
f_k^{(J)}=\frac{1}{N}\sum_{j=1}^N f^{(J)}_{k,j}
\neq f_k(\langle A\rangle)\,,\quad\text{where}\quad f^{(J)}_{k,j}= f_k(A^{(J)}_j).
\]
It can easily be checked that the single elimination
jackknife error of $A$ is identical to its standard deviation, where
for the moment being we neglect autocorrelations:
\begin{equation}
\Delta A^2=\frac{N-1}{N}\sum_j^N{\left(A^{(J)}_j-\langle A\rangle\right)}^2
=\frac{1}{N(N-1)}\sum_n^N{\left(A_n-\langle A\rangle\right)}^2.
\end{equation}
For secondary quantities no error propagation is needed but the jackknife error
\begin{equation}
{\left(\Delta^{(J)} f_k\right)}^2=\frac{N-1}{N}\sum_j^N
{\left(f^{(J)}_{k,j}-f_k^{(J)}\right)}^2
\end{equation}
or the covariance matrix
\begin{equation}
\label{eq:covar}
C^{(J)}_{k\ell}=\frac{N-1}{N}\sum_j^N
\left(f^{(J)}_{k,j}-f_k^{(J)}\right)
\left(f^{(J)}_{\ell,j}-f_{\ell}^{(J)}\right),
\end{equation}
where $C^{(J)}_{kk}={\left(\Delta^{(J)} f_k\right)}^2$, can be computed directly.

Instead of single elimination jackknife, one can ``block'' the $N$
original measurements into $J<N$ bins, averaging over $S=\lfloor N/J\rfloor\in\mathbb{N}$
subsequent
configurations and construct jackknifes out of these $J$ bins. This
will reduce --- or if $S\gg\tau_{\text{int}}$ (where $\tau_{\text{int}}$
is the largest integrated autocorrelation time of the $f_k$)
effectively remove --- autocorrelations.
It turns out that due to the reduced variance between jackknife
samples, the jackknife is a more stable procedure than the bootstrap
explained below. Therefore, at the initial stage of the analysis,
different bin sizes are realized, and the jackknife errors
of the $f_k$ are computed as a function of the bin size. These are
then extrapolated to infinite bin size (as will be detailed
in appendix~\ref{sec:bin} below),
giving the scale factors
\begin{equation}
\label{eq:scale}
s_k =\lim_{S\rightarrow\infty}\frac{\Delta^{(J)}f_k[S]}{\Delta^{(J)}f_k}\approx \sqrt{2\tau_{k,\text{int}}},
\end{equation}
where $\Delta^{(J)}f_k[S]$ is the jackknife error on $f_k$, obtained for
the bin size $S$, and $\tau_{k,\text{int}}$ is the integrated autocorrelation
time of $f_k$.

We wish to combine information from different ensembles
with different numbers of configurations to perform
various extrapolations and interpolations. The binning approach
would require a bin size $S$ that is larger than twice the
maximal integrated autocorrelation time that we encounter for any of
our observables. In addition, a fixed dimension of
the resampled data across ensembles is preferable.
Our strategy to achieve this is to resample all data into a
bootstrap ensemble of a fixed dimension $B=500$ but to rescale the
resulting distributions of $f_k$ by the scale factors $s_k$ introduced
above. For correlated fits, in addition the extrapolated
covariance matrices eq.~\eqref{eq:covar} are needed, see appendix~\ref{sec:bin}.

The bootstrap sets are generated,
randomly dialling $b_n\in\{1,\ldots,N\}$ for each $b=1,\ldots, B$.
The bootstrap sample $b$ then contains
$\{A_{b_1}, A_{b_2},\cdot,A_{b_N}\}$. We do not require the whole
sample but only the corresponding bootstrap averages
\begin{equation}
A^{(B)}_b=\frac{1}{N}\sum_n^N A_{b_n},
\end{equation}
which can be generated directly from the original data with a
well-defined random number sequence. This sequence is stored
so that additional observables can be added when they are computed
at a later time.

As discussed above, prior to any analysis we rescale the bootstrap
distributions of $f_k$ by the scale factor $s_k$:
\begin{equation}
\label{eq:rescale}
f^{(B)}_{k,b}=f_k+s_k\left[f_k(A^{(B)}_b)-f_k\right].
\end{equation}
We then define the bootstrap average of $f_k$ as
\begin{equation}
f_k^{(B)}=\frac{1}{B}\sum_b^B f^{(B)}_{k,b}.
\end{equation}
In the limit $N\rightarrow\infty$ and for sufficiently large
$B$: $f_k^{(B)}\rightarrow f_k$, up to $1/N$ corrections. There is
a bias between the true (but unknown) result $f_k^*$ and the estimate $f_k$,
obtained using a finite sample size:
$b_k=f_k-f_k^*$. Assuming that a similar skew exists
between the average of the resampled distribution $f_k^{(B)}$
and the original ensemble average, one obtains~\cite{Efron}
$b_k\approx f_k^{(B)}-f_k$.
Subtracting this bias gives the so-called ``unbiased estimate'':
\begin{equation}
f_k^*\approx 2f_k-f_k^{(B)}.
\end{equation}
We have decided to quote as our central values either the original
ensemble averages $f_k$ or the (statistically more robust)
median of the bootstrap histogram, since the bias can only amount
to a fraction of the statistical error. 

In the absence of autocorrelations, one can determine
the covariance matrix from the bootstrap ensemble instead of
the jackknife sample. However, the results for the off-diagonal
matrix elements would be incorrect if obtained from the
rescaled bootstrap ensemble.
Therefore, we also keep the binned jackknife
samples, from which the relevant covariance matrix elements
can be reconstructed if needed,
extrapolating the definition~\eqref{eq:covar} to infinite
bin size, as will be detailed below.

The above errors and the covariance matrix (which will
assume a block-diagonal form when data from different ensembles are combined)
are needed for subsequent fits to the secondary $f_k$ data. However,
it is also possible to quote an (in general asymmetric) error range,
by sorting the bootstrap results and discarding the upper and lower 16\%
of the distribution, resulting in upper and lower 68\% confidence limits
($\pm 1\sigma$ error band).\footnote{Note that if the unbiased estimate
$f_k^*$ rather than the
ensemble average $f_k$ is quoted, the 68\% confidence range should be
shifted accordingly.} We monitor the differences
$b_k$ between $f_k^{(B)}$ and $f_k$ as well as the deviation of
the width of the confidence band from $2\Delta f_k^{(B)}$, as a measure
of the non-Gaussianity of the resampled distribution.

The analysis described above factorizes into two stages.
The output of the first stage is the ensemble average
$f_k$, the resampled (and rescaled, see eq.~\eqref{eq:rescale})
bootstrap distribution $\{f^{(B)}_{k,b}, b=1,\ldots B\}$ and
jackknife ensembles for different bin sizes.
From this the errors and covariance matrix can be reconstructed and
further fits carried out. In principle, one could compute
sub-bootstraps on the individual bootstrap ensembles, aiming at
constructing separate covariance matrices for each bootstrap.
These will be subject to larger statistical fluctuations, potentially
resulting in non-positive or unusually
small eigenvalues on some of the samples.
Instead, we implement the more robust frozen covariance matrix
approach, employing one and the same matrix on all the bootstraps
in the subsequent fits to the secondary data. This approximation
is justified by the fact that differences between the two methods should be
of order $1/N$ in the sample size $N$.

\subsection{Binning and autocorrelation times}
\label{sec:bin}
We consider the autocorrelation function of the covariance matrix
element between two primary observables $A$ and $B$, i.e.\
in the notation of the previous subsection $A=A^1$ and $B=A^2$:
\begin{align}
\Gamma_{AB}(t)&=\frac{1}{N-t}\sum_{t_0=1}^{N-t}\left(A_{t+t_0}-\langle A\rangle\right)
\left(B_{t_0}-\langle B\rangle\right),\nonumber\\
\Gamma_{BA}(-t)&=\Gamma_{AB}(t)\quad \text{for}\quad t\geq 0,
\end{align}
where $t\in\mathbb{Z}$,
$|t|\ll N$ and the Monte Carlo time series
is defined by $\{A_n,B_n\}$, $n=1,\ldots, N$.
$\langle A\rangle$ is the ensemble average defined in
eq.~\eqref{eq:ensemble}. Above, we assume phase conventions such that
$\langle A\rangle, \langle B\rangle\in\mathbb{R}$ and we have taken
into account that the length of the time series available for a given
value of $t$ is $N-t$. Note that $B=A$ is included as a special case.
The discussion for the primary observables below
can easily be generalized to secondary observables.
We remark that estimates of the errors of (normalized) autocorrelation
functions (which are errors of errors), including the case of
secondary observables, can, e.g., be
obtained following the procedure detailed in Appendix C of
ref.~\cite{Madras:1988ei}, see also refs.~\cite{Wolff:2003sm,Ramos:2018vgu}.

Next we define the normalized autocorrelation function,\footnote{
    This equation is justified by the fact that
    the HMC algorithm is ultra-local in the
    Monte Carlo time, i.e.\ only the
    configuration at the time $t$ is used to obtain the $t+1$ result.
    Note that our convention differs for $A\neq B$ from
    $\rho_{AB}(t)=\Gamma_{AB}(t)/\Gamma_{AB}(0)$, which is sometimes found in the
    literature.}
\begin{equation}
\label{eq:expand}
\rho_{AB}(t)=\frac{\Gamma_{AB}(t)}{\sqrt{\Gamma_{AA}(0)\Gamma_{BB}(0)}}=\sum_k c_{A,k}c_{B,k}e^{-|t|/\tau_k},
\end{equation}
where $\tau_k$ is the autocorrelation time of the mode $k$
of the system
and the largest exponential autocorrelation time
is defined as $\tau_{\exp}=\max_k\{\tau_k\}$.
The coefficients $c_{A,k}$ and $c_{B,k}$
that describe the couplings of the observables
$A$ and $B$ to the mode $k$ are normalized: $\sum_k c_{A,k}^2=\sum_k
c_{B,k}^2=1$.

Extending the autocorrelation function to infinite times,
the integrated autocorrelation time is defined as
\begin{align}\nonumber
\tau_{AB,\text{int}}&=\frac12\int_{-\infty}^{\infty}\!{\dd t}\,\rho_{AB}(t)
=\int_0^{\infty}\!{\dd t}\,\frac12\left[\rho_{AB}(t)+\rho_{BA}(t)\right]
\\&=\sum_k c_{A,k}c_{B,k}\int_0^{\infty}\!{\dd t}\,
e^{-t/\tau_k}=\sum_k c_{A,k}c_{B,k}\tau_k\,:
\label{eq:tau}
\end{align}
It corresponds to the weighted average of autocorrelation times
of modes contributing both to $A$ and $B$. We remark that
for $A\neq B$ in principle $\tau_{AB,\text{int}}$ can be negative.
The Cauchy--Schwarz inequality implies that
\begin{equation}
\label{eq:ineq}
\tau_{AB,\text{int}}^2\leq\tau_{A,\text{int}}\tau_{B,\text{int}},
\end{equation}
where $\tau_{A,\text{int}}=\tau_{{AA},\text{int}}$ is the
integrated autocorrelation time of the observable $A$,
needed to compute the scale factor $s_A\approx 2\tau_{A,\text{int}}$
of eqs.~\eqref{eq:scale} and~\eqref{eq:rescale}.

The variance of $A$ can be obtained from $\Gamma_{AA}(t)$:
\begin{align}
\text{var}(A)&=
\int_{-\infty}^{\infty}\!\!\!{\dd t}\,\Gamma_{AA}(t)
=\left\langle{\left(A-\langle A\rangle\right)}^2\right\rangle\,\int_{-\infty}^{\infty}\!\!
{\dd t}\,\rho_{AA}(t)\nonumber\\
&=\left(2\tau_{A,\text{int}}\right)(N-1)\sigma_{A}^2[1],
\end{align}
where the argument ``$[1]$'' denotes the bin size $S=1$ (no binning) and
\begin{align}
\sigma_{A}^2[1]=\frac{1}{N-1}\left\langle{\left(A-\langle A\rangle\right)}^2\right\rangle=C_{AA}[1]
\end{align}
is the naive squared standard deviation, ignoring autocorrelation effects.
In appendix~\ref{sec:boot} we referred to it as
$\Delta A^2$.

It is well known that autocorrelation times can also be estimated
employing a binning procedure: we define $J$ bins of length
$S=\lfloor N/J\rfloor$ with $1\leq S\ll N$ and
we will only average over the first
$N'=JS$ measurements. We define $\delta A_n=A_n-\langle A\rangle$ and
compute the standard deviation
on the binned ensemble:
\begin{align}
\sigma^2_A[S]&=\frac{1}{J(J-1)}\sum_{j=0}^{J-1}
{\left(\frac{1}{S}\sum_{s=1}^S\delta A_{jS+s}\right)}^2\nonumber\\
&=\frac{1}{N'(N'-S)}\sum_{j=0}^{J-1}\sum_{s,t=1}^S\delta A_{jS+s}\delta A_{jS+t}
\approx\frac{(N-1)\sigma^2_A[1]}{S(N'-S)}\,2\int_{0}^S\!\!\!{\dd s}\,\int_s^S\!\!\!{\dd t}\,\rho_{AA}(t-s)\nonumber\\
&=\frac{2(N-1)\sigma^2_A[1]}{N'-S}\sum_k c_{A,k}^2\tau_k\left[1-\frac{\tau_k}{S}\left(1-e^{-S/\tau_k}\right)\right]\nonumber\\
&=\frac{(N-1)\sigma^2_A[1]}{N'-S}\,2\tau_{A,\text{int}}\left[1-
  \sum_kc_{A.k}^2\frac{\tau_k}{\tau_{A,\text{int}}}\frac{\tau_k}{S}\left(1-e^{-S/\tau_k}\right)\right].
\label{eq:binn}
\end{align}
In the approximate step\footnote{This approximation is valid for
  $S, \tau_k\gg 1$.
  The most general case can be obtained by employing incomplete geometric sums:
  $\sum_{t={s+1}}^S e^{-t/\tau}=(e^{-s/\tau}-e^{-S/\tau})/(e^{1/\tau}-1)\approx
  \tau(e^{-s/\tau}-e^{-S/\tau})$.}
we replaced the autocorrelation function $\Gamma_{AA}(t)$ by the normalized
autocorrelation function $\rho_{AA}(t)$ times $(N-1)\sigma^2_A[1]$,
whereas in the second last step we employed
eq.~\eqref{eq:expand}. Using
$N-1\approx N'-S$ gives the leading order expectation
\begin{equation}
  \frac{\sigma_A^2[S]}{\sigma_A^2[1]}\approx 2\tau_{A,\text{int}}\left(1-\frac{c_A}{S}\right),
  \label{eq:binnn}
\end{equation}
where $c_A=\sum_k c_{A,k}^2\tau_k^2/\tau_{\text{int}}\geq 0$.
Therefore, up to $1/S$ corrections in the bin length, the ratio of
binned over unbinned squared standard deviations
approaches twice the integrated autocorrelation time.
This suggests, in a first step, to compute this ratio in a
self-consistent way, e.g., at $S_0\approx 2\tau_{\text{int}}$ and then to
estimate the remaining ratio $\sigma_A^2[\infty]/\sigma_A^2[S_0]$
via a fit to the data obtained with
bin sizes $S\geq S_0$
We remark that for $\tau=\tau_k=\tau_{\text{int}}$ one can write:
\begin{equation}
  \frac{\sigma^2[S]}{\sigma^2[1]}\approx 2\tau\left[1-\frac{\tau}{S}
    \left(1-e^{-S/\tau}\right)\right].
  \label{eq:binnn2}
\end{equation}
Taking into account the exponential corrections to eq.~\eqref{eq:binnn}, we obtain
\begin{equation}
  \frac{\sigma_A^2[S]}{\sigma_A^2[1]}\approx 2\tau_{A,\text{int}}\left[1-\frac{c_A}{S}+\frac{1}{S\tau_{A,\text{int}}}\sum_k \left(c_{A,k}\tau_k\right)^2e^{-S/\tau_k}\right],
\end{equation}
which suggests the three-parameter fit
\begin{equation}
  \frac{\sigma_A^2[S]}{\sigma_A^2[1]} \approx 2\tau_{A,\text{int}}\left(1-\frac{c_A}{S}+\frac{d_A}{S}e^{-S/\tau_{A,\text{int}}}\right),
  \label{eq:binnn3}
\end{equation}
where $c_A> d_A\geq 0$. The rationale for this parametrization
is that on the one
hand large values of $\tau_{A,\text{int}}$ imply large couplings to slow modes,
in which case one may be unable to resolve the exponential correction to the
coefficient of the leading $1/S$ decay. In this case
eq.~\eqref{eq:binnn} should be adequate, unless of course only one mode
dominates, see eq.~\eqref{eq:binnn2}.
If, however, $\tau_{A,\text{int}}$ is dominated by the faster modes
then effectively replacing the smaller $\tau_k$-values by $\tau_{A,\text{int}}$
in eq.~\eqref{eq:binnn3},
thereby reducing the number of fit parameters, should be a sensible
approximation.

The above equations also hold for jackknifed primary data and therefore,
within the jackknife approach, generalize identically to secondary observables
such as hadron masses. Therefore, this method can
directly be used to determine the corresponding integrated autocorrelation
times from binned jackknife estimates.
We remark that the $\Gamma$-method of ref.~\cite{Wolff:2003sm}
has the advantage of exponential as opposed to power-law corrections.
However, binning provides smaller errors at a given value
of $S$ and is more robust in the case of limited statistics.
For a recent discussion of the $\Gamma$-method
and its application to secondary observables, see ref.~\cite{Ramos:2018vgu}.
As pointed out above, the generalization to secondary observables is trivial.

Assuming a normal distribution of independent measurements,
i.e.\ $S\gtrsim 2\tau_{\text{int}}$, the error on $\sigma_A^2[S]$, that is
needed to fit the $1/S$ tail,
can be estimated as~\cite{kenney}
\begin{equation}
\Delta\left(\sigma_A^2[S]\right)=\sqrt{\frac{2(J-1)}{J^2}}\sigma_A^2[S],
\end{equation}
where $J=N'/S\approx J-1$ is the number of bins.

The same binning can be performed for off-diagonal elements of the
covariance matrix. We define unbinned covariance matrix elements
$C_{AB}[1]$, see eq.~\eqref{eq:covar}
for the jackknife definition of the covariance matrix, where $A$ and $B$
can also be secondary data.
For observables that are simple ensemble averages the relation
$C_{AB}[1]=\Gamma_{AB}(t=0)/(N-1)=\rho_{AB}(0)\sigma_A[1]\sigma_B[1]$
holds. The autocorrelation time-corrected
covariance matrix is given as $C_{AB}[\infty]$, where
$C_{AB}[S]$ denotes the covariance matrix, computed for the
bin size $S$. The resulting expression reads:
\begin{equation}
C_{AB}[S]\approx 2\tau_{AB,\text{int}}\sigma_A[1]\sigma_B[1]\left(1-\frac{c_{AB}}{S}+
  \sum_kc_{A.k}c_{B,k}\frac{\tau_k}{\tau_{AB,\text{int}}}\frac{\tau_k}{S}e^{-S/\tau_k}\right),\label{eq:covaut}
\end{equation}
where, for $A\neq B$,
$c_{AB}$ is not necessarily semi-positive and $\tau_{AB,\text{int}}$ may be
negative.

Naively reconstructing the autocorrelation time-corrected
covariance matrix from the rescaled
bootstrap data eq.~\eqref{eq:rescale} would amount to
assuming $\tau_{AB,\text{int}}>0$ and
$\tau^2_{AB,\text{int}}C_{AB}[1]
=\tau_{A,\text{int}}\tau_{B,\text{int}}\sigma_A[1]\sigma_B[1]$,
which in general does not hold. Therefore, to facilitate
correlated fits to secondary data, in addition to the
diagonal integrated autocorrelation times, giving us the
rescaling factors, we need to estimate
the off-diagonal elements of $C[\infty]$, using eq.~\eqref{eq:covaut}.
Combining the inequality~\eqref{eq:ineq} with
$C_{AB}^2[1]\leq\sigma_A^2[1]\sigma_B^2[1]$
is consistent with the necessary condition
$C_{AB}^2[\infty]\leq\sigma_A^2[\infty]\sigma_B^2[\infty]$, which we
take as an upper limit for our extrapolation.
In the end we store the normalized covariance matrices
$\overline{C}_{jk}=C_{jk}[\infty]/(\sigma_{f_j}[\infty]\sigma_{f_j}[\infty])$
for the secondary data $\{f_j\}$. Since we also keep the
binned jackknife distributions, this matrix can be enlarged to incorporate
further observables at a later stage.

\begin{figure}[htp]
  \centering
  \resizebox{0.32\textwidth}{!}{\includegraphics[width=\textwidth]{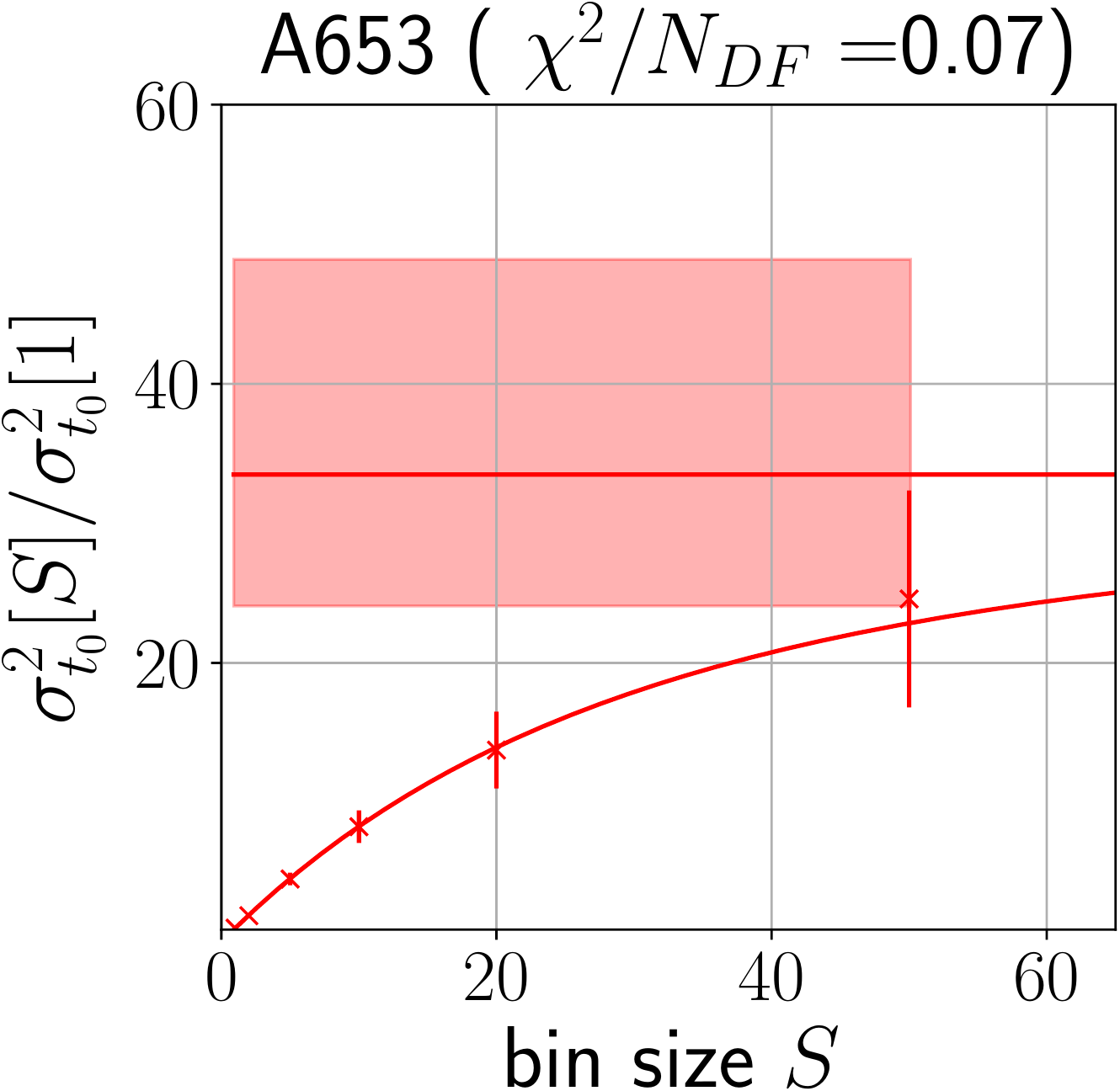}}
  \resizebox{0.32\textwidth}{!}{\includegraphics[width=\textwidth]{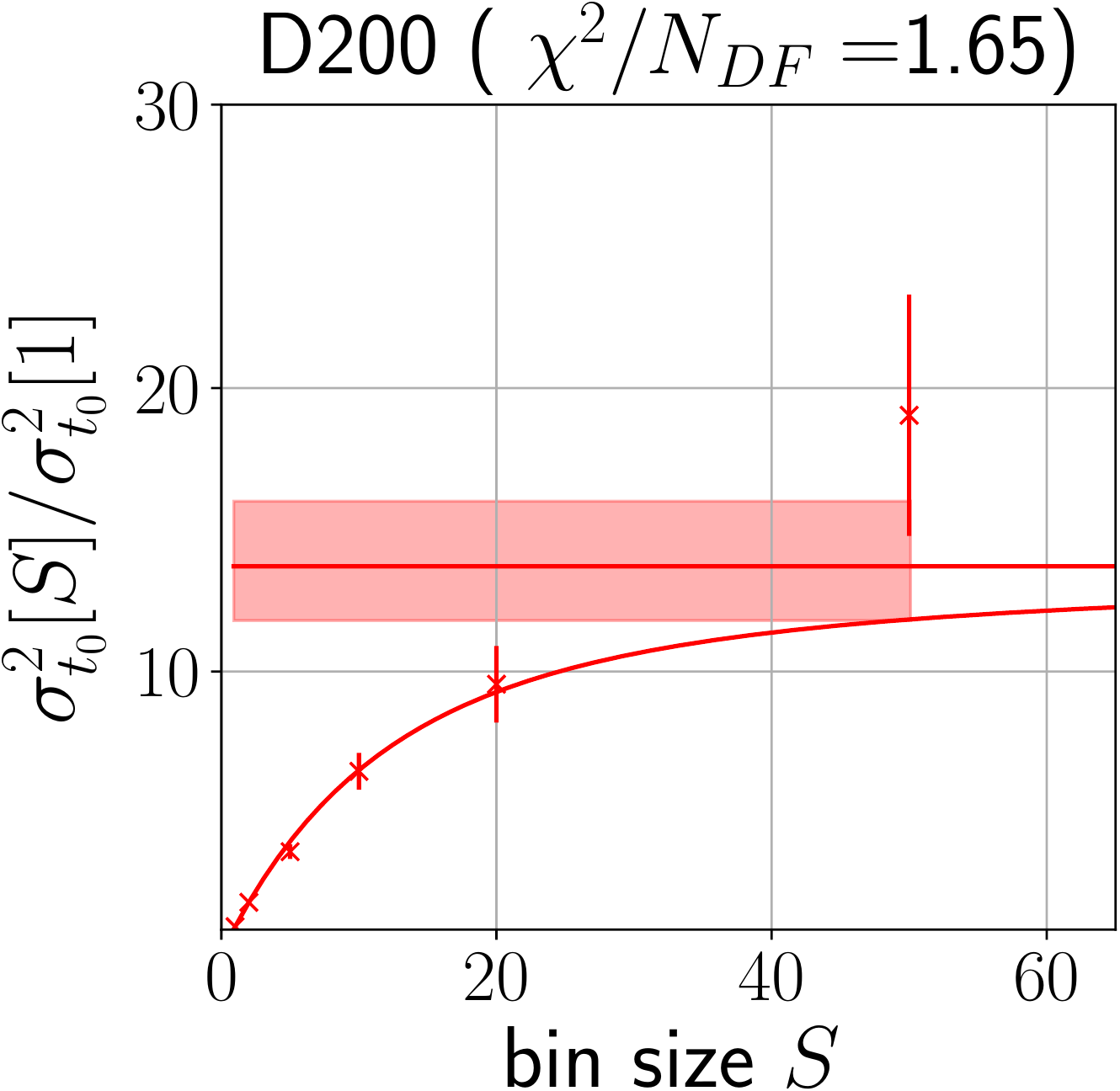}}
  \resizebox{0.32\textwidth}{!}{\includegraphics[width=\textwidth]{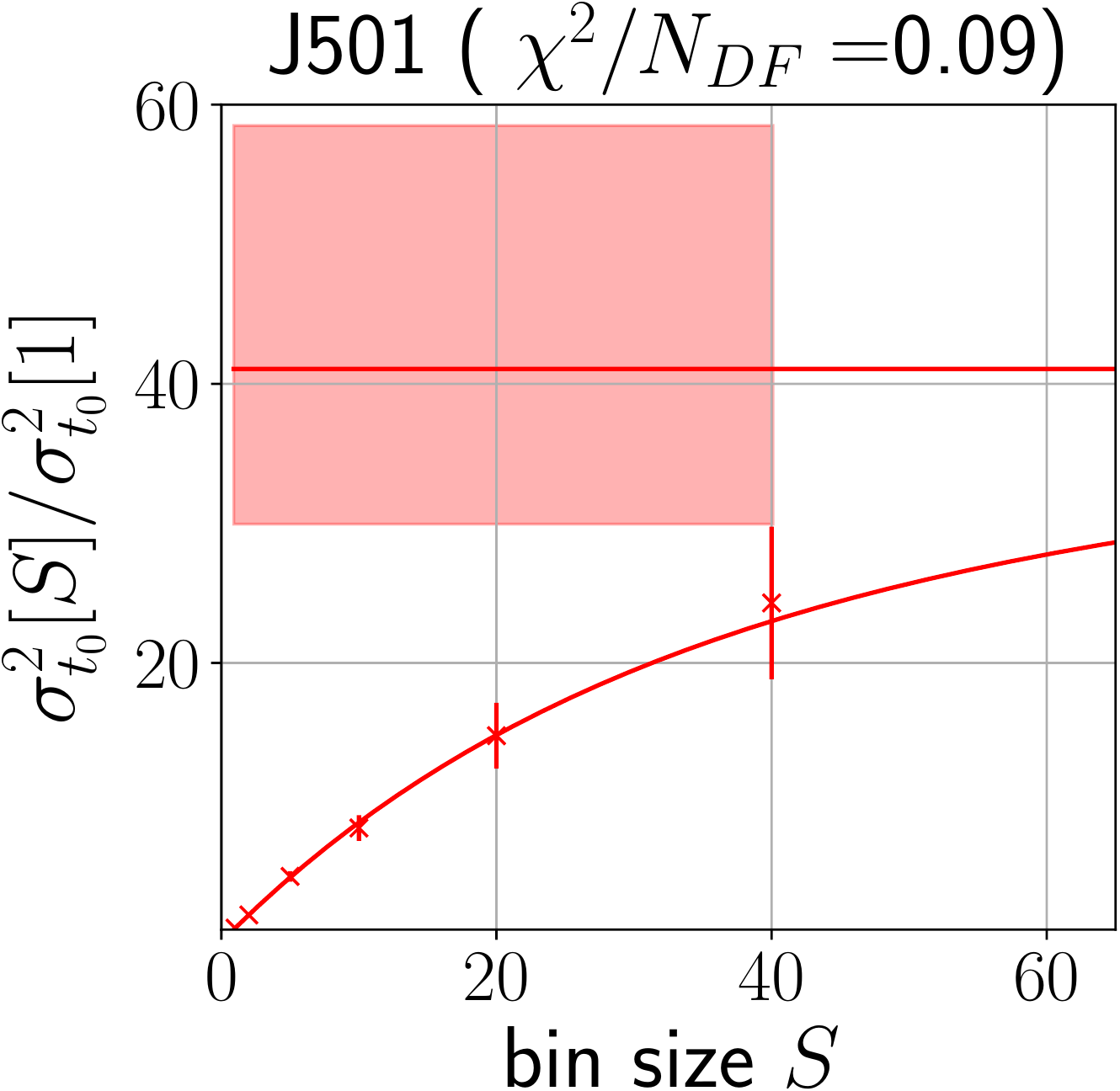}}
  \caption{Bin size extrapolation of the variance of $t_0/a^2$ according
    to the one-parameter fit eq.~\eqref{eq:binnn2} for the ensembles
    A653 (left), D200 (centre) and J501 (right). The bin size $S$ is in units
    of 4~MDU. The error bands indicate the extrapolated values.
    Note that the central plot has a different scale on the $y$-axis than the
    other two panels. The respective simulation parameters can be found in
    tables~\ref{tab:parameters},~\ref{tab:physparams} and~\ref{tab:paramsNewSim}
    (for D200 in table~2 of ref.~\cite{Bruno:2014jqa}). An overview in terms
    of the pion masses and lattice spacings is also
    provided in the middle panel of figure~\ref{fig:ensembles}.
    \label{fig:auto1}}
\end{figure}

\subsection{Examples of infinite bin size extrapolations}
\label{sec:autocorr}
We show some examples for the extrapolation discussed above
of statistical errors to infinite bin sizes, starting with $t_0/a^2$.
In the case of this observable our data, including the
smallest bin size $S=1$, are well-described by
the one-parameter fit eq.~\eqref{eq:binnn2}, indicating that
this quantity basically only couples to one --- possibly the slowest ---
mode of the system. In figure~\ref{fig:auto1} we show the
extrapolation for three representative examples:
A653, at the coarsest lattice spacing
($\beta=3.34$, $a\approx 0.098\,\textmd{fm}$)
and a large pion mass
$M_{\pi}\approx 429\,\textmd{MeV}$,  D200, a finer lattice
($\beta=3.55$, $a\approx 0.064\,\textmd{fm}$) at a small pion mass
$M_{\pi}\approx 202\,\textmd{MeV}$ and J501, at the finest lattice spacing
($\beta=3.85$, $a\approx 0.039\,\textmd{fm}$) at an intermediate pion mass
$M_{\pi}\approx 336\,\textmd{MeV}$. A653 has periodic boundary conditions in
time while the other two ensembles have open boundary conditions.
Note that the data for different bin sizes are highly correlated.
Therefore, the uncorrelated $\chi^2/N_{\text{DF}}$-values are all much smaller
than one.

The integrated autocorrelation times (which in this case
seem to coincide with the exponential autocorrelation times)
in units of four MDUs can be read off the figures by dividing the extrapolated
results (error bands) by a factor of two. The general trend for the
autocorrelation times is to increase with $\beta$ towards smaller
lattice spacings, with the exception of $\beta=3.34$,
which, being about the coarsest lattice
spacing that we can simulate with our action, also suffers from
large autocorrelations. Some time series are depicted in
figure~\ref{fig:wilson}. We list the
extracted autocorrelation times for $t_0$ and their errors in
the last column of table~\ref{tab:parameters} in section~\ref{sec:ensembles}.
In some cases our statistics are insufficient
for a reliable estimation of the error of the autocorrelation time
(indicated in Italics). Note that the ensembles that are labelled as
``rqcd0{\tt mn}'' have been
generated using {\sc BQCD}~\cite{Nakamura:2010qh} instead of
{\sc openQCD}~\cite{Luscher:2012av}.

\begin{figure}[htp]
  \centering
  \resizebox{0.49\textwidth}{!}{\includegraphics[width=\textwidth]{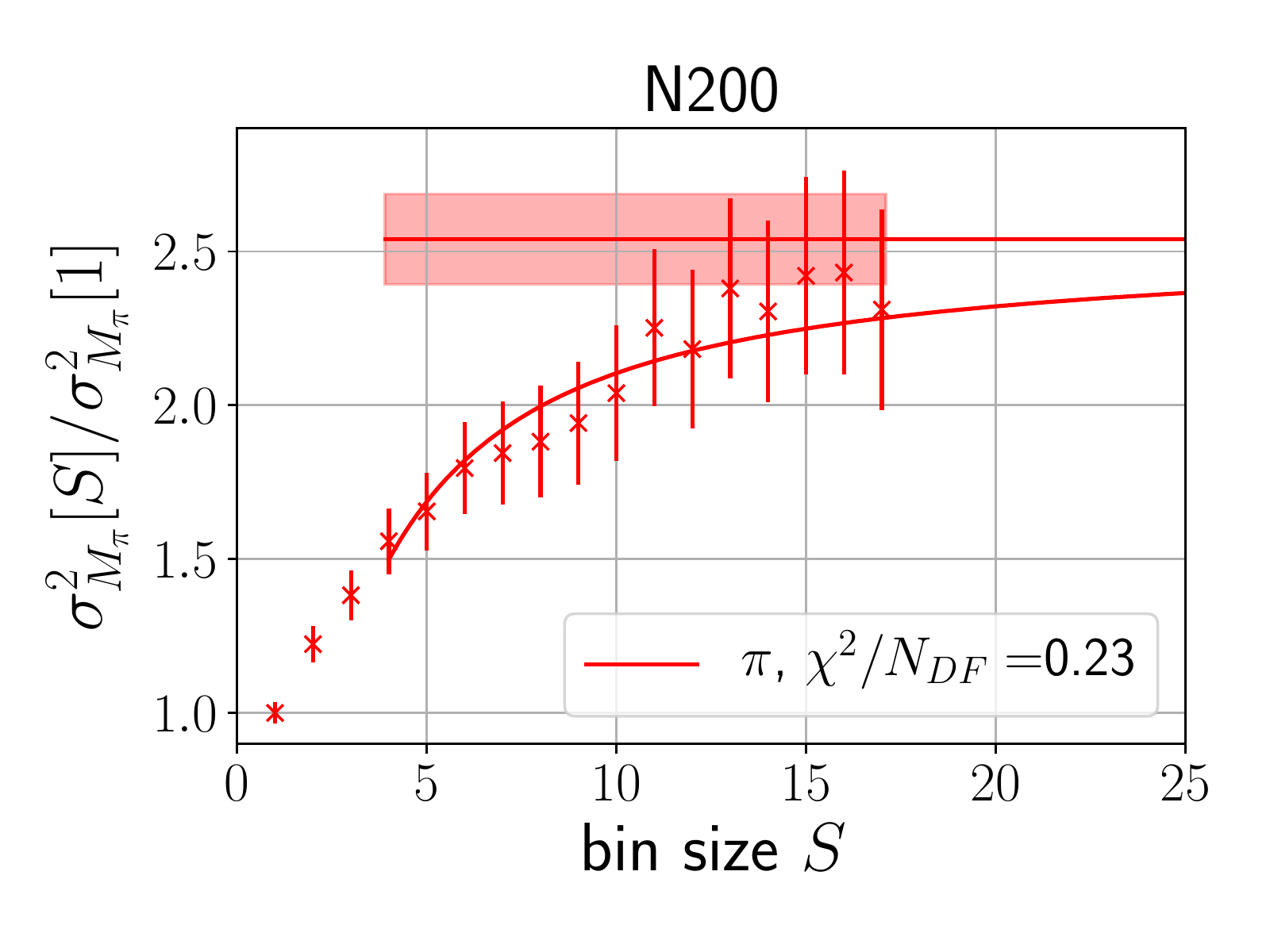}}
  \resizebox{0.49\textwidth}{!}{\includegraphics[width=\textwidth]{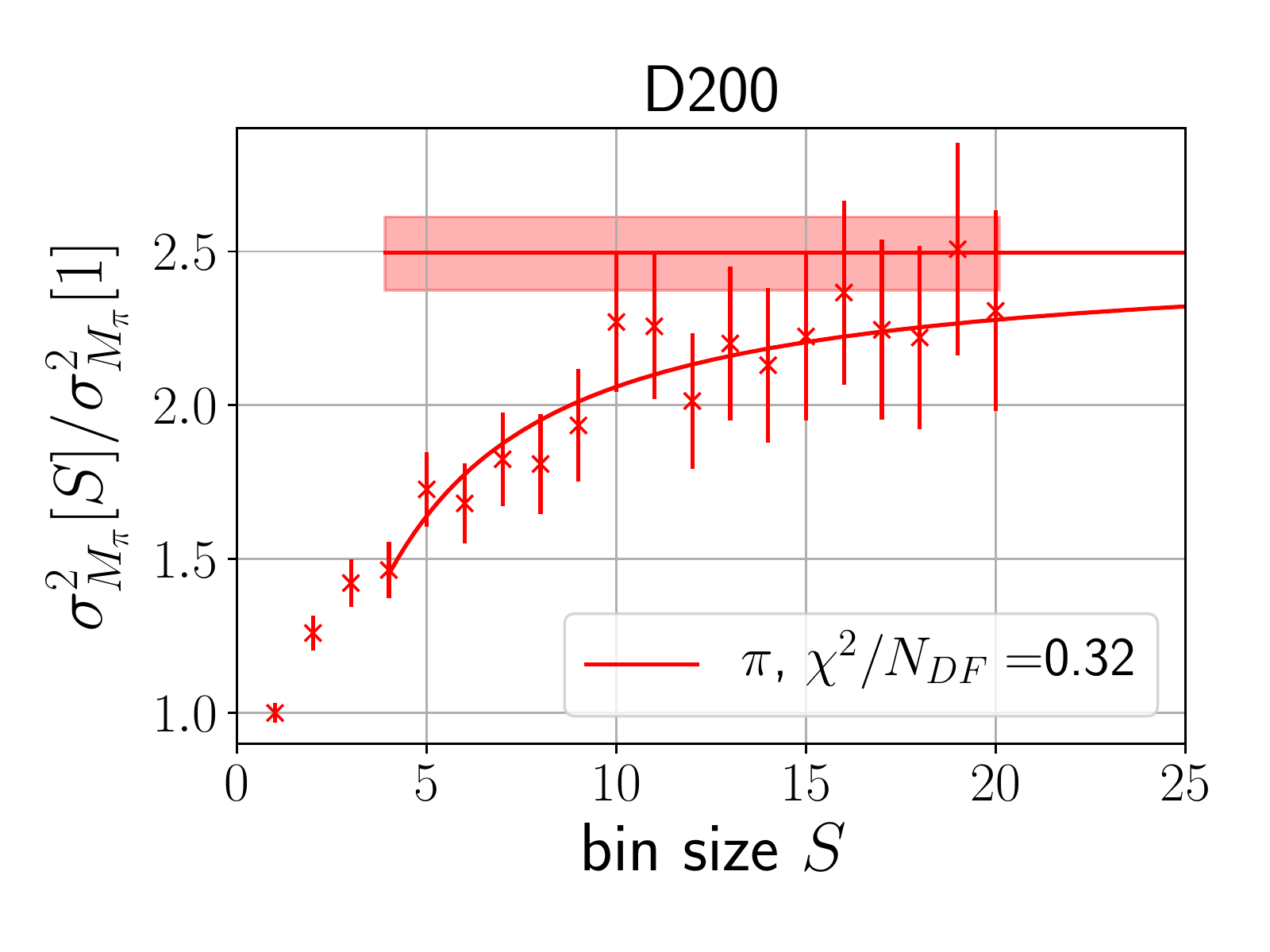}}
  \caption{Bin size extrapolation of the variance of the pion mass
    $M_{\pi}$ according
    to the three-parameter fit eq.~\eqref{eq:binnn3} for the ensembles
    N200 (left, $M_{\pi}\approx
    280\,\textmd{MeV}$) and D200 (right, $M_{\pi}\approx 200\,\textmd{MeV}$),
    both at $a\approx 0.064\,\textmd{fm}$. Successive measurements are separated
    by 4~MDUs and the fit starts at thrice the integrated autocorrelation time.
    The red error band indicates the fit range and the extrapolated value.
    \label{fig:auto2}}
\end{figure}

\begin{figure}[htp]
  \centering
  \resizebox{0.49\textwidth}{!}{\includegraphics[width=\textwidth]{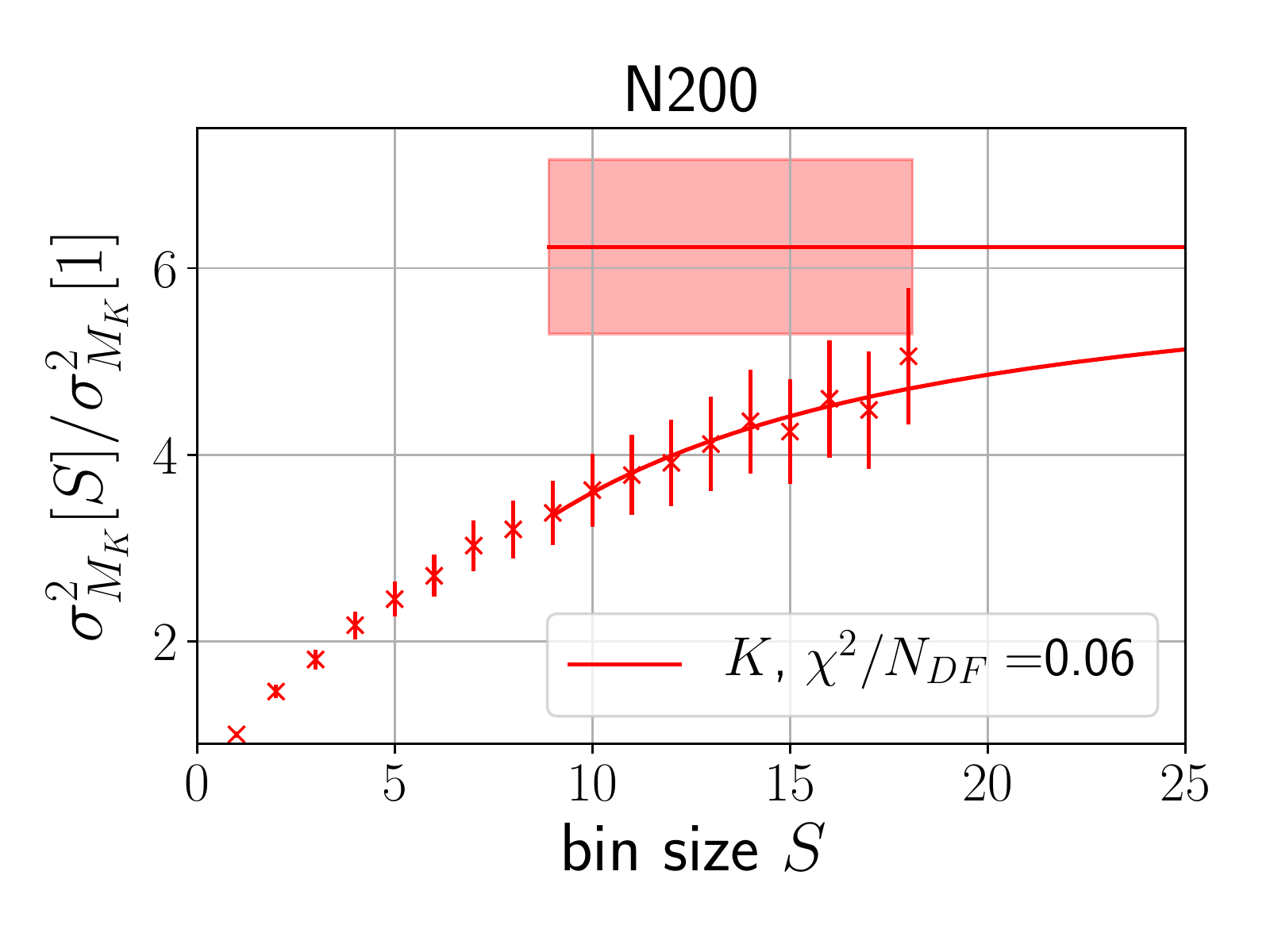}}
  \resizebox{0.49\textwidth}{!}{\includegraphics[width=\textwidth]{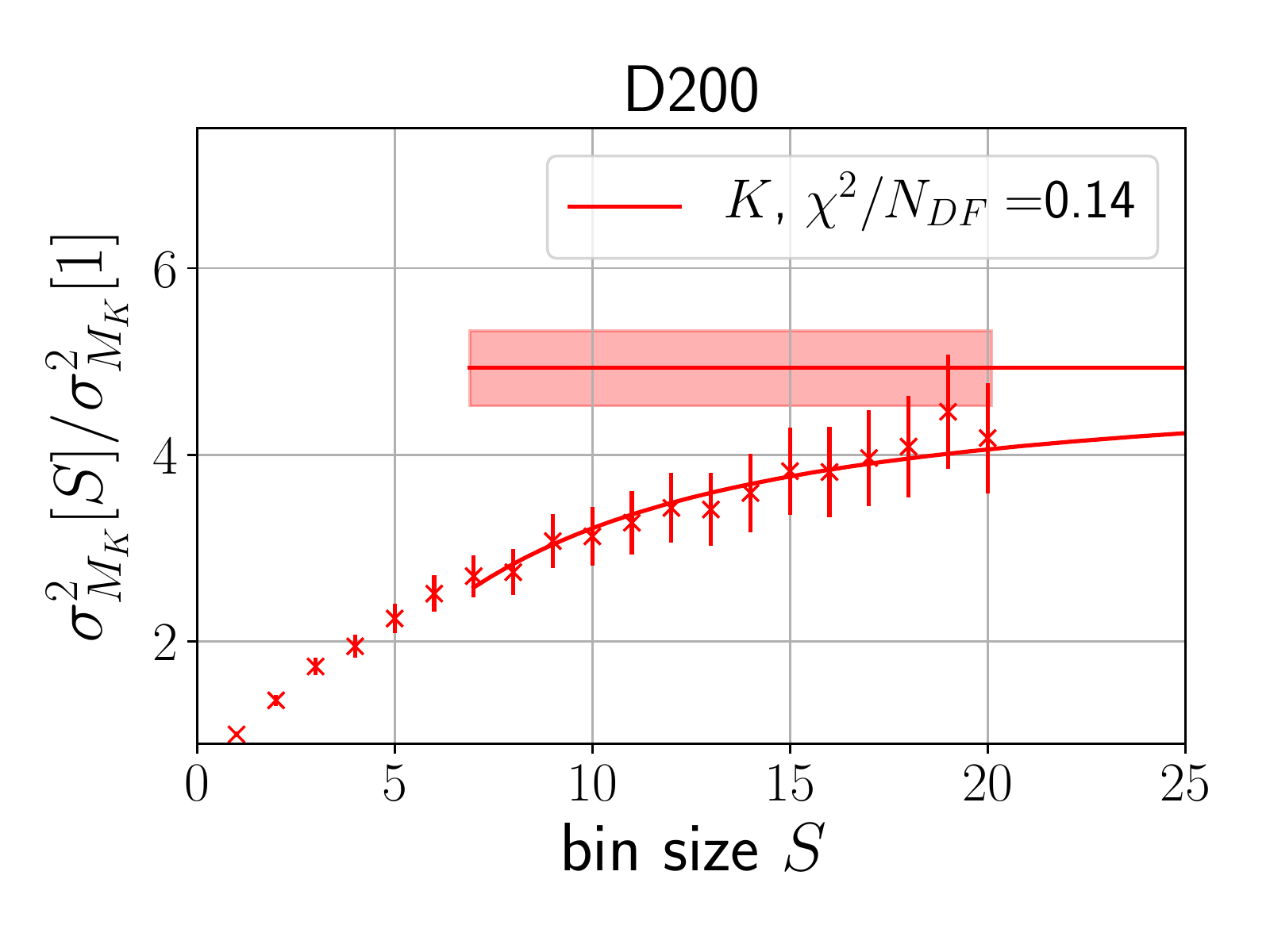}}
  \caption{The same as figure~\ref{fig:auto2} for the kaon mass $M_{K}$.
    \label{fig:auto3}}
\end{figure}

\begin{figure}[htp]
  \centering
  \resizebox{0.49\textwidth}{!}{\includegraphics[width=\textwidth]{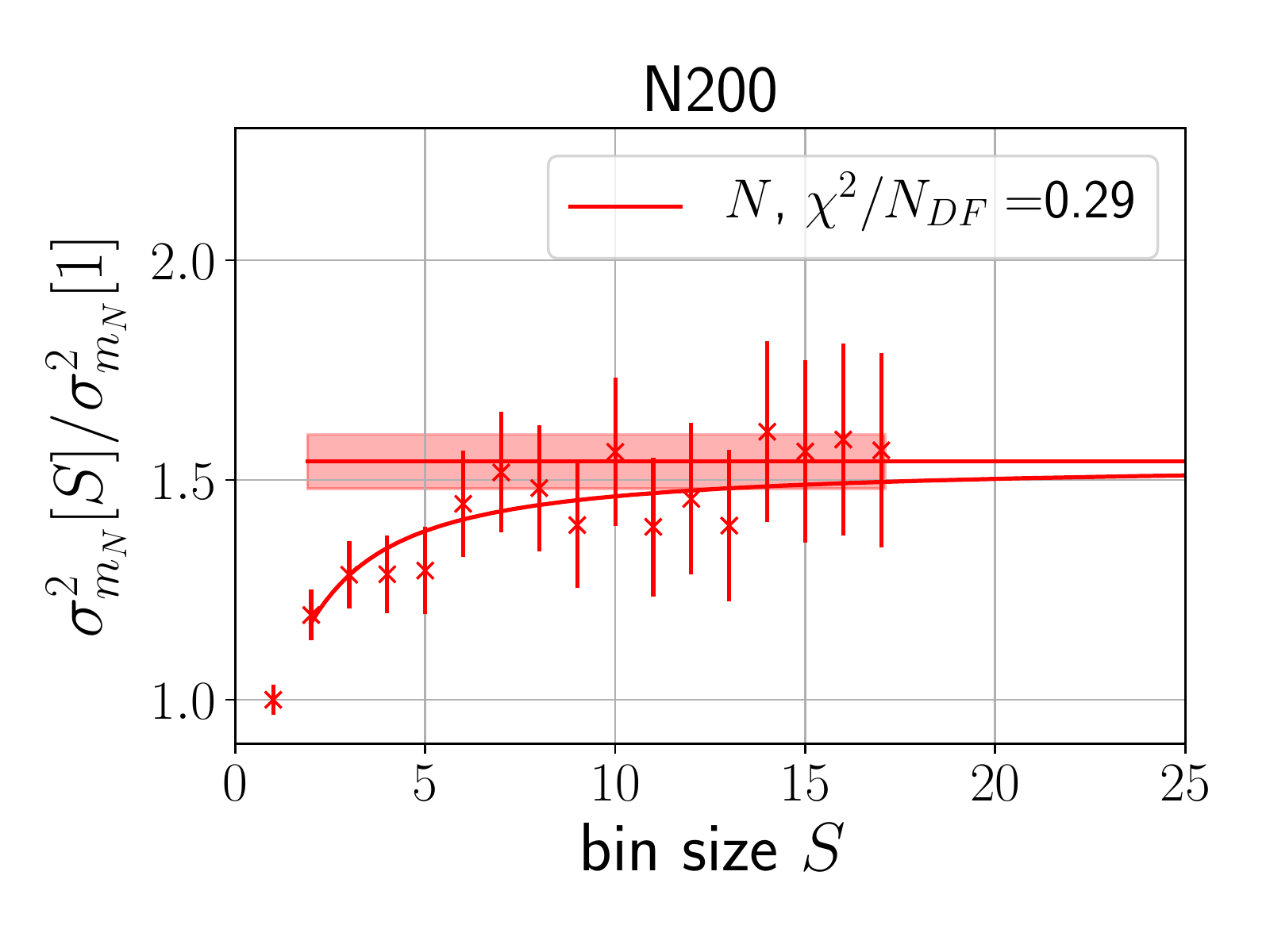}}
  \resizebox{0.49\textwidth}{!}{\includegraphics[width=\textwidth]{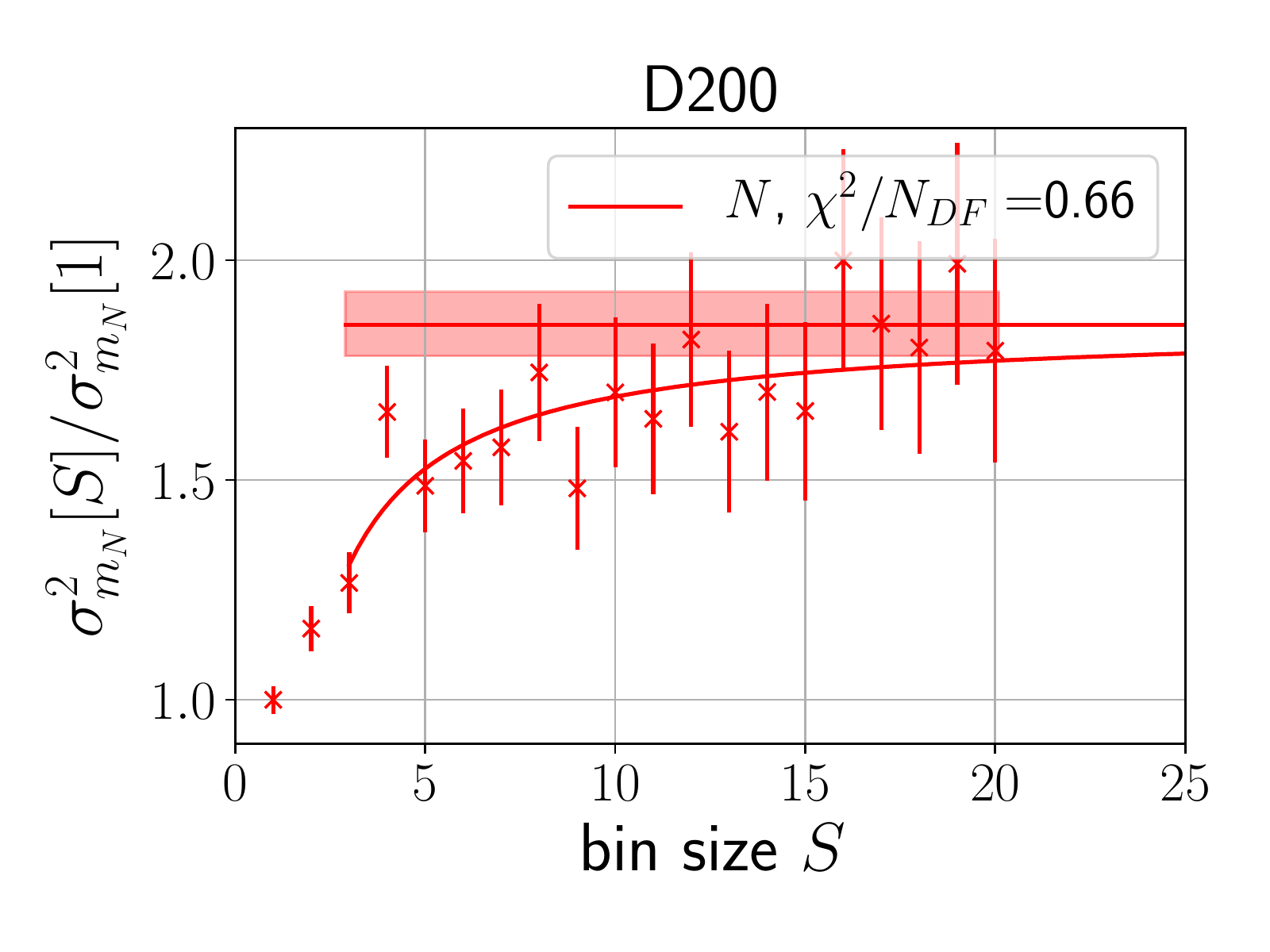}}
  \caption{The same as figure~\ref{fig:auto2} for the nucleon $m_{N}$.
    \label{fig:auto4}}
\end{figure}

\begin{figure}[h]
  \centering
  \resizebox{0.49\textwidth}{!}{\includegraphics[width=\textwidth]{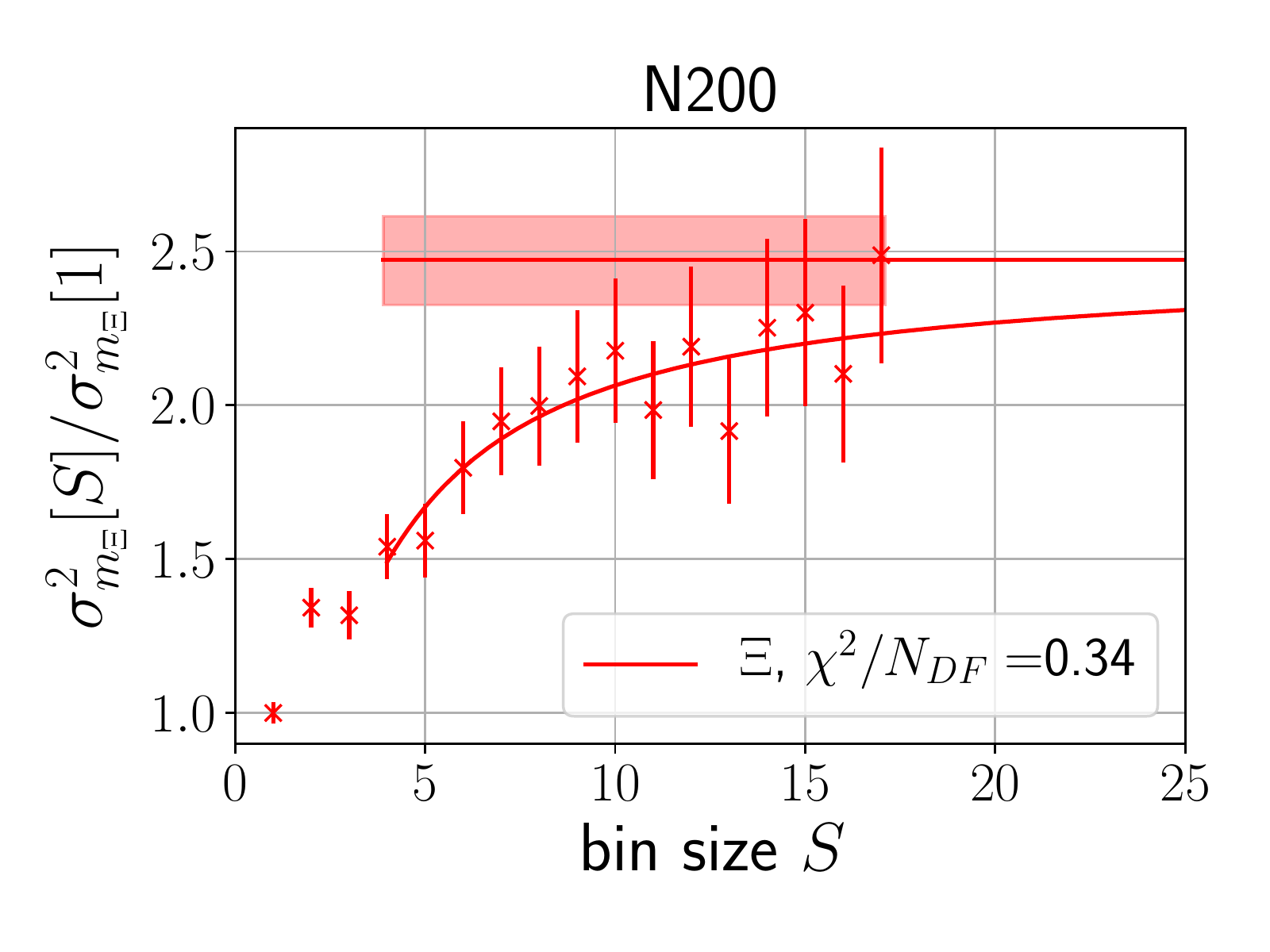}}
  \resizebox{0.49\textwidth}{!}{\includegraphics[width=\textwidth]{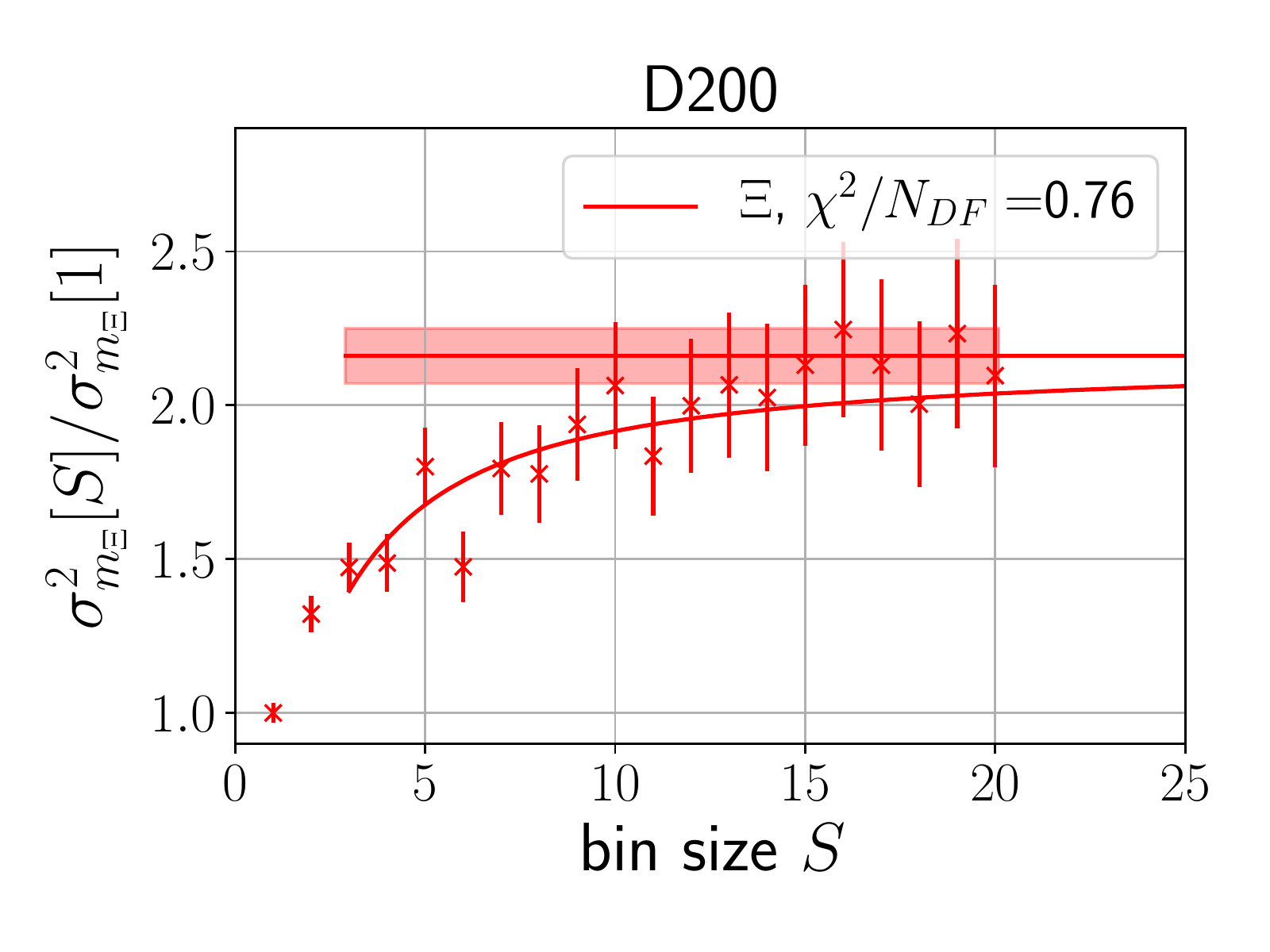}}
  \caption{The same as figure~\ref{fig:auto2} for the cascade mass $m_{\Xi}$.
    \label{fig:auto5}}
\end{figure}

\begin{figure}[htp]
  \centering
  \resizebox{0.49\textwidth}{!}{\includegraphics[width=\textwidth]{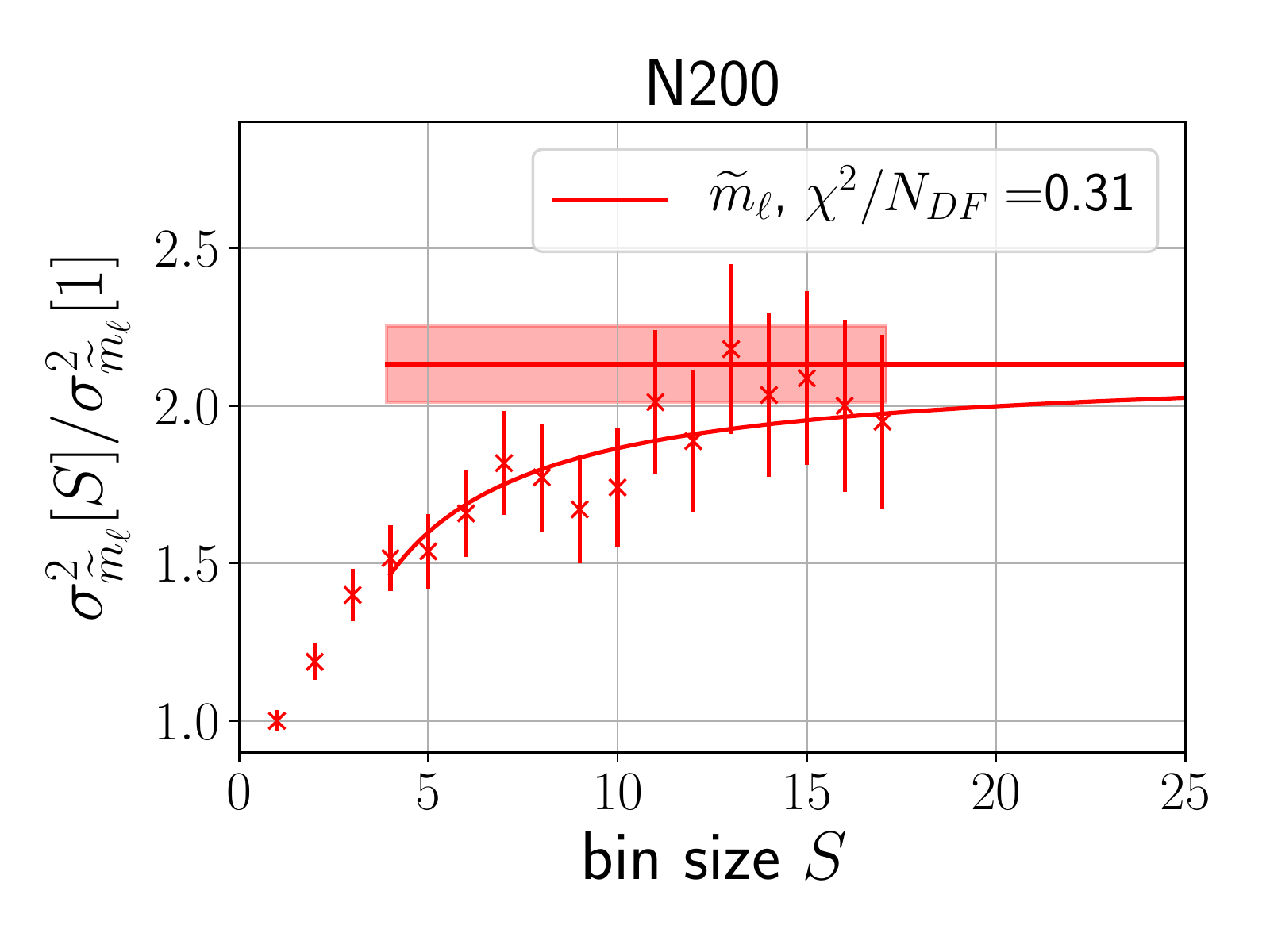}}
  \resizebox{0.49\textwidth}{!}{\includegraphics[width=\textwidth]{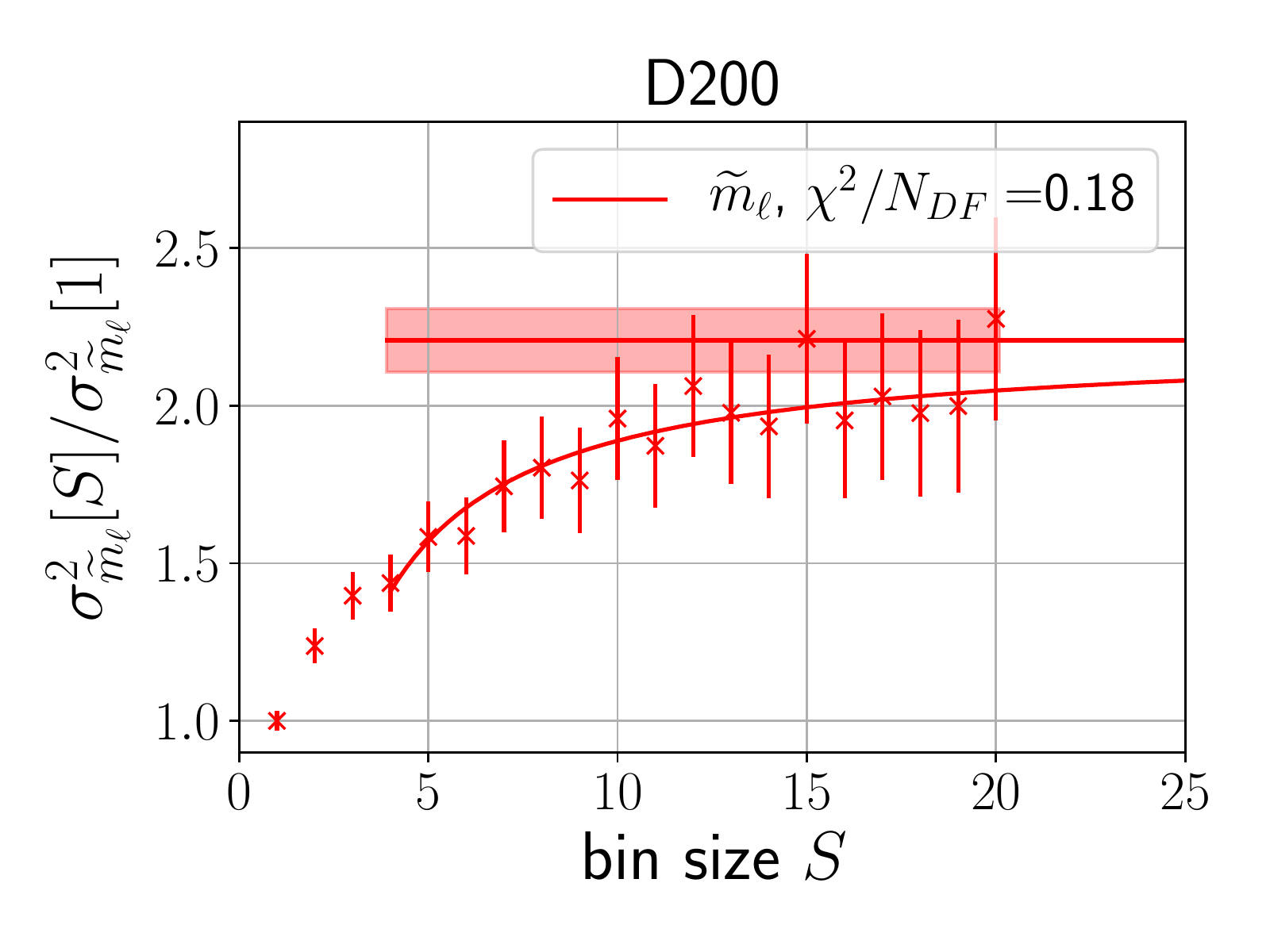}}
  \caption{The same as figure~\ref{fig:auto2} for the light AWI quark mass
    $\widetilde m_{\ell}$. However, the
    two-parameter fit~\eqref{eq:binnn} was employed, starting at
    four times the integrated autocorrelation time.
    \label{fig:auto6}}
\end{figure}

\begin{figure}[htp]
  \centering
  \resizebox{0.49\textwidth}{!}{\includegraphics[width=\textwidth]{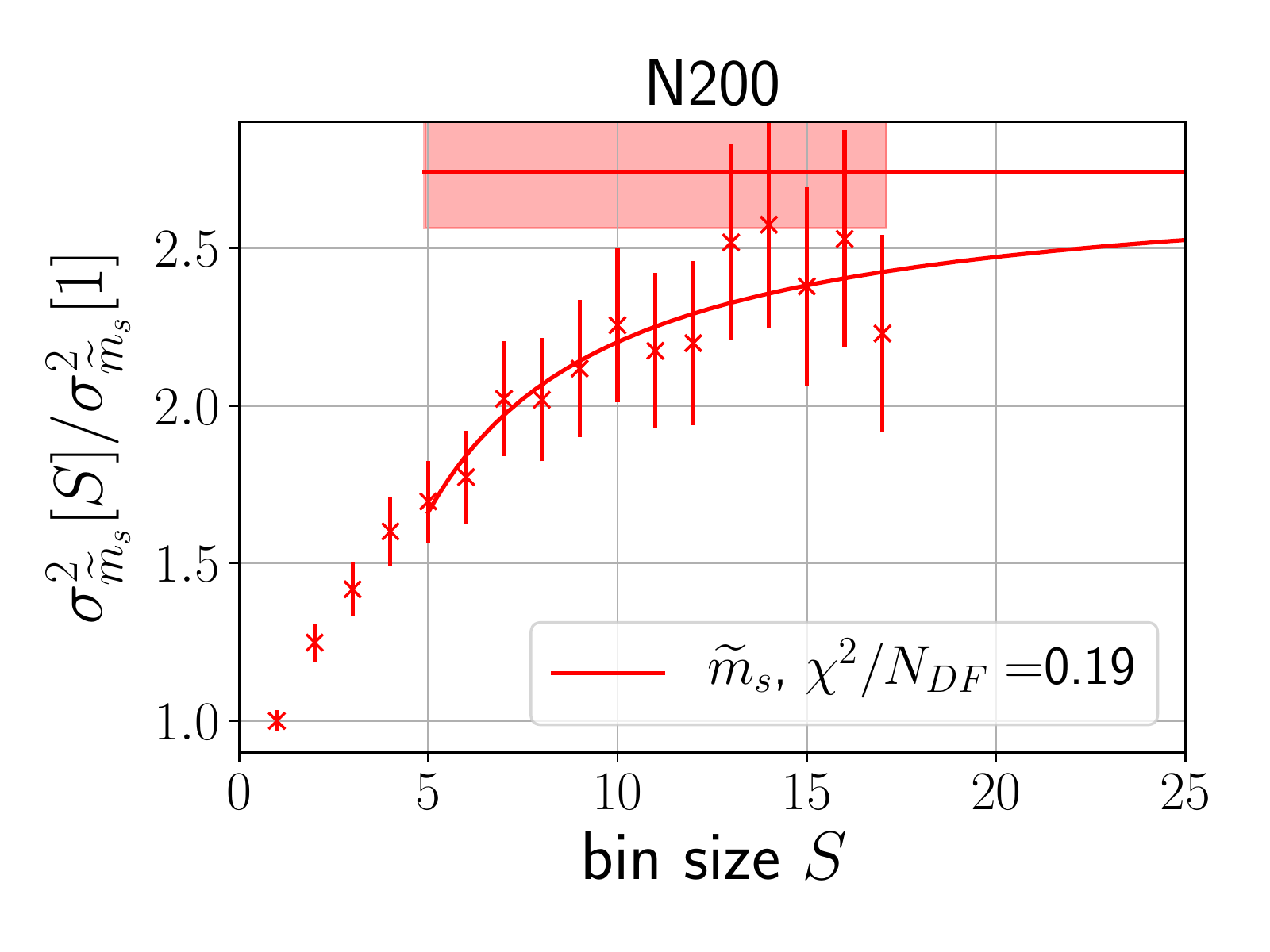}}
  \resizebox{0.49\textwidth}{!}{\includegraphics[width=\textwidth]{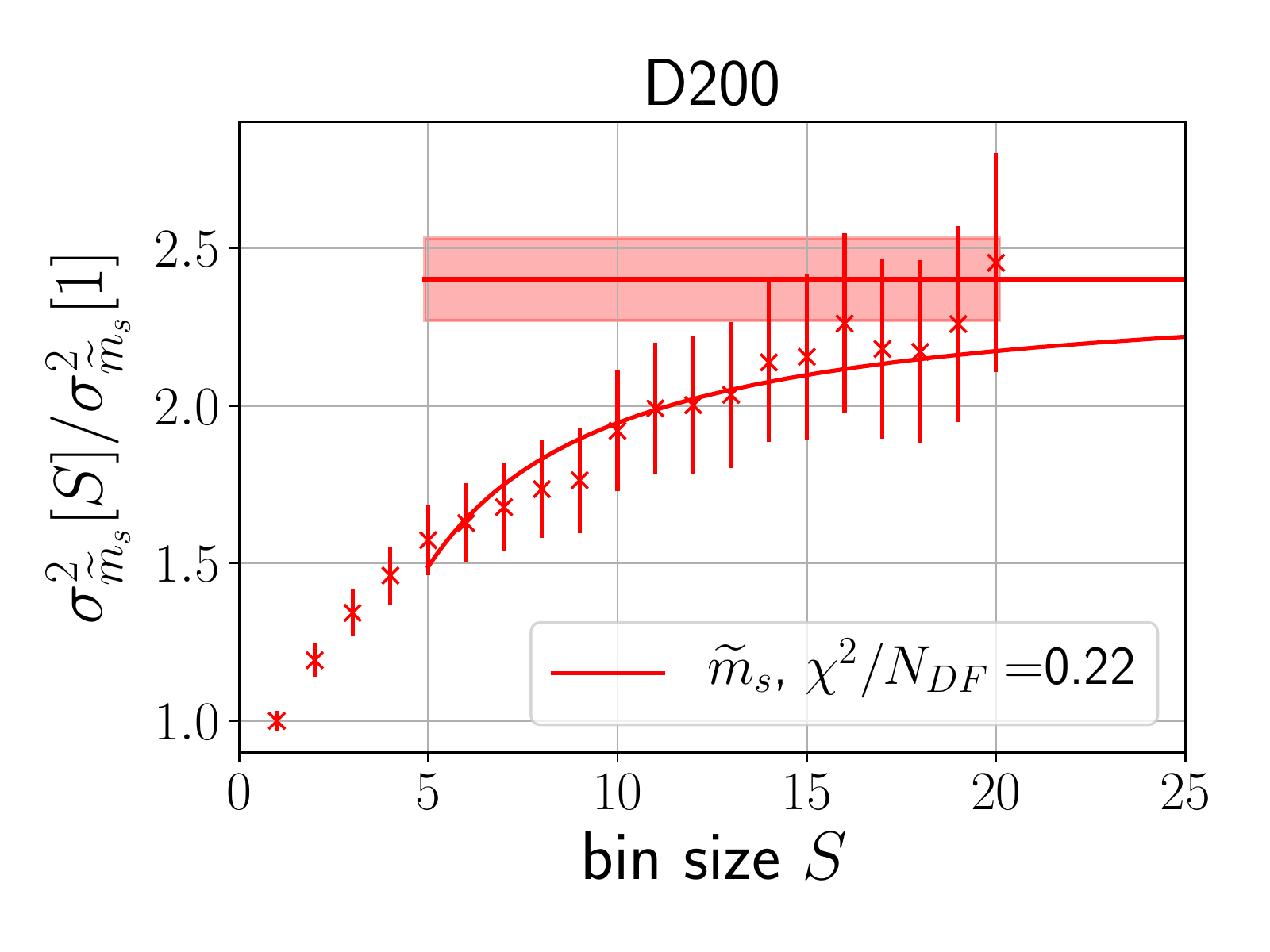}}
  \caption{The same as figure~\ref{fig:auto6} for the strange AWI quark mass.
    \label{fig:auto7}}
\end{figure}
\begin{figure}[htp]
  \centering
  \resizebox{0.49\textwidth}{!}{\includegraphics[width=\textwidth]{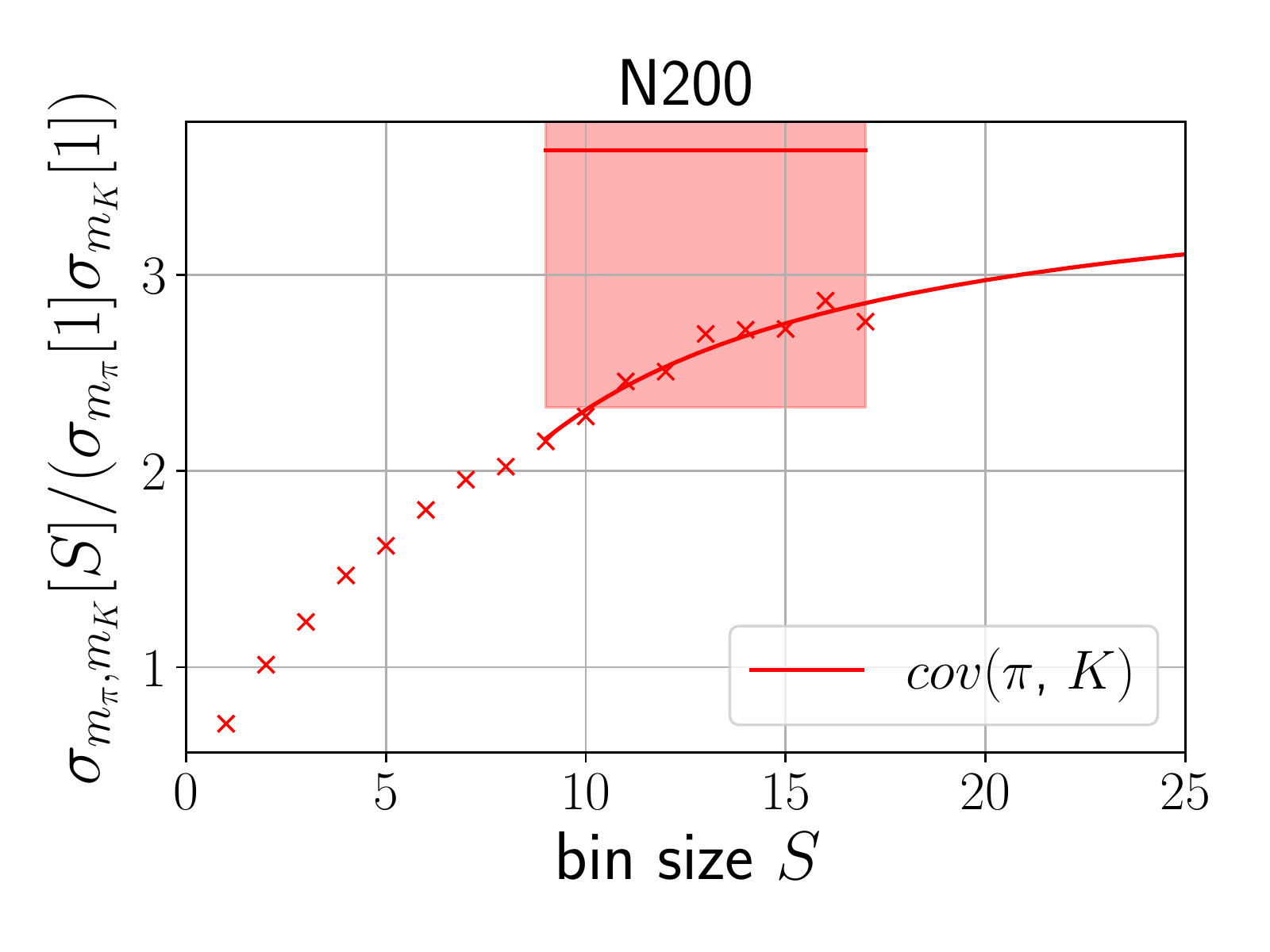}}
  \resizebox{0.49\textwidth}{!}{\includegraphics[width=\textwidth]{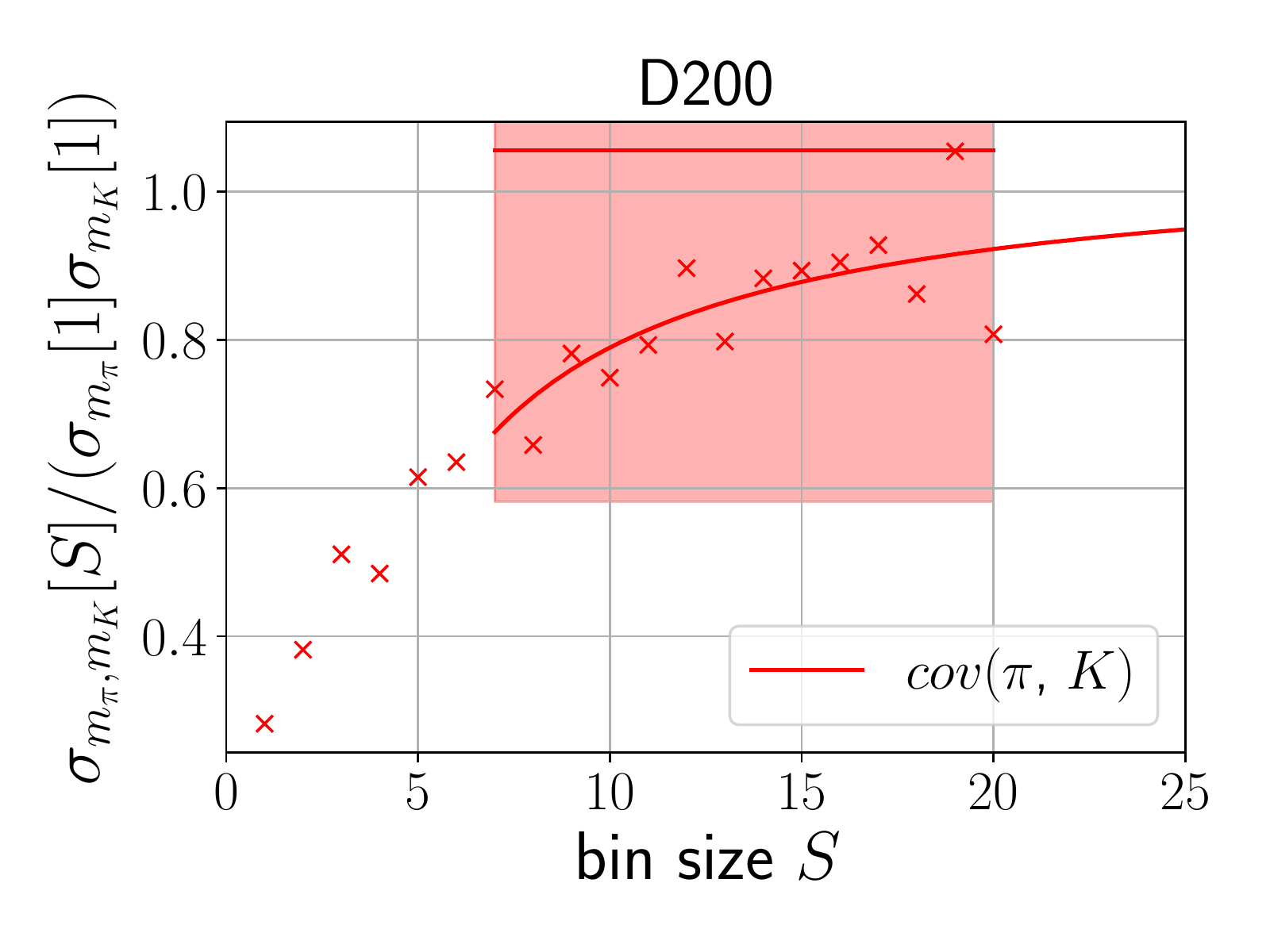}}
  \caption{Bin size extrapolation of the covariance of the pion mass
    $M_{\pi}$ with the kaon mass $M_K$ according
    to the three-parameter fit eq.~\eqref{eq:binnn3} for the ensembles
    N200 (left, $M_{\pi}\approx
    280\,\textmd{MeV}$) and D200 (right, $M_{\pi}\approx 200\,\textmd{MeV}$),
    both at $a\approx 0.064\,\textmd{fm}$.
    Successive measurements are separated
    by 4~MDUs.
    \label{fig:auto8}}
\end{figure}
\begin{figure}[h]
  \centering
  \resizebox{0.49\textwidth}{!}{\includegraphics[width=\textwidth]{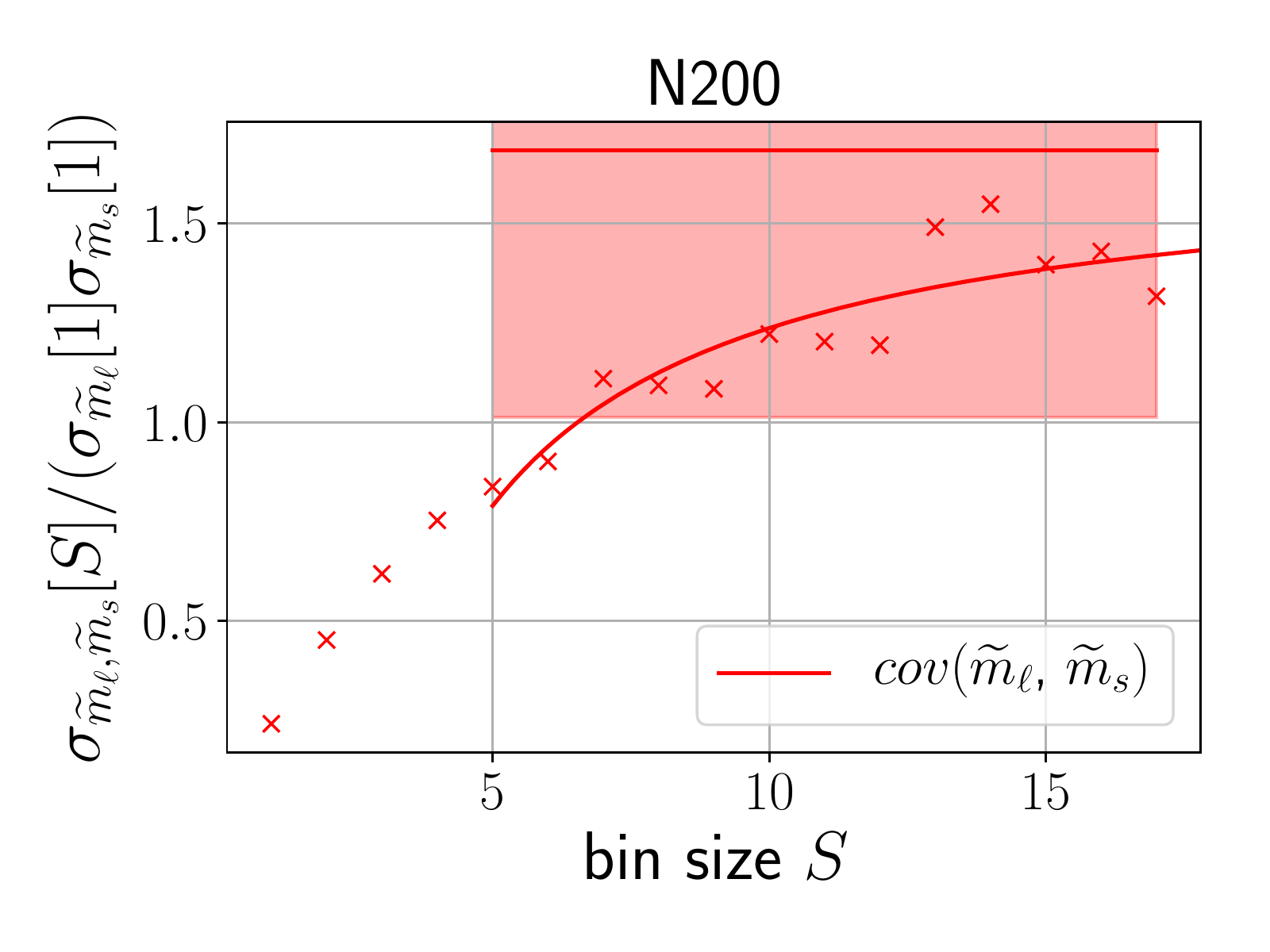}}
  \resizebox{0.49\textwidth}{!}{\includegraphics[width=\textwidth]{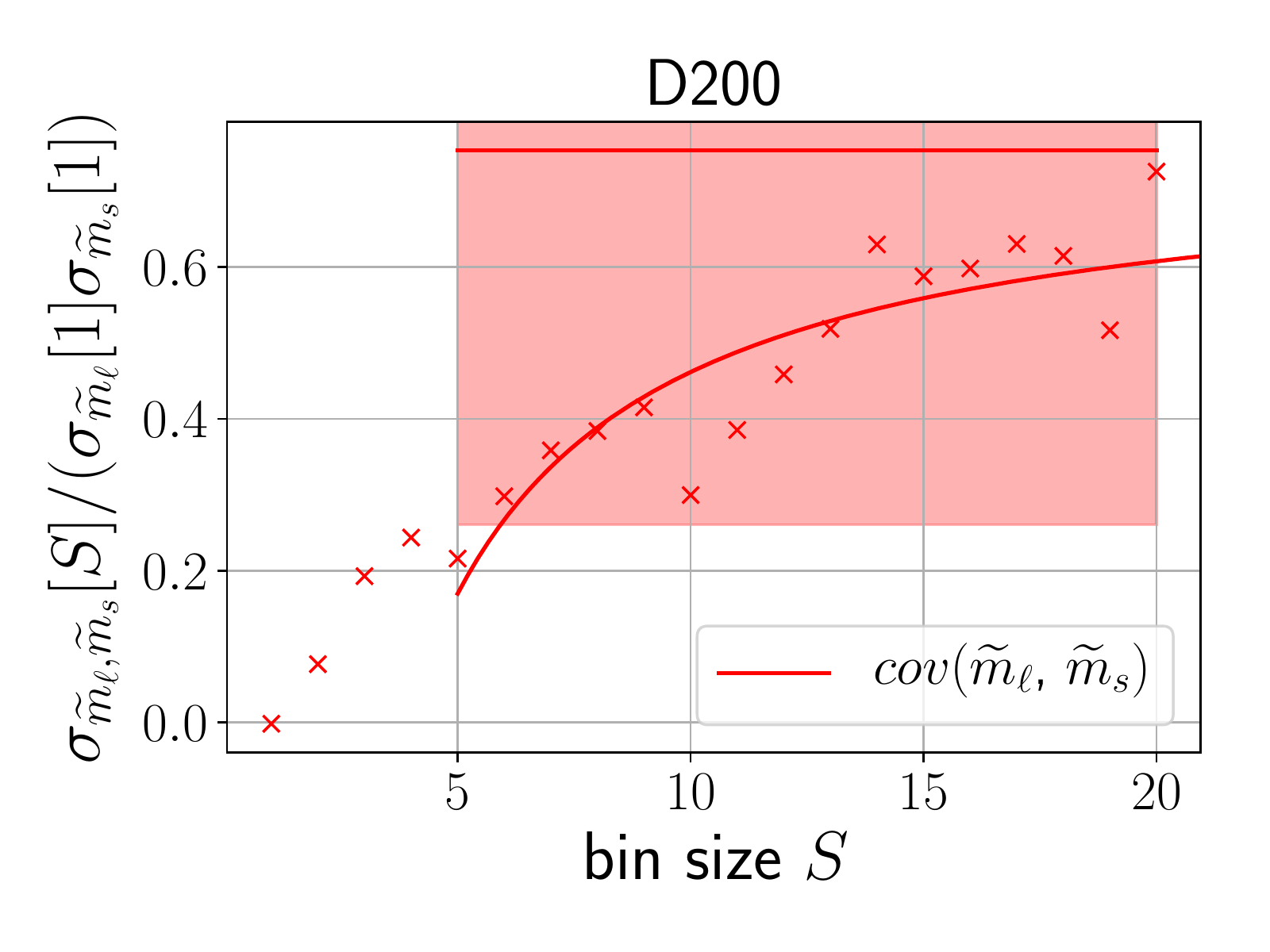}}
  \caption{The same as figure~\ref{fig:auto8} for the strange AWI quark mass $\widetilde m_{s}$. However, the two-parameter fit~\eqref{eq:binnn} was employed.
    \label{fig:auto9}}
\end{figure}

The hadron masses couple to more than just one mode and cannot be
described in terms of eq.~\eqref{eq:binnn2}. However,
the data are in agreement with two- and three-parameter fits according to
eqs.~\eqref{eq:binnn} and~\eqref{eq:binnn3}, starting from
some minimum bin size. We find that the
results from these parametrization usually agree, however, the
three-parameter fits turn out to be more stable. Therefore,
we extrapolate the error according to eq.~\eqref{eq:binnn3}, where we
self-consistently start the fit range at a bin size that is larger
than thrice the extracted integrated autocorrelation time.
As a general pattern, we find larger autocorrelation times for hadrons
that contain
strange quarks and also the autocorrelation times for pseudoscalar mesons
are larger than those for baryons. In all the cases the integrated
autocorrelation times turn out to be much smaller than for $t_0/a^2$.

In figures~\ref{fig:auto2}--\ref{fig:auto7}
we illustrate the extrapolations for the examples of N200 ($M_{\pi}\approx
280\,\textmd{MeV}$) and D200 ($M_{\pi}\approx 200\,\textmd{MeV}$),
both at $a\approx 0.064\,\textmd{fm}$, with successive measurements
separated by 4~MDUs. Regarding the AWI quark masses (figures~\ref{fig:auto6}
and~\ref{fig:auto7}) the data show a preference for
the two-parameter fits~\eqref{eq:binnn}, which we carry out using the
$S\gtrsim 4\tau_{\text{int}}$ data.
Finally, in figures~\ref{fig:auto8} and~\ref{fig:auto9} we extrapolate
the covariance matrix elements between the pion and the kaon masses
and between the light and strange quark AWI masses, respectively.
The starting bin size of the fit always corresponds to the maximum of the starting
bin sizes for the corresponding fits to the two corresponding
diagonal elements.
In the first case the data are well described by the three-parameter
fit~\eqref{eq:binnn3}, while for the AWI masses we employ
the two-parameter fit~\eqref{eq:binnn}. In some cases (not shown) we
also encountered covariances that became smaller or changed sign
when the bin size was increased, which is to be expected.
These were then parameterized
according to eq.~\eqref{eq:covaut}.

\subsection{Least squares fits with errors on the arguments and priors}
\label{sec:leastsquares}
We wish to fit expectation values $y_e$, e.g., a baryon masses
determined on an ensemble $e$, $e\in\{1,\ldots,n_e\}$, to a parametrization
$f(x_e;\{a^{\ell}\})$, where $x_e$ could be the expectation value
of the pion mass on ensemble $e$. The function depends
on a set of parameters $\{a^{\ell}\}$, $\ell\in\{1,\ldots,n_p\}$.
The baryon mass has an error $\Delta y_e$, the pion mass an
error $\Delta x_e$ and, for the moment being, we neglect correlations
between the two.
The result can be obtained using Orear's ``effective variance
method''~\cite{Barker:1974,Orear:1981qt} by minimizing the functional
\begin{equation}
  \chi^2[\{a^{\ell}\}]=\sum_{e=1}^{n_e}
  \frac{{\left[y_e-f(x_e;\{a^{\ell}\})\right]}^2}
  {\Delta y_e^2+{\left[\dd f(x_e;\{a^{\ell}\})/\dd x_e\,\Delta x_e\right]}^2}
\end{equation}
with respect to the parameters $\{a^{\ell}\}$. This is easily generalizable
to simultaneous fits to several baryon masses, which may also depend on
more than one parameter, e.g., on the pion as well as on the kaon mass.
Correlations can easily be taken into account too. The drawback of
this method is that derivatives of the function(s) with respect to the
argument(s) have to be computed for each update of a parameter value
$a^{\ell}$ and, in the case of correlated fits,
the resulting new covariance matrix needs to be inverted each time.

While the $x$-errors affect the weights within the Orear $\chi^2$-functional,
it is the $y$-difference that is minimized. It appears more natural
to minimize the shortest (weighted) distance between the curve and a
data point instead. We remark that within the Gaussian approximation
both views are equivalent.
A more efficient and stable class of algorithms that are based on
the latter minimization strategy was suggested
in ref.~\cite{Marshall:2005}. Here we describe
a related approach, which we employ in our study,
in a form that is as general as needed within the present
analysis. This method places the expectation value(s) that appear
as argument(s) of the fit function(s) on an equal footing with
the expectation value(s) that are to be approximated
by the fit function(s). The $\chi^2$-functional is then constructed
such as to minimize the $\ell^2$-distance between the fitted
parametrization and the data, where the distance is defined through
a scalar product involving the inverse covariance matrix. We remark that
the uncorrelated case of this generalized least squares method is actually
the starting point of Orear's derivation~\cite{Orear:1981qt} of his effective
variance method.

We define expectation values $x^i_e$, e.g., the mass of the
baryon $i$, $i\in\{1,\ldots,n\}$ on ensemble $e$.\footnote{These $x_e^i$
correspond to the expectation values of secondary observables
$f_k$ for an ensemble $e$, as introduced in section~\ref{sec:boot}.}
These are
parameterized in terms of functions
\begin{equation}
  {f_e}^i(\{a^{\ell}\})\coloneqq
  f^i(x^{n+1}_e,\ldots,x^N_e;\{a^{\ell}\})
\end{equation}
that depend on a set of parameters $\{a^{\ell}\}$ as well as on the arguments
$x_e^j$, $j\in\{n+1,\ldots,N\}$, e.g., the pion and the kaon mass
on ensemble $e$. So the first $n$ elements of $x_e$ are functions
of the subsequent $N-n$ elements. We define the $N\times N$
covariance matrix $C_e$ on ensemble $e$
whose elements $C_e^{ii'}$ are computed according to eq.~\eqref{eq:covar}
(and extrapolated to infinite bin size, e.g., via eq.~\eqref{eq:covaut}).
There exist no correlations between different ensembles.

Instead of minimizing the differences
$|x^i-{f_e}^i(\{a^{\ell}\})|$ using, e.g., the Orear effective variance
method outlined above, we
define a parametrization $p^k_e$, $k\in\{1,\ldots,N-n\}$, both of the
functions ${f_e}^i$ and of the original arguments $x^j_e$, $j>n$. The
functional forms in terms of the $p^k_e$ are defined as follows:
\begin{equation}
  {f_e}^i(\{a^{\ell}\})\coloneqq\left\{\begin{array}{ccc}
  f^i(p^1_e,\ldots,p^{N-n}_e;\{a^{\ell}\})&,&i\leq n\\
  p^{i-n}_e&,&n<i\leq N
  \end{array}\right..
\end{equation}
Then the differences ${\delta f_e}^i=x_e^i-{f_e}^i(\{a^{\ell}\})$,
$i\in\{1,\ldots, N\}$, are minimized via the $\chi^2$-functional
\begin{equation}
  \chi^2[\{a^{\ell},p^k_e\}]=\sum_{e=1}^{n_e}\sum_{i,i'=1}^N
  {\delta f_e}^i{\left(C_e^{-1}\right)}^{ii'}{\delta f_e}^{i'}
\end{equation}
with respect to $\{a^{\ell}\}$ and $\{p^k_e\}$.
Note that the ${f_e}^i$ for $i\leq n$ contain products of the $p^k_e$ and
$a^{\ell}$ parameters. Therefore, this is not a linear fit. However, the
$n_e(N-n)$ additional parameters are not overly problematic in terms
of the fit stability or the algorithmic efficiency
since these directions alone have a well defined minimum
at the start values $p^k_e=x^{n+k}_e$. (For $i>n$:
$\delta f_e{}^i=x_e^i-p_e^{i-n}$.) The additional multipliers
do not alter the number of degrees of freedom $N_{\text{DF}}=n_en - n_p$.
We remark that
in our case, for the ensembles along the symmetric line where $M_K=M_{\pi}$
and all octet (or decuplet) baryons collapse to one mass value, the sums
over $i$ and $i'$
only run up to two. Therefore, the number of degrees of freedom is
$N_{\text{DF}}=(n_{e}-n_{\text{symm}})n+n_{\text{symm}}-n_p$, where $n_e$ is the total
number of ensembles, $n_{\text{symm}}$ is the number
of ensembles along symmetric lines, $n$ is the number of
different baryons included in the fit and $n_p$ is the number of fit parameters.
In the case of joint octet and decuplet fits the second
$n_{\text{symm}}$ in the above formula has to be multiplied by two.

We rescale all dimensionful data and fit parameters into units of $t_0/a^2$,
set the lattice spacing using $t_0^*/a^2$ separately for each
$\beta$-value and use $m_\Xi/\textmd{GeV}$ to set the overall scale.
We incorporate these additional measurements and information as
``priors'' into our fits.
We are not able to resolve any correlations between $t_0/a^2$
and our masses. This is not surprising, given the fact that
$t_0/a^2$ has a much larger autocorrelation time, coupling to
very different modes. Therefore, we may treat $t_0/a^2$ as
independent measurements. For each of the priors
$q_e^k$ we add a term $(r_e^k-q_e^k)^2/(\Delta q_e^k)^2$ to the
$\chi^2$-functional. The central values $q_e^k$ are drawn on a bootstrap
by bootstrap basis according to a Gaussian pseudo-bootstrap distribution
with variance $\Delta q_e^k$, while the $r_e^k$ are additional
fit parameters that modify the fit functions $f_e{}^i$.
Since the number of additional parameters equals the
number of additional measurements/priors, this procedure leaves
the number of degrees of freedom invariant. After the minimization
one may compare the bootstrap histograms for the $r_e^k$ and the
priors $q_e^k$. In our case these turn out to be fairly similar
which means that the additional knowledge of the hadron masses
has little impact on the favoured values for $t_0/a^2$ or
$t_0^*/a^2$.

\section{HMC simulation parameters and reweighting towards the target action}
\label{sec:reweight}
The CLS simulations are carried out with an action that differs
from the target action in terms of the rational approximation
made for the $N_f=1$ strange quark contribution and a twisted mass term
that is introduced to stabilize the $N_f=2$ light fermion part of the
simulation. The difference is corrected for
by reweighting the observables. For details see ref.~\cite{Bruno:2014jqa}.
We list the simulation parameters for some of the ensembles
in appendix~\ref{sec:technical} and explain aspects of the computation
of the reweighting factors in appendix~\ref{sec:wewe}.

\label{sec:technical}
\begin{table}[h]
  \caption{\label{tab:paramsNewSim}Simulation parameters for selected
    ensembles, which (to the best of our knowledge) have not been reported
    elsewhere. The \textit{ensemble id} consists of the id used for the
    ensemble plus a suffix (r000, \ldots) to distinguish different replica
    runs. $a\mu_0$ and $a\mu_i$ are the final and intermediate masses used
    in the twisted mass reweighting and factorization, respectively,
    where $N_{\text{mf},2}$ of the lightest twisted mass values are
    integrated on the coarsest time scale. $N_{\text{p}}$ is the number of
    poles in the range $[r_a,r_b]$, used for the rational approximation
    for the single quark flavour. $N_{\text{p}}^\prime$ poles are represented
    as single pseudo-fermions of which $N_{\text{p},2}$ are integrated
    on the coarsest time scale using $N_{\text{s},2}$ steps at this level
    in the MD integrator. $N_{\text{MD}}$ is the number of MDUs produced for
    this replica and $\langle P_{\text{acc}}\rangle$ is the average
    acceptance rate.
  }
  \begin{center}
    {\scriptsize
      \begin{tabular}{*{11}{c}}
        \hline
        ensemble id & $a\mu_0$ & $a\mu_i$ & $N_{\text{mf},2}$ & $N_{\text{p}}$ & $[r_a,r_b]$ & $N_{\text{p}}^\prime$ & $N_{\text{p},2}$ & $N_{\text{s},2}$ & $N_{\text{MD}}$ & $\langle P_{\text{acc}}\rangle$\\\hline
        A651r000 & 0.0 & (0.0005,0.005,0.05,0.5) & 1 & 11 & $[0.01,6.5]$ & 5 & 2 & 7 & 4152 & 0.94\\
        A651r001 & 0.0 & (0.0005,0.005,0.05,0.5) & 1 & 11 & $[0.01,6.5]$ & 5 & 2 & 7 & 16252 & 0.94\\
        A652r000 & 0.0 & (0.0005,0.005,0.05,0.5) & 1 & 11 & $[0.01,6.5]$ & 5 & 2 & 6 & 3988 & 0.87\\
        A652r001 & 0.0 & (0.0005,0.005,0.05,0.5) & 1 & 11 & $[0.01,6.5]$ & 5 & 2 & 6 & 15992 & 0.85\\
        A650r000 & 0.0 & (0.0005,0.005,0.05,0.5) & 1 & 12 & $[0.005,6.5]$ & 5 & 2 & 7 & 2520 & 0.94\\
        A650r001 & 0.0 & (0.0005,0.005,0.05,0.5) & 1 & 12 & $[0.005,6.5]$ & 5 & 2 & 7 & 17728 & 0.94\\
        A653r000 & 0.0   & (0.0005,0.005,0.05,0.5) & 1 & 11 & $[0.01,6.5]$ & 5 & 2 & 7 & 20200 & 0.92\\
        A654r000 & 0.001 & (0.005,0.05,0.5)        & 1 & 11 & $[0.01,6.5]$ & 5 & 2 & 7 & 20722 & 0.95\\
        D150r000 & 0.003 & (0.00045,0.0006, & 3 & 16 & $[0.001,7.8]$ & 9 & 3 & 20 & 1616 & 0.80\\
                 &       & \;0.0055,0.06,0.7)\\
        D150r001 & 0.003 & (0.00045,0.0006, & 3 & 16 & $[0.001,7.8]$ & 9 & 3 & 20 & 796 & 0.77\\
                 &       & \;0.0055,0.06,0.7)\\
        X450r001 & 0.00065 & (0.005,0.05,0.5) & 1 & 15 & $[0.001,7.2]$ & 8 & 3 & 8 & 1604 & 0.95\\
        B450r000 & 0.001   & (0.005,0.05,0.5) & 1 & 13 & $[0.002,7.5]$ & 7 & 4 & 7 & 6448 & 0.95 \\
        S400r000 & 0.00065 & (0.005,0.05,0.5) & 1 & 12 & $[0.01,7.3]$ & 6 & 3 & 7 & 3488 & 0.92\\
        S400r001 & 0.00065 & (0.005,0.05,0.5) & 1 & 12 & $[0.01,7.3]$ & 6 & 3 & 7 & 8004 & 0.93\\
        N401r000 & 0.00065 & (0.005,0.05,0.5) & 1 & 14 & $[0.002,7.5]$  & 6 & 3 & 8 & 4400 & 0.87\\
        B451r000 & 0.001 & (0.005,0.05,0.5) & 1 & 13 & $[0.002,7,5]$  & 7 & 4 & 7 & 8032 & 0.97\\
        B452r000 & 0.001 & (0.005,0.05,0.5) & 1 & 13 & $[0.002,7,5]$  & 7 & 4 & 7 & 7776 & 0.98\\
        N450r000 & 0.00065 & (0.005,0.05,0.5) & 1 & 14 & $[0.002,7.5]$ & 6 & 3 & 8 & 3232 & 0.91\\
        N450r001 & 0.00065 & (0.005,0.05,0.5) & 1 & 14 & $[0.002,7.5]$ & 6 & 3 & 8 & 1548 & 0.87\\
        D451r000 & 0.0003 & (0.0007,0.007,0.07,0.5) & 1 & 13 & $[0.006,7.8]$ & 6 & 5 & 12 & 4112${}^*$
 & 0.87\\
        N305r000 & 0.001 & (0.005,0.05,0.5) & 1 & 13 & $[0.008,7.0]$ & 6 & 3 & 6 & 4004 & 0.95\\
        N305r001 & 0.001 & (0.005,0.05,0.5) & 1 & 13 & $[0.008,7.0]$ & 6 & 3 & 6 & 4000 & 0.95\\
        N304r000 & 0.001 & (0.005,0.05,0.5) & 1 & 13 & $[0.008, 7.0]$ & 6 & 3 & 6 & 3896 & 0.95 \\
        N304r001 & 0.001 & (0.005,0.05,0.5) & 1 & 13 & $[0.008, 7.0]$ & 6 & 3 & 6 & 3008 & 0.94 \\
        J304r000 & 0.00075 & (0.002625,0.009187, & 1 & 13 & $[0.008, 7.0]$ & 7 & 3 & 6 & 3320 & 0.90\\
                 &         & \;0.032156,0.112547,0.5) \\
        J304r001 & 0.00075 & (0.002625,0.009187, & 1 & 13 & $[0.008, 7.0]$ & 7 & 3 & 6 & 3216 & 0.90\\
                 &         & \;0.032156,0.112547,0.5) \\
        N500r000 & 0.0005 & (0.01,0.05,0.5) & 1 & 13 & $[0.0038, 7.0]$ & 6 & 3 & 6 & 3912 & 0.97\\
        J500r004 & 0.0005 & (0.01,0.05,0.5) & 1 & 13 & $[0.0038,7.0]$ & 6 & 3 & 6 & 6312 & 0.94 \\
        J500r005 & 0.0005 & (0.01,0.05,0.5) & 1 & 13 & $[0.0038,7.0]$ & 6 & 3 & 6 & 2020 & 0.94 \\
        J501r001 & 0.0005 & (0.01,0.05,0.5) & 1 & 13 & $[0.0038,7.0]$ & 6 & 3 & 6 & 6540 & 0.93 \\
        J501r002 & 0.0005 & (0.01,0.05,0.5) & 1 & 13 & $[0.0038,7.0]$ & 6 & 3 & 6 & 1588 & 0.94 \\
        \hline
      \end{tabular}
  }\end{center}
{\footnotesize ${}^*$300 MDUs at the beginning were excluded since the run was insufficiently thermalized.}
\end{table}

\subsection{Technical HMC parameters}

In table~\ref{tab:paramsNewSim} we
list the technical parameters for some of the recently performed
simulations. This table is a continuation of table~2 of
ref.~\cite{Bruno:2014jqa}.
For details on the simulation of ensemble E250 we refer to
ref.~\cite{Mohler:2017wnb}. The table includes the
value of the twisted mass parameter $a\mu_0$ used to stabilize the
two flavour part of the simulation.
The generated configurations are reweighted accordingly when expectation values
of observables are computed. The table also includes the intermediate
twisted mass values $a\mu_i$ ($i>0$) employed in the corresponding factorization
of the fermion determinant.
$N_{\text{mf},2}$ denotes the number of (lightest) pseudo-fermion flavours that
are integrated on the coarsest timescale
in the multi-scale integration scheme. The rational approximation of
the one flavour part of the fermionic action utilizes $N_{\text{p}}$ poles
in the range $[r_a,r_b]$. $N_p^\prime$ of these poles are represented as
single pseudo-fermions in the simulation of which
$N_{\text{p},2}$ are integrated on the coarsest timescale,
using $N_{\text{s},2}$ steps (at this timescale).  Also the effect of the rational
approximation is corrected for by a reweighting factor during the
measurement. The trajectory length is set
to two molecular dynamics units (MDUs).
These notational conventions are the same as those used
in ref.~\cite{Bruno:2014jqa} (in particular, in table~2).
The ensembles named ``rqcd0{\tt mn}'' (with anti-periodic fermionic and
periodic gluon field boundary
conditions in time) are not part of the CLS project as these
have been generated using the {\sc BQCD} code~\cite{Nakamura:2010qh} on
the (now decommissioned) QPACE computer.

\subsection{Reweighting}
\label{sec:wewe}
We define the difference between the target action and the simulated action
on a given configuration $i$ as $s_i$.
The strange quark reweighting factor usually does not vary significantly.
However, there is an issue regarding the strange quark reweighting
which is addressed in ref.~\cite{Mohler:2020txx} and has been
taken into account in the results presented in this article too.
Here we restrict the discussion to the
factor that is associated with the twisted mass reweighting,
which can fluctuate considerably, in particular, at coarse lattice spacings
and light quark masses. By definition $s_i\geq 0$ for this
latter contribution.
The weight factors $w_i=\exp(-s_i)$ are stochastically estimated
for each configuration. We remark that it is
also possible to estimate $w_i$ in a multi-step procedure. The
discussion below trivially generalizes to this case.

Once the reweighting factors are known, expectation values of an observable
$A$ are given as
\begin{align}
  \langle A\rangle=\frac{\sum_i w_iA_i}{\sum_i w_i}.
\end{align}
In the special case $w_i=w$ this corresponds to the usual ensemble average.
It is straightforward to incorporate the reweighting into the construction
of the jackknife and bootstrap ensembles discussed above.

\begin{figure}[t]
  \centering
  \resizebox{0.49\textwidth}{!}{\includegraphics[width=\textwidth]{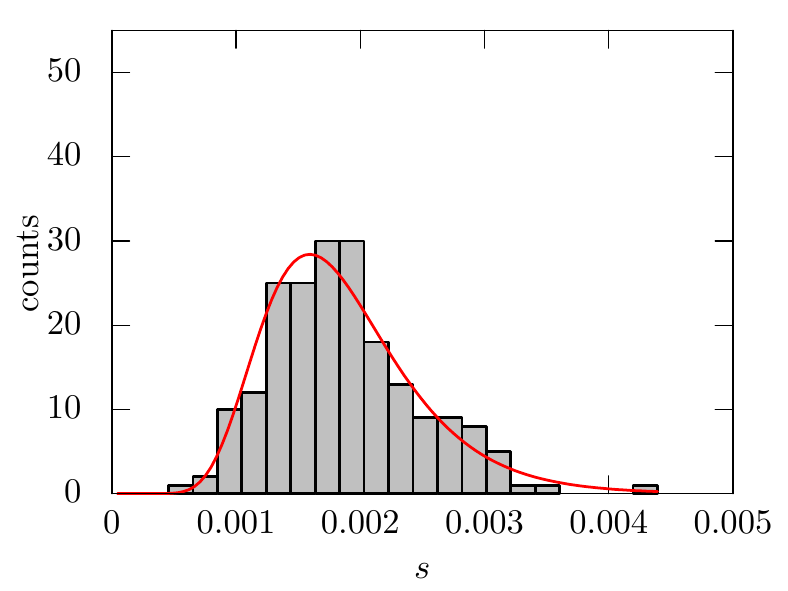}}
  \resizebox{0.49\textwidth}{!}{\includegraphics[width=\textwidth]{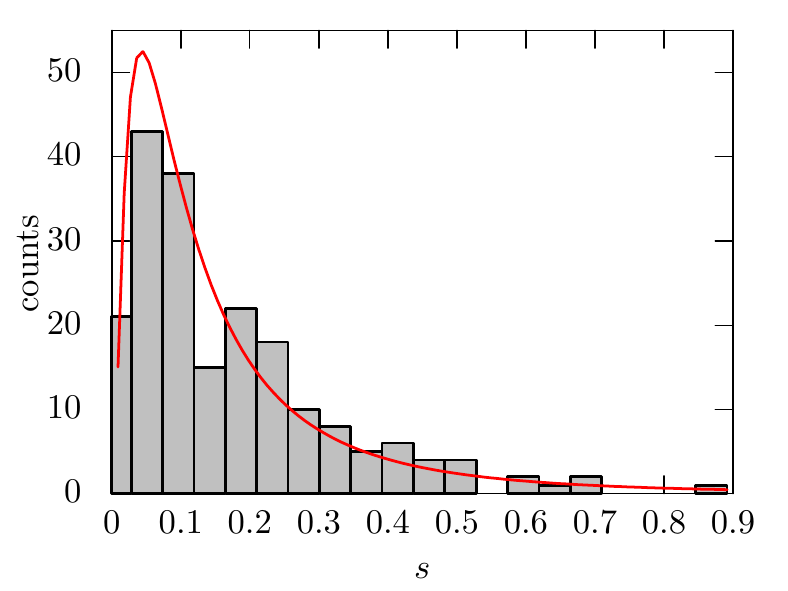}}\\
  \resizebox{0.49\textwidth}{!}{\includegraphics[width=\textwidth]{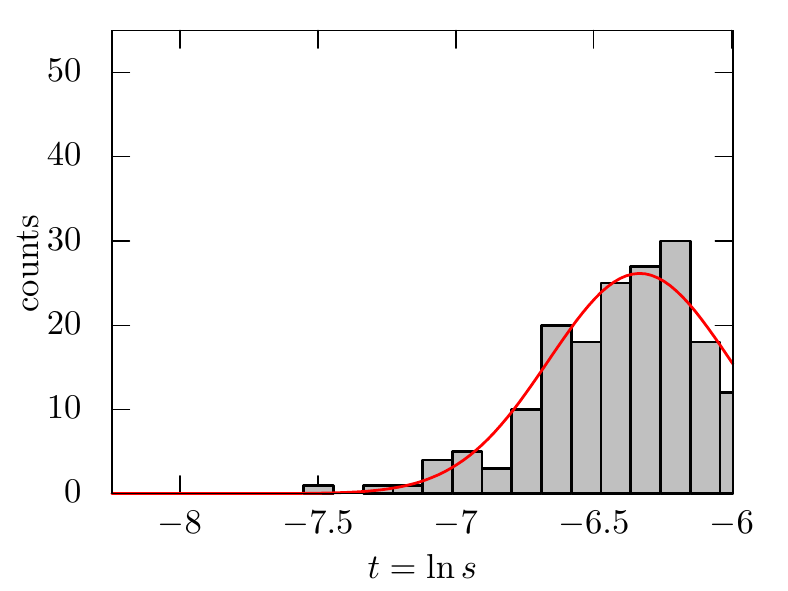}}
  \resizebox{0.49\textwidth}{!}{\includegraphics[width=\textwidth]{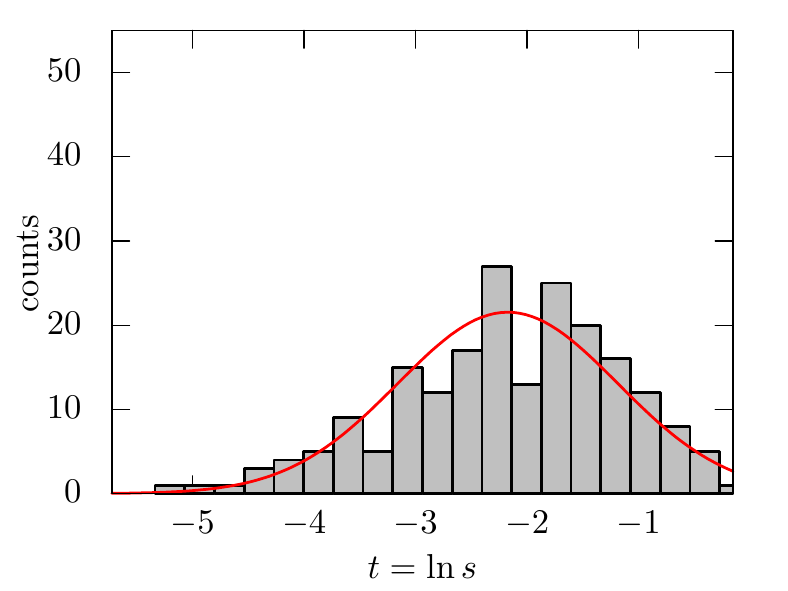}}\\
  \makebox[0.49\textwidth]{S100r003n81}
  \makebox[0.49\textwidth]{S100r003n381}
  \caption{Examples of histograms of $s$ (top row) and
    $t=\ln s$ (bottom row) for S100r003n81 (left) and
    S100r003n381 (right). To study the distribution 200 estimates
    were computed. While the distribution of $s$
    for reweighting factors close to one
    (S100r003n81: $w=\langle e^{-s} \rangle \approx 0.998$, left)
    typically resembles a Gaussian, for only slightly smaller
    reweighting factors (S100r003n381, $\langle e^{-s} \rangle\approx 0.851$,
    right) the distribution is in fact closer to a log-normal distribution.
    The curves in the bottom row show a normal distribution
    $\phi(t)$, using mean and variance of $t$. 
    The curves in the top row $\phi(\log s)/s$ correspond to this
    assumption.\label{fig:rwfhist}}
\end{figure}

The potential problem that we face is that the inclusion of some
configurations with small but imprecisely determined weights can
significantly affect ensemble average.
While this is more frequently observed for
three-point functions than for two-point functions,
we shall also address this issue here. One (expected)
observation is that
the statistical error $\sigma_{w_i}$ of the estimate $w_i$
is much bigger for small reweighting factors than it is for
those that are close to unity. In some of these cases the error on
the reweighting factor even exceeds its estimated value.
Our strategy is two-fold.
First, we introduce a more robust estimate. Then, rather than keeping
the number of estimates fixed, we increase it
on a given configuration $i$ until we would expect $\sigma_{w_i}<0.01$.
This is achieved by extrapolating the squared error
$\sigma_{w_i}^2$ for large $n_i$ linearly in the inverse number of
estimates $1/n_i$, where
$n_i\leq n_{\max}$ with $n_{\max}$ fixed.
In the cases where these target values are larger than
$n_{\max}$, additional estimates are generated. In view of avoiding
a bias, it is important that the targeted number of estimates
is predicted beforehand, based on existing measurements, rather than
continuing until the error has reached a certain threshold.

In the usual definition, $w$ (where we drop the index $i$) is estimated
on each configuration:
\begin{align}
  \label{eq:naiverwf}
  w\coloneqq\langle e^{-s}\rangle=\int\!\dd\mu(s)\,e^{-s}
  \approx \frac{1}{n}\sum_{j}^ne^{-s(j)},
\end{align}
where $s(j)$ is the $j$th estimate of $s$.
$\dd\mu(s)$ denotes the measure associated with the
probability distribution of the random variable $s$.
We define the probability density $p_s(s)$: $\dd\mu(s)=p_s(s)\dd s$,
$\int\!\dd s\,p_s(s)=1$.
One can easily change the random variable. The substitution
$s=f(t)$ implies that $p_s(s)=p_t(t)/|f'(t)|$, i.e.\
$\dd\mu(s)=|f'(t)|\dd\mu(t)$. Ideally, $p_w(w)$ with
$s=-\ln w$ would be a Gaussian of width $\Delta w$. This would guarantee
uniform convergence of the average, with an error
$\sigma_w=\Delta w/\sqrt{n-1}$ .
Clearly, this is not the case for our distribution as $s\geq 0$.
So one may anticipate a slow onset of the asymptotic
convergence behaviour, in particular, in the cases
where $w$ is not close to unity and the variance may be large.

\begin{table}
  \caption{\label{tab:momentsRwf}Skewness, $\mu_3/\mu_2^{3/2}$, and kurtosis,
    $\mu_4/\mu_2^2$, of the distribution of the logarithm of the estimates
    used to compute the
    reweighting factors, $t=\ln s$ at two different values of the coupling,
    $\beta=3.4$ and $\beta=3.7$, for ensembles along the
    $\Tr M=\mathrm{const}$ line. We
    also display the approximate pion mass and the
    volume. For a Gaussian distribution, one would expect $\mu_3/\mu_2^{3/2}=0$
    and $\mu_4/\mu_2^2=3$.}
  \begin{center}
    {
      \scriptsize
      \begin{tabular}{lcclllccll}
        \toprule
\multicolumn{5}{c|}{$\beta=3.4$} & 
\multicolumn{5}{c}{$\beta=3.7$} \\\midrule
        id & $M_{\pi}$/$\textmd{MeV}$ & $N_t\cdot N_s^3$ & $\mu_3 / \mu_2^{3/2}$ & $\mu_4 / \mu_2^2$ & id &$M_{\pi}$/$\textmd{MeV}$ & $N_t\cdot N_s^3$ & $\mu_3 / \mu_2^{3/2}$ & $\mu_4 / \mu_2^2$ \\\midrule
U103 & 420 & $128\cdot 24^3$ & 0.142(7)  & 2.629(5)  & N300 & 425 & $128\cdot 48^3$ & 0.071(9)  & 2.769(31) \\
H101 & 423 & $96\cdot 32^3$ & 0.163(10) & 2.574(26) &  & & & & \\\midrule
U102 & 357 & $128\cdot 24^3$ & 0.120(18) & 2.619(18) & N302 & 348 & $128\cdot 48^3$ & 0.071(9)  & 2.731(16) \\
H102a& 359 & $96\cdot 32^3$ & 0.132(18) & 2.524(52) &  & & & & \\\midrule
U101 & 271 & $128\cdot 24^3$ & 0.032(14) & 2.669(12) & J303 & 259 & $192\cdot 64^3$ & 0.119(17) & 2.777(26) \\
H105 & 281 & $96\cdot 32^3$ & 0.094(17) & 2.665(24) &  & & & & \\
N101 & 281 & $128\cdot 48^3$ & 0.201(12) & 2.757(15) & & & & & \\\midrule
S100 & 214 & $128\cdot 32^3$ & 0.018(16) & 2.578(23) &  & & & & \\
C101 & 222 & $96\cdot 48^3$ & 0.109(7)  & 2.684(15) &  & & & & \\
D101 & 222 & $128\cdot 64^3$ & 0.162(10) & 2.683(15) &  & & & & \\\midrule
D150 & 127 & $128\cdot 64^3$ & $-$0.078(21)& 2.701(35) &  & & & &
\\\bottomrule 
      \end{tabular}
  }\end{center}
\end{table}

One possibility would be that $p_w(w)$ is log-normal distributed, i.e.\
that $p_s(s)$ is a Gaussian distribution centred about $s_0$, with
a width $\Delta s$:
\begin{align}
  p_s(s)=\frac{1}{\sqrt{2\pi\Delta s^2}}
  \exp\left[-\frac{(s-s_0)^2}{2\Delta s^2}\right].
\end{align}
It is easy to see that in this case
\begin{align}
  w=\langle e^{-s}\rangle=\exp\left(-s_0+\frac{\Delta s^2}{2}\right),
\end{align}
i.e.\ we could obtain $w$ from the average $s_0=\langle s\rangle$ and its
second moment $\Delta s^2=\langle(s-s_0)^2\rangle$.
It turns out that also this is not the correct statistical model.
Instead, the distribution of $s$ itself appears to be close
to log-normal. We demonstrate this for two configurations of ensemble S100
in figure~\ref{fig:rwfhist}, where in the second row we show
histograms of $t=\ln s$.
To further quantify the approximate log-normality, we compute
$t_0=\langle t\rangle$ and moments $\mu_n=\langle(t -t_0)^n\rangle$
for $n=2,3,4$ ($\mu_1=0$ by definition). For a log-normal
distribution of $s$ one would expect
$\mu_3/\mu_2^{3/2}\approx 0$ and $\mu_4/\mu_2^2\approx 3$.
The averages for some of our ensembles are shown in table~\ref{tab:momentsRwf}.
Skewness and kurtosis are not universal but somewhat vary between
ensembles. However, we are unable
to detect any patterns, regarding the pion mass, the volume
or the twisted mass parameter (not shown). Mostly the skewness
is slightly positive while the kurtosis is a bit smaller than
three. While the distributions are not perfectly log-normal, they are
close to it.

\begin{figure}[htp]
  \centering
  \resizebox{0.95\textwidth}{!}{\includegraphics[width=\textwidth]{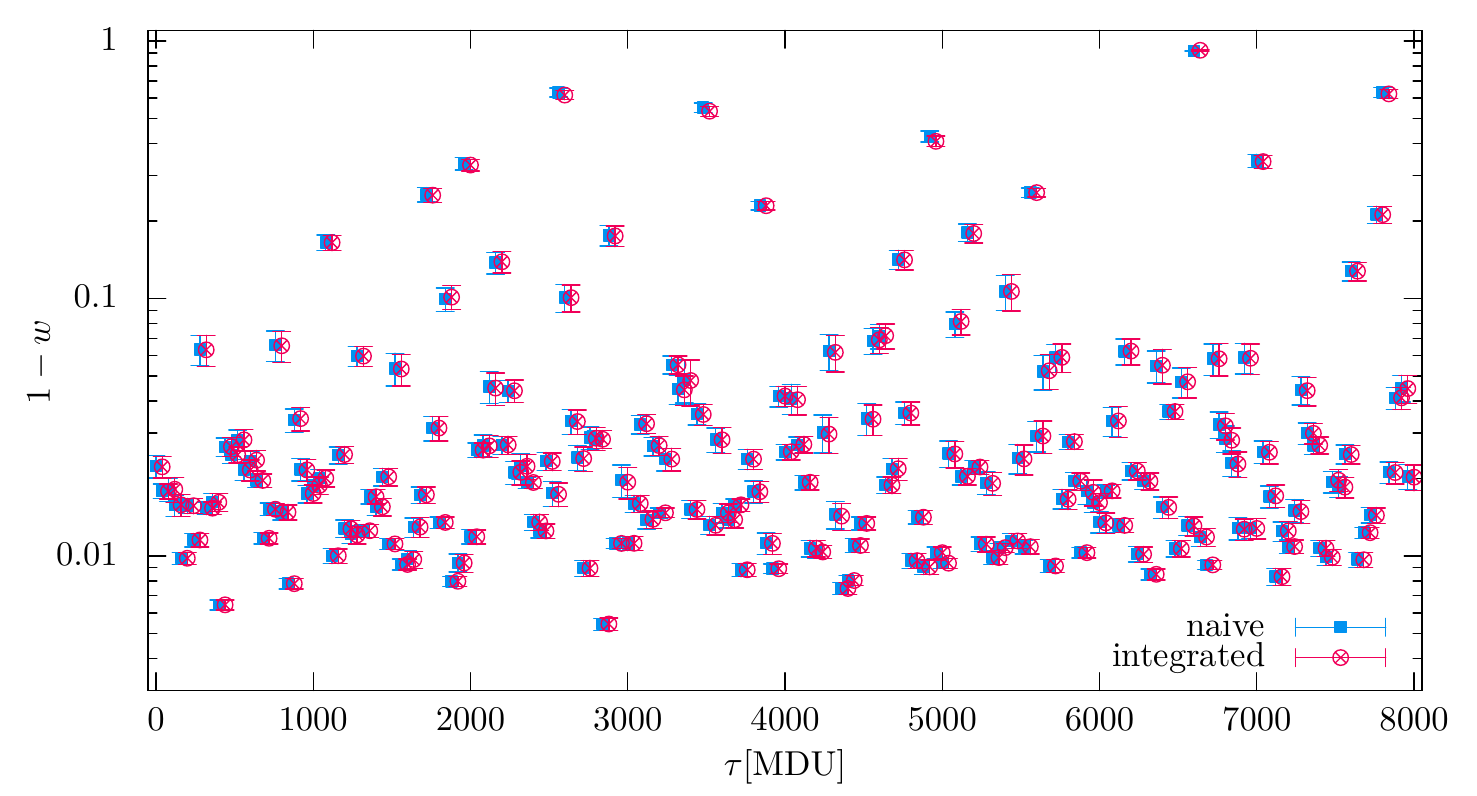}}
  \caption{Deviations of the reweighting factors from one for ensemble
    C101 ($M_\pi\approx 222\,\textmd{MeV}$, $a\approx 0.085\,\textmd{fm}$),
    computed using the naive
    definition, eq.~\eqref{eq:naiverwf} and the definition assuming a
    log-distribution of the estimates, eq.~\eqref{eq:logdistrwf}.
    The reweighting factors have been computed with the target
    precision $\sigma_w<0.01$, as predicted employing a smaller number of
    estimates. For better readability we shifted the points horizontally
    and plotted only every 40th MDU.
    \label{fig:rwfcomparison}}
\end{figure}

The approximate log-normality of $s$ suggests to
estimate $w$ by computing
\begin{align}
  t_0\approx\frac{1}{n}\sum_j^n\ln s(j),\quad
  \Delta t^2=\mu_2\approx\frac{1}{n}\sum_j^n\left[t_0-\ln s(j)\right]^2\label{eq:5}
\end{align}
and then numerically integrating
\begin{align}
  w&=\frac{1}{\sqrt{2\pi\Delta t^2}}
  \int_{-\infty}^{\infty}\!\!\dd t\,\exp\left[-\frac{(t-t_0)^2}{2\Delta t^2}-e^t\right]\nonumber\\
  &\approx\frac{1}{\sqrt{2\pi\Delta t^2}}
  \int_{-6\Delta t}^{\min(3-t_0,6\Delta t)}\!\!\!\!\!\!\dd t\,\exp\left[-\frac{t^2}{2\Delta t^2}-e^{(t+t_0)}\right].
\label{eq:logdistrwf}
\end{align}
Note that this exponential integral cannot be solved
in closed form. In the last step above, subsequent to
the substitution $t-t_0\mapsto t$, we neglected contributions
from the regions $t>3-t_0$, which gives a numerical precision
better than $10^{-8}$. We remark that for $t<-(3+t_0)$, it is
safe to approximate the integrand by a Gaussian.

The error of this improved estimate (which we coin the
``integrated'' method) is determined by
resampling over the stochastic estimates.
In figure~\ref{fig:rwfcomparison} we compare the results of the
two methods for ensemble C101 where a particularly large number
of reweighting factors is small. As expected, the two sets of
reweighting factors agree within errors.
The integrated procedure should be more reliable in terms
of the error estimate, in particular, whenever $w$ is close to zero.
However, with the increased number of estimates that we employ,
only for very few configurations are differences visible
between the two definitions on the scale of the figure.

\providecommand{\href}[2]{#2}\begingroup\raggedright\endgroup
\end{document}